\documentstyle[amssymb,graphicx,cite]{elsart}

\begin{document}

\begin{frontmatter}

\title{Effects of Fluctuations on Propagating Fronts}

\author{Debabrata Panja}

\address{Institute for Theoretical Physics, Universiteit van
Amsterdam,  Valckenierstraat 65, 1018 XE Amsterdam, The Netherlands}

\begin{abstract}
Propagating fronts are seen in varieties of non-equilibrium pattern
forming systems in Physics, Chemistry and Biology. In the last two
decades, many researchers have contributed to the understanding of the
underlying dynamics of the propagating fronts. Of these, the
deterministic and mean-field dynamics of the fronts were mostly
understood in late 1980s and 1990s. On the other hand, although the
earliest work on the effect of fluctuations on propagating fronts
dates back to early 1980s, the subject of fluctuating fronts did not
reach its adolescence until the mid 1990s.  From there onwards the
last few years witnessed a surge in activities in the effect of
fluctuations on propagating fronts. Scores of papers have been written
on this subject since then, contributing to a significant maturity of
our understanding, and only recently a full picture of fluctuating
fronts has started to emerge. This review is an attempt to collect all
the works on fluctuating (propagating) fronts in a coherent and cogent
manner in proper perspective. It is based on the idea of making our
knowledge in this field available to a broader audience, and it is
also expected to help to collect bits and pieces of loose thread-ends
together for possible further investigation.
\end{abstract}

\end{frontmatter}

\tableofcontents \vfill

\section{Front Propagation in Far from Equilibrium Systems} 

\subsection{Propagating Fronts into Unstable States: Deterministic
Systems \label{sec1.1}}

In pattern forming systems quite often situations occur where patches
of different bulk phases get separated by fronts or interfaces. In
such cases, the relevant dynamics of the system is usually dominated
by the dynamics of these fronts. When the interface separates two
thermodynamically stable phases, as in crystal-melt interfacial growth
problems, the width of the interfacial zone is usually of atomic
dimensions. For such systems, one often has to resort to a moving
boundary description, in which the boundary conditions at the
interface are determined phenomenologically or by microscopic
considerations. A question that naturally arises for such interfaces
is the influence of stochastic fluctuations on the motion and scaling
properties of such interfaces.

\begin{figure}[ht]
\begin{center}   
\includegraphics[width=0.7\linewidth]{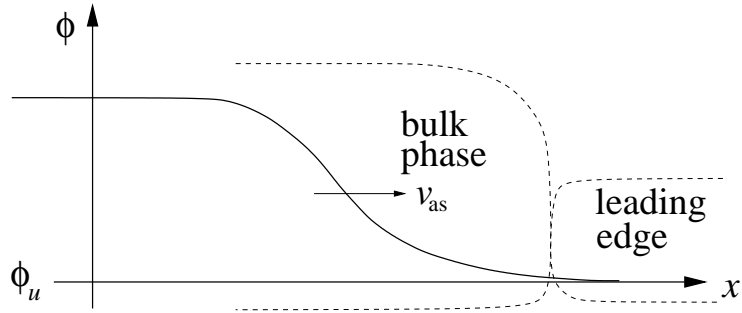}
\end{center}
\caption{A typical deterministic front propagating into a linearly
unstable state $\phi_u$ and moving with speed $v_{\mbox{\scriptsize
as}}$: at the leading edge, the deviation of the front field
$\phi(x,t)$ from its unstable state value is infinitesimal and the
nonlinear terms in the deterministic equation that describes the front
can be neglected. For pulled fronts, the perturbation around the
unstable state at the leading edge ahead of the front grows linearly
and spreads with speed $v^*$, thereby pulling the rest of the front
along with it at $v_{\mbox{\scriptsize as}}=v^*$ (hence the
nomenclature). For pushed fronts, it is the bulk phase that drives the
front propagation mechanism, as if the bulk phase (whose description
requires nonlinear terms in the equation describing the front) pushes
the leading edge from behind to achieve a front speed $>v^*$.}
\label{fig2}
\end{figure}
At the other extreme is a class of fronts that arise in pattern
forming systems, where the occurrence of fronts or transition zones is
{\it fundamentally related to their inherent nonequilibrium nature},
as they do not connect two thermodynamic equilibrium phases that are
separated by a first order phase transition. In such cases --- for
example, chemical fronts \cite{meron}, the temperature and density
transition zones in thermal plumes \cite{tabeling}, the domain walls
separating domains of different orientations in rotating
Rayleigh-B\'enard convection \cite{tucross}, streamer fronts in
discharges \cite{streamers}, or the aggregate fronts in diffusion
limited aggregation \cite{Levine1} --- the fronts are relatively wide
and are therefore described by the same continuum equations that
describe nonequilibrium bulk patterns. The lore in nonequilibrium
pattern formation is that when the relevant length scales are large,
(thermal) fluctuation effects are relatively small \cite{ch}. For this
reason, in the first approximation, the dynamics of many pattern
forming systems can be understood in terms of the deterministic
dynamics of the basic patterns and coherent structures. The first
questions to study for them are properties like the existence and the
asymptotic speed of propagation of the front solutions in
deterministic equations. For example, in one dimension, the existence
of an asymptotic propagation speed $v_{\mbox{\scriptsize as}}$ for
deterministic fronts means that in the comoving co-ordinate
$x-v_{\mbox{\scriptsize as}}t$, moving w.r.t. to the laboratory
co-ordinate $x$ at a speed $v_{\mbox{\scriptsize as}}$, the front
profile approaches a fixed shape as $t\rightarrow\infty$.  Given the
dynamical equations, the interest is in the magnitude of
$v_{\mbox{\scriptsize as}}$, and as well as in how the asymptotic
front shape and speed are approached in time from a given initial
configuration. In most cases of deterministic fronts, these equations
are partial differential equations.

Instances of such fronts, where fluctuation effects are small (and
therefore they can be neglected in favour of a deterministic
description) are abundant in physics
\cite{physics1,physics2,physics3,physics4,physics5,physics6,physics7,physics8,physics9,physics10,physics11,physics12},
chemistry \cite{chemistry1,chemistry21,chemistry22,chemistry3} and
biology \cite{biology1,biology2,biology3,biology4}.\footnote{On
account of the fact that this article is about fluctuating fronts,
these citations are representative and by no means complete. For a
more comprehensive set of references, see Ref. \cite{wimreview}, a
review article on deterministic fronts.\label{fn1}} As a result of
detailed studies carried out in the last decade, it has emerged that
in these systems, the dynamics is described by {\it propagation of
(deterministic) fronts into unstable states}, i.e., the state of the
system in the region far ahead of the front is {\it linearly
unstable\/} \cite{wimreview}. These studies have classified
deterministic fronts propagating into unstable states in two
categories in a broad sense: the so-called {\it pulled\/} and {\it
pushed\/} fronts (see Fig. \ref{fig2}). {\it Pulled fronts\/} are the
fronts that propagate into a linearly unstable state, and for which
the asymptotic front speed $v_{\mbox{\scriptsize as}}$ is the {\it
linear spreading speed\/} $v^*$ of infinitesimal perturbations around
the unstable state \cite{dee,bj,vs2,ebert}. The name pulled fronts
refer to the intuitive picture that at the leading edge of these
fronts,\footnote{The term leading edge, which will appear in this
review article over and over again, is (and will be) used to denote
the front region where the value of the front field $\phi$ is very
close to its value at the the state it propagates into. In other
words, in this region, the front evolution equation can be linearized
around the value of $\phi$ at the the state it propagates
into. \label{leading}} the perturbation around the unstable state
ahead of the front grows and spreads with speed $v^*$, while the rest
of the front gets ``pulled along'' by the leading edge. On the other
hand, fronts that propagate into a linearly unstable state and whose
asymptotic speed is $ >v^*$ are referred to as {\it pushed fronts}, as
it is the nonlinear growth in the region behind the leading edge that
pushes their front speeds to higher values.\footnote{By definition,
therefore, the asymptotic speed of a pushed front can be obtained only
by solving the full nonlinear equation; in general it is not possible
to do so except for special sets of parameter values.\label{pushed}}
If the state is not linearly unstable, then $v^*$ is trivially zero,
and in addition, there can also be fronts propagating into unstable
states with $v^*=0$; in such cases front propagation is always
dominated by the nonlinear growth in the front region itself, and
hence fronts in these cases are in a sense pushed too.\footnote{At
this point, I must warn the reader that I am being much more than just
a little bit naive and simplistic in describing the broad
classification of fronts propagating into unstable states in this
manner. The actual issue of how $v^*$ comes out naturally for fronts
propagating into unstable states and there onwards when one can expect
a pulled front (possibly a further subclassification of coherent or
incoherent pulled fronts) or a pushed front is a fairly involved
subject in itself. For a proper analysis of it, the reader is
encouraged to go through Sec. 2 of Ref. \cite{wimreview}. The point,
once again, is that this is a review article on fluctuating (and
propagating) fronts. From this point of view, it is enough to
understand the basic issues that involve the classification in terms
of pulled or pushed fronts in a broad sense. Such an attitude will be
reflected all along Sec. \ref{sec1.1}, where I will highlight these
basic features of pulled and pushed fronts in terms of examples and
pictorial representations. Furthermore, in Sec. \ref{sec1.1}, I would
completely leave out any discussion on deterministic pattern forming
fronts such as Complex Ginzburg-Landau equation or Swift-Hohenberg
equation etc., simply because literature on the effect of fluctuations
in such pattern forming fronts do not exist in the
literature.\label{warn}}

For a front propagating into linearly unstable state, the linear
spreading speed $v^*$ is obtained from the time evolution of a
localized initial perturbation around the unstable state. A beautiful
analysis of this can be found in Sec. 2.1 of
Ref. \cite{wimreview}. The knowledge of the existence or the magnitude
of $v^*$ alone however does not answer the important questions like
how and under what conditions one can expect a pulled front in the
system, or when to expect a pushed front in the a given
model. Addressing these questions satisfactorily is an involved
process, and as already mentioned in footnote \ref{warn}, from the
point of view of this review article, in Sec. \ref{sec1.1} we present
the basic necessary results by considering an example system. The
equation that we choose for the illustration of the properties of
pulled fronts is the so-called Fisher-Kolmogorov-Petrovsky-Piscounov
equation (we will refer to it as Fisher-Kolmogorov equation
hereafter), which was at first used to model the spreading of
advantageous genes in a population \cite{fisher,kpp}. In this model,
the density of the advantageous genes is denoted by $\phi(x,t)$, a
non-negative quantity, whose dynamics is described by the equation
\begin{eqnarray}
\frac{\partial\phi}{\partial t}\,=\,\frac{\partial^2\phi}{\partial
x^2}\,+\,\phi\,-\,\phi^n.\quad\quad n>1,\quad\mbox{for example 2 or 3}
\label{e1}
\end{eqnarray}
Equation (\ref{e1}) has two stationary states, of which $\phi_u=0$ is
(linearly) unstable and $\phi_s=1$ is stable. Therefore, if the system
is prepared in a way such that these two states coexist in a certain
region of space, then the stable state invades the unstable one and
propagates into it. The profile of the resulting front is similar to
that of Fig. \ref{fig2}.

To obtain the front solution admitted by Eq. (\ref{e1}), we rewrite it
in a frame that moves w.r.t. the laboratory frame at a constant speed
$v$, by means of a change of variables from $x$ to the comoving
co-ordinate $\xi\equiv x-vt$ as
\begin{eqnarray}
\frac{\partial\phi}{\partial
t}\,-\,v\,\frac{\partial\phi}{\partial\xi}\,=\,\frac{\partial^2\phi}{\partial
\xi^2}\,+\,\phi\,-\,\phi^n,
\label{e2}
\end{eqnarray}
and look for a stationary solution of $\phi$ in this comoving
frame. The crucial relevance of the growth and spreading of
infinitesimal perturbations enters naturally in this front solution,
as the propagating infinitesimal perturbations around the unstable
state in the leading edge ahead of the front sets on the instability
making way for further growth. At the leading edge of the front, the
$\phi$-values are very close to the unstable state value, i.e.,
$\phi\ll1$, and one can neglect the nonlinear term $\phi^n$ compared
to $\phi$ in Eq. (\ref{e2}). The stationary solution of the resulting
linear equation can then be solved by using
$\phi(\xi)\sim\exp[-\lambda\xi]$, yielding the relation
\begin{eqnarray}
v(\lambda)\,=\,\lambda\,+\,\frac{1}{\lambda}
\label{e3}
\end{eqnarray}
\begin{figure}[ht]
\begin{center}
\includegraphics[width=0.5\linewidth]{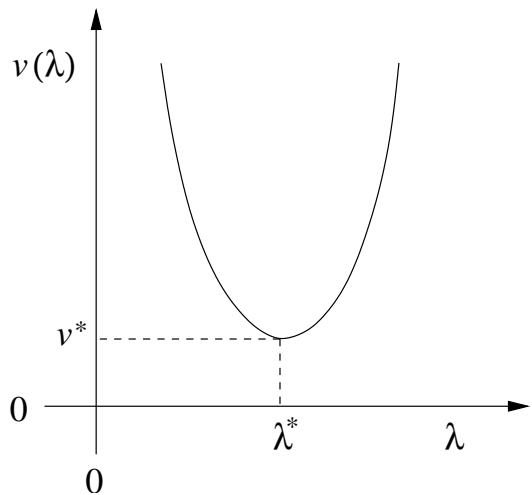}
\end{center}
\caption{The dispersion relation (\ref{e3}) is schematically shown
above. \label{figdisperse}}
\end{figure}
between $v$ and $\lambda$. The curve for the dispersion relation
between $v(\lambda)$ and $\lambda$ is schematically shown in
Fig. \ref{figdisperse}. It has a minimum at $\lambda^*$, and
$v^*=v(\lambda^*)=2$.

For any front propagating into a linearly unstable state, there is a
solution of the leading edge of the form $\exp[-\lambda\xi]$, and
consequently, there exists a dispersion relation between $\lambda$ and
the front speed $v(\lambda)$. Of course, the actual dispersion
relation depends on the model that one studies. However, irrespective
of the model, the universality of pulled fronts lies in the fact that
although Fig. \ref{figdisperse} indicates that one has a front
solution for all values of $\lambda$ (and correspondingly all possible
front speeds), from which it might a priori seem that the quantities
$\lambda^*$ and $v^*$ are not special in any way, the actual selection
of the asymptotic front speed is obtained only after a proper
stability analysis of the front profile in the comoving frame. Such a
stability analysis yields the result that with an initial condition
that $\phi(x,t)|_{t=0}$ that decays faster than $\exp[-\lambda^*x]$
for $x\rightarrow\infty$\footnote{this condition is also known as
``sufficiently steep initial condition''.\label{fn5}}, for long times,
the front speed converges uniformly\footnote{Uniform convergence means
that the convergence behaviour (\ref{e4}) of the front speed is the
same irrespective of the value of $\phi$ at which the speed is being
measured.\label{fn6}} to $v^*$ as \cite{ebert,bramson}
\begin{eqnarray}
v(t)\,=\,v^*\,-\,\frac{3}{2\lambda^* t}\,+\,{\cal O}(t^{-3/2})\,,
\label{e4}
\end{eqnarray}
as the front shape relaxes to its asymptotic configuration
$\phi^*(x-v^*t)$. However, if the initial shape of the front is such
that the leading edge is given by $\phi(x,t)|_{t=0}\sim\exp[-\lambda
x]$ for $x\rightarrow\infty$ with $\lambda<\lambda^*$, then the front
speed $v$ remains fixed at $v(\lambda)$, given by the dispersion
relation, while its shape remains unchanged at $\exp[-\lambda(x-vt)]$.

\begin{figure}[ht]
\begin{center}   
\includegraphics[width=\linewidth]{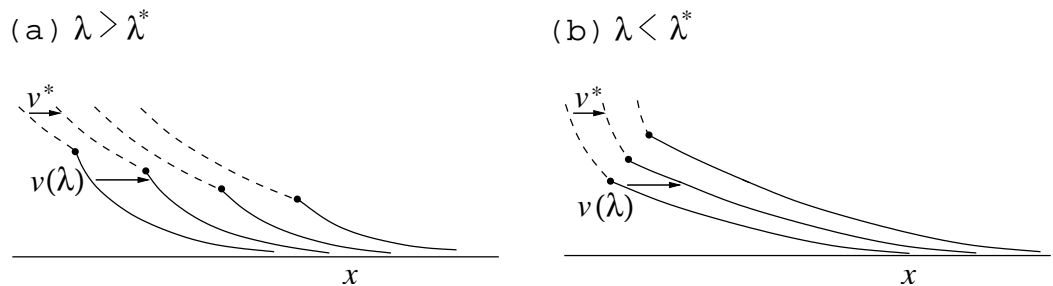}
\end{center}
\caption{An intuitive illustration of front speed selection: (a) for
$\lambda>\lambda^*$, and (b) for $\lambda<\lambda^*$. The dashed and
the solid profile propagate with their respective speeds. The kink (or
the discontinuity in slope is only a symbolic representation of the
crossover region \cite{vs2}. A more precise analysis can be found in
Sec. 2 of Ref. \cite{wimreview}.}
\label{fig4}
\end{figure}
Although obtaining the power law (\ref{e4}) is mathematically quite
involved, a beautiful qualitative illustration of the different speed
selection behaviour of a pulled front, for $\lambda<\lambda^*$ and
$\lambda>\lambda^*$ with an initial configuration of the leading edge
$\phi(x,t=0)\sim\exp[-\lambda x]$ as $x\rightarrow\infty$, has been
provided in Ref. \cite{vs2} (a more precise analysis can be found in
Sec. 2 of Ref. \cite{wimreview}). Imagine that we start with an
initial configuration of the leading edge, shown in the leftmost curve
of Fig. \ref{fig4}(a): an exponential decay $\exp[-\lambda x]$ with
$\lambda>\lambda^*$ on the right denoted by the solid curve, and the
exponential decay $\exp[-\lambda^*x]$ on the left denoted by the
dashed curve. The two curves are joined together at a kink shown by a
small filled circle. One should keep in mind that a kink in the
initial front profile is not propagated as it is for fronts that
involve higher than first order spatial derivative, but nevertheless,
one can get a clear intuitive picture of the evolution of the leading
edge by evolving the solid curve $\exp[-\lambda^*x]$ and dashed curve
$\exp[-\lambda x]$ profile separately and by following the position of
the kink in a pictorial representation. The U-shape of the dispersion
relation $v(\lambda)$ vs. $\lambda$ dictates that the speed of dashed
curve profile $v(\lambda)$ is higher than the solid curve profile
$v^*$, as shown in three later snapshots of the evolving leading edge,
taken at times $t_2<t_3<t_4$. As a result of the differences in speeds
of the dashed and the solid curves, the height of the kink from the
$x$-axis keeps decreasing, which shows that front profile at the
leading edge is being taken over by the dashed curve, corresponding to
an exponential decay with exponent $\lambda^*$. A similar picture is
shown in Fig. \ref{fig4}(b), where the solid and dashed curves
correspond to $\exp[-\lambda x]$ and $\exp[-\lambda^* x]$
respectively, and $\lambda<\lambda^*$. The situation is reversed in
this case --- the front profile at the leading edge is being taken
over by the solid curve profile.\footnote{The discussion in the above
two paragraphs may, at first glance, seem to downplay any special
significance of $v^*$ --- after all, any other front speed $>v^*$ can
be reached with an appropriately chosen $\lambda$ in the initial
configuration of the front. However, in most physical situations, one
is interested in the dynamics of ``localized initial conditions'', for
which the spatial decay of the leading edge is steeper than that of
$\exp[-\lambda^* x]$. From this point of view, the $\exp[-\lambda x]$
(with $\lambda<\lambda^*$) initial configuration of the leading edge,
which leads to the eventual front speed $v(\lambda)$, is very special,
and therefore is not interesting. \label{importance}}

The discussions below Eq. (\ref{e3}), so far, is relevant only with an
underlying understanding that we are considering a front (propagating
into a linearly unstable state) that asymptotically yields a pulled
front. In that sense, the usage of Fisher-Kolmogorov equation as an
example is very well-placed, since it asymptotically admits only a
pulled front solution. As for the question whether an equation
describing a propagating front into an unstable state gives rise to a
pulled front or not, a necessary condition is not known, but for a
sufficient condition, it is known that if all the nonlinear terms in
the time evolution of the front suppress growth for a front that
propagates into a linearly unstable state, then the resulting front is
a pulled front. In particular, for the nonlinear diffusion equation
$\partial_t\phi=D\partial^2_x\phi+f(\phi)$ (an equation that has been
extensively studied in the front propagation literature), the above
sentence implies that one can expect pulled fronts if $f(u)/u\leq
f'(0)$ \cite{wimreview}.

A similar general analysis [i.e., Eqs. (\ref{e2}-\ref{e3}) and related
discussions] for the asymptotic speed selection mechanism for pushed
fronts does not exist. This is not a surprise in itself, as for pushed
fronts, the front speed is really determined in the nonlinear bulk
phase of the front (which one cannot solve except for special sets of
parameter values). As a result, the leading edge of the front does not
play any role in the front speed selection. Nevertheless, it is known
that for sufficiently steep initial conditions, as opposed to the
$1/t$ convergence to the asymptotic front speed $v^*$ for pulled
fronts, the convergence to asymptotic front speed $v^\dagger$ for
pushed fronts\footnote{Just like $v^*$ is a standard notation for the
linear spreading speed, $v^\dagger$ is also a standard notation for
the asymptotic speed of pushed fronts.\label{pushednot}} is {\it
exponential\/} in time. For the front speed, a more mathematical
representation of the exponential convergence for pushed fronts
vs. the non-exponential convergence for pulled fronts can be traced to
the stability criteria of the asymptotic front solutions
\cite{wimreview,vs1,ebert1}. The idea is that at long times, an
intermediate front profile evolving towards its final asymptotics can
be decomposed as the asymptotic front profile and infinitesimal
localized perturbations around it. The convergence properties to the
asymptotic front speed is then determined by how fast, in the comoving
frame of the asymptotic front profile, these infinitesimal
perturbations decay in time.

To elucidate the relation between the stability of the asymptotic
pulled or pushed front solutions and their convergence (in time)
behaviour, let us consider a front solution $\Phi_v(\xi)$  propagating
with a speed $v$, and decompose an intermediate front profile
$\phi(x,t)$ as
\begin{eqnarray}
\phi(\xi,t)\,=\,\Phi_v(\xi)\,+\,\eta(\xi,t)\,,
\label{e5}
\end{eqnarray}
such that $\eta(\xi,t)$ is of infinitesimal magnitude everywhere. At
the leading order, the dynamics of $\eta(\xi,t)$ is described
by\footnote{In this example, we use an equation that involves only a
first order derivative in time. A similar analysis can also be worked
out for higher time derivatives.\label{fn7}}
\begin{eqnarray}
\frac{\partial\eta}{\partial t}\,=\,{\cal L}_v\,\eta\,+\,{\cal
  O}(\eta^2)\,,
\label{e6}
\end{eqnarray}
where the linear operator ${\cal L}_v$ is obtained by linearizing the
equation for $\phi(\xi,t)$ in the comoving frame of the asymptotic
solution.\footnote{E.g., for Eq. (\ref{e1}), ${\cal
L}_v=D\partial^2_\xi+v\partial_\xi+(1-\Phi^2_v)$.\label{fn8}} In
Eq. (\ref{e6}), $\eta(\xi,t)$ can be expanded as a linear combination
of the eigenfunctions of the operator of ${\cal L}_v$.\footnote{In
general, the operator ${\cal L}_v$ is not a Hermitian operator, and
therefore its left and right eigenvectors are different.\label{fn9}}
In this expansion, there is an eigenfunction with zero eigenvalue that
corresponds to the translational invariance of the asymptotic front
solution, while the nature of its lowest-lying eigenvalues  decides
the decay properties of time convergence properties of the asymptotic
front speed \cite{vs1,ebert1}. For pushed fronts with asymptotic front
speed $v^\dagger$, the eigenspectrum of ${\cal L}_{v^\dagger}$ for
these lowest-lying eigenfunctions is gapped --- this indicates that
the convergence to the asymptotic front solution
$\phi^\dagger(x-v^\dagger t)$ is exponential in time. On the other
hand, for pulled fronts, the spectrum of ${\cal L}_{v^*}$ is gapless,
indicating that the convergence to the corresponding asymptotic front
solution $\phi^*(x-v^*t)$ has to be non-exponential in nature
\cite{vs1}.\footnote{The continuous spectrum of ${\cal L}_{v^*}$ for
pulled fronts is known to be responsible for the breakdown of the
so-called solvability analysis; for a detailed study, see
Ref. \cite{ebert1}. Moreover, the continuity of the spectrum of ${\cal
L}_{v^*}$ is a mathematical representation of the so-called marginal
stability criterion of pulled fronts, see for example
Refs. \cite{wimreview,bj,vs1,ebert1,oono1,oono2,oono3} etc. and
references cited therein. \label{fn10}}

\subsection{Fluctuating (and propagating) Fronts: A Separate
Parad\-igm \label{sec1.2}}

From the point of view that the thermal fluctuations are rather small
for the pattern forming systems that relate to the inherent
non-equilibrium nature of the bulk phases, it is less of a surprise
that the effect of fluctuations on propagating fronts has not
attracted the attention of physicists until relatively recently. The
understanding that the fluctuations are indeed important was not only
motivated by the realization that matter is composed of discrete
particles, but also from the fact that there are situations, where
fronts are naturally made of discrete constituents on a lattice (see
for example,
Refs. \cite{kerstein,ds1,breuer,krug1,doering1,vanzon,ellak1,ellak2}). In
addition, it is also of general interest to see the effect of
externally added noise on the otherwise deterministic fronts, to see
how severely this properties are affected, or if these noise terms
give rise to new phenomena. Although in some cases, studies concerning
the effect of externally added fluctuations on the fronts were
motivated by such theoretical interests, e.g. in Refs.
\cite{lutz1,lutz2,lutz3,pasquale,armeroprl,armero,jaume1,jaume2,jaume3,santos},
in some others, the noise terms were added to model the effect of
discreteness of constituent particles and the lattice, e.g. in
Refs. \cite{lemar1,lemar2,levine,doering21,doering22}. In this review
article, we will review both of these cases. In Sec. 2, our focus will
be on propagating fronts made of discrete particles on a lattice,
there we will review all the known results of such fronts. In Sec. 3,
we will review propagating fronts where fluctuations are introduced by
means of externally added noise terms. The discussion regarding to
what extent the externally added fluctuations correspond to discrete
particle models will be discussed subsequently.

The presence of fluctuations, be it as a result of discreteness of
particles and the lattice, or be it as a result of the externally
noise terms to otherwise deterministic equations, immediately implies
that (i) at a contrast to the deterministic fronts propagating with a
fixed shape at a fixed speed $v_{\mbox{\scriptsize as}}$ at
$t\rightarrow\infty$, strictly speaking, different snapshots of one
particular realization of a fluctuating front, taken at different
times, will be microscopically different from each other; although in
an average sense, each of these realizations will have the {\it
same\/} well-defined shape. The consequences of this is the following:
unlike the deterministic fronts, the front position defined by the
location $x_0$ of the point, where the number of particles per
correlation volume is a fixed number, say $N_0$ (or by the point where
the front field reaches a fixed value $\phi_0$), does not move with a
constant speed for a given realization of a fluctuating front {\it
even for large times}. It is only when the movement of this point over
a long period of time is considered (at large times), the speed of
this point, calculated as a long time average, can then be defined as
the asymptotic front speed.\footnote{Clearly, for the asymptotic front
speed to be well-defined, it should be independent of $N_0$ (or
$\phi_0$), which in itself is a consequence that in an average sense,
the front realization has a well-defined shape.\label{fn11}} (ii)
Moreover, just like any stochastic quantity having a variance around
its average, if one follows $x_0$ as a function of time, then one also
expects $\langle[x_0(t)-v_{\mbox{\scriptsize as}}t]^2\rangle$ to
increase over time. The angular brackets denote an averaging over an
ensemble, each individual member realization of which has reached
their steady shapes in an average sense. This means that the
displacements of individual front realizations w.r.t. each other {\it
as a whole front\/} keeps increasing with time --- this a phenomenon
known as the wandering of fluctuating fronts.

\begin{figure}[ht]
\begin{center}   
\includegraphics[width=0.55\linewidth,angle=0.5]{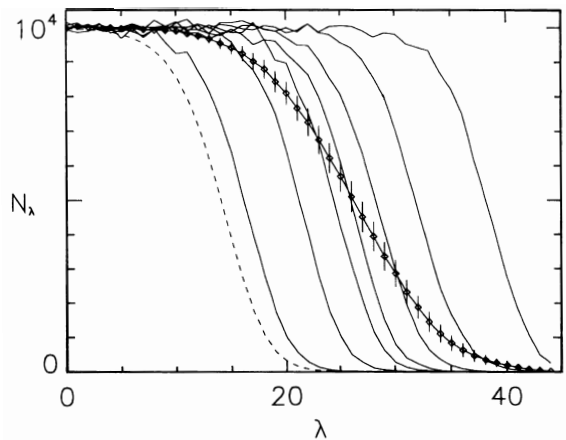}
\end{center}
\caption[]{Adopted from Ref. \cite{breuer}, this figure depicts the
front shapes for the reaction-diffusion process X $\leftrightharpoons$
2X for discrete particles on a lattice, indexed by $\lambda$. The rate
of forward reaction X $\rightarrow$ 2X is $N/2$ with $N=10^4$, while
rate for the backward reaction 2X $\rightarrow$ X has been normalized
to unity. Particles diffuse to their nearest neighbour sites with a
unit rate. Solid curves: front profiles for seven different
realizations at a certain time. Solid curve with error bars: front
profile obtained by averaging over number of particles on each lattice
site for 66 realizations.\label{breuer}}
\label{fig5}
\end{figure}
An example of the phenomena in (i) and (ii) of the previous paragraph
for fluctuating fronts is shown in Fig. \ref{fig5}. It has been
adopted from Ref. \cite{breuer}, and it depicts the shapes of fronts
made of discrete particles in a reaction-diffusion process X
$\leftrightharpoons$ 2X on a lattice indexed by $\lambda$. The rate of
forward reaction X $\rightarrow$ 2X is $N/2$ with $N=10^4$, while the
rate for the backward reaction 2X $\rightarrow$ X has been normalized
to unity. The seven solid curves are the front shapes obtained by
taking simultaneous snapshots of seven different realizations at a
certain time. All these realizations started with the same initial
microscopic configuration. At a microscopic level, the shapes of these
snapshots are different, but in an average sense, each of them has a
shape identical to that of the dashed curve. The displacements of the
individual front realizations as a whole w.r.t. each other demonstrate
front wandering. On the other hand, the solid line with error bars in
it depicts the front shape obtained by averaging over number of
particles on the lattice sites for 66 different front realizations. As
one can expect, the front obtained by such an averaging process is
much wider than any of the seven realizations, since such an averaging
fails to filter out the effect of front wandering --- in future, we
will refer to this averaging process as ``simple averaging''.

The above discussion, therefore, points us to the following
conclusions: (a) to obtain the average asymptotic shape of fluctuating
fronts, one has to separate out the wandering effects, i.e., one has
to first realign the simultaneous realizations and then calculate the
average shape --- from now on, we will refer to such an averaging as
``conditional (ensemble) averaging'',\footnote{It is precisely this
conditionally averaged front profile that is described by an
appropriate deterministic mean-field equation; in future, these
conditionally averaged front profile will be denoted by
$\phi^{(0)}$. To find out how to align different snapshots of
fluctuating fronts in the sense of (a) above, see Eq. (\ref{e81}) and
footnote \ref{fnconditional2}. For now, the mere assumption that
$\phi^{(0)}$ exists will be enough.\label{phi0}} and $\langle\langle
N_k\rangle\rangle$ will denote the conditionally averaged number of
particles in the $k$-th lattice site. (b) For measuring the front
speed, a conditional averaging is not necessary; one can obtain the
front speed by tracking how fast a given value of the simply averaged
front profile $\phi$, say $\phi_0$, moves. However, we have seen
before, for deterministic fronts, that the front shape and the speed
are related to each other, and therefore, from that angle, for a
theoretical prediction/analysis of front speed, one does need the
conditionally averaged front profile. A good description of
fluctuating fronts must take both (a) and (b) into account. In
addition, (c) a comprehensive description of fluctuating fronts must
not only include the expression of the asymptotic front speed and as
well as its approach in time towards a steady (conditionally averaged)
front shape starting from an initial configuration, but also must it
identify the nature of its wandering (i.e., whether it is diffusive,
sub- or super-diffusive), along with the relevant characterizing
exponents.\footnote{In one spatial dimension, fluctuating fronts are
always diffusive; in higher dimensions, the situation is more
complicated.\label{fn12}} {\it Such a comprehensive description for
fluctuating fronts propagating into unstable states is the central
theme of this review article. In one spatial dimension, our convention
will be to consider fronts propagating from the left to the right, and
keeping in line with the propagation direction of the front, we will
often use forward for rightward and backward for leftward
interchangeably}.

Although I claim to provide a comprehensive description for
fluctuating fronts in this review article, its precise meaning has to
be properly laid out at this point --- after all, a front is nothing
but an interface that demarcates the boundary between two different
phases within a system; and the properties of such interfaces for,
e.g., fluid flows in porous media, flame fronts in a burning paper,
atom deposition processes in molecular beam epitaxy experiments,
aggregate fronts in diffusion-limited-aggregation etc. have already
been very well-studied in the literature. Analyzing the properties of
fronts in such vastly different systems is by no means the intention
of this review. Instead, my purpose here is the following: in spite of
the existence of a large variety of dynamical equations that describe
moving interfaces in different systems, there is an overall underlying
framework for the propagation mechanism of deterministic fronts; thus,
I want to review the effect of fluctuations on propagating fronts
under the same framework in a comprehensive manner.

In general, providing such a comprehensive description for fluctuating
fronts is by no means an easy task. As mentioned in the first
paragraph of Sec. \ref{sec1.2}, two major approaches have been used to
this end --- the first one being that of noisy dynamical equations,
where one simply adds external noise terms to the otherwise
deterministic equations to introduce fluctuations. In the second
approach, many researchers in the last few years have taken  the route
of studying fluctuating fronts that are constituted of discrete
particles on a lattice, where stochasticity of the microscopic
dynamics of the particles themselves gives rise to fluctuations. In a
way, the philosophy of both these approaches is to first have a
deterministic front as the underlying structure, and then to study the
effect of fluctuations superimposed thereupon: there exists a
parameter in these approaches that controls the fluctuation
strength. For example, in the second approach, this parameter is
$1/N$, where $N$ is the conditionally averaged number of particles per
lattice site in the stable phase of the front, and the deterministic
mean-field results are obtained when $N\rightarrow\infty$. In the
recent past, there has also appeared a new approach (along the lines
of quite the opposite philosophy of the two approaches discussed
above) to extract the mean-field behaviour at small $N$
\cite{moron}. Therein, one defines a variable $N^*$ that measures the
number of particles per correlation volume in the stable phase of the
front, and depending on the system parameters, one can have
$N^*\rightarrow\infty$ for correlation volume $\rightarrow\infty$ and
small $N$; in this limit, one also recovers the deterministic
mean-field description for the front.

Nonetheless, it has to be noted that the majority of the studies on
fluctuating fronts have been carried out on fluctuating
``pulled''\footnote{The use of the quotes is motivated by the fact
that fluctuating fronts, whose deterministic limit yield pulled
fronts, are actually {\it weakly pushed}. See points IV and V of
Sec. \ref{sec2.1} for more details.\label{fn13}} fronts, i.e., on
fronts whose deterministic limit yield pulled fronts, as opposed to on
fluctuating pushed fronts\footnote{In view of footnote \ref{fn13},
what we mean by fluctuating pushed fronts is that the deterministic
limit of these fronts yield pushed fronts.\label{fn14}} . This is not
very surprising --- first of all, for the deterministic counterparts,
it is the pulled fronts, for which there is a solid theoretical
understanding. Secondly (and more importantly), it is the sensitivity
of the pulled fronts to the dynamics of the leading edge, which is
severely affected by the discreteness of particles and the lattice, or
by the presence of external noise terms --- as a result, fluctuating
``pulled'' fronts exhibit many surprising characteristics. \vfill

\subsection{A Note on the Notations \label{sec1.3}}

Before we proceed further on, it will be very helpful to have the
convention of the notations used summarized. Below we provide a table
for this purpose. Any notation that does not appear in Table I is
explained before its usage in this review article.
\begin{center}
\begin{tabular}{||p{0.07\linewidth}|p{0.85\linewidth}||}
\hline \hline $x,k$ & Notations for continuum space and for lattice
index respectively.\\ \hline $v^*$ & Notation for the linear spreading
speed for {\it deterministic} fronts propagating into unstable states,
and also for the selected asymptotic front speed for (deterministic)
pulled fronts with a steep enough initial condition. Has already been
introduced in Sec. \ref{sec1.1}. Applies to deterministic pulled
fronts both in continuum space and on a lattice.\\ \hline $v^\dagger$
& Notation for the selected asymptotic front speed for (deterministic)
pushed fronts. Also introduced already in Sec. \ref{sec1.1}. Applies
to deterministic pushed fronts both in continuum space and on a
lattice.\\ \hline $v_{\mbox{\scriptsize as}}$ & General notation for
the asymptotic front speed of any front (pulled and pushed) in any
system (deterministic and stochastic; discrete and as well as
continuum space).\\ \hline $v_N$ & Notation for front speed for a
discrete particle and lattice system of fronts, where there are $N$
particles on (conditional) average per lattice site.\\ \hline $\xi$ &
Notation for comoving co-ordinate for pulled ($\xi=x-v^*t$, or
$\xi=k-v^*t$), or fluctuating ``pulled'' fronts ($\xi=k-v_Nt$), or
even for fronts whose deterministic mean-field limit yield pulled
fronts but whose speeds have not been determined yet (in these cases,
$\xi$ will be used to denote $x-v_{\mbox{\scriptsize as}}t$ or
$k-v_{\mbox{\scriptsize as}}t$ or $k-v_Nt$ as appropriate).\\ \hline
$\zeta$ & Notation for comoving co-ordinate for pushed
($\zeta=x-v^\dagger t$, or $\zeta=k-v^\dagger t$), or fluctuating
pushed fronts ($\zeta=k-v_Nt$), or even for fronts whose deterministic
mean-field limit yield pushed fronts but whose speeds have not been
determined yet (in these cases, $\zeta$ will be used to denote
$x-v_{\mbox{\scriptsize as}}t$ or $k-v_{\mbox{\scriptsize as}}t$ or
$k-v_Nt$ as appropriate).\\ \hline $\phi^*(\xi)$ & Notation for the
front profile for pulled fronts; either $\xi=x-v^*t$ or
$\xi=k-v^*t$. Already introduced in Sec. \ref{sec1.1}.\\ \hline
$\phi^\dagger(\zeta)$ & Notation for the front profile for pulled
fronts; either $\zeta=x-v^\dagger t$ or $\xi=k-v^\dagger t$. Already
introduced in Sec. \ref{sec1.1}.\\ \hline \hline
\end{tabular}
\end{center}
{\footnotesize Table I: Convention behind the notations used in this
review article.}  \vfill

\section{Fluctuating Fronts in Discrete Particle and Lattice Systems
\label{sec2}} 

\subsection{Summary of Known Results \label{sec2.1}}

In Sec. \ref{sec2.1}, we summarize all the known results for
fluctuating (propagating) fronts, made of discrete particles on a
lattice. Among the physical systems, where front propagation has been
studied under this scheme, various sorts of reaction-diffusion systems
constitute the majority
\cite{kerstein,breuer,doering1,ellak1,ellak2,goutam,moro,bramson1,kerstein1,ba,doering3,kns,deb11,deb12,lemar2,lemar32,sokolov1,sokolov2}.
In addition, propagating fronts were also studied in the context of
directed polymers in random media
\cite{ds1,cook,derrida1,bd1,bd2,bd3}, and the calculation of the
largest Lyapunov exponent in a gas of hard spheres
\cite{vanzon,vanzonarticles1,vanzonarticles2,vanzonarticles3,vanzon1}. In
all these models, fronts propagate from a stable to an unstable
state. The unstable state is constituted of empty lattice sites,
whereas in our notation, the (conditionally) averaged number of
particles on the $k$-th lattice site $\langle\langle
N_k\rangle\rangle$ in the stable state and the corresponding
asymptotic front speed are respectively denoted by $N$ from now
on. Also, unless otherwise stated, in Sec. \ref{sec2}, fronts will be
assumed to propagate in one spatial dimension; fronts in higher than
one spatial dimensions will be considered only in Sec. \ref{sec2.6.6}.

For these models, the following results have been obtained in the last
decade:

\begin{itemize}

\item[]{I.} The limit of deterministic pulled front propagating with
speed $v^*$ is reached for $N\rightarrow\infty$. However, the
convergence of $v_N$ to its deterministic limit $v^*$ for increasing
value of $N$ is extremely slow. The approach of $v_N$ to $v^*$ is from
below, and for asymptotically large $N$ values, the leading order form
of $v_N$ is {\it model independent\/}
\cite{vanzon,kns,deb11,deb12,bd1,bd2,bd3,vanzonarticles1,vanzonarticles2,vanzonarticles3},
given by
\begin{eqnarray}
v_N\,=\,v^*\,-\,\frac{d^2v(\lambda)}{d\lambda^2}{\Bigg{|}}_{\lambda^*}\,\frac{\pi^2\,{\lambda^{\!*}}^{\,2}}{\ln^2N}\,.
\label{e7}
\end{eqnarray}

\item[]{II.} For intermediately large values of $N$, the subdominant
corrections to the leading order result (\ref{e7}) are very strong,
and they depend on the details of the model. For some models, these
corrections can be so strong that one does not observe the asymptotic
scaling (\ref{e7}) unless $N$ is extremely large
\cite{breuer,kns,deb11,deb12,lemar2,lemar32}.
\vspace{4mm}

\item[]{III.} Unlike the semi-infinite leading edge for deterministic
pulled fronts, in any snapshot of these discrete particle and lattice
systems for any value of $N$ at any time, there exists a ``foremost
occupied lattice site (f.o.l.s.)'', on the right of which no lattice
site has ever been occupied before. In the hopes of providing a theory
that would yield front speeds at intermediately large values of $N$
for these fronts, a stochastic dynamics of the f.o.l.s., and a
(standard) deterministic and uniformly translating front solution
reasonably far behind has been proposed in
Ref. \cite{deb11,deb12}. The two solutions are then matched in the
``tip region'' of the front, spatially extended over a few lattice
sites behind the f.o.l.s. (and also including the f.o.l.s.). Despite
the fact that this formalism yields a consistent (and verifiable)
picture of these fronts, it is not predictive, and a first-principle
based predictive theory for $v_N$ at intermediately large values of
$N$ is still lacking.
\vspace{4mm}

\item[]{IV.} Unlike pulled fronts, where the asymptotic front speed
$v^*$ is reached only if the initial front profile decays faster than
$\exp[-\lambda^*x]$ for $x\rightarrow\infty$, it can be argued that
the conditionally averaged front profile for fluctuating ``pulled''
fronts is reached {\it uniquely\/}, {\it independent\/} of the initial
microscopic configuration. This observation has been made in
Ref. \cite{bd3}, and it is consistent with all the known simulation
results so far.
\vspace{4mm}

\item[]{V.} In the limit of $N\rightarrow\infty$, the spectrum of the
stability operator ${\cal L}_{v_N}$ [c.f. Eq. (\ref{e6})] is gapped,
and for the lowest-lying eigenvalues, the gap is $\propto1/\ln^2N$
\cite{kns,deb3}. Although we will get into more details later on, here
we briefly mention that points IV and V together bear the signature of
these fronts being weakly pushed.
\vspace{4mm}

\item[]{VI.} In the limit of $N\rightarrow\infty$, the diffusion
coefficient of the front $D_f$ that characterizes front wandering,
approaches zero. For fluctuating ``pulled'' front in a model that
closely resembles the so-called clock model \cite{vanzon}, $D_f$ has
been shown to scale as $1/\ln^3N$ in the large $N$ asymptotic regime
\cite{bd3}.
\vspace{4mm}

\item[]{VII.} As $N$ is reduced ``too much'', the fluctuations in the
number of particles per lattice site in the stable state become
stronger and stronger. Perhaps in view of II above, in the sense that
due to the lack of first-principle based predictive theory for
finite-$N$ corrections, works on such values of $N$ are somewhat rare
\cite{ellak1,ellak2,sokolov1}. The extreme limit of this,  namely the
case when there is at most one particle is allowed per lattice site,
however, is very well studied (we will refer to this scenario as
``$N\leq1$''). In view of the fact that there are simply too many
results to summarize here, the readers are referred directly to
Sec. \ref{sec2.6} for further details.
\vspace{4mm}

\item[]{VIII.} As far as the discrete particle (and lattice) systems
of fluctuating pushed fronts are concerned (c.f. footnote \ref{fn14}),
at the time of writing this review article, I have not seen any work
on the behaviour of $v_N$ for $N\rightarrow\infty$. Nevertheless,
having  combined the knowledge that one musters from points I through
VI above  and the solvability analysis for pushed fronts
\cite{ebert1}, for  $N\rightarrow\infty$, the convergence of the
asymptotic front speed to  $v^\dagger$ (c.f. footnote \ref{pushednot})
for fluctuating pushed  fronts to their deterministic limit behaves as
a power law of $N$.
\end{itemize} 
\vspace{1mm}

In the rest of Sec. \ref{sec2}, we will elaborate points I-VIII, and
provide a unified picture of discrete particle systems of fluctuating
fronts. Details of which point is elaborated where are given below:
point I in Sec. \ref{sec2.2}, points II and III in Sec. \ref{sec2.3},
points IV and V in Sec. \ref{sec2.4}, point VI in Sec. \ref{sec2.5},
point VII in Sec. \ref{sec2.6}, and point VIII in
Sec. \ref{sec2.7}. Finally, in Sec. \ref{sec2.8} we put the results of
Secs. \ref{sec2.2}, \ref{sec2.4}-\ref{sec2.5} and \ref{sec2.7} in a
unified perspective.

\subsection{Derivation of $1/\ln^2N$ Convergence of the Asymptotic Front
Speed to $v^*$ for Fluctuating ``Pulled'' Fronts as
$N\rightarrow\infty$
\label{sec2.2}}

From its definition in Sec. \ref{sec1.1}, pulled fronts can be clearly
seen to be realized in a continuum description, for otherwise the
``infinitesimal perturbation around the unstable state'' does not
quite make sense. Through its definition (and propagation mechanism),
pulled fronts and infinitesimal perturbations around the unstable
state are intimately intertwined --- ahead of the leading edge of a
pulled front, the perturbation around the unstable state is
infinitesimal, as it grows and propagates itself with speed $v^*$ and
pulls the rest of the front along with it. From this perspective, the
discrete particle and lattice systems of fluctuating ``pulled'' fronts
belong to a different category --- at the leading edge of these
fronts, due to discreteness of the particles, the smallest amount of
the ``infinitesimal perturbation'' for growth to start on any lattice
site cannot be less than that of one quantum of particle, a fact that
is more than likely to yield a front speed different from $v^*$.

In hindsight, the above argument sounds simple, and it is appreciated
easily when the conditionally averaged number of particles per lattice
site is translated into the language of $\phi$ of
Sec. \ref{sec1.1}. Notice that in the stable phase of the front,
$\langle\langle N_k\rangle\rangle=N$ corresponds to $\phi=1$. As a
result, the lowest amount of $\phi$ corresponding to one quantum of
particle, needed on any lattice site ahead of the leading edge for
growth to begin is $1/N$. The pulled front limit in these discrete
particle systems is therefore expected to be reached only for
$N\rightarrow\infty$.

These subtleties were first realized by Brunet and Derrida \cite{bd1},
shortly after simulations revealed that for models of fluctuating
``pulled'' fronts the deviations of the asymptotic front speed from
$v^*$ can be significant \cite{breuer}, even for
$N\gg1$.\footnote{Severe deviations of the front speeds from $v^*$ for
$N$ not so larger than $1$ have also been observed in simulations, but
we leave those aside until we discuss point VII at length in
Sec. \ref{sec2.6}.\label{fn2.1}} To mimic the lowest quantum of
particle that is needed on any lattice site ahead of the leading edge
for the growth to begin, Brunet and Derrida conjectured that the
dynamics of a fluctuating ``pulled'' front at large $N$ should be
described by an equation with a growth cutoff for
$\phi\leq\varepsilon$, where $\varepsilon=1/N$.

To verify their ideas, they used modified Eq. (\ref{e1}):
\begin{eqnarray}
\frac{\partial\phi}{\partial t}\,=\,\frac{\partial^2\phi}{\partial
x^2}\,+\,(\phi\,-\,\phi^n)\,\Theta\left(\phi\,-\,\varepsilon\right)\,.
\label{e8}
\end{eqnarray}
The stationary front solution of Eq. (\ref{e8}), propagating with
speed $v_{\varepsilon}$, was then divided into three regions: I, II
and III. In region II and III, $\phi\ll1$ and hence the nonlinear term
$\phi^n$ can be ignored; in addition, in region III,
$\phi<\varepsilon$, which leads to a further simplification of
Eq. (\ref{e8}). On the other hand, in region I, one has to consider
the full nonlinear equation. This leads one to
\begin{eqnarray}
 -\,v_\varepsilon\,\frac{d\phi}{d\xi}\,=\,\frac{d^2\phi}{d\xi^2}\,+\,\phi\,-\,\phi^n\quad\quad\,\mbox{in
 region
 I,}\nonumber\\&&\hspace{-7.75cm}-\,v_\varepsilon\,\frac{d\phi}{d\xi}\,\simeq\,\frac{d^2\phi}{d\xi^2}\,+\,\phi\quad\quad\quad\quad\,\,\,\,\,\mbox{in
 region II,
 and}\nonumber\\&&\hspace{-7.75cm}-\,v_\varepsilon\,\frac{d\phi}{d\xi}\,=\,\frac{d^2\phi}{d\xi^2}\quad\quad\quad\quad\quad\quad\quad\mbox{in
 region III} \,.
\label{e9}
\end{eqnarray}
At the boundary between regions II and III, the continuity of the
value of the field $\phi$ as well as of its comoving derivative
$\phi'(\xi)$ has to be maintained. With the understanding that
$\varepsilon\ll1$, the leading order solutions that obey these
boundary conditions were shown to be \cite{bd1}\footnote{Notice the
notation $\phi^{(0)}$: the idea is motivated by the fact that as
$N\rightarrow\infty$, the conditionally averaged front shape is given
by the effective deterministic mean-field equation (\ref{e9}). See
footnote \ref{phi0} in this context.\label{effective}}
\begin{eqnarray}
\phi^{(0)}(\xi)\,=\,\varepsilon\,e^{-\,v_\varepsilon(\xi\,-\,\xi_0)}\quad\quad\quad\quad\quad\quad\quad\quad\quad\quad\quad\,\,\,\,\,\mbox{in
region III,
and}\nonumber\\&&\hspace{-10.65cm}=\,\frac{A}{\pi\lambda^*}\,|\ln\varepsilon|\,\sin\left[\frac{\pi\lambda^*(\xi-\xi_1)}{|\ln\varepsilon|}\right]e^{-\lambda^*(\xi-\xi_1)}\,\,\,\mbox{in
region II,}
\label{e10}
\end{eqnarray}
such that the boundary between regions II and III lies at
$\xi_0\simeq|\ln\varepsilon|/\lambda^*$, whereas the solution in
region II crosses over to that in region I at $\xi_1$, and $A$ is a
quantity of ${\cal O}(1)$; yielding
\begin{eqnarray}
v_\varepsilon\,=\,v^*\,-\,\frac{d^2v(\lambda)}{d\lambda^2}{\Bigg{|}}_{\lambda^*}\,\frac{\pi^2\,{\lambda^{\!*}}^{\,2}}{\ln^2\varepsilon}\,.
\label{e11}
\end{eqnarray}
Notice that to obtain the expression (\ref{e11}) of $v_\varepsilon$,
the solution of Eq. (\ref{e9}) in region I is not required. Moreover,
although from solution (\ref{e10}) $\phi^{(0)}(\xi)$ might seem to go
to zero at $\xi=0$ violating the monotonicity of the front profile, in
reality, there exists no sharp boundary between regions I and II, like
the one at $\xi_0$. Instead, the solution (\ref{e10}) crosses over to
that of region I, and the crossover happens in a way that maintains
the monotonicity of the front profile. At the leading order, the
characteristic position $\xi=\xi_1$ for the crossover is given by
$\xi_1\simeq0$.

Few points in Brunet and Derrida's conjecture must be noted: (a) One
arrives at the leading order large $N$ asymptotic behaviour of the
front speed (\ref{e7}) by setting $\varepsilon=1/N$. The form of $v_N$
is independent of the details of the microscopic dynamics of the
constituent particles, although the actual dispersion relation
$v(\lambda)$ vs. $\lambda$ is obtained from the equation obeyed by the
deterministic mean-field limit of the microscopic model at hand. (b)
Region III is characterized by an effective (finite) width
$1/v_\varepsilon$ due to the exponential decay of the front
profile. On the other hand, with the substitution $\varepsilon=1/N$
for asymptotically large $N$, the length $(\xi_0-\xi_1)$ of region II
increases logarithmically with $N$, while the length of region I
remains finite --- this yields an effective front width, scaling as
$\ln N$. (c) Furthermore, the leading order form of $v_N$ is
independent of whether one substitutes $\varepsilon=1/N$ or
$\varepsilon=c/N$, with $c$ of ${\cal O}(1)$. In addition to amplitude
$A$, $c$ simply provides a subdominant correction to the r.h.s. of
Eq. (\ref{e7}), if $c\neq1$.

\begin{figure}[ht]
\begin{center}
\includegraphics[width=0.7\linewidth]{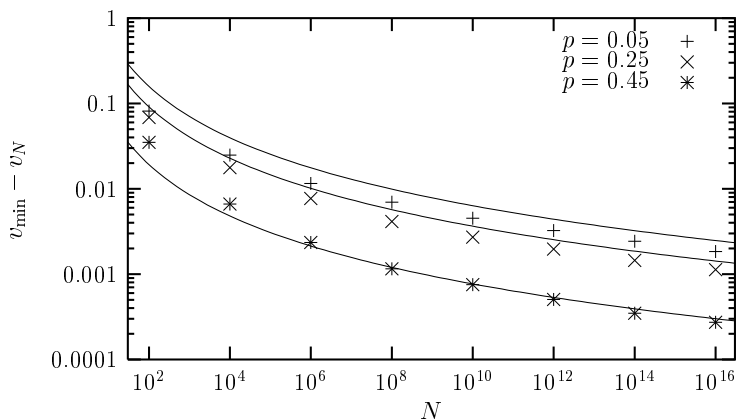}
\end{center}
\caption{The plot of $v^*-v_N$ vs. $N$ for the discrete particle and
lattice model of Refs. \protect{\cite{bd1,bd2,bd3}} relating to the
study of directed polymers in random media. In the above figure,
$v_{min}$ should be read as $v^*$. The solid curves represent
Eq. (\ref{e7}) for various $p$ values \protect{\cite{bd2}}. A
corresponding graph for the same model for much higher values of $N$
for $p=0.25$ can be found in Fig. 2 of
Ref. \protect{\cite{bd3}}.\label{bdfig}}
\end{figure}
\begin{figure}[ht]
\begin{center}
\includegraphics[width=0.5\linewidth,angle=270]{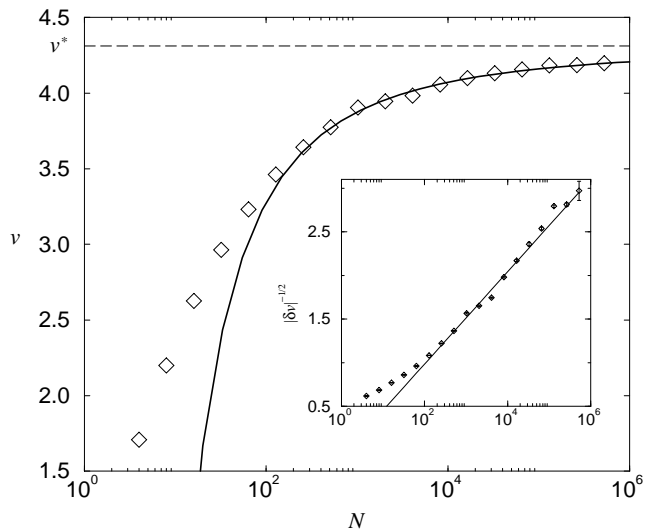}
\end{center}
\caption{The plot of $v^*-v_N$ vs. $N$ for the so-called clock model
\protect{\cite{vanzon,vanzonarticles1,vanzonarticles2,vanzonarticles3}}. Symbols:
simulation data; solid curve in the main graph: expression (\ref{e7});
inset: the same data as of the main graph plotted differently, with
$\delta v=v^*-v_N$.\label{ramsesfig}}
\end{figure}

In the history of this subject, Brunet and Derrida's insight has
turned out to have had an enormously significant impact. Their
prediction of the $N$-dependence of the front speed was immediately
confirmed in a fluctuating ``pulled'' front model (where time was also
discretized) \cite{bd1,bd2,bd3}\footnote{The deterministic mean-field
limit for this model is described by the equation
$\phi(k,t+1)=1-p[1-\phi(k-1,t)]^2-(1-p)[1-\phi(k,t)]^2$, where $k$ is
the lattice site index and $p$ is a fixed number in the  interval
$(0,0.5)$. The corresponding U-shaped dispersion relation $v(\lambda)$
vs. $\lambda$ has the form
$v(\lambda)=\ln\left[2pe^\lambda+2(1-p)\right]/\lambda$.\label{fn2.2}}
relating to directed polymers in random media (see Fig. \ref{bdfig})
\cite{cook,derrida1}. Subsequently, in another model of a fluctuating
``pulled'' front --- namely, the so-called clock model relating to the
calculation of the largest Lyapunov exponent in a gas of hard spheres
\cite{vanzon,vanzonarticles1,vanzonarticles2,vanzonarticles3}\footnote{The
deterministic mean-field limit of the so-called clock model is
described by $\displaystyle{\frac{\partial\phi_k}{\partial
t}}=-\phi_k+\phi_{k-1}^2$, leading to the dispersion relation
$v(\lambda)=\left[2e^\lambda-1\right]/\lambda$. \label{fn2.3}} --- the
idea of a cutoff was employed; and the sinusoidal profile of
Eq. (\ref{e10}) at the leading edge, along with the result that
$v^*-v_N\propto1/\ln^2N$ for asymptotically large $N$, was obtained
(see Fig. \ref{ramsesfig}).

Although a continuum-space equation was used in Ref. \cite{bd1} for
the deviation of Eq. (\ref{e7}), it turns out that for asymptotically
large $N$, one can motivate the sinusoidal profile of the leading edge
along the lines of the discussion in the first two paragraphs of
Sec. \ref{sec2.2} and the positivity of $\phi$ (in these models
$\phi_k$ represents the scaled number of particles on the lattice site
$k$, and hence it can never be
negative)\cite{deb11,deb12,vanzonarticles1,vanzonarticles2,vanzonarticles3},
from which one arrives at Eq. (\ref{e7}) very simply for any given
microscopic model. The idea is the following: since the growth of
particles ahead of the leading edge of these fronts does not start
unless there is at least one particle on any lattice site, one can
expect the discreteness of particles to {\it inhibit\/} the spreading
of the leading edge, resulting in a front speed $v_N$ smaller than
$v^*$. However, $v^*$ is the minimum value of $v(\lambda)$ for any
real $\lambda$ in the deterministic mean-field description of the
microscopic model at hand, and therefore the only way to  obtain
$v_N<v^*$ from the dispersion relation $v(\lambda)$ vs. $\lambda$ is
to consider $\mbox{Im}(\lambda)\neq0$.

Equation (\ref{e10}) originates from the consideration of non-zero
$\mbox{Im}(\lambda)$, which enters the solution as the inverse of the
wavelength of the sin-function. The positivity of $\phi$ then
restricts the value of its argument to lie within $(0,\pi)$ (modulo a
constant phase factor, which can be set to zero by shifting the
position of the origin in the comoving frame). For
$N\rightarrow\infty$, $\mbox{Re}(\lambda)\rightarrow\lambda^*$,
$\mbox{Im}(\lambda)\rightarrow0$ and $v_N\rightarrow v^*$, and as a
result, for asymptotically large $N$, one can expand the front
solution around these asymptotic values. Such an expansion confines
the values of $\mbox{Re}(\lambda)$, $\mbox{Im}(\lambda)$ and $v_N$
close to the minimum of the dispersion relation $v(\lambda)$
vs. $\lambda$ on the $\mbox{Re}(\lambda)$-plane, and at the leading
order, it yields the following results: (a) $\mbox{Im}(\lambda)=\ln
N/\pi\lambda^*$, just like in the argument of the sin-function in
Eq. (\ref{e10}), and (b) the large $N$ asymptotics (\ref{e7}), where
the second derivative $v^{''}(\lambda)\big|_{\lambda^*}$ appears due
to the fact that near the extremum of $v(\lambda)$ on the
$\mbox{Re}(\lambda)$-plane, the variation of the front speed for a
small variation of $\mbox{Re}(\lambda)$ from $\lambda^*$ is $\propto
v^{''}(\lambda)\big|_{\lambda^*}$.

We conclude Sec. \ref{sec2.2} with a short note on the subdominant
corrections to the $\propto1/\ln^2N$ deviation of $v_N$ from $v^*$. As
mentioned earlier, according to the derivations in Ref. \cite{bd1},
along with other explicit dependence on $N$, both $A$ and $c$ enter
the expression of the subdominant corrections. For asymptotically
large values of $N$, the behaviour of the subdominant corrections is
dominated by these explicit $N$-dependent terms rather than the ones
involving $A$ and $c$. For the microscopic model relating to directed
polymers in random media, a $1/\ln^{\beta}N$ form has been suggested
for the leading order behaviour of the subdominant corrections in
Ref. \cite{bd3}, where $\beta$ has been estimated to lie somewhere
between $2.5$ and $3$. However, for the reaction-diffusion system
X$\leftrightharpoons2$X in Ref. \cite{deb11,deb12}, at the leading
order, the subdominant corrections seem to behave as $\ln[\ln
N]/\ln^3N$. Although this issue has not yet been settled, it is only a
theoretical and a marginal one --- after all, at $N$ values when the
subdominant corrections start to become stronger, a very different
description of fluctuating ``pulled'' fronts is called for, as we will
find out next.

\subsection{The Case of Intermediately Large Values of $N$, Foremost
Occupied Lattice Site, Tip Region of the Front and All That
\label{sec2.3}} 

\begin{figure}[ht]
\begin{center}
\includegraphics[width=0.6\linewidth]{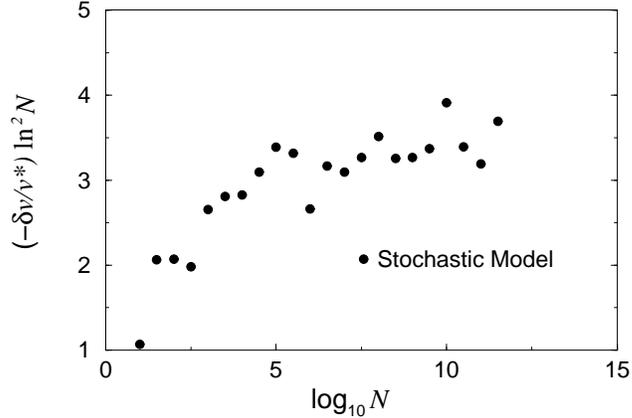}
\end{center}
\caption{Simulation data from Ref. \protect{\cite{kns}} for the
reaction-diffusion system X$+$Y$\rightarrow2$X. The rate of diffusion
$D=0.5$, the reaction rate is $0.1/N$, and $\delta v=v^*-v_N$. Note
the increasing trend of the data with increasing $N$.  For these
parameter values, according to Eq. (\ref{e7}), $\delta v\ln^2N/v^*$
should be $\simeq-9.6$. The figure is an altered version of Fig. 5 in
Ref. \protect{\cite{kns}}. The alteration was carried out for greater
clarity. \label{kesslerfig}}
\end{figure}

The large $N$ asymptotics (\ref{e7}) describes $v_N$ as an explicit
function of $N$ at the leading order. However, from comparison with
the simulation data, it seems that to observe this asymptotic result,
one often has to go to very high values of $N$. In support of their
theoretical prediction (\ref{e7}) in Ref. \cite{bd1}, Brunet and
Derrida took $N$ up to $10^{16}$ (Fig. \ref{bdfig}) \cite{bd1,bd2} in
the lattice model relating to directed polymers in random media; and
in a later work \cite{bd3}, by means of a clever simulation algorithm,
they further took $N$ to as high as $10^{150}$ in the same model. In
these studies, the range where $N$ is intermediately large but not
asymptotically large, was not examined carefully. Amid this backdrop,
a reinterpretation of the numerical results of Ref. \cite{breuer} for
the reaction-diffusion system X$\leftrightharpoons2$X suggests that
there are strong deviations of the data from the prediction (\ref{e7})
when $N$ is of ${\cal O}(10^4$-$10^6)$. Such deviations were also
noted by Kessler and co-workers \cite{kns} --- their simulations for
fluctuating ``pulled'' fronts in the reaction-diffusion system
X$+$Y$\rightarrow2$X on a lattice suggested that even for $N$ as large
as $10^{12}$, differences between simulation data and the large $N$
asymptotics (\ref{e7}) remain significant (see
Fig. \ref{kesslerfig}). In view of these, one's worry naturally shifts
to the subdominant corrections to the r.h.s. of Eq. (\ref{e7}).

The common denominator of all the derivations of Eq. (\ref{e7})
\cite{deb11,deb12,vanzonarticles1,vanzonarticles2,vanzonarticles3,bd1}
for fluctuating ``pulled'' fronts is that they were carried out to
describe a {\it conditionally averaged uniformly translating\/}
profile {\it all over the front\/}. In such a description, for the
large $N$ asymptotics, the leading order expression of $v_N$ is
independent of specific details of the microscopic model. The specific
model dependence\footnote{The dependence of $v_N$ on the specific
details of the model is apparent from a comparison of
Figs. \ref{ramsesfig} and \ref{kesslerfig}. While in
Fig. \ref{kesslerfig}, even at $N$ as large as $10^{12}$, the
$1/\ln^2N$ scaling of $v^*-v_N$ seems rather far off for the
reaction-diffusion system X$+$Y$\rightarrow2$X, Fig. \ref{ramsesfig}
finds confirmation of the $1/\ln^2N$ scaling at $N\sim10^3$ for the
so-called clock model. \label{fn2.4}} of $v_N$ enters in the
subdominant corrections through two parameters, $A$ and $c$
($\varepsilon=c/N$). In that sense, Brunet and Derrida's derivation of
Eq. (\ref{e7}) is quite instructive --- it shows that the speed of a
fluctuating ``pulled'' front is actually fully determined from the
{\it overall\/} shape of the front (thereby taking into account the
specific details of the model in a conditionally averaged manner), and
not only from the property of the leading edge --- after all, to
obtain the value of $A$, one needs to know the overall front shape,
even in the region where the nonlinearities are
important.\footnote{This stands in sharp contrast to pulled fronts. To
obtain the front speed $v^*$ for pulled fronts, one only needs the
dispersion relation $v(\lambda)$ vs. $\lambda$. The dispersion
relation is obtained from the consideration of the leading edge of the
front, where the nonlinearities are not important. \label{fn2.5}} On
the other hand, although $c$ was set equal to $1$ by Brunet and
Derrida, one might wonder what its true value is, and to what extent
it influences the subdominant corrections for a given value of $N$ in
a microscopic model.\footnote{In this context, we draw the reader's
attention to the quantity $a$ in the derivation of the large $N$
asymptotics (\ref{e7}) in Ref. \cite{deb11,deb12}. The quantities $a$
and $c$ are in fact equivalent, both are introduced by hand, and the
values of none of them can be obtained from a first principle
derivation.\label{fn2.6}} It is in fact conceivable that these
parameters have further dependence on $N$; nevertheless, the
feasibility of ever obtaining an explicit $N$-dependent prediction of
$v_N$ for any value of $N$ \cite{kns} remains an open question.

In view of these difficulties, if one were to provide a description of
fluctuating ``pulled'' fronts to yield front speeds $v_N$ for
moderately large values of $N$ as well, one must certainly look
elsewhere to address a pressing concern; namely {\it the effect of
stochasticity in the microscopic dynamics of the individual particles
on the propagating front\/}, an effect that disappears when the
conditional averaging of a large number of front  realizations is
performed all over the front region.

\begin{figure}[ht]
\begin{center}
\includegraphics[width=0.75\linewidth]{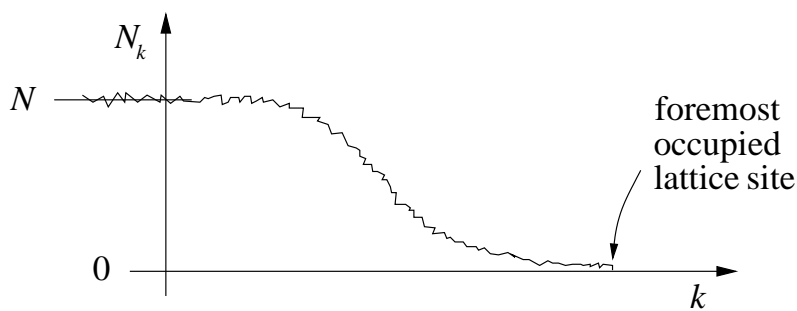}
\end{center}
\caption{An instantaneous snapshot of a front realization at large
times. Note the existence of the foremost occupied lattice site, on
the right of which no lattice site has been previously
occupied. \label{snapshot}}
\end{figure}
Such a description, put forward in Ref. \cite{deb11,deb12}, is
motivated by the microscopic picture of the front dynamics at the far
end of the front. The idea is based on the observation that {\it in
every discrete particle and lattice system of a fluctuating front, at
all times there exists a foremost occupied lattice site (f.o.l.s.),
which is defined as the one, on the right of which no lattice site has
ever been occupied before\/} (see Fig. \ref{snapshot}). As front
propagation in this framework is effectively tantamount to forward
(rightward) movement of the f.o.l.s., the crucial role played by the
discreteness effects of the particles and the lattice is reflected in
the mechanism of the front propagation at the f.o.l.s., where the
mechanism is {\it not\/} that of uniform translation, but instead, is
of {\it ``halt-and-go''}. A lattice site, which has never been
occupied before, attains the status of the f.o.l.s. as soon as one
particle hops into it from behind (left). In reference to the lattice,
the position of the f.o.l.s. remains fixed at this site for some time,
i.e., after its creation, an f.o.l.s.  remains the f.o.l.s. for some
time. During this time, however, the number of  particles on and
behind the f.o.l.s. continues to grow, and as a consequence, the
chance of one of them making a hop on to the right of the present
f.o.l.s. also increases. At some instant, a particle from behind hops
over to the right of the present f.o.l.s.: as a result of this hop,
the position of the f.o.l.s. advances, or, viewed from another angle,
a new f.o.l.s. is created on the right of the present one.

\begin{figure}[ht]
\begin{center}
\includegraphics[width=0.75\linewidth]{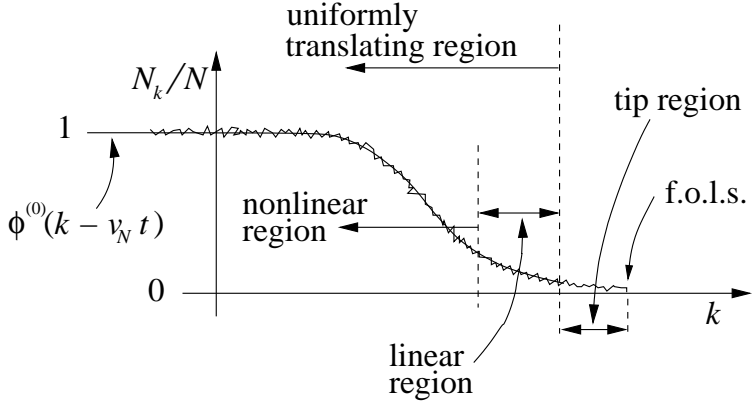}
\end{center}
\caption{A snapshot of a realization of a fluctuating front (jaggered
curve) at large times, and how a theorist might picture such a front
(smooth curve). In this picture, as the  smooth curve shows, a
uniformly translating solution travelling with speed $v_N$, given by
$\phi_k(t)=\phi^{(0)}(k-v_Nt)$ and obeying the deterministic
Eq. (\ref{e14}). However, the uniformly translating profile is
suitable everywhere but few lattice sites at the tip of the front
leading up to the f.o.l.s. The uniformly translating region is further
subdivided into two parts --- in the ``linear region'', the nonlinear
term $\displaystyle{\left[\phi^{(0)}\right]^2}$ can be neglected. In
the ``nonlinear region'', however, all the terms of Eq. (\ref{e5})
have to be taken into account.\label{haltgotheory}}
\end{figure}
Having taken this into consideration, the following scenario was
proposed in Ref. \cite{deb11,deb12}: the lattice and finite particle
effects lead to a halt-and-go dynamics at the f.o.l.s., while
reasonably far behind, the average front ``crosses over'' to a
uniformly translating (conditionally averaged) solution (see
Fig. \ref{haltgotheory}). In this formulation, the effect of
stochasticity on the asymptotic front speed is coded in the
probability distribution of the times required for the advancement of
the f.o.l.s. The goal therein has been to develop a separate
probabilistic theory for the particle hops to create the new f.o.l.s.,
and then to demonstrate by matching the front solution on the
f.o.l.s. with the more standard one (of growth and uniform
translation) reasonably far behind it in a mean-field type
approximation at the ``tip region'' of the front,\footnote{The tip
region of the front spans a few lattice sites right behind the
f.o.l.s. and also includes the f.o.l.s. (see
Fig. \ref{haltgotheory}).\label{fn2.7}} one obtains a self-consistent
description of the stochastic and discreteness effects on front
propagation. Although this formalism is not predictive, it allows one
to deal with much smaller values of $N$ than it is required for the
$\ln^{-2}N$ asymptotics to be applicable. Moreover, it has the right
asymptotics --- it shows that as the value of $N$ is increased, one
does approach to a uniformly translating front profile all the way to
one lattice site behind the f.o.l.s., justifying the validity of the
use of the conditionally averaged uniformly translating profile {\it
all over the front region\/}, used in
Refs. \cite{bd1,deb11,deb12,vanzonarticles1,vanzonarticles2,vanzonarticles3}
for the large-$N$ asymptotics (\ref{e7}). Below we provide few
mathematical details of this procedure and present the key results,
while the full details can be found in Ref. \cite{deb11,deb12}. We
note here that the formalism therein was developed for the
reaction-diffusion system X$\leftrightharpoons2$X, but in principle it
can be worked out for any other microscopic model of fluctuating
``pulled'' fronts.

The difficulty in prediction of $v_N$ in this formalism is the
following. Microscopically, the selection process for the length of
the time span between two consecutive f.o.l.s. creations is
stochastic, and having followed the movements of the f.o.l.s. of {\it
one single front realization\/} over a long time at large times, and
thereafter having denoted the $j$ successive values of the duration of
halts of the f.o.l.s. by $\Delta t_1, \Delta t_2,\ldots,\Delta t_j$
($\Delta t_{j'}\geq0\,\forall j'$), one defines the front speed as the
inverse of the long time average of this time span, i.e.,
\begin{eqnarray}
v_N\,=\,\lim_{j\rightarrow\infty}\,j\,\left[\sum_{j'=1}^{j}\Delta
t_{j'}\right]^{-1}\,.
\label{e12}
\end{eqnarray}
Simultaneously, the amount of growth of particle numbers on and behind
the f.o.l.s. itself depends on the time span between two consecutive
f.o.l.s. creations (the longer the time span, the longer the amount of
growth). As a consequence, on average, the selection mechanism for the
length of the time span between two consecutive f.o.l.s. movements,
which determines the asymptotic front speed, is nonlinear.

This inherent nonlinearity makes the prediction of the asymptotic
front speed difficult. One might recall the difficulties associated
with the prediction of pushed fronts due to nonlinear terms in this
context, although the nature of the nonlinearities in these two cases
is {\it completely different\/}. In the case of pushed fronts, the
asymptotic front speed is determined by the mean-field dynamics of the
fronts, and the nonlinearities originate from the {\it nonlinear
growth terms in the partial differential equations\/} that describe
the mean-field dynamics. On the other hand, for fronts consisting of
discrete particles on a discrete lattice, the corresponding mean-field
growth terms in the tip region of the front are {\it
linear\/}. However, since the asymptotic front speed is determined
from the probability distribution of the time span between two
consecutive f.o.l.s. movements, on average, it is the relation between
this probability distribution and the effect of the linear growth
terms that the nonlinearities stem from (and makes it very difficult
to theoretically deal with).

If we now denote the probability that a f.o.l.s. remains the
f.o.l.s. for time $\Delta t$ by ${\cal P}(\Delta t)$, then the
asymptotic front speed of Eq. $(\ref{e12})$ is given by\footnote{In
this form, one starts the clock at zero as soon as a lattice site
attains the status of the f.o.l.s. Thereafter, as soon as the next new
f.o.l.s. is created, the clock reading is reset to zero.\label{fn2.8}}
\begin{eqnarray} 
v_N\,=\,\langle\Delta t\rangle^{-1}\,=\,\left[\,\int_0^\infty d(\Delta
t)\,\,\Delta t\,\,{\cal P}(\Delta t)\,\right]^{-1}\,.
\label{e13}
\end{eqnarray}
As outlined before, the theoretical expression of ${\cal P}(\Delta t)$
is obtained from the crossover of the mean-field type front solution
at the tip region to the uniformly translating front solution
behind. The front solution in the uniformly translating region (of
Fig. \ref{haltgotheory}) is given by $\phi^{(0)}(k-v_N\Delta t)$, such
that
\begin{eqnarray} 
\frac{\partial\phi^{(0)}_k}{\partial (\Delta
t)}\,=\,D\left[\phi^{(0)}_{k+1}\,+\,\phi^{(0)}_{k-1}\,-\,2\phi^{(0)}_k\right]\,+\,\phi^{(0)}_k\,-\,\left[\phi^{(0)}_k\right]^2\,,
\label{e14}
\end{eqnarray}
where $D$ is the rate of diffusion of particles to their nearest
neighbour lattice sites in the reaction-diffusion system
X$\leftrightharpoons2$X. The uniformly translating region can be
further subdivided into two parts, a ``linear region'' and a
``nonlinear region'', where the nonlinear
$\displaystyle{\left[\phi^{(0)}\right]^2}$ term can and cannot be
neglected respectively. On the other hand, in the mean-field type
description of the tip region, the front solution is written as
$\phi^{(0)}(k-v_N\Delta t)+\delta\phi_k(\Delta t)$. The quantities
$\delta\phi_k(\Delta t)$ are expected to be {\it strictly positive\/}
due to the halt-and-go motion of the f.o.l.s. --- during the time the
f.o.l.s. has halted at a given lattice site, the accumulation of
particles flowing in from behind should result in an {\it excess\/} of
``particle density buildup'' over the uniformly translating solution
$\phi^{(0)}(k-v_N\Delta t)$ at the tip region. Finally, in this
formalism, the crossover point from the tip region to the uniformly
translating region is located at the site, where $\delta\phi_k(\Delta
t)/\phi^{(0)}(k-v_N\Delta t)$ remains negligibly small $\forall\Delta
t$.

\begin{figure}[ht]
\begin{center}
\includegraphics[width=0.9\linewidth]{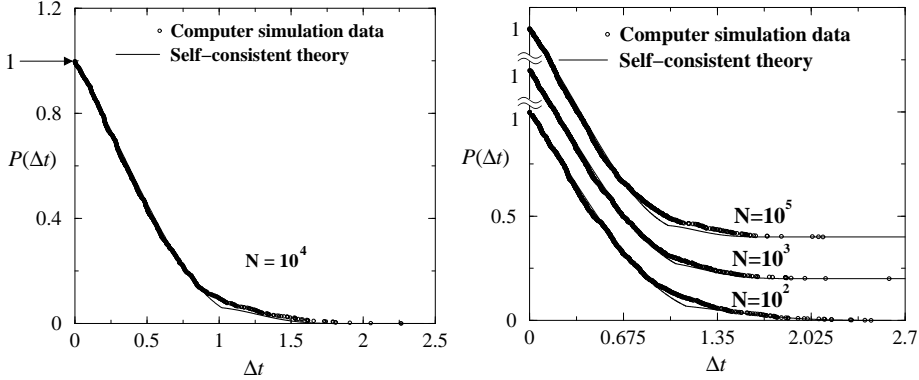}
\end{center}
\caption{The agreement between the self-consistent theoretical curve
of $P(\Delta t)$ {\protect\cite{deb11,deb12}} and computer simulation
results.\label{2345}}
\end{figure}
It is clear that to obtain the expression of ${\cal P}(\Delta t)$, one
needs the expressions of $\phi^{(0)}(k-v_N\Delta t)$ and
$\delta\phi_k(\Delta t)$. On the other hand, the expression of $v_N$
itself is needed to solve for $\phi^{(0)}(k-v_N\Delta t)$, and $v_N$
can be determined only from ${\cal P}(\Delta t)$ as Eq. (\ref{e9})
shows. This indicates that the only way to obtain the expression of
${\cal P}(\Delta t)$ is to solve a whole system of equations
self-consistently \cite{deb11,deb12}. We also note here that in this
self-consistent formalism, there is an effective parameter.

\begin{figure}[ht]
\begin{center}
\includegraphics[width=\linewidth,angle=0]{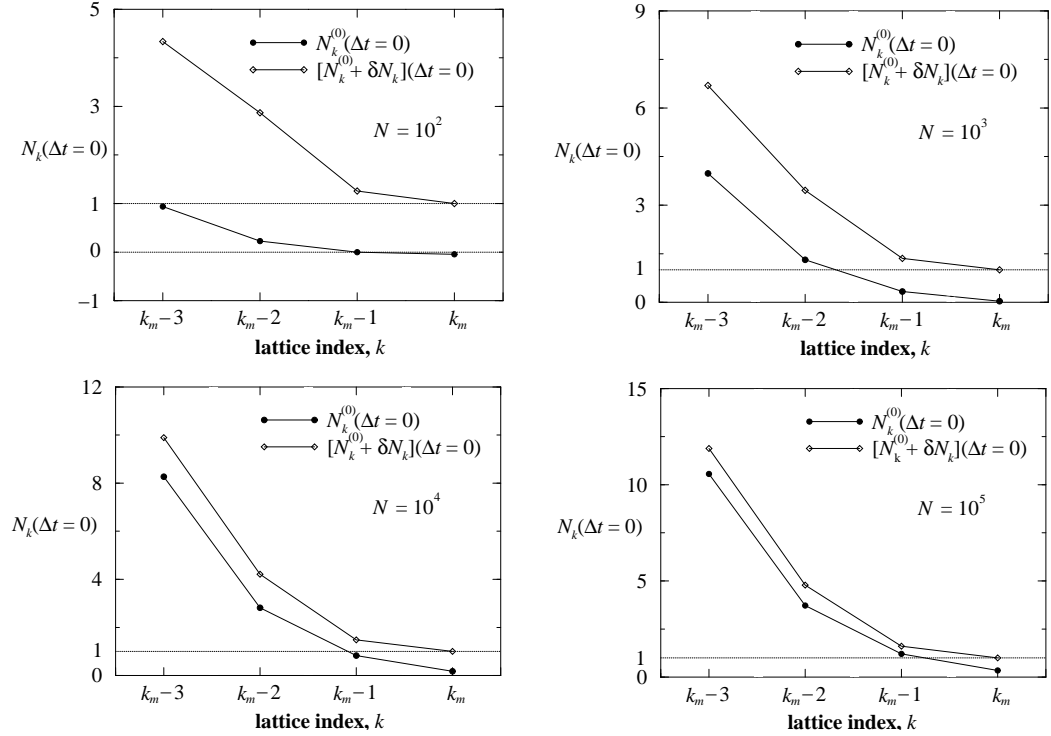}
\end{center}
\caption{Comparison between the $\delta N_k(\Delta
t=0)=N\delta\phi(\Delta t=0)$ and the $N^{(0)}_k(\Delta
t=0)=N\phi^{(0)}(k)$ values obtained from the mean-field type theory
of \protect\cite{deb11,deb12} for four foremost occupied lattice sites
(the f.o.l.s. is indexed by $k_m$) and for $N=10^2$, $10^3$, $10^4$
and $10^5$. Note that as $N$ increases, the $\delta N_k(\Delta t=0)$
values compared to the $N^{(0)}_k(\Delta t=0)$ profile become less and
less important.\label{comb2345}}
\end{figure}
The corresponding self-consistent set of equations are highly
nonlinear and complicated, but due to the presence of the effective
parameter in this formalism, the procedure to obtain the ${\cal
P}(\Delta t)$ in this formalism is {\it not predictive\/}. However,
the fact that it generates a probability distribution that agrees so
well with numerical simulations is indicative of the essential
correctness of such a description of a fluctuating ``pulled''
front. The self-consistent theoretical curves of $P(\Delta
t)=\displaystyle{\int_{\Delta t}^\infty dt'\,{\cal P}(t')}$ for $D=1$
and $N=10^4, 10^2, 10^3$ and $10^5$ (in that order) are shown in
Fig. \ref{2345}. The corresponding numerical comparison of front
speeds are shown in Table I. First, we notice that in the graphs of
Fig. \ref{2345}, the theoretical curves lie below the simulation
histograms at $\Delta t\simeq2/v_N$ --- this is an artifact of the
matching that one has to carry out for the expressions of $P(\Delta
t)$ below and above $\Delta t\simeq2/v_N$. This difference occurs due
to certain fluctuation and correlation effects
\cite{deb11,deb12}. Secondly, the agreement between the simulation
data and the self-consistent formalism is not very good for $N=10^5$
--- at this value of $N$, the simulation gets very slow and one has to
continuously remove particles from the stable phase of the front to
gain program speed, which affects the $P(\Delta t)$ histograms
significantly for large $\Delta t$ values.
\begin{center}
\begin{tabular}[ht]{||@{}c|c@{}|c@{}|c@{}||}
\hline\hline
\rule[-2mm]{0pt}{4ex}\,\,\,\,$N$\,\,\,\,&\,\,\,\,$v_N\mbox{(simulation)}$
\,\,\,\,&\,\,\,\,$v_N\mbox{(theoretical)}$\,\,\,\,&\,\,\,\,$v_N\mbox{[Eq.
(\ref{e7})]}$\,\,\,\,\\  \hline
\rule[-2mm]{0pt}{4ex}$10^2$&\,\,\,\,$1.778$\,\,\,\,&
\,\,\,\,$1.808$\,\,\,\,&$1.465$\\ \hline
\rule[-2mm]{0pt}{4ex}$10^3$&\,\,\,\,$1.901$\,\,\,\,&
\,\,\,\,$1.899$\,\,\,\,&$1.803$\\ \hline
\rule[-2mm]{0pt}{4ex}$10^4$&\,\,\,\,$1.964$\,\,\,\,&
\,\,\,\,$1.988$\,\,\,\,&$1.925$\\ \hline
\rule[-2mm]{0pt}{4ex}$10^5$&\,\,\,\,$2.001$\,\,\,\,&
\,\,\,\,$2.057$\,\,\,\,&$1.976$\\ \hline\hline
\end{tabular}
\end{center}
{\footnotesize Table II: Comparison of $v_N$ values: simulation,
self-consistent formalism \cite{deb11,deb12} [denoted by
$v_N$(theoretical)], and that of the large $N$ asymptotics (\ref{e7}).}
\begin{figure}[ht]
\begin{center}
\includegraphics[width=0.6\linewidth]{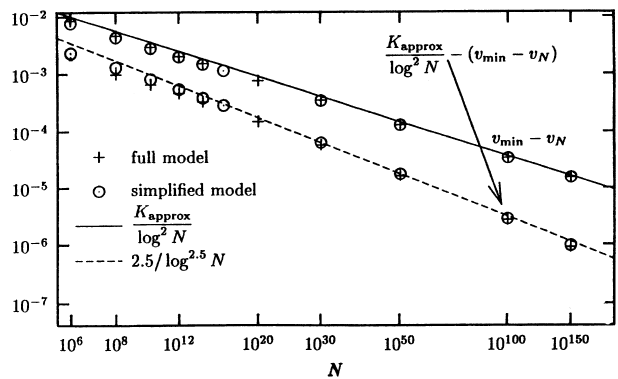}
\end{center}
\caption{Comparison between the actual stochastic model and the
corresponding ``simplified model'' relating to directed polymers in
random media \protect{\cite{bd3}}. For the large $N$ asymptotics of
$v_N$ and the leading order $N$-dependent behaviour of the subdominant
corrections to it, note the agreement between the results of the two
models. Here, $v_{min}$ is $v^*$, and $K_{approx}$ is the coefficient
of $1/\ln^2N$ in Eq. (\ref{e7}). The figure has been modified from the
original for greater clarity. \label{bdfig2}}
\end{figure}

In addition to good agreements between the self-consistent formalism
and simulations for the $P(\Delta t)$ curves in Fig. \ref{2345}, a
significant observation of the formalism is the following: as the
value of $N$ is increased in the self-consistent formalism, it is
found that the $\delta\phi_k/\phi^{(0)}_k$ values at the tip region of
the front gradually reduces \cite{deb11,deb12} (see
Fig. \ref{comb2345}). The stochastic halt-and-go character of the
movement of the f.o.l.s., which is usually occupied by ${\cal O}(1)$
number of particles, however, continues to remain valid for any value
of $N$. This implies that for very large $N$, one approaches the
picture of a fluctuating ``pulled'' front, where a uniformly
translating mean-field description (\ref{e5}) holds all the way up to
one lattice site behind the f.o.l.s., while only the dynamics of the
f.o.l.s. remains a stochastic halt-and-go process. Such a scenario was
numerically investigated in Ref. \cite{bd3} by means of constructing a
``simplified model'', in which stochasticity was introduced only on
the f.o.l.s., while the rest of the front was described by a
deterministic equation. A comparison of the simulation data of the
actual stochastic model (the ``full model'') and the simplified model
from Ref. \cite{bd3} is shown in Fig. \ref{bdfig2}. The agreement
between the two cases --- not only at the level of large $N$
asymptotics (\ref{e7}), but also the leading order $N$-dependent
behaviour of the subdominant corrections (c.f. the last paragraph of
Sec. \ref{sec2.2}) --- suggests that the ``simplified model'' is a
very good representation of the true stochastic model at
asymptotically large $N$ values.\footnote{Such a statement goes beyond
the consideration of simply the front speed, as the front diffusion
coefficients for the actual stochastic model and the ``simplified
model'' show no difference for asymptotically large $N$ values. We
will address the issue of front diffusion later in
Sec. \ref{sec2.5}.\label{fn2.9}}

\subsection{The weakly pushed nature of fluctuating ``pulled''
fronts\label{sec2.4}}

\subsubsection{(In)dependence of the front shape and speed on the
initial configuration\label{sec2.4.1}}

One common aspect of propagating fronts in the microscopic models of
fluctuating fronts is that at a given value of $\Delta t$, the more
particles there are on the f.o.l.s., the more likely it is for the
f.o.l.s. to move forward. This indicates that if there are more
particles put on the f.o.l.s. in these models at short values of
$\Delta t$, the faster the f.o.l.s. moves.

In reaction-diffusion models, the number of particles on the
f.o.l.s. can increase by means of two processes: flow of particles
from the other lattice sites in the tip region behind the f.o.l.s.,
and growth of particles on the f.o.l.s. by themselves without having
any input of particles from other lattice sites. In the so-called
clock model
\cite{vanzon,vanzonarticles1,vanzonarticles2,vanzonarticles3,vanzon1}
or the discrete particle and lattice model related to directed
polymers in random media \cite{ds1,cook,derrida1,bd1,bd2,bd3}, the
number of particles on the f.o.l.s. can increase only by means of flow
of particles from the other lattice sites behind. For the two latter
models, when a conditionally averaged steady shape of the front
emerges at long times, a complicated balance develops between
$\langle\Delta t\rangle=v_N^{-1}$ (i.e., on average for how long the
f.o.l.s. halts at a given lattice site) and the average flow of
particles on to the f.o.l.s. from behind, and the balance is
established at $v_N<v^*$. Such a balance also develops for the
reaction-diffusion models, but it is more complicated in nature ---
the particles that flow on to the f.o.l.s. can start their own growth,
thereby further contributing to the increasing number of particles on
the f.o.l.s.

The observation above demonstrates the pushed\footnote{The above
argument is not quantitative, but we have seen earlier that the
asymptotic front speed $v_N$ differs from $v^*$ by only a small amount
for large values of $N$. This is why we call them weakly pushed. We
will see later that such a nomenclature is also supported by the
spectrum of ${\cal L}_{v_N}$.\label{fn2.10}} nature of these fronts,
as their speed of propagation depends on the front solution far behind
the f.o.l.s. The front solution far behind the f.o.l.s. contributes to
the movements of the f.o.l.s. in the following way: a uniformly
translating solution exists behind the tip region (see
Fig. \ref{haltgotheory}), and for the time the f.o.l.s. halts at a
given lattice site, the uniformly translating region of the front
brings in particles into the tip region --- ultimately that
contributes to how fast the f.o.l.s. moves forward. The supply of
particles from behind is therefore a very crucial ingredient for the
front (along with the f.o.l.s.) to propagate at a speed $v_N$,
indicating that if the supply of particles on to the f.o.l.s. from
behind were cut off, the speed of the f.o.l.s. would reduce
drastically (this is already illustrated by the discussion of the
first two paragraphs of Sec. \ref{sec2.2}, where the ``inertia'' of
the f.o.l.s. against its forward movement due to the discreteness of
particles and the lattice effects are mimicked by a cutoff). It also
indicates that if the number of particles on the f.o.l.s. are
increased artificially in these models at small values of $\Delta t$,
the front should propagate faster. Such a model was investigated in
Ref. \cite{deb2} in a reaction-diffusion system with enhanced growth
rate of particles in the tip region, resulting in a front  that
propagates with a speed significantly {\it higher\/} than
$v^*$.\footnote{As readers can find out, Ref. \cite{deb2} studies a
continuum space Fisher-Kolmogorov equation type model, where the
cutoff in the growth function at $\phi=\varepsilon$ is supplemented by
an enhancement in growth between $\varepsilon<\phi\leq\varepsilon/r$
for $r<1$. Of course, as $\varepsilon\rightarrow0$, the evolution
equation of the front reaches its pulled front limit
(Fisher-Kolmogorov equation). The analysis however shows that first
taking the $\varepsilon\rightarrow0$ limit to obtain the
Fisher-Kolmogorov equation and then obtaining the front speed ($v^*$)
is not right --- one has to first calculate the front speed as a
function of $\varepsilon$ and $r$ and only then take the limit
$\varepsilon\rightarrow0$. Interestingly enough, it turns out that
when $r\leq r_c=0.283833\ldots$, the front speed converges to $v^*$,
otherwise the front speed can be significantly higher than
$v^*$.\label{limits}}

On the basis of the above two paragraphs, the independence of the
front speed and the conditionally averaged front shape on the initial
configuration can be argued in the following manner, although no
formal proof exists in the literature.\footnote{As we have seen in
Sec. \ref{sec2.3}, the balance between how long the f.o.l.s. halts at
a given lattice site and the average flow of particles on to the
f.o.l.s. from behind is very complicated. In view of this, it is
unlikely that a formal proof can be obtained.\label{fn2.11}} When a
discrete particle and lattice system of a fluctuating ``pulled'' front
has not reached its conditionally averaged steady shape, the balance
between the average value of $\Delta t$ and the average flow of
particles on to the f.o.l.s. from behind does not hold. For example,
consider an initial microscopic configuration, when the front shape at
the leading edge resembles that of the functional form $\exp[-\lambda
k]$, with $\lambda=\lambda_>>\lambda^*$. With such a configuration,
``linear region'' initially propagates with a speed
$v(\lambda_>)>v^*$, and bring in more particles on to the tip region
(and eventually on the f.o.l.s.) than what it does at the steady
state, speeding up the f.o.l.s. Subsequently, the ``inertia'' of the
f.o.l.s. also gets communicated to the uniformly translating front
solution behind, which then starts to slow down. A combination of
these two phenomena makes the leading edge less and less steep, until
the front finally settles to its steady shape (an illustration of it
can be found in Fig. 1 of Ref. \cite{breuer}). Contrast this situation
with an initial microscopic configuration that resembles the
functional form $\exp[-\lambda k]$, with
$\lambda=\lambda_<<\lambda^*$, such that
$v(\lambda_<)=v(\lambda_>)$. In this case, the ``linear region''
initially propagates with the same speed as before, but since it is
less steep than $\exp[-\lambda_>k]$, it does not bring in as many
particles on to the f.o.l.s. as it does for $\lambda=\lambda_>$, and
as a result, the f.o.l.s. does not speed up as much as it did
before. By means of a simple comparison of couple of diagrams
contrasting these two cases, one can easily convince oneself that the
front profile at the leading edge in this case continues to becomes
steeper and steeper. Although there does not exist any specific
simulation data showing that for arbitrary initial configuration, the
front ultimately settles down to the same conditionally averaged front
shape (and speed), the arguments above strongly suggest that it is
indeed the case.

\subsubsection{The Large $N$ Asymptotic Spectrum of ${\cal
L}_{v_N}$\label{sec2.4.2}}

To obtain the large $N$ asymptotic spectrum of ${\cal L}_{v_N}$, we
follow the procedure developed in Refs. \cite{kns,deb3}, which in turn
use the standard route of transforming the linear eigenvalue problem
into a Schr\"odinger eigenvalue problem
\cite{bj,ebert,oono1,oono2,oono3}. To represent a discrete particle
and lattice system of a fluctuating ``pulled'' front'' as
$N\rightarrow\infty$, the idea is to consider the cutoff picture
(\ref{e8}) in continuum space, developed by Brunet and Derrida.

In the comoving co-ordinate $\xi=x-v_N t$, Eq. (\ref{e8}) in the
steady state reads
\begin{eqnarray}
-\,v_N\,\frac{d\phi^{(0)}(\xi)
 }{d\xi}\,=\,\frac{d^2\phi^{(0)}(\xi)}{d\xi^2}\,+\,f[\phi^{(0)}(\xi)]\,,
\label{e15}
\end{eqnarray}
with
$f[\phi^{(0)}(\xi)]=\left\{\phi^{(0)}(\xi)-\left[\phi^{(0)}(\xi)\right]^n\right\}\Theta[\phi^{(0)}(\xi)-1/N]$.
Now consider a function $\phi(x,t)$ infinitesimally different from
$\phi^{(0)}(\xi)\equiv\phi^{(0)}(x-v_N t)$ in the comoving frame,
i.e., $\phi(x,t)=\phi^{(0)}(x-v_N t)+\eta(\xi,t)$. Upon linearizing
the dynamical equation in the comoving frame, one finds that the
function $\eta(x,t)\equiv\eta(\xi,t)$ obeys the following equation:
\begin{eqnarray}
\frac{\partial\eta}{\partial t}\,=\,{\cal
L}_{v_N}\,\eta\,=\,\left[v_N\,\frac{\partial}{\partial\xi}\,+\,\frac{\partial^2}{\partial\xi^2}\,+\,\frac{\delta
f(\phi)}{\delta\phi}\bigg|_{\phi\,=\,\phi^{(0)}}\right]\,\eta\,.
\label{e16}
\end{eqnarray}
In this formalism, the substitution
$\eta(\xi,t)=e^{-Et}e^{-v_N\xi/2}\psi_E(\xi)$ converts Eq. (\ref{e16})
to the following one-dimensional Schr\"odinger equation for a particle
in a  potential (with $\hbar^2/2m=1$)
\cite{bj,ebert,oono1,oono2,oono3}:
\begin{eqnarray}
\left[\,-\,\frac{d^2}{d\xi^2}\,+\,\frac{v^2_N}{4}\,-\,\frac{\delta
f(\phi)}{\delta\phi}\bigg|_{\phi\,=\,\phi^{(0)}}\right]\psi_E(\xi)\,=\,E\,\psi_E(\xi).
\label{e17}
\end{eqnarray}
In Eq. (\ref{e17}), the quantity
$\displaystyle{V(\xi)=\left[\,\frac{v^2_N}{4}\,-\,\frac{\delta
f(\phi)}{\delta\phi}\bigg|_{\phi\,=\,\phi^{(0)}}\right]}$ plays the
role of the potential. Having denoted the the coordinate of the point
where $\phi^{(0)}(\xi)=1/N$ by $\xi_0$ as before, we have
\begin{eqnarray}
V(\xi)\!=\!\left[\frac{v^2_N}{4}-\!1\!+\!n\left[\phi^{(0)}\right]^{n-1}\right]\!\Theta(\xi_0-\xi)-\!\frac{1}{v_N}\delta(\xi-\xi_0)\!+\!\frac{v^2_N}{4}\Theta(\xi-\xi_0).
\label{e18}
\end{eqnarray}

\begin{figure}[ht]
\begin{center}
\includegraphics[width=0.495\linewidth]{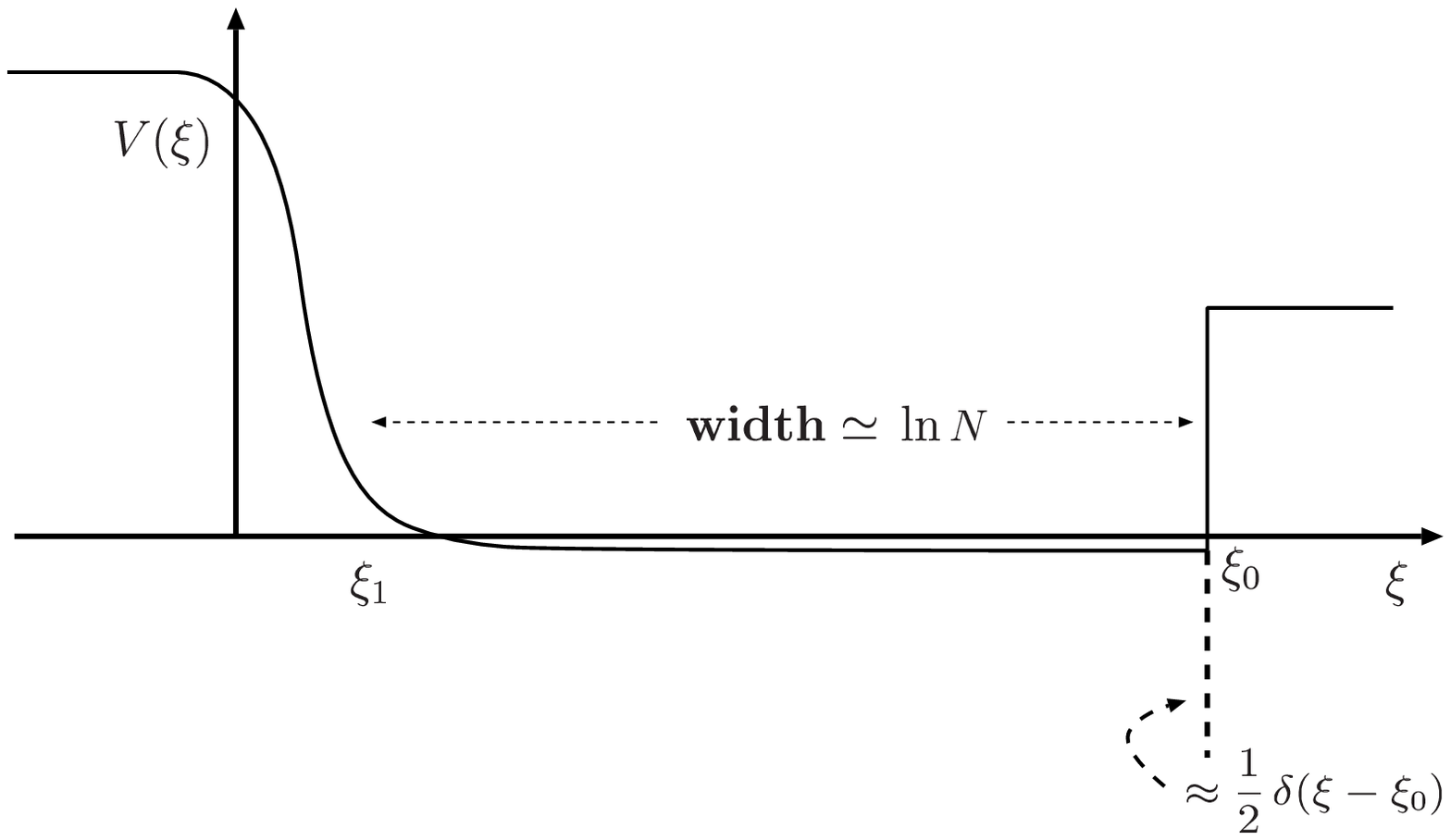}\hspace{0.1cm}\includegraphics[width=0.495\linewidth]{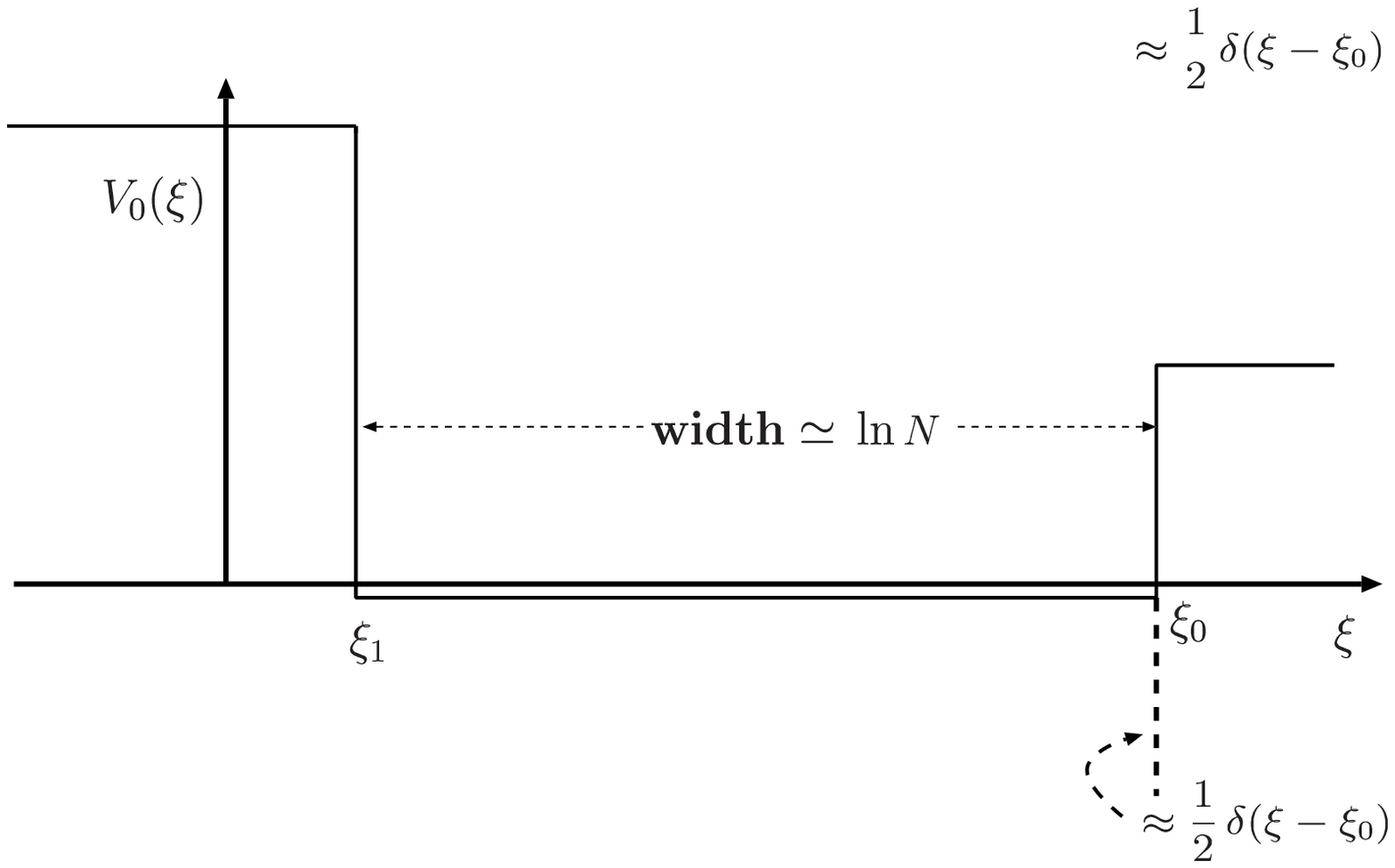}
\end{center}
\caption{Left: The potential $V(\xi)$ in the Schr\"odinger operator
obtained in the stability analysis. Right: The approximate potential
$V_0(\xi)$ that can be used for calculating the low-lying modes for
large widths of the bottom well, i.e., for large
$N$. \label{simpschro}}
\end{figure}

The form of the potential $V(\xi)$ is sketched in the left figure of
Fig. \ref{simpschro}. Notice that $\phi^{(0)}(\xi)$ is a monotonically
increasing function from $1/N$ at $\xi_0$ {\it towards the left\/},
asymptotically reaching the value $1$ as $\xi\rightarrow-\infty$. As a
result, for $\xi<\xi_0$, $V(\xi)$ also increases monotonically towards
the left, from $v^2_{N}/4 -1+nN^{1-n}\simeq -\pi^2/\ln^2N$ at
$\xi=\xi_{0-}$, to $(n-\pi^2/\ln^2N)\approx n$ as
$\xi\rightarrow-\infty$. At $\xi_0$, there is an attractive
$\delta$-function potential of strength $1/v_{N}\approx 1/2$ and a
finite step of height 1. The crucial feature for the stability
analysis below is the fact that $V(\xi)$ stays remarkably flat at a
value $-\pi^2/\ln^2N$ over a distance $(\xi_0-\xi_1)\simeq\ln N$, and
then on the left of $\xi_1$, it increases to the value $\approx n$,
over a distance of order unity. This is a consequence of the nature of
the solution (\ref{e10}).

The question of stability of the front solution can be answered by
studying the spectrum of the temporal eigenvalues. If there are
negative eigenvalues of the above Schr\"odinger equation, then
according to Eq. (\ref{e16}) $\eta(\xi,t)$  grows in time in the
comoving frame, i.e.,  the front solution $\phi^{(0)}(\xi)$ is
unstable. On the other hand, if there are no negative eigenvalues,
then the asymptotic front shape is stable, and the spectrum of the
eigenvalues then determines the nature of the relaxation of
$\phi(x,t)$ to the solution $\phi^{(0)}(\xi)$. The full spectrum in
general depends on the boundary conditions imposed on the
eigenfunctions $\psi_E$. Here we consider only localized
perturbations, for which we need to have $\eta(\xi,t)\to 0$ as $\xi
\to \pm \infty$. Because of the exponential factor in Eq. (\ref{e16}),
any eigenfunction $\psi_E$ which vanishes as $\xi \to \infty$ is
consistent with vanishing $\eta$ towards the right\footnote{The fact
that eigenfunctions $\psi_E$ which diverge as $\xi\to \infty$ are
allowed, means that there are admissible eigenfunctions which are not
in the Hilbert space of the Schr\"odinger operator. See e.g.,
Ref. \cite{ebert} for further discussion of this
point.\label{fn2.12}}. However, for $\xi\to -\infty$, the
eigenfunctions $\psi_E$ need to vanish exponentially fast with a
sufficiently large exponent, so that when it is combined with the
exponentially diverging term $e^{-v_N/2}$, they are still consistent
with the requirement that $\eta $ vanishes for $\xi\to-\infty$. For
the lowest ``energy'' eigenvalues investigated below, we demonstrate
that these requirements are obeyed.

From the form in the potential, it is clear that the lowest ``energy''
eigenmodes, i.e., the slowest relaxation eigenmodes, are the ones that
are confined to the bottom of the potential. We notice that among
these modes, invariably there is a zero mode of the stability operator
that is associated with the uniformly translating front solution of a
dynamical equation like Eq. (\ref{e8}): since  $\phi^{(0)}(\xi)$ and
$\phi^{(0)}(\xi+a)$ are solutions of Eq. (\ref{e14}) for any arbitrary
$a$, we find by expanding to first order in $a$ that
$\psi_0(\xi)=e^{v_N\xi/2}\displaystyle{\frac{d\phi^{(0)}}{d\xi}}$ is a
solution of Eq. (\ref{e17}) with eigenvalue $E=0$.\footnote{We will
later see in Sec. \ref{sec3.1.2} that
$\displaystyle{\frac{d\phi^{(0)}}{d\xi}}$ is also known as the
so-called Goldstone mode
\cite{schlogl1,schlogl2,pasquale}.\label{goldstone}} From the result
(\ref{e19}) for the asymptotic front solution, we then immediately
get, {\em to dominant order},
\begin{eqnarray}
\psi_0 \sim \sin[\pi\lambda^*(\xi-\xi_1)/\ln
N]\quad\quad0\simeq\xi_1\lesssim\xi\leq\xi_0.
\label{e19}
\end{eqnarray}
Furthermore, since $\phi^{(0)}(\xi)$ is a monotonically decreasing
function of $\xi$, the solution
$\psi_0(\xi)=e^{v_N\xi/2}\displaystyle{\frac{d\phi^{(0)}}{d\xi}}$ is
nodeless. Since we know from elementary quantum mechanics that the
nodeless eigenfunction has the lowest eigenvalue, this implies that
all the {\it other\/} eigenvalues of Eq. (\ref{e17}) are positive,
i.e., the solution $\phi^{(0)}(\xi)$ is stable.

The spectrum of eigenvalues of Eq. (\ref{e17}) for $E>0$ determines
the decay property of localized perturbations $\eta(\xi,t)$ in
time. We notice that for $E>v_N^2/4 \approx 1$ (the value of the
potential on the far right), the spectrum of eigenvalues are
continuous. However, we are particularly interested in the smallest
eigenvalues $E_m>0$ for small $m$, since these are  the eigenmodes
that decay the slowest in time. These are the eigenvalues associated
with bound states in the potential well. As $N\rightarrow\infty$, the
bottom well of the potential becomes very wide: its width diverges as
$\ln N$. As we know from elementary quantum mechanics, the lowest
``energy'' eigenfunctions then become essentially sine or cosine waves
in the potential well with small wave numbers $k$ and correspondingly
small ``energy'' eigenvalues. Based on the fact that the potential
$V(\xi)$ on the left rises over length scales of order unity, we now
make an approximation. In the limit that the bottom well is very wide
and the $k$ values of the bound state eigenmodes very small, it
becomes an increasingly good and an asymptotically correct
approximation to view the left wall of the well as a steep step,
sketched in the right figure of Fig. \ref{simpschro} --- one can
therefore approximate the potential by
\begin{eqnarray}
V_0(\xi)=n\left[1-\Theta(\xi)\right]-\frac{\pi^2}{\ln^2N}\Theta(\xi)\left[1-\Theta(\xi-\xi_0)\right]\nonumber\\&&\hspace{-3cm}-\frac{1}{2}\delta(\xi-\xi_0)+\Theta(\xi-\xi_0).
\label{e20}
\end{eqnarray}

On the right hand side, there is an attractive delta-function
potential at the point where the potential shows a step to a value
close to 1. It is easy to check that the prefactor of the
delta-function of 1/2 is not strong enough to give rise to bound
states with $E<0$, and as a result, for very large values of $N$, the
low-lying eigenmodes  approach  sine waves with nodes at the position
of the walls of the potential,\footnote{More explicitly, if we write
the solution within the well as $A\sin [q (\xi-\xi_1) +B]$, and for
$\xi > \xi_0$ as $A_+ e^{-(\xi-\xi_0)}$ (which is correct to lowest
order in $\epsilon$), then we get from the boundary conditions at
$\xi_0$: $2q\,\mbox{cotg} [q(\xi_0-\xi_1)+B] = -1$; likewise, at the
left boundary, of the well, we get $q\, \mbox{cotg}\,B = const.$,
where the constant is determined by the size of potential step. For
$\xi_0-\xi_1\gg1$, the small-$q$ solutions are those with $B \to
0$.\label{fn2.13}} $\psi_m \simeq \sin\left[ q_m (\xi-\xi_1)\right]$.
The condition that these solutions have nodes at the right edge of the
well then yields
\begin{eqnarray}
q_m \simeq \frac{(m+1) \pi}{\xi_0-\xi_1} \simeq \frac{(m+1)\pi}{\ln N},
\label{e21}
\end{eqnarray}
implying that the corresponding eigenvalues are given
by\footnote{Notice that for the eigenvalues in Eq. (\ref{e22}), the
corresponding eigenmodes $\psi_E(\xi)$ decay as
$\exp[-\sqrt{n}\,\,|\xi|]$ for $\xi\rightarrow-\infty$ and as
$\exp[-v_N\xi/2]$ for $\xi\rightarrow\infty$, which make
$e^{v_N\xi/2}\psi_E(\xi)$ to go to zero for $\xi\rightarrow\pm\infty$,
satisfying the boundary conditions discussed previously.\label{fn2.14}}
\begin{eqnarray}
E_m\,\simeq \,\frac{\left[
(m+1)^2\,-\,1\right]\,\pi^2}{\ln^2N}\,\quad\quad m\,=\,0,1, 2,\ldots
\label{e22}
\end{eqnarray}
Here, the first term between square brackets comes from the ``kinetic
energy'' term $q^2$, while the second term originates from  the value
of the potential at the bottom. Note that for $m=0$, the eigenmode
$\sin q_0$ with eigenvalue $E_0$ is indeed the same as the zero
eigenmode of Eq. (\ref{e19}) with $q_0=\pi\lambda^*/\ln N$, which we
calculated from the shape of the front solution $\phi^{(0)}$ in the
leading edge. Besides verifying the consistency of our approach, this
also confirms that there are no corrections to Eq. (\ref{e22}) for
$m=0$: for $m=0$, it will yield an eigenvalue zero to all orders in
$N$. Therefore, the smallest nonzero eigenvalue, which governs the
relaxation of the front velocity and profile to the asymptotic ones is
$E_1$ with relaxation time $\tau_1$ given by
\begin{eqnarray}
\tau_1^{-1} = E_1 \simeq \,\frac{3 \,\pi^2}{\ln^2N}\,.
\label{e23}
\end{eqnarray}
\begin{figure}[ht]
\begin{center}
\includegraphics[width=0.42\linewidth]{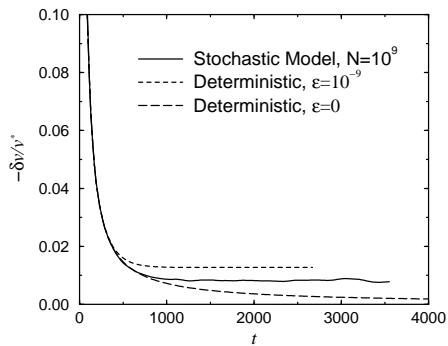}
\end{center}
\caption{The time convergence of the asymptotic front speeds: for
Eq. (\ref{e8}) [short dashed line], and for the stochastic
reaction-diffusion system X$+$Y$\rightarrow2$X (solid line)
\protect{\cite{kns}}. They follow corresponding curve for the
deterministic equation (\ref{e1}) with $n=2$ (long dashed line) for a
while, and then they suddenly break off to settle down to their
asymptotic values. As the long dashed line represents the power law
convergence $t^{-1}$ for pulled fronts, such a behaviour for the solid
and the short dashed curve is expected for a very slow exponential
decay of Eq. (\ref{e23}). Note that the notations have been changed
from the original to suit the present article. Here $\delta
v=v^*-v$. \label{kesslerfig2}}
\end{figure}
The time convergence of the asymptotic front speed for the
reaction-diffusion system X$+$Y$\rightarrow2$X is shown in the
simulation data from Ref. \cite{kns} (Fig. \ref{kesslerfig2}), in
support of Eq. (\ref{e23}). As can be seen therein, as a function of
time, the front speed for Eq. (\ref{e8}) [short dashed line] and for
the stochastic reaction-diffusion system X$+$Y$\rightarrow2$X (solid
line) follow corresponding curve for the deterministic equation
(\ref{e1}) with $n=2$ (long dashed line) for a while, and then they
suddenly break off to settle down to their asymptotic values. As the
long dashed line represents the power law convergence $t^{-1}$ for
pulled fronts, such a behaviour for the solid and the short dashed
curve is expected for a very slow exponential decay of Eq. (\ref{e23}).

We have noted earlier in the introduction that for a pushed front, the
lowest lying eigenspectrum is gapped. For discrete particle and
lattice systems of a fluctuating ``pulled'' front with
$N\rightarrow\infty$, according to the above analysis, the spectrum is
also gapped, but the gap is very small, as it is
$\propto\ln^{-2}N$. This is another confirmation of these fronts being
weakly pushed. Moreover, Eq. (\ref{e22}) also confirms that as
$N\rightarrow\infty$, the gap between the spectral lines decreases as
$\ln^{-2}N$, which is consistent with the fact that for a pulled front
$N=\infty$ and the spectrum becomes gapless (see footnote \ref{fn10}
in this context).

In connection to the gapless spectrum in the limit
$N\rightarrow\infty$, a curious observation was made in
Ref. \cite{kns}. Put simply, the observation is the following: in the
time convergence of the front speed to its asymptotic value $v^*$, the
leading edge increases in length $\sim\sqrt{t}$ leading to the uniform
$t^{-1}$ convergence (\ref{e4}). For a discrete particle and lattice
system of fluctuating ``pulled'' fronts with $N\rightarrow\infty$, the
length of the leading edge in Eq. (\ref{e10}) increases as $\ln N$,
which ultimately leads to the $1/\ln^2N$ deviation of the asymptotic
front speed $v_N$ from $v^*$ (although in this similarity the
prefactors do not match correctly). Based on these observation,
Kessler and co-workers suggested \cite{kns} that the initial
conditions, which are responsible for the front speed selection of a
pulled front, act as an {\it effective time-dependent cutoff\/} beyond
which the steady state equation no longer holds, and the
time-dependent cutoff goes to zero with increasing time\footnote{A
similar observation was also made therein about pushed fronts with
cutoffs.\label{fn2.15}}. To what extent there is a deep physics
underlying this observation, however, is not clear.

\subsection{The Large $N$ Asymptotic Scaling of Front Diffusion
Coefficient $D_f$ for Fluctuating ``Pulled'' Fronts\label{sec2.5}}

So far, we have only dealt with (and observed) how the discreteness of
particles and the lattice sites affect the asymptotic front speed of
fluctuating ``pulled'' fronts. With the idea behind random wandering
of fronts has been introduced already in Sec. \ref{sec1.2}, we now
address the issue of their diffusive wandering.

It turns out that the theoretical concept of random wandering of
fronts is actually an extremely tricky issue, the full flavour of
which will only be clear in Secs. \ref{sec3} and
\ref{sec4.2}. However, operationally, i.e., in a computer simulation,
the front diffusion for a front consisting of discrete particles on a
lattice is rather straightforward to measure. Consider for example the
fluctuating ``pulled'' front in the so-called clock model
\cite{vanzon,vanzonarticles1,vanzonarticles2,vanzonarticles3}. This
model, which arose in the calculation of the largest Lyapunov exponent
for a gas of hard spheres
\cite{vanzon,vanzonarticles1,vanzonarticles2,vanzonarticles3}, one
considers a set of $N$ clocks with integral clock readings
$k=0,1,2,\ldots$. The number of clocks with a certain reading $k$ at
time $t$ is $n_k(t)$. Any two clocks can ``collide'' continuously in
time; the post-collisional readings of a pair of mutually
``colliding'' clocks with pre-collisional readings $k_i$ and $k_j$ are
{\it both\/} updated to $\mbox{max}(k_i,k_j)+1$. In terms of
$\phi_k(t)=\sum_{k'=k}^{\infty}n_{k'}(t)/N$ (with time $t$ redefined
to have a unit mean single clock collision frequency), the relevant
asymptotic solution corresponding to an initial state in which all
clock readings are finite is a front  propagating into the state
$\phi_k=0$ for large $k$. In the mean-field limit, the dynamics is
that of a pulled front propagating with speed $v^*=4.31107\ldots$
\cite{vanzon,vanzonarticles1,vanzonarticles2,vanzonarticles3}.
\begin{figure}[ht]
\begin{center}
\includegraphics[width=0.6\linewidth]{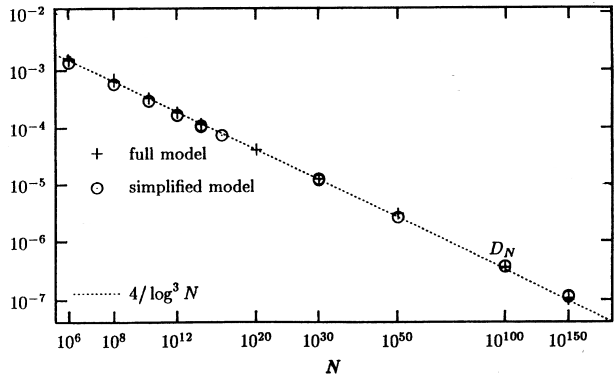}
\end{center}
\caption{Simulation data from Ref. \protect{\cite{bd3}} for the front
diffusion coefficient, for both the microscopic stochastic model
relating to directed polymers in random media (``full'' model) and the
``simplified'' model. In the above figure, $D_N$ should be read as
$D_f$. The figure has been modified from the original for greater
clarity.\label{bdfig3}}
\end{figure}

A straightforward way to measure the front speed and the front
diffusion coefficient in a computer simulations of the clock model is
by tracking the position of the centre of mass of the clock
distribution in individual realizations as a function of time. With
the centre of mass location for the clock distribution in an actual
realization r at time $t$ given by
\begin{eqnarray}
S_{\mbox{\scriptsize r}}(t)\,=\,N^{-1} \sum_k k n_k(t)\equiv\sum_k
\phi_k(t)\,,
\label{srt}
\end{eqnarray} 
one has $v_N=\langle S_{\mbox{\scriptsize r}}(t)\rangle$, and
$D_f=\lim_{T\rightarrow\infty}d\langle[\int_t^{t+T}dt\dot{S}_{\mbox{\scriptsize
r}}(t)-v_NT]^2\rangle/dT$ [the angular brackets denote first an
average over all possible evolution sequence for a given initial
realization (at time $t$), and then a further averaging over the
ensemble of initial realizations]. This was indeed the same method
used by Brunet and Derrida to obtain the front speed and front
diffusion coefficient \cite{bd3} for a model which is slightly
different from the clock model. The superb cleverness in their
measurement of the front diffusion lies in the simulation algorithm
they used --- this algorithm allowed them to take $N$ as high as
$10^{150}$, and they demonstrated that the $N\rightarrow\infty$
asymptotic scaling of front diffusion coefficient $D_f$ behaves as
$1/\ln^3N$ \cite{bd3}. Moreover, a ``simplified model'', where the
fluctuations are {\it randomly\/} generated {\it only\/} on the
instantaneous foremost occupied lattice site
(i.f.o.l.s.),\footnote{The i.f.o.l.s. is simply the location of the
lattice site on the right of which all lattice sites are
instantaneously empty. It is different from the f.o.l.s. of
Sec. \ref{sec2.3}.\label{ifols}} was found to exhibit the {\em same}
$1/ \ln^3N$ asymptotic scaling of $D_f$ as the full stochastic model
\cite{bd3}. However, we note that in the absence of any studies on the
diffusion coefficient of fluctuating ``pulled'' fronts on other
microscopic models, to what extent this scaling is independent of the
model is not known.

For the clock model, we will provide a argument of the $1/\ln^3N$
asymptotic scaling of $D_f$ in Sec. \ref{sec4.2}, while presently we
simply obtain the formal Green-Kubo expression for $D_f$. Needless to
say, dividing the derivation in this manner is complicated for the
readers. Nevertheless, the reason behind postponing the derivation of
the $1/\ln^3N$ asymptotic scaling of $D_f$ till Sec. \ref{sec4.2.2} is
the following: it will require certain inputs and comparisons with the
stochastic differential equation used in the Langevin-type
field-theoretical approaches of fluctuating fronts, which we will
discuss in Secs. \ref{sec3} and \ref{sec4}.

The central theme to obtain $D_f$ is that there is stochasticity in
the front evolution at two levels: first, in any snapshot of the clock
model realization r at time $t$, its shape
$\{n_k(t)\}_{\mbox{\scriptsize r}}$ fluctuates around
$\{n^{(0)}_k(t)\}\equiv\{n^{(0)}(\xi)\}$, where
$n^{(0)}(\xi)=\phi^{(0)}(\xi)-\phi^{(0)}(\xi-1)$. In the subsequent
front evolution, {\it in an average or mean-fieldish sense}, the
clocks are chosen from a ``distorted distribution''
$\{n_k(t)\}_{\mbox{\scriptsize r}}$ rather than from
$\{n^{(0)}_k(t)\}$, and this introduces fluctuations in the front
speed $\delta v_{\mbox{\scriptsize r}}(t)$ [measured by tracking the
centre of mass of the clocks]. Let us denote, by $\delta
v_{\mbox{\scriptsize r,mf}}(t)$, this {\it average\/} fluctuations in
the instantaneous front speed around $v_N$ at time $t$ due to the
shape fluctuation of the front at time $t$. From the actual clocks
that constitute the front, at time $t$, we have
\begin{eqnarray}
v_N\,+\,\delta v_{\mbox{\scriptsize
r,mf}}(t)=\sum_{i,j}\,\,[2-\delta_{k_i,k_j}]\,(|k_i-k_j|+2)(n_{k_i}[n_{k_{j}}-\delta_{k_i,k_j}])/2N^2\,.
\label{clock1}
\end{eqnarray}
Equation (\ref{clock1}) is obtained by applying a probabilistic
argument in the microscopic dynamics of the clock model --- in a
collision between two clocks with readings $k_i$ and $k_j$, the mean
clock reading increases by an amount $(|k_i-k_j|+2)/N$, the
probability of having such a collision is
$[2-\delta_{k_i,k_j}][n_{k_i}(n_{k_{j}}-\delta_{k_i,k_j})]/[N(N-1)]$,
and finally in a unit time, on average, $(N-1)/2$ collisions occur.
The $\delta_{k_i,k_j}$'s arise due to the indistinguishability of the
clocks with the same readings. When we write
$\{n_k(t)\}_{\mbox{\scriptsize r}}=n^{(0)}_k(t)+\{\delta
n_k(t)\}_{\mbox{\scriptsize r}}$, the expression of $v_N$ is obtained
from Eq. (\ref{clock1}) by replacing $\{n_k(t)\}_{\mbox{\scriptsize
r}}$ by $\{n^{(0)}_k(t)\}$ in the first step of Eq. (\ref{clock1}),
implying that at the leading order
\begin{eqnarray}
\delta v_{\mbox{\scriptsize
r,mf}}(t)\,=\,\sum_{i,j}\,[2-\delta_{k_i,k_j}]\,\frac{|k_i-k_j|+2}{N}\,\frac{n^{(0)}_{k_i}[n^{(0)}_{k_j}-\delta_{k_i,k_j}]}{2N}\times\nonumber\\&&\hspace{-5.9cm}\times\left[\frac{\delta
n_{k_i}}{n^{(0)}_{k_i}}\,+\,\frac{\delta
n_{k_j}}{n^{(0)}_{k_j}-\delta_{k_i,k_j}}\right]\,.
\label{clock2}
\end{eqnarray}
The idea behind writing $\delta v_{\mbox{\scriptsize r,mf}}(t)$ in
this manner is to emphasize the fact that despite several factors of
$N$ involved in the expression of $v_N$, its magnitude is of ${\cal
O}(1)$; and therefore, the order of magnitude estimate of $\delta
v_{\mbox{\scriptsize r,mf}}(t)$ must behave as the {\it ratio\/} of
$\delta n_k(t)$ and $n^{(0)}_k(t)$ in realization r.

Secondly, beyond (the mean-fieldish) $\delta v_{\mbox{\scriptsize
r,mf}}(t)$, there is another source of the instantaneous front speed
fluctuation that depends on what is the precise sequence of random
numbers that updates the clock distribution in an infinitesimally
short time window after $t$, i.e., between times $t$ and
$t+\varepsilon$ for an infinitesimally small $\varepsilon$. Let us
denote the front speed fluctuation due to the stochasticity in the
front evolution at time $t$ by $\delta v_{\mbox{\scriptsize
r,s}}(t_+)$ [this depends on the updating sequence of random numbers s
immediately after $t$]. The total instantaneous front speed
fluctuation $\delta v_{\mbox{\scriptsize r}}(t)$ around $v_N$ is
therefore given by
\begin{eqnarray}
\delta v_{\mbox{\scriptsize r}}(t)\,=\,\dot{S}_{\mbox{\scriptsize
r}}(t)\,-\,v_N\,=\,\delta v_{\mbox{\scriptsize r,mf}}(t)\,+\,\delta
v_{\mbox{\scriptsize r,s}}(t_+)\,.
\label{diff1}
\end{eqnarray}
Notice that the definition of $\delta v_{\mbox{\scriptsize r,s}}(t_+)$
in this manner automatically warrants the condition that the average
of $\delta v_{\mbox{\scriptsize r,s}}(t_+)$ [but not $\delta
v_{\mbox{\scriptsize r}}(t)$] over all possible collision sequences
after $t$ is identically zero. A further average of $\delta
v_{\mbox{\scriptsize r}}(t)$ over an ensemble of realizations at time
$t$, however, must yield zero.

The formal expression of $D_f$ is then finally obtained using the
Green-Kubo formula
\begin{eqnarray}
D_f=\frac{1}{2}\lim_{T\rightarrow\infty}\int_{0}^{T}dt'\,\langle\langle\delta
v_{\mbox{\scriptsize r}}(t)\,\delta v_{\mbox{\scriptsize
r}}(t+t')\rangle_{\mbox{\scriptsize s}}\rangle_{\mbox{\scriptsize
r}}\,.
\label{kubo}
\end{eqnarray}
In Eq. (\ref{kubo}), one first needs to average over the random number
sequences chosen to update the front after time $t$ (subscript s), and
then average over the initial realizations at time $t$ (subscript r).

\subsection{The Case of Small Values of $N$ \label{sec2.6}}

By now we have seen that the extent of dependence of the front speed
on the details of the models increases with decreasing $N$. This in
itself is not surprising when one realizes that the relative strength
in fluctuations in the number of particles increases with deceasing
$N$ and that the fluctuation characteristics are indeed controlled by
the specific details of the model. In Sec. \ref{sec2.6}, we discuss
propagating fronts in discrete particle and lattice systems where the
average number of particles $N$ per lattice site in the stable phase
of the front is ``small''. Studies on front propagation at small
values of $N$, however, have been limited to only the
X$\leftrightharpoons2$X and the X$+$Y$\rightarrow$2X
reaction-diffusion systems
\cite{kerstein,doering1,ellak1,ellak2,goutam,moro,bramson1,kerstein1,ba,doering3,sokolov1,sokolov2,sokolov1a,deb5,sokolov3}.\footnote{In
general, any model of discrete particle on a lattice with a small
number of particles per lattice site, relatively strong fluctuations
render mean-field descriptions inapplicable, and there exist numerous
accounts of it in the literature. In the context of reaction-diffusion
systems, on general issues of suitability of mean-field descriptions
of discrete particle and lattice systems at small number of particles
per lattice site, readers may be interested to consult
Refs. \cite{ba11,ba12,ba13,ba14,ba15,ba16,ba17} and the references
cited therein, and also references cited in Ref. \cite{sokolov1}. The
reaction-diffusion systems in these references are of
X$\leftrightharpoons2$X, X$+$Y$\rightarrow$2X, X$+$X$\rightarrow$X,
X$+$X$\rightarrow$inert and X$+$Y$\rightarrow$inert
type.\label{fn2.20}} The mean-field description of all the
one-dimensional models studied in these references yield the
Fisher-Kolmogorov equation on a lattice, given by (with a redefined
time scale and in full generality of parameters)
\begin{eqnarray} 
\frac{\partial\phi_k}{\partial
t}\,=\,D\left[\phi_{k+1}\,+\,\phi_{k-1}\,-\,2\phi_k\right]\,+\,r(N\phi_k\,-\,\phi_k^2)\,,
\label{e29}
\end{eqnarray}
where $r$ is the rate of reaction, $D$ is the rate of diffusion of the
particles to their nearest neighbours, and $\phi_k=\langle\langle
N_k\rangle\rangle/N$. In the stable phase of these fronts there are
$N$ particles per lattice site on average, and in the mean-field
limit, the front becomes a pulled front with speed $v^*$ (when the
space continuum limit is taken, then $v^*=2\sqrt{DrN}$).

\subsubsection{Summary of the Models Studied and Known
Results\label{sec2.6.1}}

In the various models studied, the value of $N$ varies between $N\ll1$
to $N\simeq20$. Section \ref{sec2.6} is divided into several
subsections with progressively reduced value of $N$. One common trend
in all these (one-dimensional) lattice models is to identify the front
position with the {\it instantaneous\/} position of the foremost
particle --- do notice that this is identical to the
i.f.o.l.s. defined in Sec. \ref{sec2.5}, and therefore, it is not the
same as the f.o.l.s. defined in Sec. \ref{sec2.3}.

\begin{itemize}
\item[]{(i)} Reaction-diffusion model X$+$Y$\rightarrow2$X on a
lattice, where both X and Y particles can diffuse to their next
nearest  neighbour lattice sites with the same diffusion rate $D$; any
Y particle coming into contact with an X particle on the same lattice
site is {\it instantaneously\/} converted to an X particle
\cite{sokolov1}: In this model, initially, the entire lattice contains
only Y particles, and one X particle is introduced at the left end of
the lattice at $t=0$. Soon afterwards, a front consisting of X
particles propagates into a region full of Y particles. The mean-field
description of this model yields a reaction rate $r\rightarrow\infty$,
resulting in an infinite mean-field front speed. However, in the
actual lattice model for $N\rightarrow0$, this is not the case:
dimensional arguments, confirmed by simulation shows that $v_N\propto
DN$, while the front diffusion coefficient $D_f$ is $\propto D$, with
proportionality constants of ${\cal O}(1)$ \cite{sokolov1}. In a
discrete-time equivalent of this model, where time is measured in
discrete units of $D$ (thereby rendering $D=1$) and {\it the system is
parallelly updated}, the front speed has been shown to behave as
\cite{ellak1,ellak2}
\begin{eqnarray}
v_N\,=\,1\,-\,e^{-N/2}\,,
\label{e30}
\end{eqnarray} 
with $v_N$ saturating at $1$ for large $N$.
\vspace{4mm}
\item[]{(ii)} Reaction-diffusion model X$+$Y$\rightarrow2$X, where on
a lattice both X and Y particles can diffuse to their next nearest
neighbour lattice sites with the same diffusion rate $D$, and a Y
particle coming into contact with an X particle on the same lattice
site is converted to an X particle with a finite rate $r$: although
one might expect the front speed at small $N$ to have no resemblance
with the mean-field expression $v^*$, it turns out that for some small
value of $N$ and $r\ll1$, surprisingly the mean-field front speed
$v^*$ is recovered again \cite{sokolov2}.
\vspace{4mm}
\item[]{(iii)} A reaction-diffusion model similar to the one in (i)
above was considered in Ref. \cite{sokolov1a}, with the X and Y
particles have different diffusion rates, say $D_X$ and $D_Y$, then
for $D_X\neq0$, the front speed depends only on $D_X$, and not on
$D_Y$. At $D_X=0$ however, there is no uniformly translating front
solution \cite{sokolov1a}.
\vspace{4mm}
\item[]{(iv)} With only one particle allowed per lattice site, one of
the major variants of the model that has been considered in the
literature is the reaction-diffusion system X$\leftrightharpoons2$X.
In this system, only three basic moves are allowed (see
Fig. \ref{depsw}): (a) an X particle can diffuse to any one of its
neighbour lattice sites with a diffusion rate $D$, provided this
neighbouring site is empty; (b) an X particle can give birth to
another one on any one of its empty neighbour lattice site with a
birth rate $\varepsilon$; (c) any one of two X particles belonging to
two neighbouring filled lattice sites can get annihilated with a
annihilation rate $W$.

For all values of the parameters $D$, $\varepsilon$ and $W$, there
exists an exact relation between the front speed $v_N$ and $D_f$,
namely \cite{deb5}
\begin{eqnarray}
\frac{v_N}{2}\,+\,D_f\,=\,D\,+\,\varepsilon\,.
\label{e31}
\end{eqnarray}
For $W=0$ and $D/\varepsilon\rightarrow\infty$ \cite{bramson1}, and
for $W\neq0$ and
$\sqrt{D\varepsilon}/(\varepsilon+W)\rightarrow\infty$ \cite{moron},
the front speed is given by its mean-field limit,
$v^*=2\sqrt{2D\varepsilon}$, but when $D$ and $\varepsilon$ are of the
same order of magnitude, then there are tremendous deviations from
this mean-field theory results \cite{bramson1,kerstein,kerstein1}. In
Ref. \cite{kerstein}, these differences were studied numerically, and
it was shown in \cite{kerstein1} that the differences from the
mean-field theory results can be explained by means of a
``self-consistent two-particle approach''. For $W\neq0$, the $W=D$
case was considered in Refs. \cite{ba,doering3} in continuum space,
where the front speed is obtained as $v=\varepsilon$ and front
diffusion coefficient $D_f=D$.
\begin{figure}[ht]
\begin{center}
\includegraphics[width=0.6\linewidth]{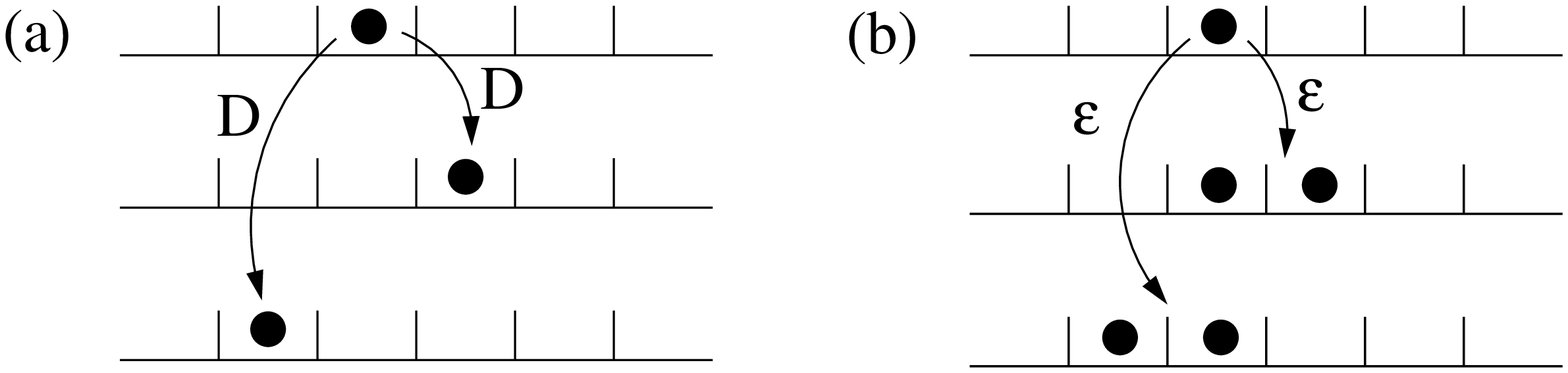}\hspace{5mm}
\includegraphics[width=0.3\linewidth]{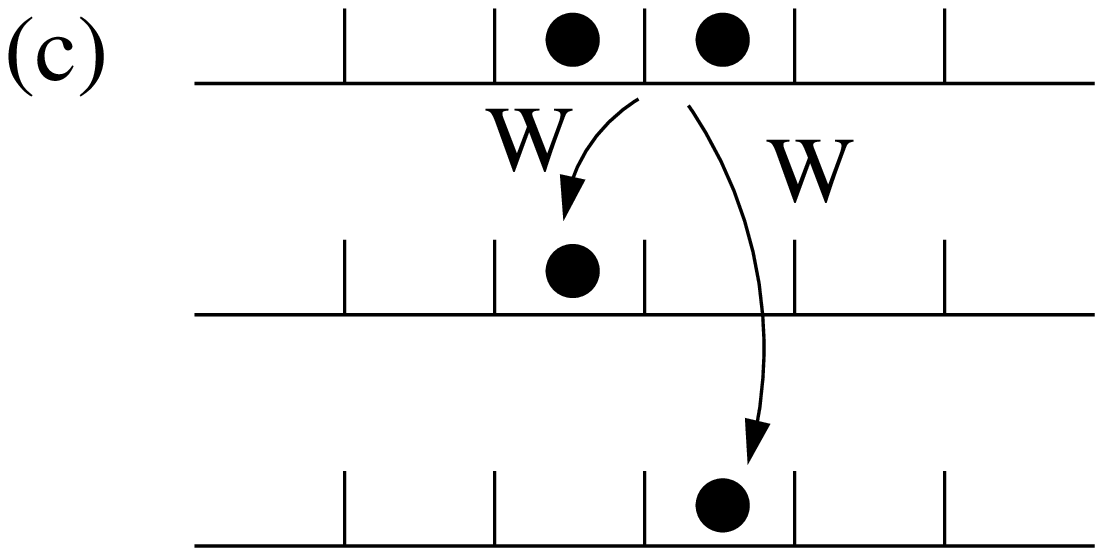}
\end{center}
\caption{The microscopic processes that take place inside the system:
(a) a  diffusive  hop with rate $D$ to a neighbouring empty site; (b)
creation  of a new particle on a site neighbouring an occupied site
with rate $\varepsilon$; (c) annihilation  of a particle on a site
adjacent to an occupied site at a rate $W$.\label{depsw}}
\end{figure}

The full lattice model where $D$, $W$ and $\varepsilon$ are all
different, and $D/W$ and $D/\varepsilon$ are of the same order of
magnitude was considered in Ref. \cite{deb5}; in this case, one has a
good approximate solution \cite{deb5}
\begin{eqnarray}
N\simeq\frac{\varepsilon}{\varepsilon+W}\,\quad\quad
v_N\simeq\frac{\varepsilon(\varepsilon+D)}{\varepsilon+W}\,\quad\mbox{and}\quad
D_f\simeq\frac{(\varepsilon+2W)(\varepsilon+D)}{2(\varepsilon+W)}\,.
\label{e32}
\end{eqnarray}
\item[]{(v)} Fluctuating fronts in higher than one spatial dimensions:
all the known studies on propagating fronts in discrete particle and
lattice realizations in higher than one spatial dimensions have been
on X$+$X$\leftrightharpoons2$X reaction-diffusion system
\cite{doering1,goutam,moro}, or on X$+$Y$\rightarrow2$X. We have seen
earlier that for fluctuating fronts in one spatial dimension, the
front position suffers {\it diffusive\/} wandering around its mean. In
higher than one spatial dimensions, the front wandering properties
change: it is then possible to formulate the problem along the lines
of interfacial growth phenomena \cite{kpz,krug2,halpin,barabasi,krug3}
and study the associated scaling properties. However, there exists
tremendous ambiguity regarding how to properly characterize the
wandering behaviour of fronts in higher than one dimensions
\cite{goutam,moro}, although the results indicate that any generic
upper critical dimension must be higher than $4$ \cite{moro}.
\end{itemize}
 
In the following sections \ref{sec2.6.2} through \ref{sec2.6.6}, we
provide detailed descriptions of points (i) through (v) above, one
point in each section.

\subsubsection{Front Propagation in the Diffusion-Limited Irreversible
Process X$+$Y$\rightarrow2$X \label{sec2.6.2}}

In Sec. \ref{sec2.6.2}, we elaborate the model introduced in point (i)
of Sec. \ref{sec2.6.1}. It was first introduced and studied in the
$N\rightarrow0$ limit in Ref. \cite{sokolov1}. In this model, the
front position is (naturally) identified by the location of the
foremost (rightmost) X particle, and the only way the front
(consisting of X particles) can propagate into the region occupied by
the Y particles is by diffusion. Front propagation in this lattice
model is therefore diffusion-limited, which results in a front speed
$\propto N$, as opposed to an infinite front speed as the mean-field
theory predicts.

To solve for front speed and front diffusion, the following
dimensional argument was proposed in Ref. \cite{sokolov1}: there are
only two dimensionful parameters in this problem: $N$, which scales as
inverse length, and $D$. Therefore, $\alpha=v_N/DN$ is a dimensionless
quantity. On the other hand, wandering of the front region is
diffusive, and if one denotes the front width due to the diffusive
broadening by $w(t)$, then $w^2(t)\sim D_ft$, and from the definition
of $w(t)$ and using the parameters of the problem $N$ and $D$, one can
define another dimensionless quantity $\beta=w^2(t)/Dt\equiv
D_f/D$. These arguments suggest that $v_N\propto DN$ and $D_f\propto
D$. These conjectures were confirmed in simulation, and the values of
$\alpha$ and $\beta$ were numerically obtained to be equal to $1.05$
and $3.57$ respectively (with uncertainties in the last digits), using
$N$ values $0.025$, $0.05$ and $0.1$.\footnote{Notice that the front
really propagates due to the diffusion of X particles, in the model of
Ref. \cite{sokolov1}, if the two species X and Y have different
diffusion rates, then one would simply expect $v_N=\alpha D_XN$ with
the same value of $\alpha$ as deduced above. That this idea is
actually correct was verified by simulations \cite{sokolov1a}. We will
return to this point in Sec. \ref{sec2.6.4}.\label{fn2.21}}

In a continuum space formulation, in the comoving frame of the
foremost X particle ($\xi=Nx$), the probability density $p_{XY}(\xi)$
that the first Y particle is at a distance $\xi$ on the right of the
foremost X particle, and the probability density $p_{XX}(\xi)$ that
the first X particles on the left of the foremost X particle were also
calculated in Ref. \cite{sokolov1}. In this frame, the foremost X
particle is stationary (say at $\xi=0$), and the Y particles in the
region $\xi>0$ and the X particles in the region $\xi<0$ undergo
uncorrelated diffusive movements towards $\xi=0$. The density of
particles $n(\xi)$ (Y particles for $\xi>0$ and X particles for
$\xi<0$) are then obtained from\footnote{In Eq. (\ref{e33}) we need to
use the diffusion rate of X particles for $\xi<0$ (or the diffusion
rate of Y particles for $\xi>0$) in the comoving frame of the foremost
X particle. Therefore, these diffusion rates are actually {\it
relative\/} diffusion rates of a pair of already diffusive particles,
and the value for both of them is $2D$. Notice however that in
Ref. \cite{sokolov1}, $D$ was used as the relative diffusion rate as
opposed to $2D$. As a result, the solutions of Eq. (\ref{e33}) that we
obtain here, which are the correct ones, are not the same as the ones
obtained in Ref. \cite{sokolov1}. \label{fn2.22}}
\begin{eqnarray}
\frac{\partial n}{\partial t}\,-\,v_NN\,\frac{\partial
n}{\partial\xi}\,=\,2DN^2\,\frac{\partial^2 n}{\partial\xi^2}\,,
\label{e33}
\end{eqnarray}
The stationary state solution of this equation with the appropriate
boundary conditions then yields
\begin{eqnarray}
n_Y(\xi>0)\,=\,N\,[1\,-\,\exp(-v_N\xi/[2DN])]\,=\,N\,[1\,-\,\exp(-\alpha\xi/2)]\quad\mbox{and}\nonumber\\&&
\hspace{-12.75cm}n_X(\xi<0)\,=\,N\,.
\label{e34}
\end{eqnarray}
From Eq. (\ref{e34}), the probability $\rho(\xi)$ of finding a Y
particle at a distance $\xi$ on the right of the foremost X particle,
is obtained from the integral equation
$n_Y(\xi)/N=\rho(\xi)+\displaystyle{\int_0^\xi\rho(\xi')d\xi'}$,
leading to the solution
$\rho(\xi)=\alpha(e^{-\alpha\xi/2}-e^{-\xi})/(2-\alpha)$. The
probability $\theta(\xi)$ of finding a X particle at a distance $\xi$
on the left of the foremost X particle, is similarly obtained from the
integral equation
$n_X(\xi)/N=\theta(\xi)+\displaystyle{\int_\xi^0\theta(\xi')d\xi'}$,
yielding $\theta(\xi)=\alpha\,e^{-\alpha|\xi|/2}/2$. As already
mentioned, these solutions are different from the ones obtained in
Ref. \cite{sokolov1} (see footnote \ref{fn2.22}).

Notice that Eq. (\ref{e33}) does not determine the front speed in any
way, instead $v_N$ enters as an input for the solution of $\rho(\xi)$
and $\theta(\xi)$. At best, $v_N$ has to be determined from somewhere
else. In a model closely related with the the one analyzed in
Ref. \cite{sokolov1}, where time was discretized in units of $D^{-1}$,
this incompleteness was removed: in the $N\rightarrow0$ limit, $v_N$
was deduced from first principles to be equal to $N/2$ at the leading
order of $N$ \cite{ellak1,ellak2}. In this model, in a unit time, each
particle moves one lattice distance randomly towards its right or
left, and the method of ``parallel updating''\footnote{The idea behind
parallel updating in this model is that in a unit time, {\it all\/}
particles move diffusively one lattice distance randomly either
towards their left or the right, and when this update brings an X
particle in contact with any number of Y particles on a lattice site,
all the Y particles on that lattice site are instantaneously converted
to X particles.\label{parallel}} was employed in the simulation
algorithm. The argument to derive the expression of front speed in
this model in the $N\rightarrow0$ limit is the following: the front
speed is essentially determined by considering the two foremost X
particles (which are separated by an average of $1/N$ lattice
sites).\footnote{In the limit $N\rightarrow0$, the role of a third (or
more) particle near the foremost X particle can be neglected; they
only provide higher order corrections. \label{fn2.23}} In fact, the
only non-zero contribution to the front speed occurs when the second X
particle is on the lattice site just behind the foremost one, or on
the same lattice site as the foremost one (see footnote
\ref{ellaknote1} for an explanation). In the latter case, one of the X
particles at the location of the front is tagged to be the foremost X
particle and the other is then tagged as the second X particle. In the
comoving frame of the front, let us tag the location of the foremost X
particle by $k=0$, and denote the probability that the second X
particle is $k$ lattice sites behind the foremost X particle by
$p_{-k}$. It is then easily seen that\footnote{Equation (\ref{e35}) is
obtained by noticing that in a unit time, the two foremost X particles
can execute a total of four moves, each with probability $1/4$. When
the two foremost X particles are separated by one lattice site, then
the front moves forward one lattice distance  with probability $1/2$,
moves backward one lattice distance with probability $1/4$ and remains
stationary with probability $1/4$. This explains the origin of the
$p_{-1}/4$ term. On the other hand, when the two foremost X particles
on the same site, then the front moves forward one lattice distance
with a probability $3/4$, and moves backward one lattice distance with
a probability $1/4$ --- giving rise to the $p_0/2$ term. When the
second X particle is more than one lattice site behind the foremost X
particle, then the front moves forward or backward by one lattice
distance with equal probabilities $1/2$, which is why $p_{-k}$ for
$k>1$ does not contribute in Eq. (\ref{e35}).\label{ellaknote1}}
\begin{eqnarray}
v_N\,=\,\left[\frac{p_{-1}}{4}\,+\,\frac{p_0}{2}\right]\,.
\label{e35}
\end{eqnarray}
The steady state solution of $p_{-1}$ and $p_0$ are obtained from the
master equation for $p_k$'s around foremost X particle (i.e.,
$k=-2,-1,\ldots,2$ etc.), yielding\footnote{For the master equation
and its solution, we urge the readers to consult
Ref. \cite{ellak1,ellak2}; although here we point out the similarity
between Eqs. (\ref{e36}) and (\ref{e34}) --- in both of these (very
similar) models, the density of X particles behind the foremost one is
$N$.\label{lookup}}
\begin{eqnarray}
p_{-1}\,=\,N\,,\quad p_0\,=\,N/2\,\quad\mbox{and}\quad v_N\,=\,N/2\,.
\label{e36}
\end{eqnarray}

\begin{figure}[ht]
\begin{center}
\includegraphics[width=0.6\linewidth,height=0.3\textheight]{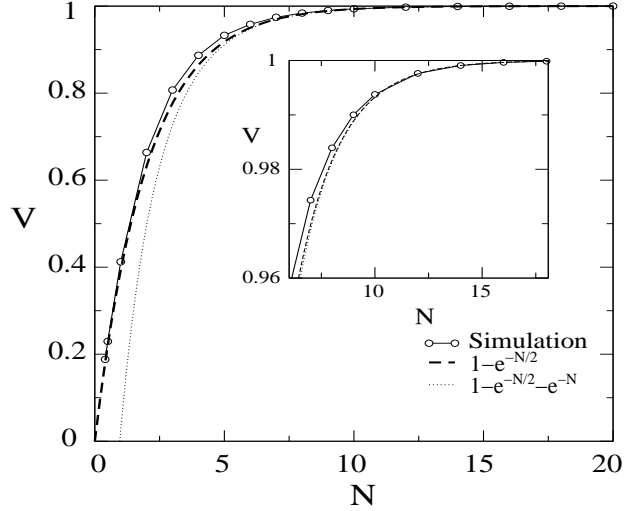}
\end{center}
\caption{Equation (\ref{e38}) vs. simulation data for front speed,
adopted from Ref. \protect{\cite{ellak1,ellak2}}. Here $v$ should be
read as $v_N$. The dotted curve shows the agreement between theory and
simulation if one uses a Poissonian distribution also for
$p(n_0)$. Notations have been modified from the original to maintain
consistency with the text.\label{sanderfig}}
\end{figure}

The above derivation clearly does not hold for higher values of $N$,
as one needs to consider the case where a lattice site is occupied by
multiple number (many more than two) of X particles. Such an analysis
was also presented in Ref. \cite{ellak1,ellak2}, in which, having
denoted the probability of having the $k$-th lattice site occupied by
$n_k$ number of particles by $p(n_k)$, one has [Eq. (\ref{e37}) is
obtained by following the logic of footnote \ref{ellaknote1}]
\begin{eqnarray}
P_-\,=\,\sum_{n_0=1}^\infty\sum_{n_{-1}=1}^\infty\frac{p(n_0,n_{-1})}{2^{n_0+n_{-1}}}\,,\quad
P_+\,=\,\sum_{n_0=1}^\infty\left[1-\frac{1}{2^{n_0}}\right]p(n_0)\,\quad\mbox{and}\nonumber\\&&\hspace{-11.5cm}P_0\,=\,\sum_{n_0=1}^\infty\sum_{n_{-1}=1}^\infty\frac{1}{2^{n_0}}\left[1-\frac{1}{2^{n_{-1}}}\right]\,p(n_0,n_{-1})
\label{e37}
\end{eqnarray}
respectively for the probability of the front to step backward one
lattice site, to remain where it presently is, and to step forward one
lattice site in one unit time. To obtain the front speed from
Eq. (\ref{e37}) however, one needs the probability distributions
$p(n_0)$ and $p(n_{-1})$ \cite{ellak1,ellak2}. In
Ref. \cite{ellak1,ellak2}, while $p(n_{-1})$ was obtained from the
simulations as a Poissonian with average $N$, a truncated
Poissonian\footnote{The truncated Poissonian distribution for the
probability of $n_0$ number of particles at $k=0$ is given by
$p_T(n_0)=(1-\delta_{n_0,0})(1-e^{-N})^{-1}n_0e^{-n_0}$.
\label{fn2.24}} was used for $p(n_{0})$, yielding the result [notice
that one recovers Eq. (\ref{e36}) for $v_N$ from Eq. (\ref{e38}) in
the limit $N\rightarrow0$]
\begin{eqnarray}
v_N\,=\,1\,-\,e^{N/2}\,.
\label{e38}
\end{eqnarray}
The agreement of the front speed with simulation results
\cite{ellak1,ellak2} is shown in Fig. \ref{sanderfig}.

\subsubsection{Reaction-diffusion Model X$+$Y$\rightarrow$2X at Small $N$
with Reaction Rate $r\rightarrow0$\label{sec2.6.3}}

In Ref. \cite{sokolov2}, the reaction-diffusion model
X$+$Y$\rightarrow$2X was numerically studied at small values of $N$,
($N=0.1,0.3,1,5$ and $10$) and reaction rates $r\ll1$
($r\leq0.02$). The details of the microscopic model are already
described in point (ii) of Sec. \ref{sec2.6.1}.

\begin{figure}[ht]
\begin{center}
\includegraphics[width=0.5\textwidth]{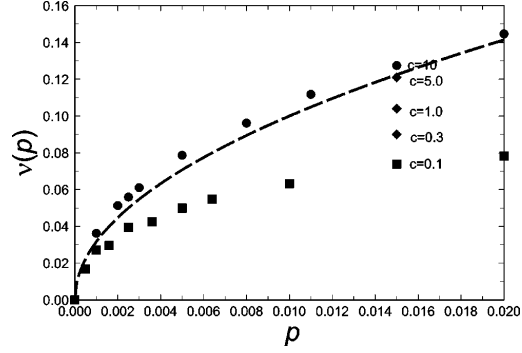}
\end{center}
\caption{Front speed at different values of $N$ and as a function of
reaction rate $r$ ($p$ and $c$ should be read as $r$ and $N$
respectively), where $\nu(r)=v_N(r)/2\sqrt{DN}$. Dashed curve:
mean-field result $\nu(r)=\sqrt{r}$. \label{sokolov2}}
\end{figure}

With the front position identified by the position of the foremost X
particle, the front speed at different values of $N$ and as a function
of reaction rate $r$ is shown in Fig. \ref{sokolov2} ($p$ and $c$
should be read as $r$ and $N$ respectively), where
$\nu(r)=v_N(r)/2\sqrt{DN}$. If the mean-field theory were applicable
in this model, then $\nu(r)$ should scale as $\sqrt{r}$ (dashed curve
of Fig. \ref{sokolov2}). Figure \ref{sokolov2} reveals that typically
most values of $N$ yield strong deviation from mean-field result for
$\nu(r)$, except for $N=10$ for all values of $r$. Reference
\cite{sokolov2} also claimed that the trend below $r\simeq0.001$
indicates that the mean-field result front speed for $r\lesssim0.001$,
however there are too few data points below $r\simeq0.001$ in
Fig. \ref{sokolov2} to arrive at a definitive conclusion.

\begin{figure}[ht]
\begin{center}
\includegraphics[width=0.45\textwidth]{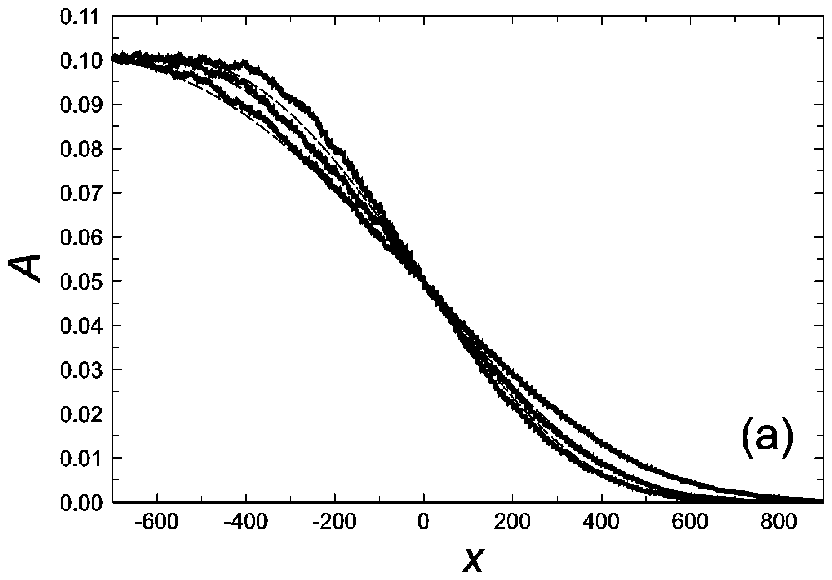}
\hspace{5mm}\includegraphics[width=0.46\textwidth]{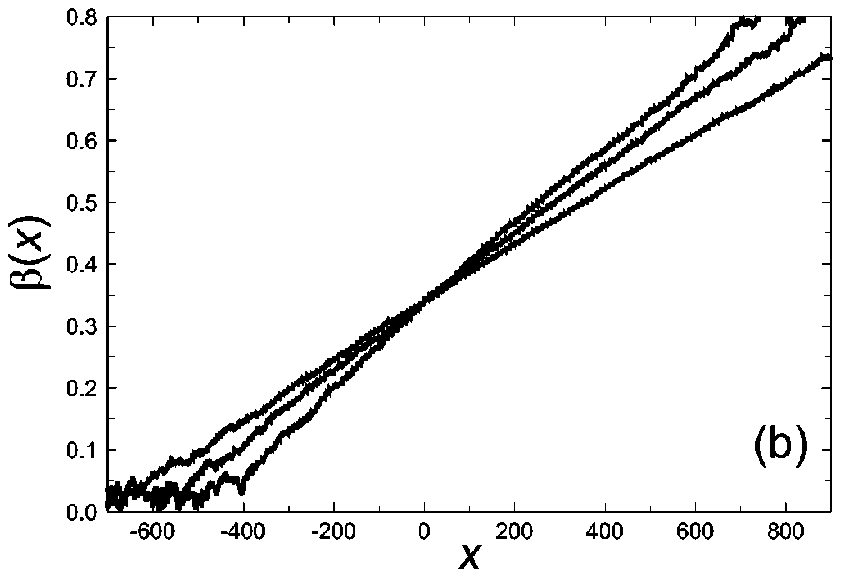}
\end{center}
\caption{Plots in support of Eq. (\ref{e39}): (a) front the front
profile (here $A$ should be read as $\phi(\xi)$ and $x$ should be read
as $\xi$) for three values of the reaction rates $r=0.0025,0.005$ and
$0.01$ (left above from left to right). The dashed lines correspond to
Eq. (\ref{e39}) with numerically determined $a$, $b$ and $\sigma$. (b)
Plots of $\beta(\xi)=\sqrt{\log_{10}|\log_{10}[\phi(\xi)]|}$ are shown
in Fig. \ref{sokolov4}(b): the right above left to right plots
supposedly correspond to $r$ values $0.0025,0.005$ and $0.01$
respectively. \label{sokolov4}}
\end{figure}

No theoretical argument for why the $N=10$ data for the front speed
follows the mean-field theory results so well was provided in
Ref. \cite{sokolov2} (neither is it clear from a simple
argument). Moreover, since Fig. \ref{sokolov2} indicates that the
mean-field results are closely followed for $N=10$, it is a natural
extension to investigate, by means of simulations, if the front
profile also follows the mean-field result. However, this point was
not investigated in Ref. \cite{sokolov2}. Instead, the front profile
was investigated for $N=0.1$, and from the plots that were provided in
Ref. \cite{sokolov2}, the front profile seems to behave as
\begin{eqnarray}
\phi(\xi)\,=\,a\,\exp\left\{-\,b\,\exp\left[\frac{(\xi-\xi_0)^2}{2\sigma^2}\right]\right\}\,.
\label{e39}
\end{eqnarray}
The plots that were supplied in Ref. \cite{sokolov2} in support of
Eq. (\ref{e39}) are shown in Fig. \ref{sokolov4}. Figure
\ref{sokolov4}(a) shows the front the front profile (here $A$ should
be read as $\phi(\xi)$ and $x$ should be read as $\xi$) for three
values of the reaction rates $r=0.0025,0.005$ and $0.01$ (left above
from left to right). The dashed lines in Fig. \ref{sokolov4}(a)
correspond to Eq. (\ref{e39}) with numerically determined $a$, $b$ and
$\sigma$. The plots of
$\beta(\xi)=\sqrt{\log_{10}|\log_{10}[\phi(\xi)]|}$ are shown in
Fig. \ref{sokolov4}(b): the right above left to right plots supposedly
correspond to $r$ values $0.0025,0.005$ and $0.01$ respectively.

\subsubsection{The Case of Different Diffusion Rates for X and Y
Particles in the Model of Point (i)\label{sec2.6.4}}

In Ref. \cite{sokolov1a}, a model similar to the one proposed in
Ref. \cite{sokolov1} was studied. In this model, the two species X and
Y particles have two different diffusion rates, namely $D_X$ and
$D_Y$. Both the cases $D_X\neq0$ and $D_X=0$ were considered in the
ensuing analysis. The lattice is initially filled with Y particles
with an average number of particles $N$ per lattice site and at $t=0$,
an X particle is introduced at the leftmost lattice site, just like
the model discussed in point (i) of Sec. \ref{sec2.6.1}. However, {\it
only one particle is allowed per lattice site in this model, and when
two neighbouring lattice sites are occupied by one X and one Y
particle, the Y particle is instantaneously converted to an X
particle}. As pointed out in Ref. \cite{sokolov1a}, the restriction
that only one particle is allowed per lattice site is not necessary,
other than when $D_X=0$.

\begin{figure}[ht]
\begin{center}
\includegraphics[width=0.5\textwidth,height=0.25\textheight]{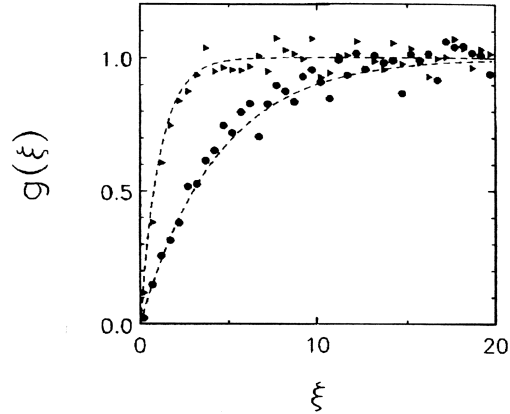}
\end{center}
\caption{Comparison of the density profile of Y particles on the right
of the foremost X particle in a continuum-space type theory
$g(\xi)=n_Y(\xi)/N=$ $1-\exp\{-v_N\xi/[(D_X+D_Y)N]\}$ and simulation
data for two different $N$ values. Triangles: $D_X=0.2$, $D_Y=0.8$,
circles: $D_X=0.8$, $D_Y=0.2$. \label{sokolov1}}
\end{figure}

In this model, the $D_X\neq0$ does not reveal any surprises (see
footnote \ref{fn2.21} \cite{sokolov1a}). Moreover, the density profile
of Y particles on the right of the foremost X particle (located at
$\xi=0)$ in a continuum-space formulation ($\xi=Nx$) turns out to be
$n_Y(\xi)=N[1-\exp\{-v_N\xi/[(D_X+D_Y)N]\}]$ \cite{sokolov1a} (see
Fig. \ref{sokolov1}) --- this is also consistent with Eq. (\ref{e34}).

The case of $D_X=0$, however, is more complicated as we explain
below. In this case, the front propagates only due to the diffusive
flux of Y particles towards the foremost X particle, and at any time
$t$, with the front position defined by the location $x_f(t)$ of the
foremost X particle in a continuum-space formulation, one has
\cite{sokolov1a}
\begin{eqnarray}
v_N(t)\,=\,-\,D_Y\,\frac{\partial n_Y(x,t)}{\partial
x}\bigg|_{x_f(t)}\,,
\label{e40}
\end{eqnarray}
while the density profile of Y particles on the right of the foremost
X particle is obtained from the analogous form of Eq. (\ref{e33}) as
(notice that the rate of diffusion of the Y particles towards the
foremost X particle is $D_Y$)\footnote{One must be aware of the fact
that the usage of $D_Y$ as the rate of diffusion of the Y particles
towards the foremost X particle is valid at a low density of Y
particles. In this model, which allows only one particle per site, the
diffusion of a Y particle is permitted only if its neighbouring
lattice sites are empty. Strictly speaking, therefore, to obtain the
the rate of diffusion of the Y particles towards the foremost X
particle one should also take into account probability of having two
next-nearest neighbour Y particles. At low density of Y particles this
probability can be neglected, permitting the use of $D_Y$ in
Eq. (\ref{e41}).\label{fn2.25}}
\begin{eqnarray}
\frac{\partial n}{\partial t}\,-\,v_N\,\frac{\partial n}{\partial
x}\,=\,D_Y\,\frac{\partial^2 n}{\partial x^2}\,.
\label{e41}
\end{eqnarray}
Equations (\ref{e40}) and (\ref{e41}) together are then solved with
the ansatz $n_Y(x,t)=Ng\{[x-x_f(t)]/\sqrt{D_Yt}\}$ and a yet unknown
function $g$ \cite{sokolov1a}. This ansatz actually suggests that
$x_f(t)\sim\sqrt{t}$, and that the front speed $v_N(t)\sim t^{-1/2}$
is actually time-dependent. At low density of particles per lattice
site $N$, the solution of $n_Y(x,t)$ turns out to be the error
function
$g\{[x-x_f(t)]/\sqrt{D_Yt}\}=\mbox{erf}\{[x-x_f(t)]/[2\sqrt{D_Yt}]\}$.
These aspects were verified in Ref. \cite{sokolov1a} (see
Fig. \ref{sokolov2}).
\begin{figure}[ht]
\begin{center}
\includegraphics[width=0.4\textwidth]{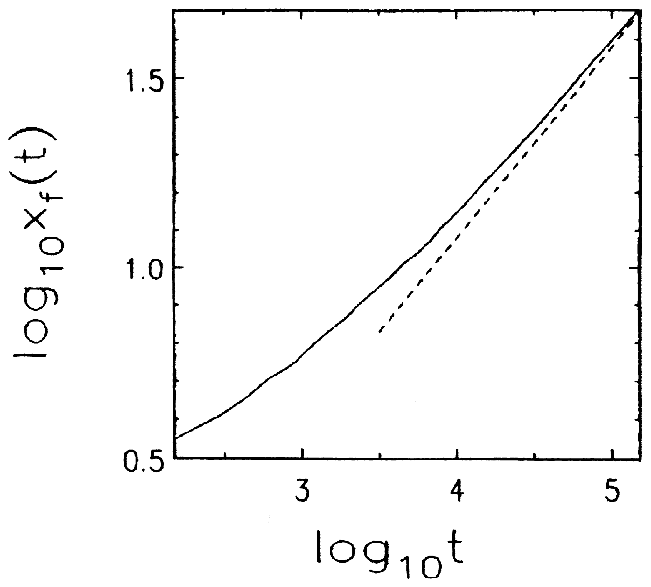}\hspace{5mm}
\includegraphics[width=0.39\textwidth,angle=0.5]{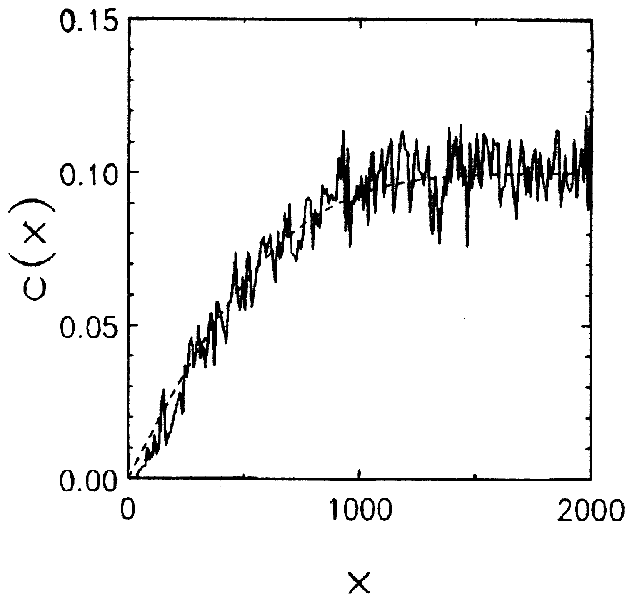}
\end{center}
\caption{The plots of $x_f(t)$ vs. $t$ in log-log scale, and that of
$n_Y(x,t)$ at $t=10000$ units. Here $c(x)$ should be read as
$n_Y(x)$. Dashed lines correspond to $x_f(t)\sim\sqrt{t}$ and
Eq. (\ref{e41}) respectively. \label{sokolov3}}
\end{figure}

\subsubsection{Reaction-diffusion Model X$\leftrightharpoons2$X with
Only One Particle Allowed per Lattice Site\label{sec2.6.5}}

In Sec. \ref{sec2.6.5}, we discuss the model of point (iv) of
Sec. \ref{sec2.6.1}. Below, up to Eq. (\ref{e46}) we present the
general features of this model following
Refs. \cite{bramson1,kerstein,kerstein1,deb5}, and then we discuss the
results in different parameter regimes.

Like in the other subsections of Sec. \ref{sec2.6}, {\it for all
values\/} of $D$, $\varepsilon$ and $W$, we define the position of the
front by the location of the foremost particle. For a given
realization, let us denote its position at time $t$ by $k_f(t)$. For
an ensemble of front realizations, having denoted the probability
distribution for the foremost particle to be at lattice site ${k_f}$
by $P_{k_f}(t)$, its evolution is then described by
\begin{eqnarray}
\frac{dP_{k_f}}{dt}\,=\,(D\,+\,\varepsilon)\,P_{{k_f}-1}\,+\,\left[D\,P^{\mbox{\scriptsize
empty}}_{{k_f}+1}\,+\,W\,P^{\mbox{\scriptsize
occ}}_{{k_f}+1}\right]\,\nonumber\\&&\hspace{-6.9cm}-\,(D\,+\,\varepsilon)\,P_{{k_f}}\,-\,\left[D\,P^{\mbox{\scriptsize
empty}}_{{k_f}}\,+\,W\,P^{\mbox{\scriptsize occ}}_{{k_f}}\right]\,.
\label{e42}
\end{eqnarray} 
Here $P^{\mbox{\scriptsize occ}}_{{k_f}}(t)$ and $P^{\mbox{\scriptsize
empty}}_{{k_f}}(t)$  denote the joint probabilities that the foremost
particle is at site $k_f$ and that the site  $k_f-1$ is occupied or
empty, respectively.  Clearly, $P_{k_f}(t)=P^{\mbox{\scriptsize
occ}}_{{k_f}}(t)+P^{\mbox{\scriptsize empty}}_{{k_f}}(t)$, and
$\sum_{k_f}P_{k_f}(t)=1$. The first term on the r.h.s. of
Eq. (\ref{e42}) describes the increase in $P_{k_f}(t)$ due to the
advancement of a foremost occupied lattice site from position
${k_f}$$-$$1$, while the second term describes the increase in
$P_{k_f}(t)$ due to the retreat of a foremost occupied lattice site
from position ${k_f}$$+$$1$. The  third and the fourth terms,
respectively, describe the decrease in $P_{k_f}(t)$ due to the
advancement and retreat of a foremost occupied lattice site from
position ${k_f}$.

From the definition of $P_{k_f}(t)$, the mean position and the width
of the distribution for the positions of the foremost occupied lattice
sites are defined  as $x(t)=\sum_{k_f}{k_f}P_{k_f}(t)$ and ${\cal
W}^2(t)=\sum_{k_f}[{k_f}-x(t)]^2P_{k_f}(t)$.\footnote{One can also
define the foremost occupied lattice site for a realization as the one
on the right of which no lattice site has ever been occupied before,
and obtain the front speed from this definition following
Sec. \ref{sec2.3}. Both of them of course yield the same result due to
the $t\rightarrow\infty$ limit.\label{fn2.26}} The mean speed and
diffusion coefficient of the front are thus given in terms of these
quantities as  the $t$$\to$$\infty $ limit of $v_N=d x(t)/dt $ and
${\cal W}^2(t)=2D_f t$ --- see Fig. \ref{depsw}{\em (c)}.  To obtain
them, we need the expressions of $P^{\mbox{\scriptsize
occ}}_{{k_f}}(t)$ and $P^{\mbox{\scriptsize empty}}_{{k_f}}(t)$. To
start with, we have
\begin{equation}
P^{\mbox{\scriptsize occ}}_{k_f}(t)=\rho_{k_f-1} P_{k_f}(t),
\label{e43}
\end{equation}
where $\rho_{k_f-1}$ is the conditional probability of having the
$({k_f}$$-$$1)$th lattice site occupied (the foremost particle is  at
the ${k_f}$th lattice site). The set of conditional occupation
densities $\rho_{k_f -m}$ for  $m \ge 1$ can be thought of as
determining the front profile in a frame moving with each front
realization. For obtaining $v_N$ and $D_f$, we simply need to know the
asymptotic long-time limit  $\rho_{k_f-1}( t\to\infty)$, which from
here on we will denote simply as $\rho_{k_f-1}$. Given $\rho_{k_f-1}$,
it is then straightforward to obtain from Eq. (\ref{e42}) and the
conditions $P_{k_f}(t)=P^{\mbox{\scriptsize
occ}}_{{k_f}}(t)+P^{\mbox{\scriptsize empty}}_{{k_f}}(t)$ and
$\sum_{k_f}P_{k_f}(t)=1$
\begin{eqnarray}
v_N\,=\,\frac{dx}{dt}\,=\,\varepsilon\,-\,\rho_{k_f-1}(W\,-\,D)\quad\quad\mbox{and}\nonumber\\&&\hspace{-7.4cm}\frac{d{\cal
W}^2}{dt}\,=\,2D\,+\,\varepsilon\,+\,\rho_{k_f-1}(W\,-\,D)\,.
\label{e44}
\end{eqnarray}
Of these, the second equation indicates that the front wandering is
diffusive, and an expression of the front diffusion coefficient $D_f$
is therefore given by
\begin{eqnarray}
D_f\,=\,\frac{1}{2}\left[2D\,+\,\varepsilon\,+\,\rho_{k_f-1}
(W\,-\,D)\right]\, .
\label{e45}
\end{eqnarray}
Note that if we use Eq. (\ref{e43}) in Eq. (\ref{e42}), the latter
equation has the form of the master equation for a single  random
walker on a chain. Thus we can think of the foremost particle as
executing a biased random walk, and $D_f$ as the effective diffusion
coefficient of this walker. Moreover, if we eliminate $\rho_{k_f-1}$
from Eqs. (\ref{e44}) and (\ref{e45}), we  get the following {\it
exact\/} relation
\begin{equation}
v_N/2\,+\,D_f\,=\,D\,+\,\varepsilon\,.
\label{e46}
\end{equation}
With this background in mind for all values of $D$, $\varepsilon$ and
$W$, we now explore the results in different parameter regimes.

The mean-field limit of this model effectively yields some form of
Fisher-Kolmogorov equation. The applicability of the mean-field
result, however, is confined only to large values of the diffusion
coefficient (while keeping the other parameters $\varepsilon$ and $W$
finite). The need to have large diffusion coefficient is not very
difficult to understand intuitively --- notice that although in
principle there are three parameters in the problem, effectively there
are only two of them, since the third one simply sets the time
scale. Let us choose these two parameters as $D/\varepsilon$ and
$D/W$. The quantity
$\ell_D=\mbox{min}\left(\sqrt{D/\varepsilon},\sqrt{D/W}\right)$
\cite{moron} then can be seen to define a correlation volume, i.e., a
length scale over which the position of a particle varies in a unit
time. When $\ell_D\rightarrow\infty$, the correlations involving two
or more particles within a finite range of lattice sites get washed
out, and then a mean-field description of the problem becomes suitable.

Having defined $x=k/\sqrt{2D}$, for $W=0$ and large values of
$D/\varepsilon$, a mathematical derivation of the mean-field limit of
this model in the form of the Fisher-Kolmogorov equation
\begin{eqnarray}
\frac{\partial\phi}{\partial
t}\,=\,\frac{1}{2}\,\frac{\partial^2\phi}{\partial
x^2}\,+\,2\varepsilon(\phi\,-\,\phi^2)
\label{e47}
\end{eqnarray}
can be found in Ref. \cite{bramson1} (also see the references cited
therein).\footnote{In a recent paper \cite{moron}, the approach to the
mean-field description of the front has been investigated in the light
of Eq. (\ref{e11}). The idea is to define a quantity
$N^*=\sqrt{D\varepsilon}/(\varepsilon+W)$ such that for
$N^*\rightarrow\infty$ (which happens for finite $N$ and
$\ell_D\rightarrow\infty)$, not only does one recover the
Fisher-Kolmogorov equation, but also for $N^*\gg1$, one expects to
observe $v^*-v_N\sim1/\ln^2N^*$ and $D_f\sim1/\ln^3N^*$. In other
words, Ref. \cite{moron} has essentially shown that $N^*$ simply plays
the same role as $N$ does in Secs. \ref{sec2.2}-\ref{sec2.5}.
\label{fnmoron}} While we leave the interested readers on their own for this
mathematical argument, here we concentrate on the Refs.
\cite{kerstein,kerstein1} that investigated, by means of a
``self-consistent two particle model'', the role of interparticle
correlations in the approach of the front speed towards its mean-field
value\footnote{The mean-field Fisher-Kolmogorov equation (\ref{e47})
has front speed $2\sqrt{\varepsilon}$. When the distance is measured
in terms of $k$ and not $x$, this means that the actual $v^*$ is
$2\sqrt{2D\varepsilon}$.\label{fn2.28}} $v^*=2\sqrt{2D\varepsilon}$
for increasing value of $D/\varepsilon$. We make an important
observation here that this approach only works for $W=0$ (we will
return to this point later).

The first step of this ``self-consistent two particle model'' is to
derive the front speed for small $D/\varepsilon$ at $W=0$, and it
appears already in \cite{bramson1}: the idea is that far behind the
foremost particle, the density of particles approach the homogeneous
equilibrium density $N$:
$\lim_{m\rightarrow\infty}\rho_{k_f-m}=N$. From the master equation of
the microscopic processes for $W=0$, it is easy to show that the
homogeneous equilibrium solution for the total probability is of
product form (so that the probability of having different sites
occupied is uncorrelated), and that the equilibrium occupation density
$N$ is simply given by $N=1$.\footnote{For $W\neq0$, $N$ becomes
$\varepsilon/(\varepsilon+W)$ \cite{deb5}; we will return to this case
later in Sec. \ref{sec2.6.5}.\label{fn2.27}} Then the crudest
approximation for the front profile $\rho_{k_f-m}$ and in particular
for $m=1$ is just to take $\rho_{k_f-1}\approx N$. Substitution of
this approximation into Eq. (\ref{e44}) then immediately yields
\cite{bramson1,kerstein,kerstein1}
\begin{eqnarray}
v_N\,=\,\varepsilon\,+\,D\,.
\label{e48}
\end{eqnarray}
Equation (\ref{e47}) agrees well with the simulations in the limit
$D/\varepsilon\rightarrow0$.

While Ref. \cite{kerstein} is mostly about numerical studies of how
the mean-field result for front speed $v^*=2\sqrt{2D\varepsilon}$ is
obtained from the first half of Eq. (\ref{e44}) in the limit
$D/\varepsilon\rightarrow\infty$, the transition from the front speed
in Eq. (\ref{e47}) to $v^*$ for increasing values of $D/\varepsilon$
in analyzed by theoretical means in Ref. \cite{kerstein1}. What
Ref. \cite{kerstein1} does is to develop a theory for evaluating
$\rho_{k_f-1}$, by means of considering the probability distribution
$p_j$ of having $j$ empty sites between the foremost particle and the
second foremost particle, and $\rho_{k_f-1}$ is then simply obtained
as $\rho_{k_f-1}=p_0$. The task of obtaining $p_0$ itself, however, is
not easy, and it is obtained from solving the master equation for the
steady state solution of $\{p_j\}$, as we outline below.

With the usage of the normalization $\sum_{j}p_j=1$, the master
equation for ${p_j}$ is given by
\begin{eqnarray}
\dot{p}_j\,=\,2D\,p_{j-1}\,+\,(2D\,+\,\varepsilon)\,p_{j+1}\,-\,(4D\,+\,3\varepsilon)\,p_j\quad\quad\mbox{for
$j>0$ and}\nonumber\\&&\hspace{-12cm}
\dot{p}_0\,=\,(2D\,+\,\varepsilon)\,p_1\,+\,2\varepsilon\,-\,2(D\,+\,\varepsilon)\,p_0\,.
\label{e49}
\end{eqnarray}
Equation (\ref{e49}) is obtained from the following considerations:
the $j>0$ state can be obtained from the $j-1$ state by a forward
diffusion of the foremost particle, or by the backward diffusion of
the second foremost particle, giving rise to the $2Dp_{j-1}$ term. The
$j>0$ state can also be obtained from $(j+1)$ state by a backward
diffusion of the foremost particle, or the forward diffusion of the
second foremost particle, or by the growth of another particle just on
the right of the second foremost particle --- this gives rise to the
$(2D+\varepsilon)p_{j+1}$ term. Any $j>0$ state contributes not only
to $j+1$ and $j-1$ states by the reverse processes, but also to the
$j=0$ state when a particle is created on either side of the foremost
particle, which finally explains the $(4D+3\varepsilon)p_j$ term. The
$(2D+\varepsilon)p_1$ is obtained from any $j>0$ state by the creation
of a particle on either side of the foremost particle. In addition,
the $j=0$ state from the $j=1$ state by a creation of a particle on
the right of the second foremost particle, by a forward diffusion of
the second foremost particle or by a backward diffusion of the
foremost particle. On the other hand, state $j=0$ contributes only to
$j=1$ state when the second foremost particle diffuses backward or the
foremost particle diffuses forward. When all these are properly taken
into account along with the normalization condition $\sum_{j}p_j=1$,
the second half of Eq. (\ref{e48}) is obtained immediately. The steady
state of Eq. (\ref{e48}) is then easily solved by the ansatz
$p_j=p_0(1-p_0)^j$ for $\forall j>0$ \cite{kerstein1}, yielding
\begin{eqnarray}
p_0\,=\,\frac{\sqrt{9\varepsilon^2\,+\,16D\varepsilon}\,-\,\varepsilon}{2(2D\,+\,\varepsilon)}\,.
\label{e50}
\end{eqnarray}

The lesson one learns from this clever approach is actually quite
instructive. Notice that Eq. (\ref{e49}) automatically yields the
proper $D/\varepsilon\rightarrow0$ limit of the front speed
(\ref{e48}). Its $D/\varepsilon\rightarrow\infty$ behaviour however
yields $v_N=\sqrt{D\varepsilon}$, which is a factor of $\sqrt{8}$
smaller than $v^*=2\sqrt{2D\varepsilon}$. This itself should not
strike one as surprising. Instead, with the observation that the
involvement of a third foremost particle on the left of the second one
is always ignored (in the sense that the site on the left of the
second foremost particle is always assumed to be empty), the fact that
even in this crude form this formalism at least provides the
$\sqrt{D\varepsilon}$ behaviour of $v_N$ at large $D/\varepsilon$ is
surprising.
\begin{figure}[ht]
\begin{center}
\includegraphics[width=0.6\textwidth,angle=1]{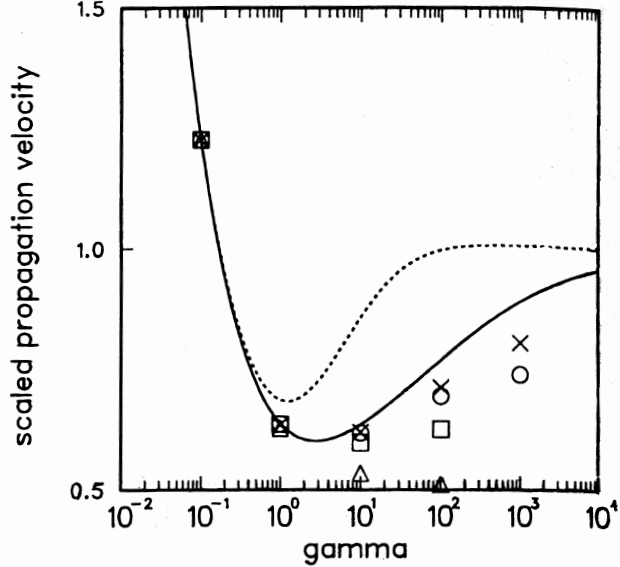}
\end{center}
\caption{$v_N/v^*$ for $\varepsilon=2$ and $\gamma=2D$: simulations
(crosses), ``correlated two-particle self-consistent approach'' (solid
curve), and ``uncorrelated $n$-particle approach'' with $n=3$ (open
triangles), $5$ (open squares) and $10$ (open circles). Also shown is
the ad hoc functional dependence $v_N=[1+\gamma(1+\gamma/8)^{-1/2}]/2$
of $v_N/v^*$ on $\gamma$ by the dotted curve.\label{kersteinfig}}
\end{figure}

For a remedy of this drawback, Ref. \cite{kerstein1} also invokes a
``correlated two-particle self-consistent approach'', in which
diffusive moves of the second foremost particle are prevented due to
the presence of the third foremost particle just next to the second
one. With $p_a$ as the probability of having the second and the third
foremost particles next to each other, the probability of the second
foremost particle making a diffusive hop towards the left is reduced
by $D(1-p_a)$. Reference \cite{kerstein1} then obtains the front speed
as a function of increasing $D/\varepsilon$ by means of an effective
functional dependence of $p_a$ on $p_0$, namely $p_a=Bp_0-(B-1)p^2_0$,
where the parameter $B$ is chosen in such a way that it leads to
$v_N\rightarrow v^*$ for $D/\varepsilon\rightarrow\infty$. However, in
this process, an analytical form of $v_N$ as a function of
$D/\varepsilon$ is not obtained, although a predicted value of $v_N$
can be numerically obtained for a given value of $D/\varepsilon$.  In
Fig. \ref{kersteinfig}, we present the graph from
Ref. \cite{kerstein1} for $\varepsilon=2$ and $\gamma=2D$. It shows
all the results for $v_N/v^*$ together: simulations (crosses),
``correlated two-particle self-consistent approach'' (solid curve),
and ``uncorrelated $n$-particle approach'' with $n=3$ (open
triangles), $5$ (open squares) and $10$ (open circles). Also shown in
Fig. \ref{kersteinfig} is the ad hoc functional dependence
$v_N=[1+\gamma(1+\gamma/8)^{-1/2}]/2$ of $v_N/v^*$ on $\gamma$
(adopted because of its simple functional dependence having proper
limits at $D/\varepsilon\rightarrow0$ and
$D/\varepsilon\rightarrow\infty$) by the dotted curve.

Having thus concluded the $W=0$ cases studied in the literature in the
above paragraph, we now embark on the $W\neq0$ case for this
model. The first case of $W\neq0$, namely $W=D$, has been analyzed in
Refs. \cite{ba} and \cite{doering3}. Notice from Eq. (\ref{e44}) that
with $W=D$, one does not require to evaluate $\rho_{k_f-1}$ any longer
--- the front speed and front diffusion are simply given by
$v_N=\varepsilon$ and $D_f=D+\varepsilon/2$ \cite{doering3,ba}.

Clearly enough, when $W\neq0$ but $W\neq D$, one needs to evaluate
$\rho_{k_f-1}$. Reference \cite{deb5} considered this case, but only
with all parameters of the same order of magnitude. In this case,
having estimated $\rho_{k_f-1}$ by
$N=\lim_{m\rightarrow\infty}\rho_{k_f-m}$ (see footnote \ref{fn2.27}),
one obtains \cite{deb5}
\begin{eqnarray}
v_N\,=\,\frac{\varepsilon\,(\varepsilon\,+\,D)}{\varepsilon\,+\,W}\quad\mbox{and}\quad
D_f\,=\,\frac{(\varepsilon\,+\,2W)(D\,+\,\varepsilon)}{2(\varepsilon\,+\,W)}\,.
\label{e51}
\end{eqnarray}
\begin{figure}[ht]
\begin{center}
\includegraphics[width=0.4\textwidth,angle=270]{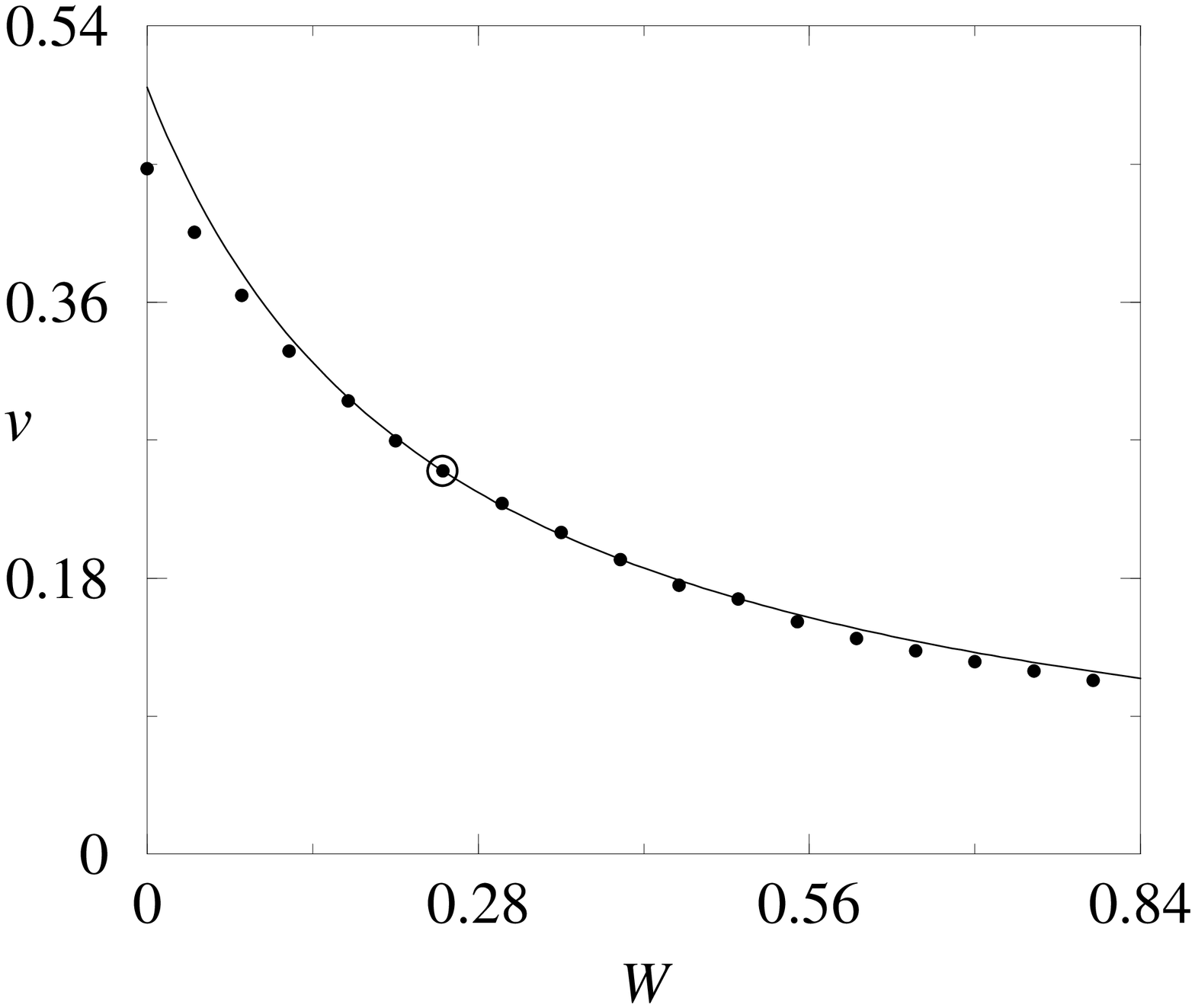}\hspace{5mm}
\includegraphics[width=0.4\textwidth,angle=270]{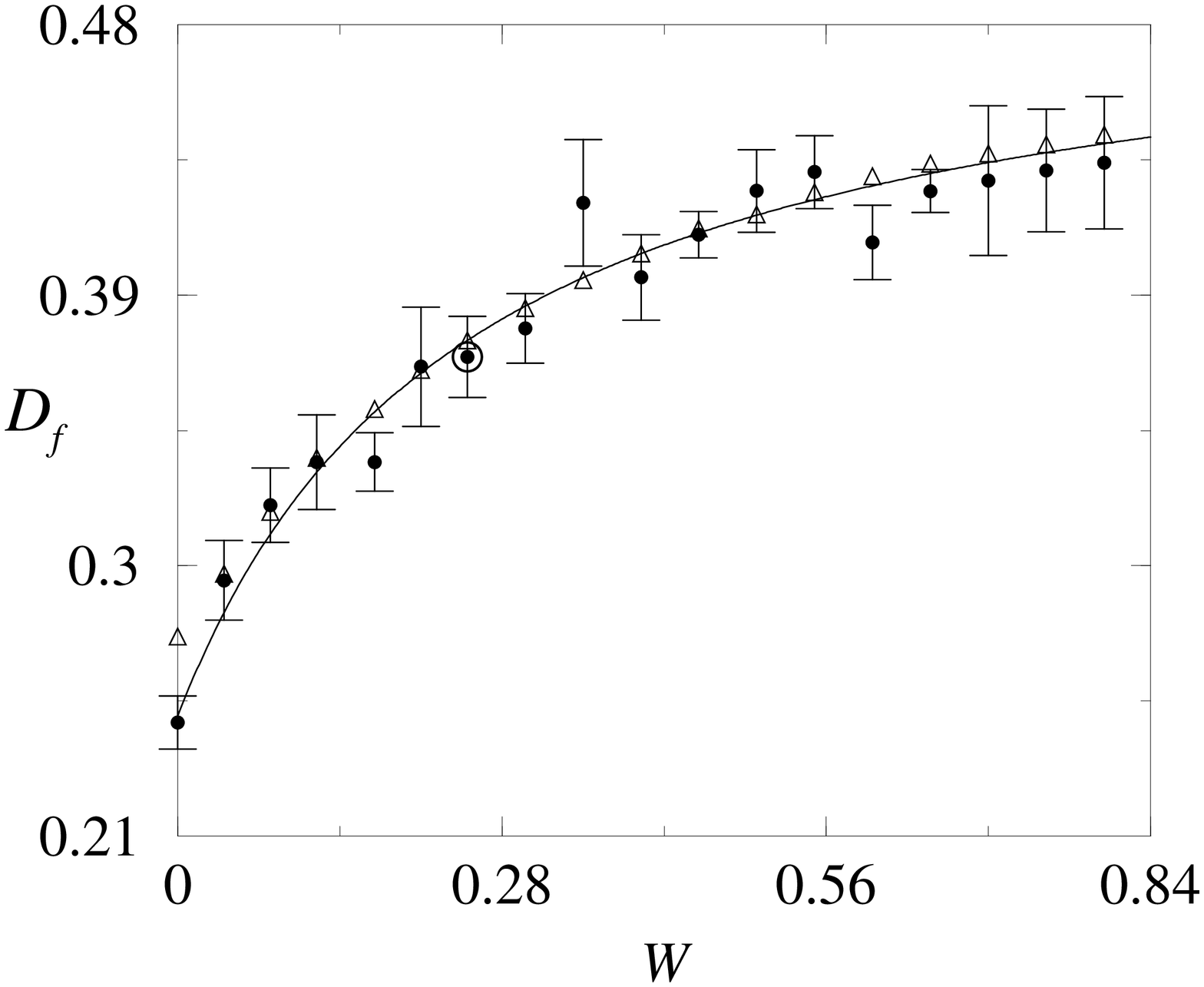}
\end{center}
\caption{Comparison of Eq. (\ref{e51}) with stochastic simulation data
for $D=\varepsilon=0.25$ are presented in Fig. \ref{deb1}  as a
function of $W$ for $D=\varepsilon=0.25$. The larger open circles in
the two graphs denote the results of $D=W$ case analyzed in
Refs. \cite{doering3,ba}.\label{deb1}}
\end{figure}
The comparison of Eq. (\ref{e51}) with stochastic simulation data for
$D=\varepsilon=0.25$ are presented in Fig. \ref{deb1}  as a function
of $W$ for $D=\varepsilon=0.25$. The larger open circles in the two
graphs of Fig. \ref{deb1} denote the results of $D=W$ case analyzed in
Refs. \cite{doering3,ba}. Notice that close to $W=0$ and at larger
values of $W$, the agreement between Eq. (\ref{e51}) with stochastic
simulation data is not good. It is in fact caused by the errors
incurred in estimating $\rho_{k_f-1}$ by $N$, as shown in
Fig. \ref{deb2} by means of the relative deviation $d=(\rho_k-N)/N$
between for a few lattice sites behind the foremost particle.
\begin{figure}[ht]
\begin{center}
\includegraphics[width=0.5\textwidth,angle=270]{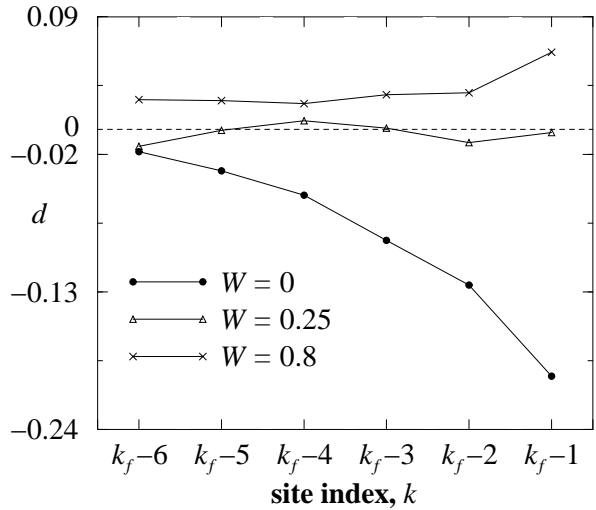}
\end{center}
\caption{Relative deviation $d=(\rho_k-N)/N$ of the average density
from $N= \varepsilon/(\varepsilon+W)$ for the first six lattice sites
to the left of the foremost occupied lattice site, $k_f$, for
$D=\varepsilon=0.25$ and three different values of $W$.\label{deb2}}
\end{figure}

In principle, for $W\neq0$, it should be possible to perform an
analysis in the spirit of Refs. \cite{kerstein,kerstein1} to get
successively more accurate expressions for $\rho_{k_f-1}$, and
correspondingly for the front speed and diffusion coefficient. In
particular, such extensions might allow one to use the results in a
wider parameter range, such as $D/W\rightarrow\infty$ while
$D/\varepsilon\sim{\cal O}(1)$, or $D/\varepsilon\rightarrow\infty$
while $W/\varepsilon\sim{\cal O}(1)$. However, inspection of the
earlier analysis suggests that such higher order analytical
expressions of $\rho_{k_f-1}$ are less trivial to obtain than one
might expect at first sight. More precisely, for $W=0$, the master
equation for the probability $p_j$ that the two foremost particles are
separated by $j$ lattice sites can be closed in a simple manner
\cite{kerstein,kerstein1}. For $W=0$ formalism, no particle gets
annihilated, and as a result, the hierarchy of equations for the joint
probability density distribution of the two foremost particles can be
closed easily at the simplest level, since in the absence of
annihilation, the third foremost particle never becomes the second
foremost particle (``uncorrelated two-particle approach''). At this
level, the expression of $\rho_{k_f-1}$ can then be analytically
solved, leading to a better approximation than what we use in this
paper for $W=0$. Of course, the master equation can be closed at a
higher level, by considering more than two foremost particles to
determine $\rho_{k_f-1}$, but then one does not obtain an analytical
expression of $\rho_{k_f-1}$ (``correlated two-particle
self-consistent approach''). As soon as $W\neq 0$, this is not true
anymore: consider the following situation where the two foremost
particles are next to each other. With annihilation of particles
allowed, one of them can annihilate the other, and then the
probability distribution function of the two foremost particles is
crucially coupled to those which involve particles further back at the
simplest level. It is therefore clear that for $W\neq 0$, the master
equation for $p_j$ involves particles that are further back. While it
is certainly possible to solve the master equation numerically, it
does not appear to lead one to an analytical expression of
$\rho_{k_f-1}$ that provides a better approximation than the one
already used in Ref. \cite{deb5}.\vfill

\subsubsection{Fluctuating ``Pulled'' Fronts in Higher than One Spatial
Dimensions\label{sec2.6.6}}

Up until now in Sec. \ref{sec2.6}, we have only considered fronts on a
one-dimensional lattice --- and we have witnessed that with $N$ being
the conditionally averaged number of particles per lattice site in the
stable phase for propagating fronts into unstable states, the front
propagates with a speed $v_N$, and simultaneously the front positions
in different realizations of an ensemble undergoes {\it diffusive\/}
wandering around its ensemble averaged position. In higher than one
spatial dimensions, the situation gets more complicated, although the
basics of front propagation in discrete particle and lattice systems
remain unchanged --- therein the interest is in the wandering
properties of the front, not so much in its speed
\cite{doering1,goutam,moro}.

\begin{figure}[ht]
\begin{center}   
\includegraphics[width=0.5\linewidth]{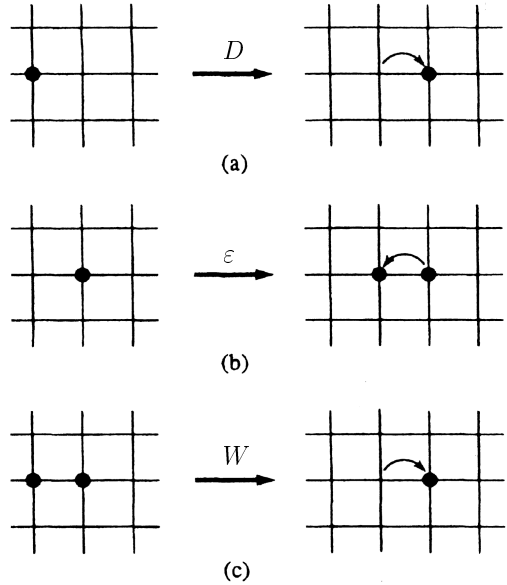}
\end{center}
\caption{Microscopic processes in the $d$-dimensional
reaction-diffusion system X$+$X$\leftrightharpoons2$X along the $k_l$
direction ($l=1,2,\ldots,d$) for a given value of $l$. On any lattice
site, at most one X particle is allowed at any time. The dynamics
along any one of the $k_l$ directions is exactly the same as that of
Fig. \ref{depsw}.}
\label{doeringfig1}
\end{figure}
Since all the studies on discrete particle and lattice system of
propagating fronts in higher than one spatial dimensions have been on
reaction-diffusion systems  X$+$X$\leftrightharpoons2$X
\cite{doering1,goutam,moro} or X$+$Y$\rightarrow2$X \cite{sokolov3},
to understand the general idea, it is best to first follow the
reaction-diffusion  system X$+$X$\leftrightharpoons2$X of
Ref. \cite{doering1} on a $d$ dimensional {\it square\/} lattice for
various values of $d$ and briefly discuss its results. The system
consists of a long lattice in the $k_1$ direction $(-L_1\leq k_1\leq
L_1)$ with a transverse length $L=L_2=L_3=\ldots=L_{d}$. Any lattice
site can hold at most one X particle at any time. The microscopic
dynamics of the particles along the $k_l$ direction ($l=1,2,\ldots,d$)
for a given value of $l$ is shown in Fig. \ref{doeringfig1} --- the
dynamics along any one of the $k_l$ directions is exactly the same as
that of Fig. \ref{depsw}.  With periodic boundary conditions applied
along all the transverse directions $k_2,k_3,\ldots,k_d$ and the
initial condition for the particle density
$\phi_{\vec{k}}(t=0)=\Theta(k_1)N$, one studies front propagation in
the $k_1$-direction.\footnote{Even in $d$ spatial dimensions,
$N=\varepsilon/(\varepsilon+W)$, just like the one-dimensional model
of Fig. \ref{depsw}. See footnotes \ref{fn2.27} and
\ref{e60}.\label{fnN}}

From this perspective, it is conceivable that the front properties can
be studied by simply projecting the quantities defined in $d$ spatial
dimensions on to the $k_1$ direction. Precisely this theme was
developed in Ref. \cite{doering1}: a mean particle density
$\rho_{k_1}(t)$ was introduced by projecting the particle density
profile in the $d$-dimensional space on to the $k_1$ direction as
$\rho_{k_1}(t)=\left[\sum_{j=2}^d\phi_{\vec{k}}(t)\right]/L^{d-1}$.\footnote{It
is then clearly seen that $\rho_{k_1}(t)$ obeys the Fisher-Kolmogorov
equation
\begin{eqnarray}
\frac{\partial\rho_{k_1}}{\partial
t}\,=\,D[\rho_{k_1+1}\,+\,\rho_{k_1-1}\,-\,2\rho_{k_1}]\nonumber\\&&\hspace{-3.5cm}+\,[\rho_{k_1\,+\,1}\,+\,2(d-1)\rho_{k_1}\,+\,\rho_{k_1-1}]\,[\varepsilon\,-\,(\varepsilon+W)\rho_{k_1}]\,.
\label{e60}
\end{eqnarray}
Notice that the corresponding one-dimensional problem in
Sec. \ref{sec2.6.5} also yields Eq. (\ref{e60}) with $d=1$ (although
we never quite discussed it in Sec. \ref{sec2.6.5}). From
Eq. (\ref{e60}), we find that $N=\varepsilon/(\varepsilon+W)$ for any
$d$, just as we found in Eq. (\ref{e32}).\label{fn50}} One of the
natural consequences of having to study the front properties by means
of this projection is that one {\it cannot\/} define a front position
based on the foremost particle as we have done so far in
Secs. \ref{sec2.6.2}-\ref{sec2.6.5}. To bypass this problem, a second
quantity
$K_1(t)=\left[\sum_{k_1=-L_1}^{L_1}\rho_{k_1}(t)\right]/N-L_1$ was
introduced in Ref. \cite{doering1} for the front position for any
single realization. While the ensemble average $\langle
K_1(t)\rangle\sim v_Nt$ at long times, it was proposed that the
wandering properties of the front around its ensemble averaged front
position $\langle K_1(t)\rangle$ as a function of time can be measured
by tracking the time development of quantity $w(t)$ \cite{doering1},
where [the angular brackets in Eq. (\ref{e61}) also denote ensemble
averaging]
\begin{eqnarray}
w^2(t)\,=\,\left\langle\frac{2}{N}\,\sum_{k_1=-L_1}^{L_1}k_1\rho_{k_1}(t)\,-\,\langle
K_1(t)\rangle^2\,-\,L^2_1\right\rangle\,.
\label{e61}
\end{eqnarray}
With these propositions, Ref. \cite{doering1} invoked the scaling
hypothesis for $w(t)$, i.e., $w(t)\sim t^{\alpha}F(t/L^\beta)$, and
then continued further on to measure $\alpha$ and $\beta$ from
simulations: for $d=2$, $\alpha=0.272\pm0.007$ and $\beta=1$; for
$d=3$, the data are indistinguishable between the two possibilities
$w(t)\sim t^{0.1}$ and $w(t)\sim\sqrt{\ln t}$; and for $d=4$,
$\alpha=0$ (all simulations were carried out with $D=0.5$ and
$\varepsilon/W=0.1$).\footnote{At this juncture we briefly discuss the
results of Ref. \cite{sokolov3}. This paper considers front
propagation in a three-dimensional reaction-diffusion system
X$+$Y$\rightarrow2$X, where initially the whole lattice is full of Y
particles with an equilibrium concentration $N$, and at time $t=0$ one
X particle is introduced at one location. Both the species diffuse
with equal rate $D$ on the three-dimensional nearest neighbour lattice
sites, and any Y particle coming in contact with any other X particle
on the same lattice site is instantaneously and irreversibly converted
to an X particle --- this is the $d$-dimensional generalization of the
one-dimensional lattice model of Sec. \ref{sec2.6.2}. The object of
study was to see the deviations of $v_N$ from the prediction of the
Fisher-Kolmogorov equation, $v^*$. The numerical finding was that
$v_N$ deviates from $v^*$. Not only that there was no study of the
broadening of the front region by means of measuring $w(t)$ therein,
but also the front speed as a function of the total average particle
density was not carefully fitted with any theoretical curve. In this
context, we refer to Fig. 2 of Ref. \cite{moro}, which shows, for the
reaction-diffusion model X$+$X$\leftrightharpoons2$X of
Fig. \ref{doeringfig1}, that for $\varepsilon/D\rightarrow0$, front
speed $v_N$ in $d>1$ does behave as $\sqrt{\varepsilon}$ as the
Fisher-Kolmogorov equation would predict.\label{fn52}}

It turns out however, that for the reaction-diffusion model
X$+$X$\leftrightharpoons2$X of Fig. \ref{doeringfig1}, invoking the
scaling relation to measure the front wandering is an extremely tricky
issue. In fact, although $w(t)$ in Eq. (\ref{e61}) is {\it a\/}
measure of the front wandering, the full measure of wandering of the
$(d-1)$-dimensional interface for fronts in $d$ spatial dimensions is
already lost when the projection $\rho_{k_1}(t)$ of
$\phi_{\vec{k}}(t)$ along the $k_1$ direction is taken {\it before\/}
the front wandering is defined. Truly speaking, if $\phi_{\vec{k}}(t)$
is not reduced to $\rho_{k_1}(t)$ before the scaling relation is
invoked, then on a $d$-dimensional lattice, the problem of front
wandering can be related to interfacial growth and associated scaling
analyses of the interfacial roughness
\cite{kpz,krug2,halpin,barabasi,krug3}. The scaling idea in these
interfacial growth phenomena is the following: for interfacial growth
in $d$ dimensions spanned by $(k_1,k_2,\ldots,k_d)$ co-ordinates
[leading to a $(d-1)$-dimensional interface spanned by
$(k_2,k_3,\ldots,k_d)$ co-ordinates], the ``height'' of the interface
in the $k_1$ direction is described by
$h_{k_2,k_3,\ldots,k_d}(t)\equiv h(t)$ at time $t$. With overline
denoting the projection of a $d$-dimensional quantity on to one (along
the $k_1$ direction) dimension [e.g.,
$\rho_{k_1}(t)=\overline{\phi_{\vec{k}(t)}}\,$], and the angular
brackets denoting ensemble average, the actual scaling relation is
given by ${\cal W}(t)=t^\alpha F(t/L^\beta)$, where ${\cal
W}^2(t)=\langle\overline{[h(t)-\overline{h(t)}]^2}\rangle$. When the
argument $\eta$ of $F(\eta)$ is $\ll1$, then $F(\eta)$ behaves like a
constant; on the other hand, for $\eta\gg1$,
$F(\eta)\sim\eta^{-\alpha}$. This implies that for $t\gg L^\beta$,
${\cal W}(t)$ saturates at ${\cal W}_{\mbox{\scriptsize sat}}\sim
L^\gamma$, where $\gamma=\alpha\beta$.

Even with this modification, the actual values of $\alpha$ and $\beta$
in different dimensions for the X$+$X$\leftrightharpoons2$X
reaction-diffusion model of Fig. \ref{doeringfig1} are {\it not} free
from ambiguity \cite{goutam,moro}.\footnote{We note here that only
$W=D$ case was considered in Ref. \cite{moro}.\label{fn54}} The
ambiguity stems from how $h(t)$ is related to $\phi_{\vec{k}}(t)$
[although it is clear that one should not use $\rho_{k_1}(t)$ for
$h(t)$ to draw analogy between the front wandering properties and the
interface growth roughness scaling relations on a $d$-dimensional
lattice]. To illustrate this point, let us consider the example case
for $d=2$. It turns out that when $h(t)$ equals a simple local
coarse-graining over the field values $\phi_{\vec{k}}(t)$ within a
cubic box of size $(2\ell+1)^d$ around $\vec{k}$, then for $\ell=1$,
one obtains $\alpha=0.29\pm0.01$ and $\gamma\simeq0.4\pm0.02$
\cite{goutam} ($\alpha=0.27\pm0.01$ and $\gamma\simeq0.41\pm0.02$
\cite{moro}). Interestingly enough, these exponents are surprisingly
close to the KPZ exponents for a two-dimensional interface
\cite{goutam} ($\alpha=0.245\pm0.003$, $\gamma=0.393\pm0.003$
\cite{marinari}). However, it has been shown that with increasing
values of $\ell$, i.e., $\ell=2,3,4$ etc., the exponents $\alpha$ and
$\gamma$ approach the KPZ exponents for a one-dimensional interface
($\alpha=1/3$, $\gamma=0.5$) \cite{moro}. So the general picture that
emerges from these analyses is that as one looks at progressively
longer length scales, the wandering properties of the front makes a
transition from a non-KPZ to a KPZ behaviour \cite{moro}.

With $\ell=4$, the KPZ scaling behaviour for the front wandering on a
three ($\gamma=0.393$ \cite{marinari}) and a four dimensional lattice
($\gamma=0.313$ \cite{marinari}) for the X$+$X$\leftrightharpoons2$X
reaction-diffusion model of Fig. \ref{doeringfig1} has been recovered
\cite{moro}. These results together indicate that in this model, any
generic upper critical dimension has to be higher than $4$. More
simulations are however needed to conclusively clarify this point.

\subsection{Convergence of the Asymptotic Front
Speed to $v^\dagger$ for Fluctuating Pushed Fronts as
$N\rightarrow\infty$\label{sec2.7}}

We had earlier pointed out that at the time of writing this review
article, I have not seen any work on how $v_N$ behaves as a function
of $N$ in the limit $N\rightarrow\infty$. Nevertheless, in
Sec. \ref{sec2.7}, we will show that if one incorporated the discrete
nature of the particles by means of an effective cutoff\footnote{The
introduction of an effective cutoff to model the discreteness of
particles, even for fluctuating pushed fronts, is not
unusual. Consider for example the following deterministic equation in
continuous space and time \cite{bj,vs2}
\begin{eqnarray}
\frac{\partial\phi}{\partial t}\,=\,D\,\frac{\partial^2\phi}{\partial
x^2}\,+\,\phi\,+\,(b^{-1}-1)\phi^2\,-\,b^{-1}\phi^3
\label{footnoteeq1}
\end{eqnarray}
for $b>0$. This system then has a linearly unstable state at $\phi=0$,
and a stable state at $\phi=1$. In a discrete particle and lattice
model of this model with $\phi_k=\langle\langle N_k\rangle\rangle/N$,
where $N$ is the (conditionally) average number of particles per
lattice site at the stable phase of the front, at the tip, the value
of $\phi$ is once again of ${\cal O}(1/N)$. The effect of discreteness
of particles on the front, once again, can then be mimicked by putting
a growth cutoff for $\phi$ at $\varepsilon\simeq1/N$.\label{fn57}}
following the ideas of Brunet and Derrida as described in
Sec. \ref{sec2.2} \cite{bd1}, then the corresponding front speed $v_N$
must behave as $v^\dagger-v_N\sim N^{-\gamma}$, where $\gamma$ is a
positive number (c.f. footnote \ref{pushednot}).

The route we follow to demonstrate this power law convergence of $v_N$
to $v^\dagger$ is by means of considering an example of a front
propagation from a stable state to a (meta)stable state in \cite{kns}
\begin{eqnarray}
\frac{\partial\phi}{\partial t}\,=\,D\,\frac{\partial^2\phi}{\partial
x^2}\,+\,(1-\phi^2)(\phi+a)\,,
\label{e63}
\end{eqnarray}
where $0<a<1$. This equation, which appears with the name of
Ginzburg-Landau equation or with the so-called Schl\"{o}gl model, has
two stable states, $\phi=\pm1$, and an unstable state at
$\phi=-a$. For a steep initial condition that connects $\phi=1$ to
$\phi=-1$, a front propagates from the $\phi=1$ to the $\phi=-1$ state
with speed $v^\dagger=a\sqrt{2}$; and the corresponding front profile
in terms of $\zeta=x-v^\dagger t$ is described by \cite{kns}
\begin{eqnarray}
\phi^\dagger(\zeta)\,=\,-\tanh\left[\lambda^\dagger(\zeta-\zeta_0)/2\right]\,,
\label{e64}
\end{eqnarray}
such that
$\lambda^\dagger=[v^\dagger+\sqrt{v^{\dagger\,2}+8D(1-a)}]/(2D)$.\footnote{It
is not surprising at all that the front solutions in
Eqs. (\ref{footnoteeq1}) and (\ref{e63}) bear strong similarities
\cite{dee} --- with appropriate redefinition of $\phi$ and rescaling
of the diffusion coefficient $D$ and time $t$, one equation can be
converted into another. However, one has to be careful here: despite
the convertibility of one form to another, the difference between the
nature of the fronts in Eqs. (\ref{footnoteeq1}) and (\ref{e63}) stems
from the fact that $b>0$ in Eq. (\ref{e63}) corresponds to $a>1$ in
Eq. (\ref{footnoteeq1}), for which the front propagation is into a
{\it linearly\/} unstable state. On the other hand, $0<a<1$ in
Eq. (\ref{footnoteeq1}) corresponds to $b<0$ in Eq. (\ref{e63}) --- in
this case, front propagates from a stable to an (meta)stable
state.\label{fn58}} Notice that the Ginzburg-landau equation
(\ref{e63}) does not fall in the category of fronts propagating into
unstable state, the procedure to obtain
$v^\dagger-v_\varepsilon\sim|\varepsilon|^{-\gamma}$ in
Eq. (\ref{e63}) is very instructive, and it can be easily applied to
pushed fronts propagating into unstable states.\footnote{Notice that
with $v^\dagger=a\sqrt{2}$, the front profile (\ref{e64}) is directly
obtained by solving Eq. (\ref{e63}). In general, in the comoving frame
for any given asymptotic front speed $v_{\mbox{\scriptsize as}}$, upon
linearizing Eq. (\ref{e63}) around $\phi=-1$ and using the front
profile ansatz $\phi(\zeta)\sim-1+Ae^{-\lambda\zeta}$ for
$\zeta\rightarrow\infty$, two roots of the exponent $\lambda$ are
obtained: $\lambda_1=[v_{\mbox{\scriptsize
as}}+\sqrt{v^2_{\mbox{\scriptsize as}}+8D(1-a)}]/(2D)>0$ and
$\lambda_2=[v_{\mbox{\scriptsize as}}-\sqrt{v^2_{\mbox{\scriptsize
as}}+8D(1-a)}]/(2D)<0$. In the corresponding front solution
$\phi(\zeta)\sim-1+A_1e^{-\lambda_1\zeta}+A_2e^{-\lambda_2\zeta}$,
$A_2$ must be zero for the solution to be physically relevant. This
condition is clearly obeyed by the front profile
(\ref{e64}).\label{fn59}}

\begin{figure}[ht]
\begin{center}   
\includegraphics[width=0.6\linewidth]{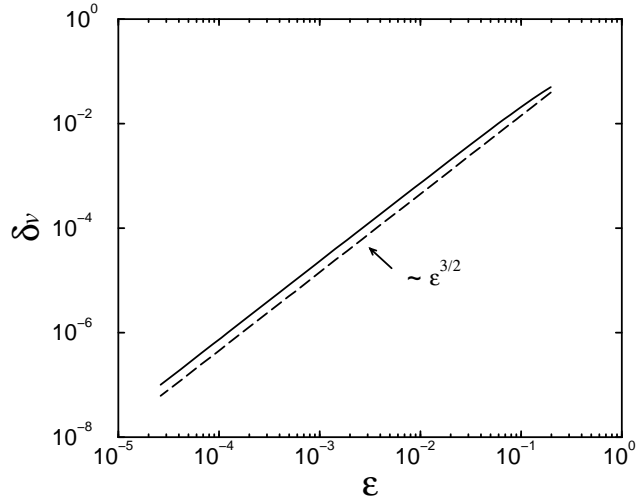}
\end{center}
\caption{The log-log curve of $\delta v=v^\dagger-v_\varepsilon$
vs. $\varepsilon$ for the Ginzburg-Landau equation (\ref{e63}) with
$D=a=1/2$ \protect{\cite{kns}}. Solid line: simulation data, dashed
line: visual aid for the $\delta v\sim\varepsilon^{3/2}$ behaviour
predicted in Eq. (\ref{e66}).}
\label{kesslerfig3}
\end{figure}

As we add a growth cutoff in the Ginzburg-Landau equation (\ref{e63})
at $\phi=-1+\varepsilon$, we can use the technique developed by Brunet
and Derrida (c.f. Sec. \ref{sec2.2} \cite{bd1}) to solve for the
asymptotic front speed $v_\varepsilon$ \cite{kns}. Let us denote the
location, in the comoving co-ordinate $\zeta=x-v_\varepsilon t$, where
the value of $\phi$ reaches $-1+\varepsilon$ by $\zeta_0$. For
$\varepsilon\rightarrow0$, Eq. (\ref{e63}) can clearly be linearized
on the right of $\zeta_0$. In addition, in the left neighbourhood of
$\zeta_0$, the value of $\phi$ is still infinitesimally close to $-1$,
which allows one to linearize Eq. (\ref{e63}) around the unstable
state once again. This yields the following form of the front solution
\cite{kns}\footnote{The reason for expressing the coefficient of
$e^{-\lambda_2\zeta}$ in Eq. (\ref{e65}) in this manner is actually
quite subtle. The idea lies in the fact that just like we linearize
Eq. (\ref{e63}) with a growth cutoff around $-1$ for large $\zeta$ to
obtain  Eq. (\ref{e65}), we can also linearize it around $1$ for
$\zeta\rightarrow-\infty$. This linearization suggests an equivalent
form of the front solution
$\phi^{(0)}(\zeta)=1-B_1e^{\lambda^\dagger_{1'}\zeta}-B_2(v^\dagger-v_\varepsilon)e^{\lambda^\dagger_{2'}\zeta}$
for $\zeta\rightarrow-\infty$, where
$\lambda^\dagger_{1'}=[-v^\dagger+\sqrt{v^{\dagger\,2}+8D(1+a)}]/(2D)>0$
and
$\lambda^\dagger_{2'}=[-v^\dagger-\sqrt{v^{\dagger\,2}+8D(1+a)}]/(2D)<0$. The
linearized equations at $\zeta\rightarrow\pm\infty$ involve
$v_\varepsilon$ as an input parameter. However, in this form of
$\phi^{(0)}(\zeta)$ at $\zeta\rightarrow\pm\infty$, $A_1$, $A_2$,
$B_{1'}$ and $B_{2'}$ all become {\it independent\/} of
$\varepsilon$.\label{fn60}}
\begin{eqnarray}
\phi^{(0)}(\zeta)\,=\,-1\,+\,A_1\,e^{-\lambda^\dagger_1\zeta}\,+\,A_2\,(v^\dagger-v_\varepsilon)\,e^{-\lambda^\dagger_2\zeta}\,\,\,\mbox{for
large $\zeta$, but
$\zeta<\zeta_0$}\nonumber\\&&\hspace{-11.2cm}=\,-1\,+\,\varepsilon
e^{-v_\varepsilon(\zeta-\zeta_0)}\quad\quad\quad\quad\quad\quad\quad\quad\quad\!\!\mbox{for
$\zeta\geq\zeta_0$}\,,
\label{e65}
\end{eqnarray}
where the expressions of $\lambda^\dagger_1$ and $\lambda^\dagger_2$
are obtained from footnote \ref{fn59} after having
$v_{\mbox{\scriptsize as}}$ substituted by $v^\dagger$. As the two
solutions of Eq. (\ref{e65}) and their derivatives are matched at
$\zeta_0$ (with the condition that $A_1$ and $A_2$ are independent of
$\varepsilon$; see footnote \ref{fn60}), it is easily seen that
$\zeta_0\sim|\ln\varepsilon|$ and \cite{kns}\footnote{Equation
(\ref{e66}) has also been obtained by Chomaz {\it et al\/}
\cite{physics11}, in a study of the influence of a fixed boundary in a
system with a convective instability.\label{chomaz}}
\begin{eqnarray}
v^\dagger-v_\varepsilon\,\sim\,\varepsilon^{1-\lambda^\dagger_2/\lambda^\dagger_1}\,.
\label{e66}
\end{eqnarray}
For $D=a=1/2$, $\lambda^\dagger_2/\lambda^\dagger_1=-1/2$, and
$v^\dagger-v_\varepsilon\sim\varepsilon^{3/2}$. This scaling of
$v^\dagger-v_\varepsilon$ has been observed in the simulation
\cite{kns}, and it is shown in Fig. \ref{kesslerfig3}.

As already pointed out, the exercise can be repeated to obtain a power
law behaviour of $v^\dagger-v_\varepsilon$ for the
Eq. (\ref{footnoteeq1}), where the front does propagate into an
unstable state. Based on Brunet and Derrida's insight that the effect
of discreteness in terms of particles and the lattice is well captured
by the choice $\varepsilon\sim1/N$ as $N\rightarrow\infty$, the above
analysis indicates that in a discrete particle and lattice model of
fluctuating pushed fronts $v^\dagger-v_N$ would indeed behave as
$N^{-\gamma}$ for some $\gamma>0$. However, at the time of writing
this review article, I have not seen any empirical verification of
this prediction at this point.

\subsection{Epilogue I \label{sec2.8}}

As we discussed front propagation in discrete particle and (mostly
one-d\-i\-m\-e\-n\-s\-i\-o\-n\-a\-l) lattice systems throughout
Sec. \ref{sec2}, we started with an asymptotically large value of $N$
in Sec. \ref{sec2.2}, and from there onwards we gradually reduced the
value of $N$ till Sec. \ref{sec2.6}. We have witnessed that pulled
fronts in discrete particle and lattice systems do not exist, at best
they are weakly pushed. With decreasing value of $N$, the weakly
pushed nature of these fronts becomes more and more prominent, and the
front properties start to depend heavily on the details of the models
under consideration. Because of these complications at intermediate
values of $N$, in general, it is very difficult to obtain a
first-principle based predictive expression of $v_N$ for fluctuating
``pulled'' fronts. In a contradictory trend to this, however, at the
limit of $N$ when there is at most one particle allowed per lattice
site, front speed and the front diffusion coefficient become tractable
again (c.f. Sec. \ref{sec2.6.5}), giving rise to exact relation like
Eq. (\ref{e46}), or approximate theoretical expressions like
Eqs. (\ref{e46}) and (\ref{e51}).

Even if we exclude the small or moderately large (but not
asymptotically large!) values of $N$ from these concluding remarks on
Sec. \ref{sec2}, so many different features of discrete particle and
lattice systems of fronts propagating into unstable states for
$N\rightarrow\infty$ discussed in Secs. \ref{sec2.2},
\ref{sec2.4}-\ref{sec2.5} and \ref{sec2.7} may at first seem a bit
puzzling to the reader. A moment's reflection, however, shows that the
origin of such versatile phenomena can be intuitively understood in a
unified perspective --- to be more precise, from the stability
spectrum of (deterministic) pulled and pushed fronts propagating into
unstable states, already briefly discussed at the end of
Sec. \ref{sec1.1}: the gapless stability spectrum of pulled fronts
leads to its power law convergence (\ref{e4}) of the front speed
$v(t)$ to $v^*$ in time for the asymptotic front speed
\cite{wimreview,vs1,ebert} has already been noted in
Sec. \ref{sec1.1}. On the other hand, stability spectrum of pushed
fronts is gapped \cite{wimreview,vs1,oono1,oono2,oono3,kns,ebert1},
which leads to exponential convergence of the front speed $v(t)$ to
$v^\dagger$ \cite{wimreview,kns}.

What we observe in fluctuating ``pulled'' fronts and as well as in
fluctuating pushed fronts for $N\rightarrow\infty$ is intimately
connected to their stability spectra. For pulled fronts, the gapless
nature of the stability spectrum is indeed another representation of
the linear marginal stability criterion. It therefore must come as no
surprise that any change in the front dynamics close to the linearly
unstable state must affect the front properties severely. In
Secs. \ref{sec2.2} and \ref{sec2.5} this is manifested by the very
slow convergence of $v_N$ and $D_f$ to $v^*$ and zero respectively,
while the stability spectrum of fluctuating ``pulled'' fronts itself
becomes gapped with a very small gap $\propto1/\ln^2N$
(c.f. Sec. \ref{sec2.4}). On the other hand, for pushed fronts, the
spectrum is gapped, and any change in the front dynamics close to the
linearly unstable [or (meta)stable] state is expected to affect the
front properties minimally. This is then reflected in the stronger
power law (in $N$) convergence of $v_N$ to $v^\dagger$.\footnote{Let
us just state here that the theoretical arguments indicate that the
diffusion coefficient of fluctuating pushed fronts made of discrete
particles on a lattice should scale as $1/N$; however, for now we will
leave this issue until Sec. \ref{sec4.2.1}.\label{fnpusheddiff}}

We finally close Sec. \ref{sec2} on discrete particle and lattice
systems of fluctuating fronts propagating into unstable states by
pointing out a few problems, whose solutions are not yet completely
settled. These are: (a) for fluctuating ``pulled'' fronts a first
principle based {\it predictive\/} theory for the front speed $v_N$ at
intermediate values of $N$, (b) for the model of Sec. \ref{sec2.6.3},
why one observes mean-field Fisher-type behaviour even at a reasonably
small value of $N$, (c) a better analytical prediction for the model
of Sec. \ref{sec2.6.5} for a larger set of parameter values, and (d)
possible existence of an upper critical dimension for fluctuating
``pulled'' fronts, when there are at most one particle allowed per
lattice site (c.f. Sec. \ref{sec2.6.6}).

\section{Field-theory of Fluctuating Fronts: External
Fluctuations\label{sec3}}

\subsection{External Fluctuations, Multiplicative Noise and Novikov's
Theorem\label{sec3.1}}

\vspace{-4mm}  We have seen earlier in Sec. \ref{sec1.1} how
propagating fronts have been observed in many systems related to
physics, chemistry and biology. In all of these cases, the evolution
equations of the propagating fronts invariably involve some parameters
that relate to experimental situations, such as parameters in chemical
or biological reaction kinetics, an externally imposed fields
etc. When these parameters correspond to externally imposed fields,
then the variation in these fields acts as a varying external
influence on the system --- such as an externally imposed electric or
magnetic field, or light intensity in a photosensitive
reaction. Similarly, when these parameters originate within the system
itself, then usually they are law-of-mass-action-based mesoscopic
mean-field theory estimates of quantities that originate in the
underlying microscopic dynamics. For the most part, (Langevin-type)
field-theoretical studies on the effects of fluctuations in systems
with propagating fronts have been motivated by these phenomenological
considerations. Other significant motivating factors for the
field-theoretical approaches have been phase transitions induced by
noise \cite{sancho1,parrondo1,becker1,parrondo2}, and the discovery
that surface roughness in various growth processes fall under a
handful number of universality classes
\cite{barabasi,krug2,krug3,kpz,edwards1}.

In my opinion, for systems admitting propagating front solutions,
external fluctuations are easier to handle than the internal ones ---
not only will we find ample evidences of this later on, but also two
simple examples at this stage will make this point clear. Consider a
front dynamics under the influence of an external electric field
$\vec{E}=E\hat{x}$ that couples linearly to the front field $\phi$,
also propagating in the $x$-direction. Fluctuations in the electric
field of magnitude $\delta E(x,t)$ around its mean value $E^{(0)}$
simply gives rise to a term $\propto\delta E(x,t)\phi(x,t)$ in the
evolution equation of the front. Contrast this situation with a
chemical reaction front: therein the deterministic front evolution
equation is already a mesoscopic (mean-field) description of the
underlying microscopic process --- these mesoscopic descriptions
collectively give rise to the very definition of the front field
$\phi$, and as well as to the mass-action based parameters in the
dynamics such as the reaction rate. When fluctuations are to be
considered in such systems, then in general,  fluctuations in both the
number of particles per correlation volume (which usually defines the
front field) and as well as in the parameters are to be considered
together. In the full description of these fluctuations, naturally,
complicated correlations between the fluctuating front field and the
parameters are unavoidable. To get a reasonable grip on the
theoretical handling of these processes, the mass-action based
mesoscopic parameters for the internal dynamics are often kept intact
in the dynamics of the system (we have seen a large number of examples
of it in Sec. \ref{sec2}) and separate fluctuation terms are added by
hand to the evolution equations to mimic the internal fluctuations.

Precisely because of these subtleties, field theories for the internal
fluctuations often prove to be very tricky. For now, we will leave it
for Sec. \ref{sec4}. As for Sec. \ref{sec3}, we will devote it to the
field-theory of external fluctuations on propagating fronts.

To begin with, let us consider the deterministic equation for a scalar
front in one spatial dimension coupled to $n$ external fields $a_i$,
$i=1,2,\ldots,n$:
\begin{eqnarray}
\frac{\partial\phi}{\partial t}\,=\,{\cal F}(\{a_i\},\phi)\,.
\label{e67}
\end{eqnarray}
The effect of the fluctuations in the fields $a_i$ around their
mean-field values $a^{(0)}_i$ on the front properties is then given by
\begin{eqnarray}
\frac{\partial\phi}{\partial t}\,=\,{\cal
F}(\{a^{(0)}_i\},\phi)\,+\,\sum_{i=1}^{n}\,\sum_k\frac{1}{k!}\,\frac{\delta^k{\cal
F}}{\delta
a_i^k}\bigg|_{a^{(0)}_i}\,\left[a_i(x,t)-a^{(0)}_i\right]^k\,.
\label{e68}
\end{eqnarray}
In Eq. (\ref{e68}), the functional derivative
$\displaystyle{\frac{\delta^k{\cal F}}{\delta
a^k_i}\bigg|_{a^{(0)}_i}}$ is a function of the front field. What is
the precise form of this function depends on how the fields are
coupled to the front dynamics, but the fact remains that in the
evolution equation of the front, the coefficients of the fluctuations
in the external fields are multiplied by functions of the front field
--- Eq. (\ref{e68}) is thus an equation with {\it multiplicative
noise}.

Before we proceed further, it is however necessary to first properly
interpret the multiplicative noise term in Eq. (\ref{e68}). This
brings us to the well-documented It\^{o} vs. Stratonovich dilemma for
multiplicative noise \cite{vankampen,gardiner}. In this review
article, we will not delve deep into the details of which
interpretation is the right one, {\it but it is extremely important to
be clear about the fact that throughout Sec. \ref{sec3}, the noise
term in Eq. (\ref{e68}) will be interpreted in the Stratonovich
sense}. Section \ref{sec3}, in fact, is a collection of works done on
the field theory of external fluctuations on propagating fronts for
the last two decades,\footnote{Field-theoretical studies on the effect
of external fluctuations (as multiplicative noise) on propagating
front in the context of reaction-diffusion systems date back to the
early $1980$s \cite{lutz1,lutz2}. Some threads of it were picked up in
the early $1990$s \cite{lutz3,pasquale}; but all of these works
considered only weak noise, i.e., the case of
$\left|[a_i-a^{(0)}_i]/a^{(0)}_i\right|\ll1$ in Eq. (\ref{e68}). The
late $1990$s till now witnessed a surge of activities in this field
\cite{armeroprl,armero,jaume2,jaume3,santos,wim2,santos1} --- they
considered multiplicative noise that is not necessarily weak.
\label{fnstrato}} and all these works interpreted the noise term
in the Stratonovich sense.

We will now see in Sec. \ref{sec3.1.1} that in the context of front
propagation, the so-called Novikov's theorem \cite{novikov} has proved
to be a very useful tool for theoretical analysis with Stratonovich
interpretation of multiplicative noise. It should however be kept in
mind that under general circumstances, Novikov's theorem is not an
automatic gateway to success in these problems. First of all, the
theorem can handle noise only with Gaussian statistics. Secondly, it
is practically useful only if the spatial and temporal part of the
noise factorize, and in the limit when both the noise correlation
length and correlation time go to zero.

\subsubsection{Novikov's Theorem in the Context of Front
Propagation in Reaction-Diffusion Systems\label{sec3.1.1}}

Consider the stochastic differential equation describing a general
reaction-diffusion process coupled to a single external
field\footnote{Novikov's theorem can be easily generalized to
arbitrary spatial dimensions and as well as to the case when the front
field is a vector field, see Ref. \cite{jaume3}. Generalizations to
more than one external field coupled to the front dynamics is also
trivial.\label{fn62}}
\begin{eqnarray}
\frac{\partial\phi}{\partial t}\,=\,D\frac{\partial^2\phi}{\partial
x^2}\,+\,f(\phi)\,+\,\tilde\varepsilon^{1/2}g(\phi)\eta(x,t)\,,
\label{e69}
\end{eqnarray}
where $\tilde\varepsilon\eta(x,t)$ is the space-time dependent
fluctuation in the external field, $\tilde\varepsilon$ is the strength
of the (multiplicative) noise, and $\eta(x,t)$ has zero mean with a
correlation time $\tau_c$ and correlation length $\lambda_c$. The
factorized temporal and spatial correlations are expressed by the
functions $C_t$ and $C_x$ respectively, such that (the angular
brackets with a subscript $\eta$ denote averaging over the noise)
\begin{eqnarray}
\langle\eta(x,t)\eta(x',t')\rangle_\eta\,=\,2\,C_t\left(|t-t'|;\tau_c\right)\,C_x\left(|x-x'|;\lambda_c\right)\quad\mbox{and}\nonumber\\&&\hspace{-10.55cm}\int
dx'\,C_x(|x-x'|;\lambda_c)\,=\,\int dt'\,C_t(|t-t'|;\tau_c)\,=\,1\,.
\label{e70}
\end{eqnarray}

The difficulty with using Eq. (\ref{e69}) directly is that although
the mean of $\eta(x,t)$ is zero, in the Stratonovich interpretation,
the stochastic term in Eq. (\ref{e69}) [$\propto g(\phi)\eta(x,t)$]
does not have zero mean. {\it This is precisely where Novikov's
theorem comes in --- its purpose is to (non-perturbatively) rewrite
Eq. (\ref{e69}) in such a manner that the fluctuating term in it does
have zero mean}. Novikov's theorem, however, is applicable only if
$\eta(x,t)$ itself is a {\it Gaussian random function\/} with zero
mean \cite{novikov}.
\begin{itemize}
\item[] The exact statement for Novikov's theorem is the following:
Suppose $h(s)$ is a Gaussian random function with zero mean value and
correlation
\begin{eqnarray}
\langle h(s)h(s')\rangle_h\,=\,{\cal H}(s,s')\,,
\label{e71}
\end{eqnarray}
then for any functional $H(\{h\})$, the following result holds
\cite{novikov}:
\begin{eqnarray}
\langle h(s)H(\{h\})\rangle_h\,=\,\int_{-\infty}^{\infty}ds'\,{\cal
H}(s,s')\left\langle\frac{\delta H(\{h\})}{\delta
h(s')}\right\rangle_h\,.
\label{e72}
\end{eqnarray}
\end{itemize}
Equation (\ref{e72}) does not involve any approximations. However, to
make any practical use of it for any analytical calculation in the
present context, N\-o\-v\-i\-k\-ov's theorem is useful only if the
correlation time $\tau_c$ and correlation length $\lambda_c$ become
much smaller than the respective time and length scales of the
propagating front\footnote{For finite values of $\tau_c$ and
$\lambda_c$, and for non-Gaussian forms of $\eta(x,t)$ [and even for
non-factorizable spatiotemporal correlations, unlike Eq. (\ref{e70})],
it is possible to formulate a Novikov-type theorem by carrying out a
cumulant expansion of the noise. See
Ref. \cite{santos1}.\label{cumulant}} --- for example when $\tau_c$ is
much smaller than the inverse reaction rate and $\lambda_c$ is much
smaller than the front width. In these cases, one can use both
$\tau_c$ and $\lambda_c\rightarrow0$ and reduce Eq. (\ref{e70}) to
\begin{eqnarray}
\langle\eta(x,t)\eta(x',t')\rangle_\eta\,=\,2\delta(t-t')\,\delta(x-x')\,.
\label{e73}
\end{eqnarray}

The limit of both $\tau_c$ and $\lambda_c\rightarrow0$, however,
causes certain intricacies in the application of Novikov's theorem, as
we now illustrate below. From Eqs. (\ref{e71}-\ref{e73}) we have
\begin{eqnarray}
\langle
g[\phi(x,t)]\eta(x,t)\rangle_\eta=2\int_{-\infty}^{\infty}dt'\int_{-\infty}^{\infty}dx'\,C_t\left(|t-t'|;\tau_c\right)\,C_x\left(|x-x'|;\lambda_c\right)\times\nonumber\\&&\hspace{-6cm}\times\,\left\langle
g'[\phi(x,t)]\,\frac{\delta\phi(x,t)}{\delta\eta(x',t')}\right\rangle_\eta\,.
\label{new1}
\end{eqnarray}
After having solved $\phi(x,t)$ from Eq. (\ref{e69}) as
\begin{eqnarray}
\phi(x,t)\,=\,\int_{-\infty}^{\infty}dt''\,\left[D\frac{\partial^2\phi(x,t'')}{\partial
x^2}+f[\phi(x,t'')]+\tilde\varepsilon^{1/2}g[\phi(x,t'')]\eta(x,t'')\right]\times\nonumber\\&&\hspace{-8cm}\times\,\Theta(t-t'')\,,
\label{new2}
\end{eqnarray}
the expression
$\displaystyle{\frac{\delta\phi(x,t)}{\delta\eta(x',t')}}$ in
Eq. (\ref{new1}) reads
\begin{eqnarray}
\frac{\delta\phi(x,t)}{\delta\eta(x',t')}\,=\,\tilde\varepsilon^{1/2}\,g[\phi(x,t')]\,\Theta(t-t')\,\delta(x-x')\,,
\label{new3}
\end{eqnarray}
which subsequently implies that
\begin{eqnarray}
\langle
g[\phi(x,t)]\eta(x,t)\rangle_\eta=2\tilde\varepsilon^{1/2}C_x\left(0;\lambda_c\right)\int_{-\infty}^{\infty}dt'\,C_t\left(|t-t'|;\tau_c\right)\,\times\nonumber\\&&\hspace{-5cm}\times\,\left\langle
g'[\phi(x,t)]\,g[\phi(x,t')]\right\rangle_\eta\,\Theta(t-t')\,.
\label{new4}
\end{eqnarray}
One can now clearly see that the replacement of
$C_t\left(|t-t'|;\tau_c\right)$ by $\delta(t-t')$ in the limit of
$\tau_c\rightarrow0$ results in a perfectly well behaved r.h.s. of
Eq. (\ref{new4}), but the spatial white noise limit in the case of
$\lambda_c\rightarrow0$ does not, since the $\lambda_c\rightarrow0$
limit sends $C_x\left(0;\lambda_c\right)$ to $\infty$. In most papers
on field-theory of front propagation with multiplicative (external)
noise, the limit $\tau_c\rightarrow0$ in Eq. (\ref{e70}) is often
taken implicitly, while the explicit dependence on
$C_x\left(x-x';\lambda_c\right)$ is left intact. In this limit,
Eq. (\ref{new4}) reduces to
\begin{eqnarray}
\tilde\varepsilon^{1/2}\,\langle
g[\phi(x,t)]\eta(x,t)\rangle_\eta\,=\,\varepsilon\,\langle
g'[\phi(x,t)]\,g[\phi(x,t)]\rangle\,,
\label{e74}
\end{eqnarray}
where $\varepsilon=\tilde\varepsilon C_x\left(0;\lambda_c\right)$ and
$\Theta(0)=1/2$.

In order to avoid the difficulty associated with the remaining
$C_x\left(0;\lambda_c\right)$ in Eq. (\ref{e74}) approaching infinity
in the limit of $\tau_c$ and $\lambda_c\rightarrow0$, operationally
(i.e., in a computer simulation) one simply discretizes both space and
time in units of $\Delta x$ and $\Delta t$. These discretization units
then act as natural ultraviolet cutoffs. So long as $\Delta t$ is
sufficiently small, the results are independent of $\Delta t$. On the
other hand, when $\Delta x$ is sufficiently small, the Gaussian
representation $C_x\left(0;\lambda_c\right)$ of $\delta$-function in
space is replaced by a square representation of $\delta$-function,
such that $\lambda_c=\Delta x$, and one uses \cite{armero,jaume2}
\begin{eqnarray}
C_x\left(0;\lambda_c\rightarrow0\right)\,=\,\frac{1}{\Delta x}\,.
\label{new5}
\end{eqnarray}
That this is the right interpretation is confirmed by very good
agreements between the theoretical and the simulation
results.\footnote{Equations (\ref{new1}-\ref{new4}) are indeed a fancy
way of deriving Eq. (\ref{e74}). Because these equations use Novikov's
theorem, they are only applicable if $\eta$ is a Gaussian random
variable. When $\eta$ is not Gaussian, then there is a simpler way to
derive Eq. (\ref{e74}) for the Stratonovich interpretation of Eq.
(\ref{e69}): see for example Ref. \cite{gardiner} on the issue of  the
equivalence of It\^{o} and Stratonovich integrals.\label{fnstrat}}

With Eq. (\ref{e74}), we can finally rewrite (\ref{e69}) as (see for
example Ref. \cite{armero})
\begin{eqnarray}
\frac{\partial\phi}{\partial t}\,=\,D\frac{\partial^2\phi}{\partial
x^2}\,+\,f_g(\phi)\,+\,\tilde\varepsilon^{1/2}R(\phi,x,t)\,,
\label{e75}
\end{eqnarray}
where $f_g(\phi)=f(\phi)+\varepsilon g(\phi)g'(\phi)$. Notice now that
\begin{eqnarray}
R(\phi,x,t)\,=\,g(\phi)\eta(x,t)\,-\,\tilde\varepsilon^{1/2}C_x\left(0;\lambda_c\right)\,g(\phi)g'(\phi)
\label{e76}
\end{eqnarray}
has zero mean and the correlation property that\footnote{We will see
in Sec. \ref{sec3.1.2} that the front speed is obtained from the
equation $\displaystyle{\frac{\partial\phi}{\partial
t}=D\frac{\partial^2\phi}{\partial x^2}+f_g(\phi)}$, and the right
parameter for the effect of the stochastic term on the front speed is
$\varepsilon$ and not $\tilde\varepsilon$. In that sense, not only is
it clear that the choice of $\Delta x$ does affect the front speed,
but also in the limit $\Delta x\rightarrow0$, Eq. (\ref{e75}) captures
the entire effect of the external fluctuations on the front speed and
not just up to some order in a perturbation expansion in powers of
$\tilde\varepsilon$. The diffusion coefficient of the so-called
Goldstone mode $D_G$ \cite{schlogl1,schlogl2,pasquale}, however,
involves the assumption that $\tilde\varepsilon\ll1$, so that one can
neglect the $O(\tilde\varepsilon^{1/2})$ term in Eq. (\ref{e77}). The
issue of $D_G$  is however a bit more complicated --- although the
$\tilde\varepsilon$-dependence of the diffusion coefficient of the
Goldstone mode is more subtle than a simple series expansion in powers
of $\tilde\varepsilon$, the fact remains that for
$\tilde\varepsilon\ll1$, the contribution of the
$O(\tilde\varepsilon^{1/2})$ term in Eq. (\ref{e77}) indeed yields a
higher order correction \cite{armero}.\label{fn64}}
\begin{eqnarray}
\hspace{-5mm}\langle R(\phi,x,t)R(\phi,x',t')\rangle_\eta=\langle
g[\phi(x,t)]\eta(x,t)g[\phi(x',t')]\eta(x',t')\rangle_\eta+O(\tilde\varepsilon^{1/2}).
\label{e77}
\end{eqnarray}
 
\subsubsection{Reaction-Diffusion Systems: The Effect of
Multiplicative Noise on the Front Speed and the Diffusion Coefficient
of the So-called Goldstone Mode\label{sec3.1.2}}

\begin{figure}[ht]
\begin{center}   
\includegraphics[width=0.55\linewidth,height=0.22\textheight]{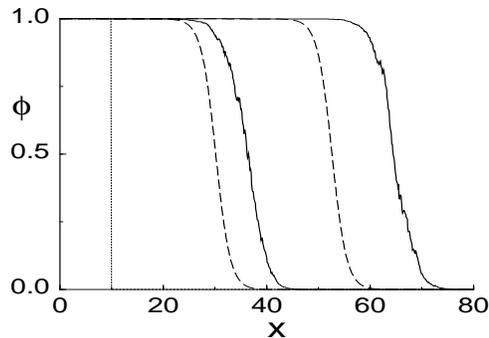}
\end{center}
\caption{Snapshots of fronts (with multiplicative noise) in the
stochastic differential equation
$\displaystyle{\frac{\partial\phi}{\partial
t}=\frac{\partial^2\phi}{\partial x^2}+\phi(1-\phi)[a(x,t)+\phi]}$ and
in the deterministic differential equation
$\displaystyle{\frac{\partial\phi}{\partial
t}=\frac{\partial^2\phi}{\partial x^2}+\phi(1-\phi)[a^{(0)}+\phi]}$ at
two different times \protect{\cite{armeroprl}}. The (dotted) shape of
a step at $x=10$ denotes initial condition for both the deterministic
and fluctuating fronts. The dashed lines correspond to the front
shapes for the deterministic differential equation with $a^{(0)}=0.3$,
and the solid lines correspond to the fronts in the stochastic
differential equation, where $a(x,t)$ fluctuates around $a^{(0)}$ with
an amplitude $\tilde\varepsilon=0.3$ ($\Delta x=0.1$, $\Delta
t=10^{-3}$). The increased gap between the deterministic and the
fluctuating front profiles at the later time is due to the increased
front speed in the presence of the noise, as described in
Eq. (\ref{e79}). Symbols have been changed from the original to keep
consistency with the ones used here.}
\label{armero1}
\end{figure}

From the discussion in Sec. \ref{sec3.1.1}, it is clear that in the
context of front propagation, application of Novikov's theorem to
rewrite Eq. (\ref{e69}) to (\ref{e75}) is {\it not\/} a series
expansion in powers of $\tilde\varepsilon$. Neither does
Eq. (\ref{e75}) involve any approximations. Instead, it is simply a
way to separate the systematic part and the fluctuating part of the
stochastic differential equation (\ref{e69}) --- as we will now see,
rewriting Eq. (\ref{e69}) in the form of Eq. (\ref{e75}) has the
advantage that {\it the systematic part gives rise to a steady shift
in the front speed}, while {\it the fluctuating part gives rise to two
other phenomena at two different time scales: the (random)
displacement of the front from its uniformly translating position at
long time scales, and shape fluctuation of the front around its
instantaneous position at short time scales\/} (see the illustration
in Fig. \ref{armero1}). A very simple and instructive illustration of
this appears in Refs. \cite{lutz1,lutz2,lutz3,pasquale,jaume2}, and
here we will follow it to make our point.

The method of separating these two different time scales goes back to
the early $1980$s \cite{lutz1,lutz2}, although it was carried out for
weak noise and without the aid of Novikov's theorem. For weak noise
one carries out a perturbative expansion in the noise strength
$\tilde\varepsilon$, but the method to extract the diffusion
coefficient of the so-called Goldstone mode arising out of the random
displacement of the front from its uniformly translating position at
long time scales, nevertheless, remains equally applicable. We will
not derive the expressions for the diffusion coefficient of the
so-called Goldstone mode for weak noise
\cite{lutz1,lutz2,lutz3,pasquale} here; instead we will summarize the
key ideas of Refs. \cite{lutz1,lutz2,lutz3,pasquale} in points
(i)-(iii) at the end of Sec. \ref{sec3.1.2}. Enthusiastic readers are
encouraged to consult these references on their own to further satisfy
their curiosity.

To analyze different phenomena taking place at different time scales,
we first express $\phi$ in Eq. (\ref{e75}) as a combination of a {\it
fixed\/} shape $\phi^{(0)}$ displaced by the amount $X(t)$ from its
uniformly translating mean position $\xi=x-v_{\varepsilon}t$ and
(time-dependent) fluctuations in the front shape $\delta\phi$ around
the instantaneous position of the front as
\begin{eqnarray}
\phi(x,t)\,=\,\phi^{(0)}[\xi-X(t)]\,+\,\delta\phi[\xi-X(t),t]\,.
\label{e78}
\end{eqnarray}
It is precisely at this point where one first separates out the
systematic and the fluctuating part of the front. As for the
fluctuating part, a further separation of the effect of fluctuating
term $\propto R(\phi,x,t)$ in Eq. (\ref{e75}) at two different time
scales also takes place in Eq. (\ref{e78}). The fluctuations at long
time scale is coded in the (random) wandering $X(t)$ of the Goldstone
mode around $\xi=x-v_\varepsilon t$, whereas the fluctuations at short
time scales in the front shape is coded in the fluctuations in the
front shape, expressed by $\delta\phi[\xi-X(t),t]$. In this form, the
shape $\phi^{(0)}$ as a function of its argument $\xi$ is obtained
from the equation
\begin{eqnarray}
D\frac{\partial^2\phi^{(0)}}{\partial\xi^2}\,+\,v_{\varepsilon}\frac{\partial\phi^{(0)}}{\partial\xi}\,+\,f_g[\phi^{(0)}]\,=\,0\,.
\label{e79}
\end{eqnarray}
As one can now see, the effective deterministic part now has a front
solution that propagates with a speed $v_\varepsilon$ --- the
important thing to notice is that $v_\varepsilon$ is different from
the speed of the deterministic front obtained by dropping the
$\propto\eta(x,t)$ term of Eq. (\ref{e69}) altogether.

Clearly, in this process, the displacement of the front $X(t)$ from
its uniformly translating mean position $\xi$ can be chosen in many
different ways. One of the convenient ways is to obtain it through the
so-called Goldstone mode. The idea is that just like in
Eq. (\ref{e15}), if $\phi^{(0)}(\xi)$ is a solution of
Eq. (\ref{e79}), then so it $\phi^{(0)}(\xi+a)$ for any $a$. This
translational invariance of Eq. (\ref{e79}) indicates that
$\Phi_{G,R}=\displaystyle{\frac{\partial\phi^{(0)}}{\partial\xi}}$ is
the right eigenvector of the linear stability operator ${\cal
L}_{v_{\varepsilon}}$ with eigenvalue zero, where
\begin{eqnarray}
{\cal
L}_{v_{\varepsilon}}\,=\,\frac{\partial^2}{\partial\xi^2}\,+\,v_{\varepsilon}\,\frac{\partial}{\partial\xi}\,+\,+\,\frac{\delta
f_g(\phi)}{\delta\phi}\bigg|_{\phi\,=\,\phi^{(0)}}
\label{e80}
\end{eqnarray}
The so-called Goldstone mode is simply the quantity $\Phi_{G,R}(\xi)$
\cite{schlogl1,schlogl2,pasquale}, which also propagates with the
speed $v_\varepsilon$ of the front solution $\phi^{(0)}(\xi)$.

As we already saw in Sec. \ref{sec2.4.2},
$\Phi_{G,L}(\xi)=e^{v_{\varepsilon}\xi/D}\,\displaystyle{\frac{\partial\phi^{(0)}}{\partial\xi}}$
is the left eigenvector of ${\cal L}_{v_{\varepsilon}}$ with
eigenvalue zero, and then based on the idea of the Goldstone mode, we
now {\it uniquely\/} define $X(t)$ from the relation
\cite{lutz1,lutz2,lutz3,pasquale,jaume2}\footnote{For fluctuating
fronts in reaction-diffusion systems, Eq. (\ref{e81}) is the
operational definition of how to extract the conditionally averaged
front profile from an ensemble of front
realizations.\label{fnconditional2}}
\begin{eqnarray}
\int_{-\infty}^{\infty}d\xi\,e^{v_{\varepsilon}\xi/D}\,\frac{d\phi^{(0)}}{d\xi}\,\delta\phi[\xi-X(t),t]\,=\,0\,.
\label{e81}
\end{eqnarray}
Thereafter, having used Eq. (\ref{e78}) in Eq. (\ref{e75}), at the
lowest order of the magnitude of the front shape fluctuation, we have,
{\it on the instantaneous comoving frame of the Goldstone mode}
\begin{eqnarray}
\frac{\partial(\delta\phi)}{\partial t}\,=\,{\cal
L}_{v_{\varepsilon}}\delta\phi\,+\,\dot{X}(t)\,\frac{\partial\phi^{(0)}}{\partial\xi}\,+\,R(\phi,\xi,t)\,,
\label{e82}
\end{eqnarray}
and then having left multiplied Eq. (\ref{e82}) with $\Phi_{G,L}(\xi)$
and integrating both sides w.r.t. $\xi$, we get
\begin{eqnarray}
\dot{X}(t)=-\tilde\varepsilon^{1/2}\left[\int_{-\infty}^{\infty}d\xi\,\Phi_{G,L}(\xi)\,R(\phi,\xi,t)\right]\!\Bigg/\!\left[\int_{-\infty}^{\infty}d\xi\,\Phi_{G,L}(\xi)\Phi_{G,R}(\xi)\right]\!\!.
\label{e83}
\end{eqnarray}

The upshot of the entire exercise in terms of the usage of Novikov's
theorem for propagating fronts is now clear --- the stochastic
differential equation (\ref{e69}) has a front speed $v_\varepsilon$,
which has to be calculated from Eq. (\ref{e79}). On the other hand,
the front speed also has (time-dependent) random fluctuations around
$v_\varepsilon$ [notice that $\langle\dot{X}(t)\rangle_\eta=0$].

One measure of the wandering of the entire front that is ubiquitous in
literature on the external fluctuation effects on propagating fronts
is the diffusion coefficient $D_G$ of the Goldstone mode. The quantity
$D_G$, as can be easily guessed, is borne out of the random
displacement $\dot{X}(t)$ at long time scales; but in order to obtain
an expression for $D_G$ from Eq. (\ref{e83}), it is necessary that one
replaces $R(\phi,\xi,t)$ by $R(\phi^{(0)},\xi,t)$. This replacement
makes $R$ $\delta$-correlated both in space and time in
Eq. (\ref{e77}), and then the front speed fluctuation becomes a Markov
process originating from the fluctuating $R$ term in
Eq. (\ref{e75}). Moreover, it also reduces the whole formalism of
Langevin-type front evolution equation (\ref{e69}) with {\it
multiplicative\/} noise in the Stratonovich interpretation to
Eq. (\ref{e75}) with {\it additive and white\/} noise, resulting in
the further reduction of the fluctuating ``Langevin force'' $R$ {\it
of variable magnitude in time\/} to a force with a {\it fixed
magnitude. This reduction is absolutely necessary in order to even
define $D_G$ [and also Eq. (\ref{e86}) in this form] for Stratonovich
noise}.\footnote{We will later see in Sec. \ref{sec4} that when
$R(\phi,\xi,t)$ is interpreted in the It\^{o} sense, $D_G$ can be
well-defined without replacing $R(\phi,\xi,t)$ by
$R(\phi^{(0)},\xi,t)$. See also Sec. \ref{sec3.5}.\label{markov}}

By using the Kubo formula for the diffusion coefficient as the time
integral of the autocorrelation function of the fluctuation in the
front speed, and with the aid of Eq. (\ref{e74}) and (\ref{e77}) [the
$O(\tilde\varepsilon^{1/2})$ term in Eq. (\ref{e77}) is dropped], the
diffusion coefficient $D_G$ of the Goldstone mode is obtained as
\begin{eqnarray}
D_G\,=\,\tilde\varepsilon\,\,\left[\,\int_{-\infty}^{\infty}d\xi\,\Phi_{G,L}^2(\xi)\,g^2[\phi^{(0)}]\right]\Bigg/\left[\,\int_{-\infty}^{\infty}d\xi\,e^{v_\varepsilon\xi/D}\,\Phi^2_{G,R}(\xi)\right]^2\,.
\label{e84}
\end{eqnarray}
 
Before we end Sec. \ref{sec3.1.2}, it is worth spending some time on
{\it three\/} important issues. First, only after having derived
Eq. (\ref{e84}), we can appreciate footnotes \ref{fn64}. The front
speed $v_\varepsilon$ is obtained from Eq. (\ref{e79}), and to obtain
$v_\varepsilon$, we have not carried out any expansion in powers of
$\tilde\varepsilon$. On the other hand, for the diffusion coefficient
$D_G$ of the Goldstone mode, we have dropped the
$O(\tilde\varepsilon^{1/2})$ term in Eq. (\ref{e77}). Despite that,
the actual $\tilde\varepsilon$-dependence of $D_G$ is {\it not\/}
simply the prefactor $\tilde\varepsilon$ on the r.h.s. of
Eq. (\ref{e84}); in fact, there are other
$\tilde\varepsilon$-dependent terms already occurring in the
expression of $v_\varepsilon$ that enter both the numerator and as
well as the denominator of Eq. (\ref{e84}) \cite{armero}.

Secondly, the nomenclature ``diffusion coefficient of the Goldstone
mode'' may {\it a priori\/} appear to be a misnomer --- after all,
since $D_G$ characterizes the wandering properties of the Goldstone
mode, one would, quite rightfully, tend to think that $D_G$ should
really be the front diffusion coefficient. Why the name ``diffusion
coefficient of the Goldstone mode'' has been chosen in this review
article is in a way related to my own personal choice of words. It
turns out that $D_f$ for the front defined earlier is in fact
conceptually totally different from the field-theory expression
(\ref{e84}). While we leave further details on this point till
Sec. \ref{sec4.2}, to differentiate these two quantities, we will
continue to use the terminology ``diffusion coefficient of the
Goldstone mode'' for Eq. (\ref{e84}).

Thirdly, the front shape fluctuations around the instantaneous
position of the front at fast time scales can also be analyzed by
defining the (mutually orthonormal) shape fluctuation modes
$\{\Psi_m(\xi)\}$ in the eigenspace of non-zero eigenvalues of the
linearized operator ${\cal L}_{v_\varepsilon}$ as
\begin{eqnarray}
\delta\phi(\xi,t)\,=\,\sum_{m\neq0}c_m(t)\,\Psi_{m,R}(\xi)\,,
\label{e85}
\end{eqnarray}
Here $\Psi_{m,R}(\xi)$ is the right eigenvector of ${\cal
L}_{v_\varepsilon}$ with eigenvalue $\tau_m^{-1}$, and
$\Psi_{m,L}(\xi)=e^{v_\varepsilon\xi/D}\Psi_{m,R}$ is the
corresponding left eigenvector. In the appropriate context,
$e^{v_\varepsilon\xi/(2D)}\Psi_{m,R}$ is basically the same as
$\psi_m(\xi)$ encountered in Eq. (\ref{e21}). Needless to say,
\begin{eqnarray}
\int_{-\infty}^{\infty}d\xi\,\Psi_{m,L}(\xi)\Phi_{G,R}(\xi)=0\,,\quad\quad\forall
m
\label{orthonormal}
\end{eqnarray} 
i.e., each of these modes is orthogonal to the Goldstone mode. The
mode expansion (\ref{e85}) then easily leads one to
\begin{eqnarray}
\dot{c}_m(t)\,=\,-\,\tau^{-1}_m\,c_m\,+\,\tilde\varepsilon^{1/2}\int_{-\infty}^{\infty}d\xi\,\Psi_{m,L}(\xi)\,R(\phi^{(0)},\xi,t)\,
\label{e86}
\end{eqnarray}
and its corresponding solution
\begin{eqnarray}
c_m(t+t')\!=\!e^{-\frac{t'}{\tau_m}}c_m(t)+\tilde\varepsilon^{1/2}\!\!\int_0^{t'}\!\!dt''\!\!\int_{-\infty}^{\infty}\!\!d\xi\,e^{-\frac{t'-t''}{\tau_m}}\Psi_{m,L}(\xi')R(\phi^{(0)},\xi,t'')\,.
\label{e97c}
\end{eqnarray}

Notice that in Eqs. (\ref{e85}-\ref{e97c}), we have first used
$\psi_m(\xi)=e^{v_\varepsilon\xi/(2D)}\Psi_{m,R}(\xi)$ to first
convert ${\cal L}_{v_\varepsilon}$ to a Hermitian operator [see
Eqs. (\ref{e16}) and (\ref{e17})], of which $\psi_m(\xi)$ is the
eigenvector with eigenvalue $\tau^{-1}_m$. We then implicitly made use
of the completeness condition for $\{\psi_m(\xi)\}$:
\begin{eqnarray}
\sum_{m\neq0}\psi_m(\xi)\,\psi_m(\xi')\,=\,\delta(\xi-\xi')\,.
\label{complete}
\end{eqnarray}
This completeness condition also means that $c_m(t)$ at any time $t$
is automatically determined from the following equation:
\begin{eqnarray}
c_m(t)\,=\,\int_{-\infty}^{\infty}d\xi\,\Psi_{m,L}(\xi)\,\delta\phi(\xi,t)\,.
\label{cm}
\end{eqnarray}

We will not need Eqs. (\ref{e85}-\ref{cm}) in Sec. \ref{sec3};
however, they will come handy for Sec. \ref{sec4.2}.

We finally end Sec. \ref{sec3.1} with the remark that to derive
Eqs. (\ref{e69}-\ref{e84}) for weak noise ($\tilde\varepsilon\ll1$)
and for the cases when $\eta(x,t)$ does not necessarily satisfy
Eq. (\ref{e73}) \cite{lutz1,lutz2,lutz3,pasquale}, using small noise
expansion formulation \cite{gardiner}, $\phi^{(0)}$ comes out to be
the same as $\phi^*$ or $\phi^\dagger$ (correspondingly with
$v_\varepsilon=v^*$ or $v^\dagger$), depending on whether the
deterministic part of Eq. (\ref{e69}) gives rise to a pulled or a
pushed front. This implies that in these approaches, while there is no
steady modification in the front speed due to the presence of the
noise, the only new phenomenon that the noise gives rise to is the
diffusive wandering of the Goldstone mode. Nevertheless, there are a
few interesting results derived in
Refs. \cite{lutz1,lutz2,lutz3,pasquale} for weak noise, and we
summarize them in points (i) through (iii) below.

(i) A derivation of the diffusion coefficient of the Goldstone mode
for g\-e\-n\-e\-r\-a\-l noise correlations can be found in
Refs. \cite{lutz1,lutz2}. An a\-p\-p\-l\-i\-c\-a\-t\-i\-o\-n to derive
an exact result for $D_G$ for front propagation in bistable systems
can be found in Ref. \cite{lutz1} when
$\langle\eta(x,t)\eta(x',t')\rangle_\eta\propto\displaystyle{\exp\left[\frac{-(t-t')^2}{\tau_c^2}\right]\exp\left[\frac{-(r-r')^2}{\lambda_c^2}\right]}$
with characteristic time and length scale $\tau_c$ and $\lambda_c$
respectively.

(ii) The probability density of $X(t)$ based on the dynamics similar
to Eq. (\ref{e83}) has been analyzed in Ref. \cite{lutz3}. An
application to derive an exact result for $D_G$ for front propagation
in a bistable system can be found therein when $\eta(x,t)$ satisfies
Eq. (\ref{e73}).

(iii) A treatment of how to calculate the correlation function
$\langle\phi(x,t)\phi(x',t)\rangle_\eta$ exists in
Ref. \cite{pasquale}. The theory was then applied to front propagation
in a bistable system and to front propagation into an unstable state
(of a Fisher equation type model).

\subsection{Application of Novikov's Theorem to Reaction-Diffusion
Systems: Effects of Multiplicative Noise on Specific Models
\label{sec3.2}} 

Equipped with the necessary groundwork in Sec. \ref{sec3.1.1}, we now
analyze the effect of multiplicative noise on specific models of
reaction-diffusion systems.

\subsubsection{An Example of (Multiplicative) Noise-Induced Front
Transitions\label{sec3.2.1}}

Consider the reaction-diffusion equation
\begin{eqnarray}
\frac{\partial\phi}{\partial t}\,=\,D\frac{\partial^2\phi}{\partial
x^2}\,+\,\phi(1-\phi)\,[a^{(0)}+\phi]\,,
\label{e87}
\end{eqnarray}
with three homogeneous stationary states, $\phi=0,-a^{(0)}$ and
$1$. To provide the state $\phi=1$ with global stability, we confine
$a^{(0)}$ within the range $[-1/2,1]$ --- with this restriction on the
value of $a^{(0)}$, one arrives back at Eq. (\ref{footnoteeq1}).

The interesting point to note in this reaction-diffusion equation is
that it admits front solutions whose characteristics are different for
different values of $a^{(0)}$, as we illustrate below in (a)-(c).

(a) $-1/2\leq a^{(0)}<0$: In this range, $\phi=0$ state is
metastable. The linear marginal stability criterion does not hold in
this range, and a front solution propagating with speed
$v^\dagger=\sqrt{D}[2a^{(0)}+1]/\sqrt{2}$ emerges {\it
uniquely\/}. Let us denote this phase of the front by M.

(b) $0\leq a^{(0)}<1/2$: In this range, $\phi=0$ is linearly unstable,
but for sufficiently localized initial conditions, the front speed is
still given by $v^\dagger$. For less localized initial conditions,
front speeds higher than $v^\dagger$ are also accessible. We denote
this phase of the front by NL.
\begin{figure}[!ht]
\begin{center}
\includegraphics[width=0.7\linewidth,height=0.3\textheight]{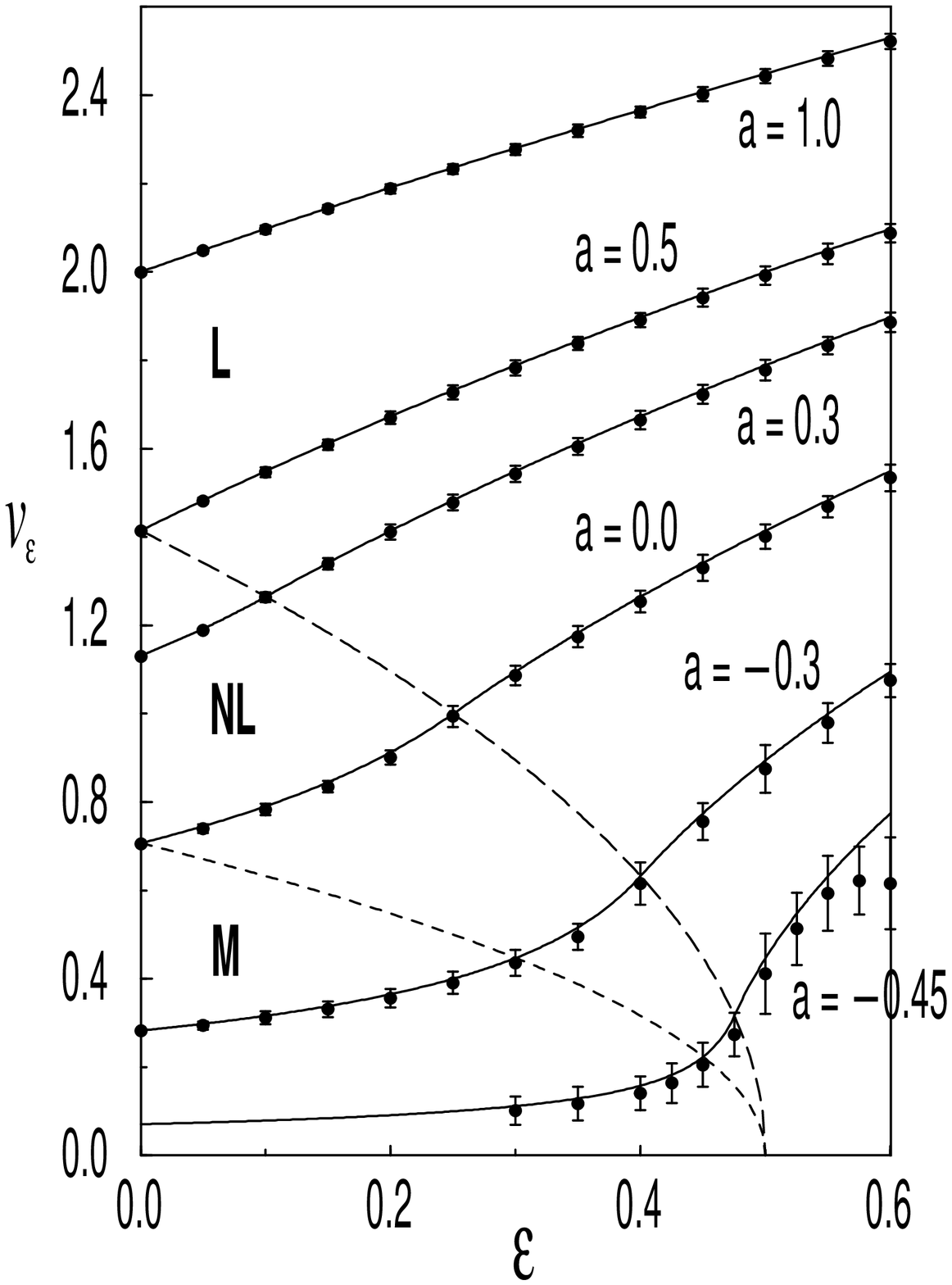}
\end{center}
\vspace{-2mm}
\caption{Noise-induced front transition in Eq. (\ref{e88}) for $D=1$
\protect{\cite{armeroprl,armero}}. Solid lines: theoretical values for
the front speed $v_\varepsilon=2\sqrt{D(a^{(0)}+\varepsilon)}$ (L
phase) and $v_\varepsilon=(2a^{(0)}+1)\sqrt{D/[2(1-\varepsilon)]}$ (NL
and M phase). Filled circles with error bars: simulation data. Symbols
have been changed from the original to keep consistency (also the
figure has been modified for greater clarity), and $a$ should be read
as $a^{(0)}$.}
\label{armero2}
\vspace{2mm}
\begin{center}   
\includegraphics[width=0.6\linewidth,height=0.35\textheight,angle=1]{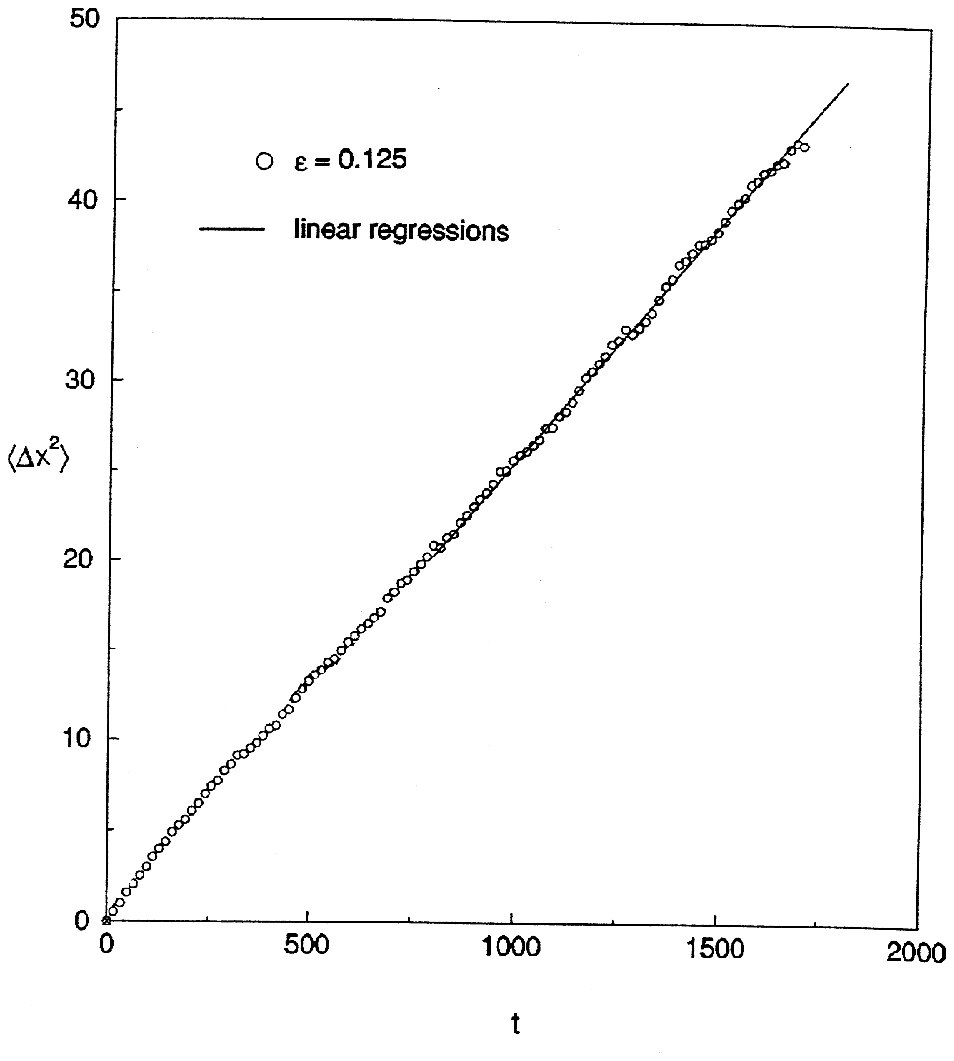}
\end{center}
\caption{Wandering behaviour of the Goldstone mode for $D=1$,
$a^{(0)}=0.1$, $\Delta x=0.5$, $\Delta t=0.1$, and $\varepsilon=0.125$
\protect{\cite{armero}}. The graph has been modified from its original
for clarity. The notation $\Delta x$ appearing in the $y$-axis should
not be confused with the discretization unit $\Delta x$.}
\label{armero3}
\end{figure}

(c) $1/2\leq a^{(0)}\leq1$: Finally, linear marginal stability holds
in this range. For sufficiently localized initial conditions, a pulled
front solution propagating with speed $v^*=2\sqrt{Da^{(0)}}$
emerges. For less localized initial conditions, front speeds higher
than $v^*$ is also accessible. This phase of the front is denoted by L.

How multiplicative noise can affect the characteristics of a front in
a reaction-diffusion system is beautifully illustrated when $a^{(0)}$
in Eq. (\ref{e87}) is replaced by a fluctuating quantity
$a(x,t)=a^{(0)}+\tilde\varepsilon^{1/2}\eta(x,t)$
\cite{armeroprl,armero}, yielding
\begin{eqnarray}
\frac{\partial\phi}{\partial t}\,=\,D\frac{\partial^2\phi}{\partial
x^2}\,+\,\phi(1-\phi)\,[a(x,t)+\phi]\,.
\label{e87a}
\end{eqnarray}
Here, $\eta(x,t)$ is a Gaussian random function --- in the limit of
correlation length and time of the noise going to zero, $\eta(x,t)$
satisfies Eq. (\ref{e73}).

The application of Novikov's theorem to this model yields the
equivalent of Eq. (\ref{e79}), expressed as \cite{armeroprl,armero}
\begin{eqnarray}
D\frac{\partial^2\phi}{\partial
x^2}\,+\,v_\varepsilon\frac{\partial\phi}{\partial
x}\,+\,\phi(1-\phi)\,[a^{(0)}+\varepsilon+(1-2\varepsilon)\phi]\,=0\,.
\label{e88}
\end{eqnarray}
In the presence of noise, this modification of the front equation from
Eq. (\ref{e87}) reorganizes the characteristics of the front ---
effectively, it is this reorganization that ends up being termed as
the {\it noise-induced front transition}. The idea behind this
transition is simple: just like the characteristics of the uniformly
translating front solution of Eq. (\ref{e87}) depend on the value of
$a^{(0)}$, the characteristics of the front solution (\ref{e88}) for a
given value of $a^{(0)}$ depends on $\varepsilon$. This implies that
by adjusting the value of $\varepsilon$, it is possible to convert a
front solution of Eq. (\ref{e87}) that belongs to one of the M, NL or
L phases to a front solution (\ref{e88}) belonging to another one of
these phases. For sufficiently steep initial conditions, in the L
phase, the front (\ref{e88}) propagates with the speed
$2\sqrt{D(a^{(0)}+\varepsilon)}$, while both in the NL and in the M
phase, the front propagates with the speed
$(2a^{(0)}+1)\sqrt{D/[2(1-\varepsilon)]}$. In fact, the boundary
between the NL and the L phase for the front solution (\ref{e88}) lies
at $\varepsilon=1/4-a^{(0)}/2$.\footnote{For example, $a^{(0)}=0.1$
and $\varepsilon=0.2$ in Eq. (\ref{e88}) lies exactly on the boundary
between the NL and the L phase. Later in Sec. \ref{sec3.3}, we will
see that this boundary between the NL and the L phase  marks a change
in the wandering properties of the Goldstone mode as
well.\label{fntransition}}

The $v_\varepsilon$ vs. $\varepsilon$ plots for Eq. (\ref{e88})
\cite{armeroprl,armero} are shown in Fig. \ref{armero2}. Simulation
results for the wandering behaviour of the Goldstone mode for $D=1$,
$a^{(0)}=0.1$, $\Delta x=0.5$, $\Delta t=0.1$, and $\varepsilon=0.125$
\cite{armero} appear in Fig. \ref{armero3}.

\subsubsection{An Example of (Multiplicative) Noise-Induced
Fronts\label{sec3.2.2}}

\begin{figure}[ht]
\begin{center}   
\includegraphics[width=0.6\linewidth,angle=-0.5]{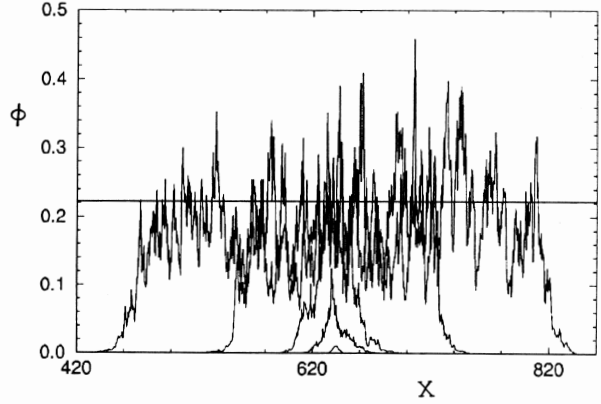}
\end{center}
\vspace{-2mm}
\caption{Snapshots of front configurations for noise-induced front
propagation in Eq. (\ref{e89}) [with $D=1$] when $a(x,t)$ fluctuates
around $a^{(0)}=0.1$ with an amplitude $\tilde\varepsilon=0.15$
\protect{\cite{santos}}. The snapshots have been taken at $t=50$,
$100$, $240$ and $450$.}
\label{santos1}
\end{figure}

In Sec. \ref{sec3.2.1}, we considered a system that already admits a
front solution in the absence of noise, and witnessed how the
characteristics of the front properties can change when multiplicative
noise is introduced in it. We now consider a system that does not
admit a propagating front solution in the absence of noise, but when
the amplitude of the external noise exceeds a threshold value, fronts
start to propagate (see Fig. \ref{santos1}). In the front propagation
literature, such a front is known as a {\it noise-induced front}. An
example system that exhibits such noise-induced fronts is \cite{santos}
\begin{eqnarray}
\frac{\partial\phi}{\partial t}\,=\,D\frac{\partial^2\phi}{\partial
x^2}\,-\,\phi[a(x,t)+\phi^2]\,.
\label{e89}
\end{eqnarray}
Clearly, when $a(x,t)$ is replaced by $a^{(0)}>0$ in Eq. (\ref{e89}),
the equation does {\it not\/} admit a propagating front solution ---
any perturbation around the state $\phi=0$ simply decays in
time. Elsewhere in the literature, this phenomenon is known as {\it
propagation failure}. Similarly, the emergence of propagating front
solutions in Eq. (\ref{e89}) when $a(x,t)$ becomes a sufficiently
strongly fluctuating quantity is known as the {\it breakdown of
propagation failure\/} (see e.g. Ref. \cite{kladko}). In fact, the
situation is comparable to the so-called stochastic resonance paradigm
\cite{bruno}.

\begin{figure}[ht]
\begin{center}   
\includegraphics[width=0.47\linewidth]{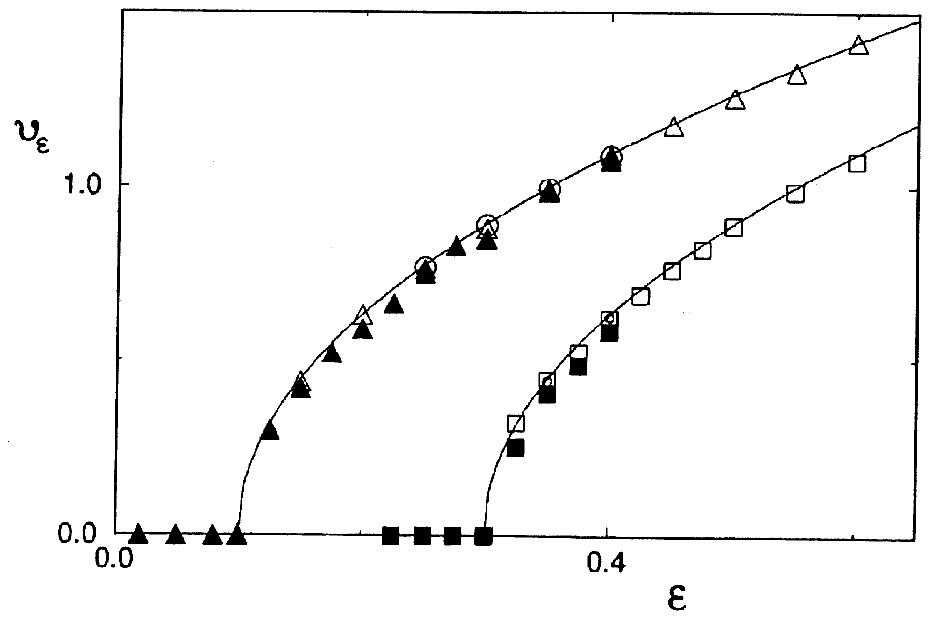}\hspace{5mm}\includegraphics[width=0.47\linewidth]{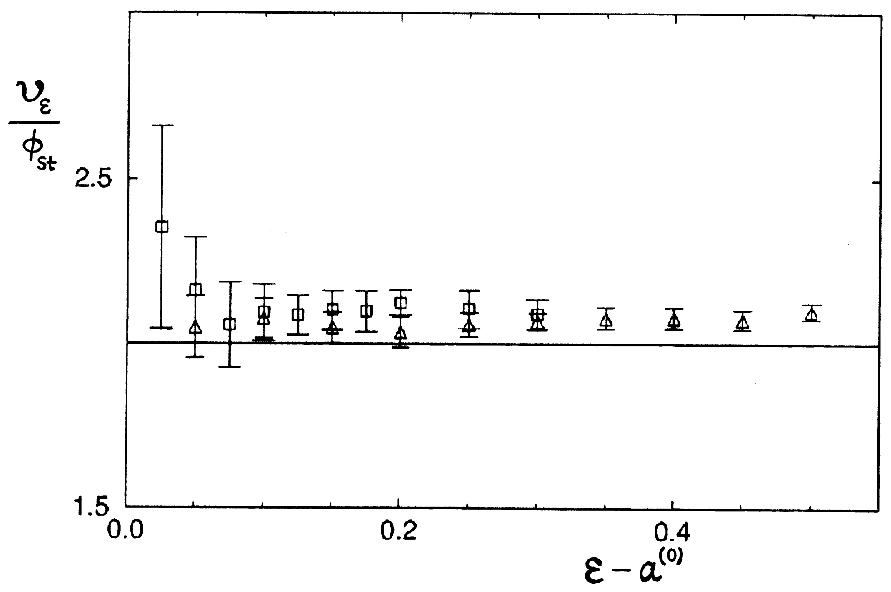}
\end{center}
\caption{Noise-induced front propagation in the reaction-diffusion
system (\ref{e89}) \protect{\cite{santos}}. Left figure:
$v_\varepsilon$ vs. $\varepsilon$ for $D=1$ and two different values
of $a^{(0)}$: namely $a^{(0)}=0.1$ (triangles), and $a^{(0)}=0.3$
(squares). Right figure: theoretically predicted value of
$v_\varepsilon/\phi_{\mbox{\scriptsize st}}$ vs. simulation for
$D=1$. Filled symbols: simulations of the stochastic differential
equation (\ref{e89}) for space and time discretization units $\Delta
x=0.5$ and $\Delta t=0.01$ respectively; open symbols: $\Delta x=0.1$
and $\Delta t=0.001$. Circles represent front solution (\ref{e90}),
and lines denote the theoretical prediction
$v_\varepsilon=2\sqrt{D[\varepsilon-a^{(0)}]}$.}
\label{santos2}
\end{figure}

With $a(x,t)=a^{(0)}+\tilde\varepsilon^{1/2}\eta(x,t)$, where
$\eta(x,t)$ is a Gaussian random function whose correlation properties
reduce to Eq. (\ref{e73}) in the limit of infinitesimal correlation
time and length, the equivalent of Eq. (\ref{e79}) obtained with the
application of Novikov's theorem is expressed as  \cite{santos}
\begin{eqnarray}
D\frac{\partial^2\phi}{\partial
x^2}\,+\,v_\varepsilon\frac{\partial\phi}{\partial
t}\,-\,\phi[a^{(0)}-\varepsilon+\phi^2]\,.
\label{e90}
\end{eqnarray}
It is clear from Eq. (\ref{e90}) that when $\varepsilon>a^{(0)}$,
fronts propagate from the stable state $\phi_{\mbox{\scriptsize
st}}=\sqrt{D[\varepsilon-a^{(0)}]}$ to the linearly unstable state
$\phi=0$ with the (pulled) speed
$v_\varepsilon=2\sqrt{D[\varepsilon-a^{(0)}]}$ (see
Fig. \ref{santos2}).

\subsection{Fluctuating Pulled Fronts with Multiplicative Noise in
Reaction-D\-i\-f\-f\-u\-s\-i\-o\-n Systems: Diffusive vs. Subdiffusive
Wandering of the Goldstone Mode\label{sec3.3}}

In footnote \ref{fntransition}, we had commented that for the
reaction-diffusion model (\ref{e87}) at $a^{(0)}=0.1$ and
$\varepsilon=0.2$, the wandering of the Goldstone mode is
subdiffusive, and also that this phenomenon is caused by the fact that
the $a^{(0)}=0.1,\varepsilon=0.2$ sits precisely on the boundary
between the L and the NL phase. Truly speaking, the above argument is
misleading --- contrary to what may appear from footnote
\ref{fntransition}, this subdiffusive behaviour has nothing to do with
noise-induced front transitions. Instead, the root of this
subdiffusive behaviour turns out to go deeper --- as soon as a
fluctuating pulled front\footnote{What we mean by fluctuating pulled
front here is a fluctuating front with multiplicative noise belonging
to the L phase. Here our usage of the phrase fluctuating pulled front
is chosen to distinguish these fronts from the fluctuating ``pulled''
fronts that we encountered in Sec. \ref{sec2}.\label{notepulled}} with
multiplicative noise belongs to the L phase, the understanding of the
wandering properties of the corresponding Goldstone mode calls for an
entirely different approach \cite{wim2}.

The reason for the requirement that a different approach is needed to
study the wandering properties of the fluctuating pulled fronts is not
difficult to trace; after all, for these fronts it is pretty clear
that one runs into trouble with Eq. (\ref{e84}). The point is that for
them the (linearized) semi-infinite leading edge of $\phi^{(0)}(\xi)$
in Eq. (\ref{e79}) behaves as $\xi\exp[-v_\varepsilon\xi/(2D)]$. This
implies that for $\xi\rightarrow\infty$, at the leading order, the
corresponding Goldstone modes
$\Phi_{G,R}\equiv\displaystyle{\frac{d\phi^{(0)}(\xi)}{d\xi}}$ and
$\Phi_{G,L}\equiv
e^{v_\varepsilon\xi/D}\displaystyle{\frac{d\phi^{(0)}(\xi)}{d\xi}}$
behave as $\sim\xi\exp[-v_\varepsilon\xi/(2D)]$ and
$\sim\xi\exp[v_\varepsilon\xi/(2D)]$ respectively, which then send the
magnitude of both the numerator and the denominator in Eq. (\ref{e84})
to $\infty$.

In the next paragraph, following Ref. \cite{wim2}, we will argue that
the wandering width ${\cal W}(t)=\sqrt{\langle X^2(t)\rangle}$ of the
Goldstone mode for fluctuating pulled fronts with multiplicative noise
scales as $t^{1/4}$ at long times. A more rigorous derivation of this
result by means of converting Eq. (\ref{e87}) for $a^{(0)}=1$ to a KPZ
equation for a one-dimensional interface in terms of the variable
$h(\xi,t)$, defined by the Cole-Hopf transformation
$\phi(x,t)=\exp[h(\xi,t)-v_\varepsilon\xi/(2D)]$, can also be found in
Ref. \cite{wim2}. We will summarize this rigorous method later, but
will not elaborate on it.

The argument for the $t^{1/4}$ scaling behaviour of ${\cal W}(t)$ at
long times \cite{wim2} is the following: notice that before even one
arrives at Eq. (\ref{e84}), a more severe problem already occurs at
the level of Eq. (\ref{e78}) for fluctuating pulled fronts --- namely
that when $\phi^{(0)}$ is pulled, then there is no inherent time scale
for the convergence of the front speed to its asymptotic value
[c.f. Eq. (\ref{e4})]. We had earlier seen that this behaviour
originate from the gapless spectrum of the stability operator
(\ref{e80}) for pulled fronts. It implies that separating the effect
of the fluctuations into a ``slow'' wandering of the Goldstone mode
from the ``fast'' modes of fluctuations in the front shape, upon which
the whole underlying philosophy of Eqs. (\ref{e78}) through
(\ref{e84}) is based, does not make sense any longer. Nevertheless,
the $t^{1/4}$ scaling behaviour of ${\cal W}(t)$ emerges beautifully
when the long-time behaviour of $\phi^{(0)}(\xi,t)$ is successfully
married to Eq. (\ref{e84}), and that is carried out in the following
manner.

With the result that the leading edge of the pulled front
$\phi^{(0)}(\xi)$ relaxes asymptotically as \cite{ebert}
\begin{eqnarray}
\phi^{(0)}(\xi,t)\,\sim\,\xi\,e^{-v_\varepsilon\xi/(2D)-\xi^2/(4Dt)}/t^{3/2}\quad\mbox{for}\,\,\xi,t\gg1
\label{e91}
\end{eqnarray}
one defines a time-dependent upper cutoff $\xi_c\sim\sqrt{4Dt}$ for
the $\xi$ integral in the denominator of Eq. (\ref{e84}). When this
upper cutoff is inserted in Eq. (\ref{e84}), one obtains a
time-dependent diffusion coefficient for the Goldstone mode \cite{wim2}
\begin{eqnarray}
D_G\,\sim\,\frac{3\tilde\varepsilon}{v^2_\varepsilon\sqrt{\pi Dt}}
\label{e92}
\end{eqnarray}
for $t\gg1$. The implication of Eq. (\ref{e92}) then is that for
$t\gg1$, ${\cal W}(t)$ could possibly scale as $t^{1/4}$.

\begin{figure}[ht]
\begin{center}   
\includegraphics[width=0.55\linewidth]{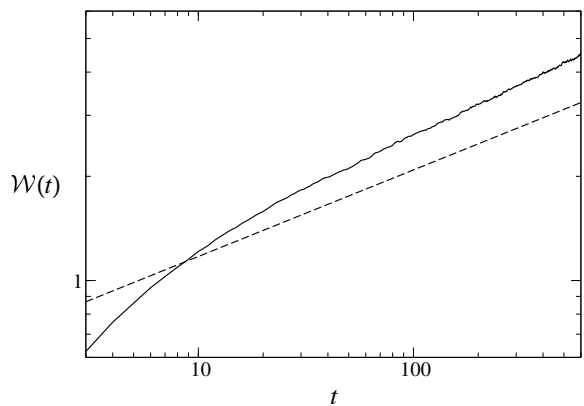}
\end{center}
\caption{Log-log plot of ${\cal W}(t)$ vs. $t$ showing subdiffusive
wandering of the Goldstone mode for fluctuating pulled fronts with
multiplicative noise (\ref{e87a}) \protect{\cite{wim2}}. The data
correspond to $a^{(0)}=1$. Solid line: simulation data, dashed line: a
line with slope $1/4$.}
\label{wim2}
\end{figure}
To obtain the $t^{1/4}$ scaling of ${\cal W}(t)$ in terms of the
rigorous Cole-Hopf transformation of Eq. (\ref{e87a}) to a
one-dimensional KPZ equation \cite{wim2}, one invokes the scaling
relation ${\cal W}(t)=t^\alpha F(t/L^\beta)$. When the argument $\mu$
of $F(\mu)$ is $\ll1$, then $F(\mu)$ behaves like a constant, and when
$\mu\gg1$, then $F(\mu)\sim\mu^{-\alpha}$. The proper way to implement
the above scaling relation is to use a  {\it time-dependent\/} length
scale $L(t)\sim\xi_c\sim\sqrt{4Dt}$. For a
one-d\-i\-m\-e\-n\-s\-i\-o\-n\-a\-l interface, the KPZ exponents are
$\alpha=1/3$ and $\beta=3/2$, which indicates that for
$t\rightarrow\infty$, the argument of $F$ also approaches
$\infty$. The long time behaviour of ${\cal W}(t)$ is then clearly
given by $t^{1/3}[t/(\sqrt{t})^{3/2}]^{-1/3}=t^{1/4}$.

The long time $t^{1/4}$ scaling behaviour of ${\cal W}(t)$ for the
model (\ref{e87}) with $a^{(0)}=1$ has been confirmed by numerical
simulations \cite{wim2} (see Fig. \ref{wim2}). We note here that
although the wandering properties of the Goldstone mode for the model
(\ref{e89}) was not studied in Ref. \cite{santos}, from the argument
based on the upper cutoff $\xi_c$ for the integrations in
Eq. (\ref{e84}), in this model too we expect the same $t^{1/4}$
scaling behaviour for ${\cal W}(t)$ at long times. A rigorous
derivation of it based on the Cole-Hopf transformation in the spirit
of Ref. \cite{wim2} is left for the enthusiastic readers.

\subsection{Fluctuating Fronts with Multiplicative Noise in
Reaction-Diffusion Systems and Kinetic Roughening\label{sec3.4}}

\begin{figure}[ht]
\begin{center}   
\includegraphics[angle=270,width=0.45\linewidth]{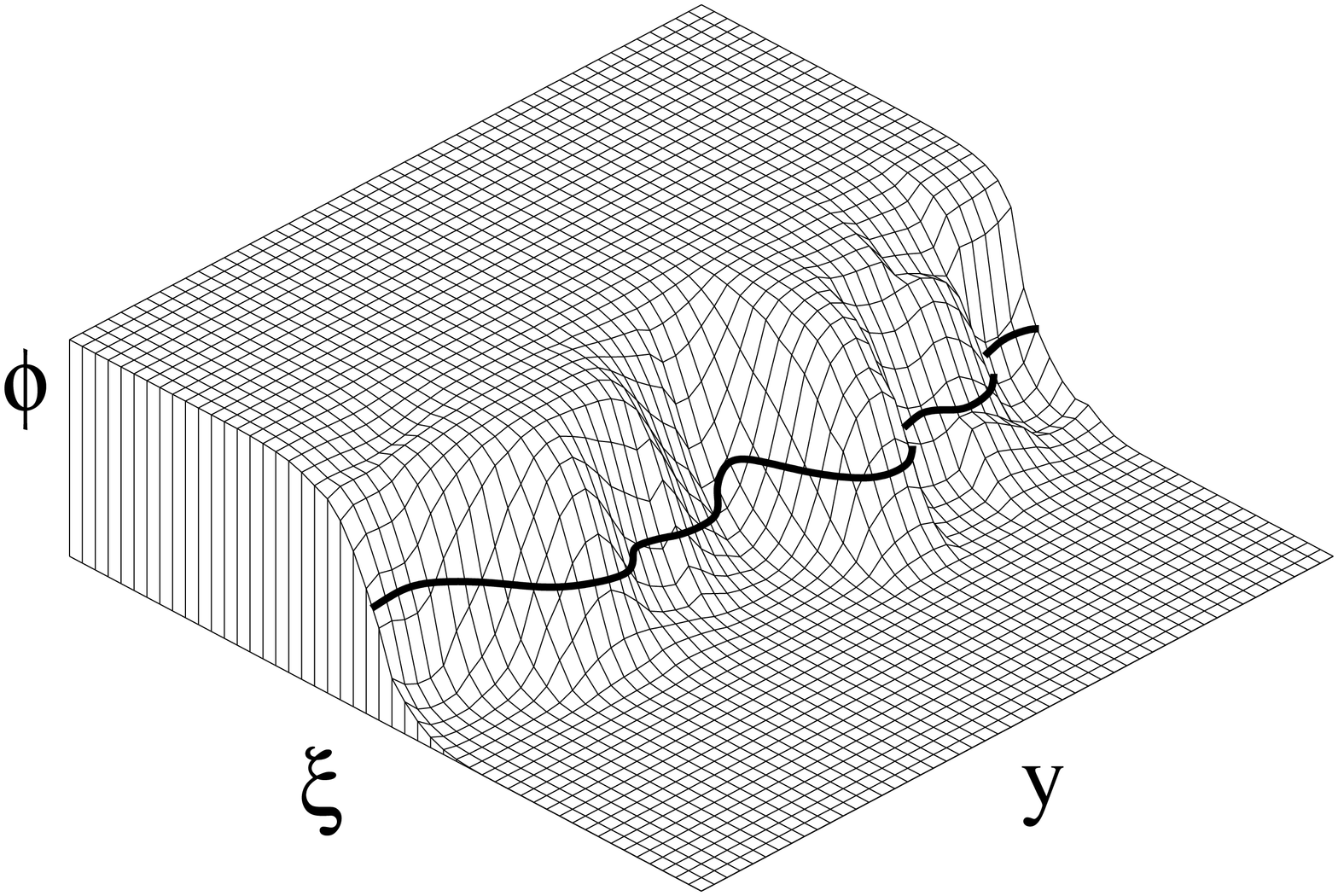}\hspace{5mm}\includegraphics[angle=270,width=0.45\linewidth]{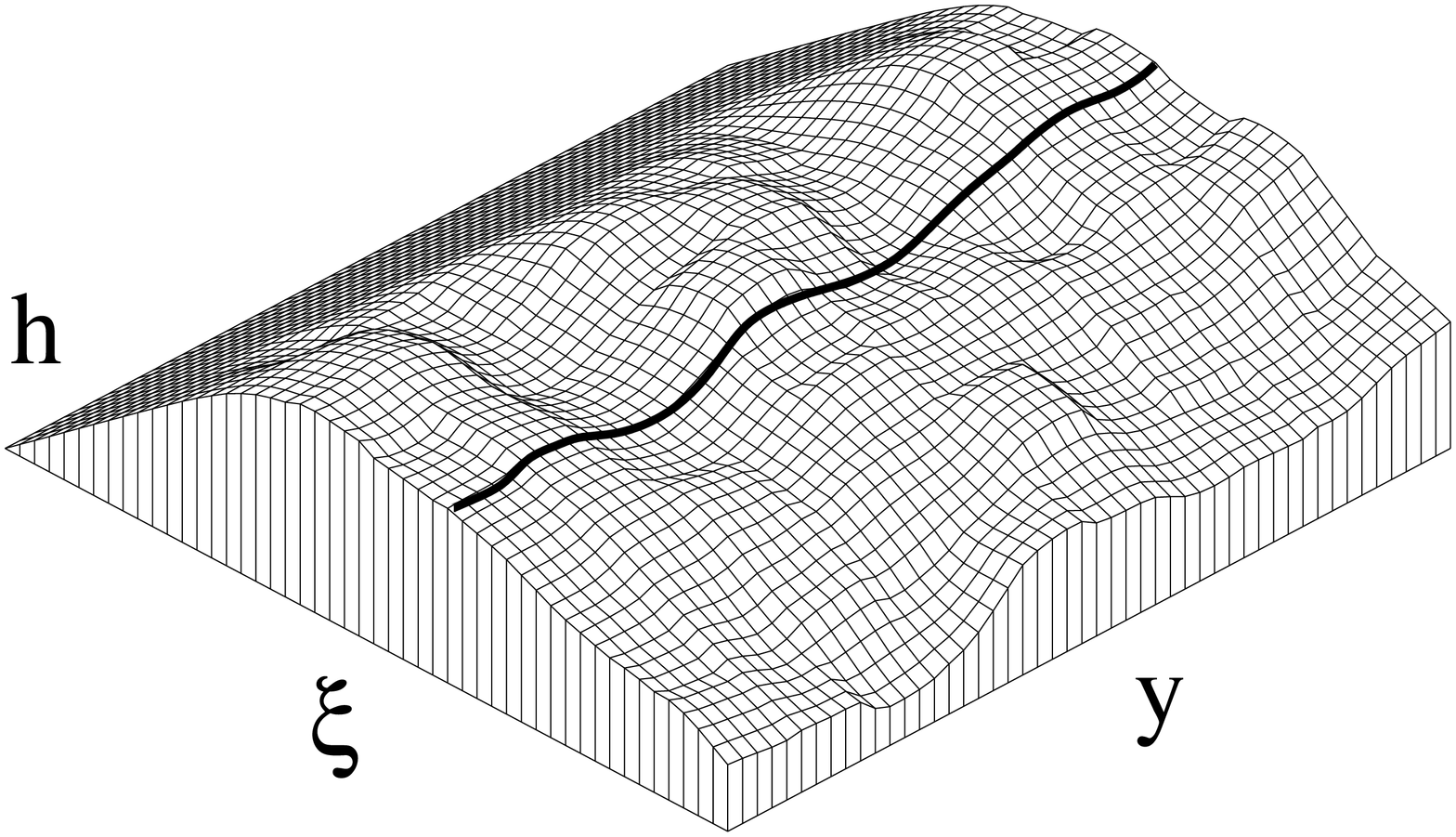}
\end{center}
\caption{Fluctuating pulled front propagation with multiplicative
noise in two spatial dimensions for the model (\ref{e93}). The front
propagates in the $x$ direction, and $\xi=x-v_\varepsilon t$. Left
figure: an instantaneous snapshot of the front configuration. Right
figure: the corresponding configuration in terms of the
two-dimensional interface ``height'' $h(\xi,t)$ obtained by the
Cole-Hopf transformation. The solid curve denotes, in both figures,
the contour for position $(\xi,y)$ where the value of $\phi$ is $1/2$.}
\label{wim3}
\end{figure}

\begin{figure}
\begin{center}   
\includegraphics[width=0.55\linewidth]{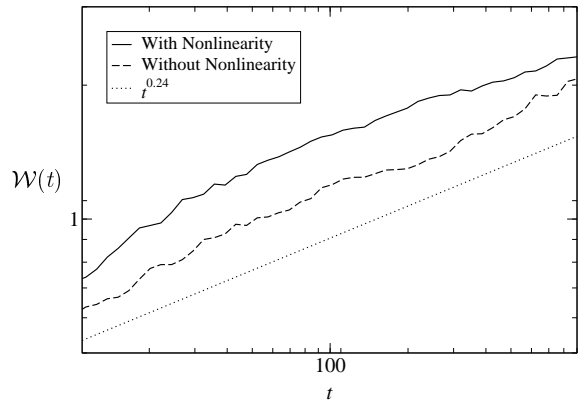}
\end{center}
\caption{Log-log plot of ${\cal W}(t)$ vs. $t$ showing the wandering
of the Goldstone mode for the fluctuating pulled front with
multiplicative noise in Eq. (\ref{e93}) \protect{\cite{jaume1}}. The
with and without nonlinearity respectively indicate whether nonlinear
terms $\propto\exp[h(\xi,t)]$ generated through the Cole-Hopf
transformation of Eq. (\ref{e93}) were retained in the simulation or
not.}
\label{wim4}
\end{figure}

While the Cole-Hopf transformation allows one to rigorously derive the
$t^{1/4}$ scaling of the wandering properties of the Goldstone modes
of fluctuating pulled fronts with multiplicative noise in one spatial
dimension, questions can certainly be raised regarding its
applicability to higher dimensions. Surprisingly enough, it turns out
that the Cole-Hopf transformation is equally successful for
reaction-diffusion systems in two spatial dimensions, only if for
$\xi\rightarrow\infty$, the multiplicative noise couples {\it
linearly\/} to the front field. Such a reaction-diffusion equation
\begin{eqnarray}
\frac{\partial\phi}{\partial
t}\,=\,D\nabla^2\phi\,+\,[1+\sqrt{\tilde\varepsilon}\eta(x,t)]\phi[1-\phi^2]\,,
\label{e93}
\end{eqnarray}
in two spatial dimensions has been considered in
Ref. \cite{jaume1}. The variable $h$ then corresponds to a
two-dimensional interface (see Fig. \ref{wim3}). The front wandering
properties in Eq. (\ref{e93}) has then been shown to obey the KPZ
scaling for a two-dimensional interface\footnote{In the scaling ${\cal
W}(t)\sim t^\alpha F(t/L^\beta)$ for two-dimensional interfaces,
existing accepted values are $\alpha=0.245\pm0.003$ and
$\alpha\beta=0.393\pm0.003$ \cite{marinari}. One has to be somewhat
careful in how to interpret ``long time'' wandering exponent in
Fig. \ref{wim4}. For fluctuating pulled front with multiplicative
noise in one spatial dimension, we earlier saw that $L$ itself scaled
as $\sqrt{t}$. Since for a one-dimensional interface, $\beta=3/2$, at
large times, the argument of $F$ then behaved as
$t^{1/4}\rightarrow\infty$, and that in turn implied that as
$t\rightarrow\infty$, $F(t/L^\beta)$ behaved as
$[t/L^\beta]^{-\alpha}$. However, it turns out that for fluctuating
pulled fronts with multiplicative noise in two spatial dimensions
(\ref{e93}), even for long times, the argument of $F$ remains small,
resulting in the behaviour of $F(t/L^\beta)\sim const.$ and
correspondingly ${\cal W}(t)$ behaves simply as $t^\alpha$
\cite{jaume1}.\label{fn69}} --- see Fig. \ref{wim4}.\footnote{At this
juncture, one cannot help but wonder whether the fluctuations in the
discrete particle and lattice model of Ref. \cite{goutam} have really
anything to do with fluctuating pulled fronts with multiplicative
noise that behaves as $\tilde\varepsilon^{1/2}\phi$ as
$\phi\rightarrow0$ --- after all both observe the same scaling for
${\cal W}(t)$ in one and two spatial dimensions. While we are leaving
the major question of how well the effect of the internal fluctuations
due to the discreteness effects of the particle and and the lattice
are represented by fluctuations in field-theory for Sec. \ref{sec4},
at this short interlude, we can say that the results of
Ref. \cite{moro} points the finger to a negative answer. We will
further see in Sec. \ref{sec4.1} that the internal fluctuations are
usually modeled by a fluctuating term in the stochastic differential
equation that behaves as $\sqrt\phi$ for $\phi\rightarrow0$, which
too, rules out an affirmative possibility.\label{howwell}}

The success of the Cole-Hopf transformation to reduce
reaction-diffusion equations with multiplicative noise in $d$ spatial
dimensions to a KPZ equation for a $d$ dimensional interface is
certainly extremely intriguing. Although in any spatial dimensions,
this transformation works only if for $\xi\rightarrow\infty$, the
fluctuations in the external parameter linearly couples to the front
field, one wonders if there exist further general connections between
the evolution equations of fronts and those describing interfacial
growth phenomena. From this perspective, a beautiful analysis for
fronts (even with vector front fields) in reaction-diffusion equations
in arbitrary dimensions has been presented in Ref. \cite{jaume3}. The
analysis is not rich in mathematical rigour, and we will not describe
its mathematical procedure here either. The idea therein is the same
as what has been already presented in Sec. \ref{sec3.1.2}; namely that
one needs to perform a separation of length and time scales. As the
short time and length scales are averaged out by means of a coarse
graining method, leaving behind the long length and time scales in a
``renormalized'' evolution equation of the front, one does recover the
dynamical equations for growing interfaces, standardly seen in
interfacial growth literature.

Applied to specific systems of Sec. \ref{sec3.2.1}, this formalism
yields the same results for the expression of the shift in front speed
and the diffusion coefficient of the Goldstone mode in one dimension,
as expected \cite{jaume3}. It also reproduces the front fluctuation
characteristics in Fig. \ref{wim4} for the reaction-diffusion system
(\ref{e93}) in two dimensions.

\subsection{Epilogue II\label{sec3.5}}

Generally speaking, (L\-a\-n\-g\-e\-v\-i\-n-type) field-theoretical
methods have been used ubiquitously by chemists to describe the effect
of external noise on chemical reaction-diffusion systems for
decades. In many of these chemical systems, one does observe front
propagation. From that point of view, it has been difficult for me to
decide, for Sec. \ref{sec3}, what to significantly include, what to
mention in appropriate context and what to leave out. In fact, in
Sec. \ref{sec3}, I have adopted the philosophy of discussing only
those papers in detail that have contributed significantly to the
field-theoretical framework on the effect of external noise on
propagating fronts.  In this process, I have strived to provide a
unified perspective of these field-theoretical methods.

While all the works that I am aware of on these field-theoretical
methods have exclusively been on reaction-diffusion systems, quite
expectedly, there exist a few scattered results that concerned weak
noise, or simply additive fluctuations. These results, in addition to
some of the real laboratory experiments that in the first place
motivated the field-theoretical studies of multiplicative noise on
propagating fronts can be traced by following the cited references in
the works that we have discussed in Sec. \ref{sec3}.

There is one further point to that I would like to bring the reader's
attention to. That is the fact that effectively majority of the works
on field-theoretical treatments of propagating fronts with
multiplicative noise considered only spatiotemporally white noise ---
after all, spatiotemporal white noise in the ensuing analytics is a
lot easier to handle than coloured or structured noise. Although we
started with a fancier form of the noise (\ref{e70}) in
Sec. \ref{sec3}, we reduced it to spatiotemporal white noise
(\ref{e73}) very soon in the limit of both $\lambda_c$ and
$\tau_c\rightarrow0$.

This however does not imply, in any way, that structured noise forms
have not been considered in the literature to study the effect of
multiplicative noise on propagating fronts. Few examples for weak
coloured noise forms have been mentioned in points (i)-(iii) at the
end of Sec. \ref{sec3.1.2}. A recent work has appeared with
spatiotemporally structured (and not necessarily weak) noise
\cite{santos1} --- wherein an approach along the lines of Novikov's
theorem\footnote{Recall that Novikov's theorem is applicable only when
the noise is Gaussian. This requirement renders it unusable for
coloured noise with non-Gaussian statistics.\label{fn74}}
\cite{santos1} has been adopted. The central theme of this line of
thought is, once again, to rewrite Eq. (\ref{e69}) in the form of
Eq. (\ref{e75}), such that the stochastic term does have a zero
mean. In the absence of the applicability of Novikov's theorem, in
this case, one carries out a cumulant expansion of the noise term to
evaluate the l.h.s. of Eq. (\ref{e74}) \cite{santos1} (the precise
mathematical procedure in this cumulant expansion is non-trivial; but
naturally, in the limit of both $\tau_c$ and $\lambda_c\rightarrow0$
one recovers the white noise results). The limitation of this
technique, however, is that it is a perturbative expansion in (small)
correlation length and time, and therefore, one should not expect more
than some limited success.

Before we end Sec. \ref{sec3}, let us return to the paragraph below
Eq. (\ref{e86}). We had commented therein that $D_G$ is defined only
when $R(\phi,\xi,t)$ is replaced by $R(\phi^{(0)},\xi,t)$ in
Eq. (\ref{e82}). While Eq. (\ref{e83}) is still well-defined without
this replacement, Eq. (\ref{e84}) is not. First of all, note that
without having $R(\phi,\xi,t)$ replaced by $R(\phi^{(0)},\xi,t)$, $R$
is no longer white both in space and time [c.f. Eq. (\ref{e77})], and
that certainly gets one into trouble. Secondly (and more importantly),
taking the average over the noise alone to obtain an expression of
$D_G$ is not enough.

To make the second point more articulate, let us take a note of the
fact that the average over the noise, such as in Eq. (\ref{e77}),
implies that for $t'>t$, one averages over all possible noise
realizations after $t$. With this average alone, the expression for
$D_G$ [without having $R(\phi,\xi,t)$ replaced by
$R(\phi^{(0)},\xi,t)$] depends on the precise front configuration at
time $t$. Thereafter, one still has to also average over the ensemble
of realizations of the front configurations at time $t$ [just like in
Eq. (\ref{kubo})] to properly define $D_G$. In that case, to obtain a
corresponding theoretical expression of $D_G$ for general noise terms
is not an easy task (we will see more of it in Sec. \ref{sec4.2.1}.

Nevertheless, whether one can replace $R[\phi(\xi,t),\xi,t]$ by
$R[\phi^{(0)}(\xi,t),\xi,t]$ under general conditions is a natural
question. {\it A priori}, it seems to be a trivial issue, because for
weak noise, one is always accustomed to think that such a replacement
is always possible. The point however is that since we are dealing
with fluctuating fronts that propagate into unstable states, i.e.,
$f_g(\phi)\sim\phi$ for $\phi\rightarrow0$, one must always bear in
mind the following note of caution: in the limit of weak noise, so
long as $R[\phi(\xi,t),\xi,t]/\phi\rightarrow0$  for
$\phi\rightarrow0$ (all the cases that we discussed in Sec. \ref{sec3}
follow this property), $R[\phi(\xi,t),\xi,t]$ can be replaced by
$R[\phi^{(0)}(\xi,t),\xi,t]$. Indeed, in Sec. \ref{sec4}, we will see
this condition violated, namely that for the sFKPP equation
(\ref{ee96}) $R[\phi(\xi,t),\xi,t]/\phi\not\rightarrow0$ for
$\phi\sim1/N$ and $N\rightarrow\infty$. In that case, if one does
replace $R[\phi(\xi,t),\xi,t]$ can be replaced by
$R[\phi^{(0)}(\xi,t),\xi,t]$, one gets a front diffusion coefficient
scaling as $1/N$, which is certainly wrong!

\section{Field-theory of Fluctuating Fronts: Internal
Fluctuations\label{sec4}}

In first three paragraphs of Sec. \ref{sec3}, we have already got a
taste of the complications associated with internal fluctuations. It
therefore should not come as a surprise that (unlike what we
experienced in the field-theory of fluctuating fronts for external
fluctuations) describing the effect of internal fluctuations on
propagating fronts is a real challenge for the field-theorists. Before
we plunge into the details in Sec. \ref{sec4.1} onwards, let us
demonstrate the major intricacies.

To start with, let us remind ourselves that all the works on the
field-theoretical treatments of the effects of internal fluctuations
on propagating fronts that I am aware of have been exclusively in the
reaction-diffusion systems. However, as far as Sec. \ref{sec4} is
concerned, the readers must not overconclude the reference to
reaction-diffusion systems in the context of field-theory for the
effects of internal fluctuations on fronts in general. To make this
point clearer, I want to draw the reader's attention to the fact that
there exists a vast literature on field-theoretical treatments of the
effect of internal fluctuations on fronts in various
reaction-diffusion systems, where two opposing currents of reactant
species of particles meet from two sides and reactions between them
take place in a reasonably localized volume of space known as the
reaction zone. In the reaction zone, one can model the fluctuations in
particle densities by means of a field-theory, and one obtains various
kinds of spatiotemporal behaviour of particle densities or
correlations between reacting particles etc (see for example,
Refs. \cite{lee1,lee2}). In these systems reaction fronts do develop,
but they do not propagate (asymptotically) with a constant speed. The
dynamics of these fronts are extremely interesting on their own merit,
and their understanding has contributed significantly to
renormalization group techniques, but as far as this review article is
concerned, they are beyond our purview.

On the other hand, as far as internal fluctuations in propagating
fronts are concerned, Langevin type field-theories of
reaction-diffusion fronts were of interest first to chemists
\cite{lemar1,lemar10} and mathematicians \cite{muller}. To be more
particular, the interest was in a Langevin description of stochastic
Fisher equation. It would still be another few years until physicists
found field-theories for propagating fronts to be of considerable
interest \cite{levine,doering21,doering22}.

So why are field-theories of internal fluctuations in propagating
fronts so difficult? The answer to this question, in fact, lies at two
levels. The difficulty at the first level stems from the complication
associated with the description of fluctuations in a real experimental
situation. In the second paragraph of Sec. \ref{sec3}, we  discussed
this briefly using the example of a chemical reaction --- that in
general, in a chemical reaction, not only are there local fluctuations
in the density of reactants, but also complicated correlations between
fluctuation in particle densities and in the mesoscopic parameters
such as reaction rate or diffusion coefficient of particles
exist. Taking all these into account in a reasonable and satisfactory
degree of detail is indeed very difficult, and other than
Ref. \cite{lemar11} in the chemistry literature, I am not aware of any
treatment where the effect of local fluctuations in the mesoscopic
reaction-diffusion parameters on the propagating front has been
considered.

To get a good theoretical grip on reaction-diffusion processes, just
like we have seen all along in Sec. \ref{sec2}, in all the
field-theoretical approaches based on Langevin type stochastic
differential equations, it is customary to consider fluctuations only
in the density of particles while the usual mesoscopic parameters such
as diffusion coefficients or reaction rates are held constant. Even
then, these field theories are not free from intricacies. A glance at
the results (a)-(c) below in the chemistry literature
\cite{lemar1,lemar10,lemar2} will make this clear:

\vspace{2mm} (a) For the reaction-diffusion process
X$+$Y$\rightarrow2$X described by the Fisher equation (\ref{e1}) with
$n=2$, where the Langevin forces have been obtained from the master
equation, a front speed {\it higher\/} than $v^*$ has been observed in
Ref. \cite{lemar1}. It appears from the results of the numerical
simulation that in the range $N\sim10^2$-$10^{3.5}$, $v_N$ behaves as
$v_N-v^*\sim N^{-1/3}$.

(b) In a Ginzburg-Landau or Schl\"{o}gl model (\ref{e63}), a similar
Langevin formalism deduced from the master equation also yields a
front speed {\it higher\/} than $v^\dagger$ \cite{lemar10}. Numerical
simulations in this case yields $v_N-v^\dagger\sim N^{-1.38}$.

(c) Direct Monte Carlo simulation techniques applied to the
reaction-diffusion process X$+$Y$\rightarrow2$X described by the
Fisher equation (\ref{e1}) with $n=2$, however yields a {\it
decrease\/} in the front speed from $v^*$, and as a function of $N$ in
the range $N\sim10$-$10^{3.5}$, $v^*-v_N\sim N^{-1/3}$ \cite{lemar2}.

Note that given our prior knowledge of how $v_N$ behaves as a function
of $N$ from Sec. \ref{sec2}, it is difficult to reconcile (a) and (b)
with anything else [even with (c), which in itself is not the
asymptotic behaviour of $v_N$ for $N\rightarrow\infty$]; and this is
precisely where the second level of difficulty comes in. The point is
that the Langevin type field-theories derived from the master equation
effectively are expansions over the ``small parameter''
$1/\langle\langle N_k\rangle\rangle$, where $\langle\langle
N_k\rangle\rangle$ is the conditionally averaged number of particles
on lattice site $k$ (see for example Chap. VII of
Ref. \cite{vankampen}, or Ref. \cite{breuer}). Clearly, the
$1/\langle\langle N_k\rangle\rangle$ expansion  is only possible if
$\langle\langle N_k\rangle\rangle$ are large quantities --- so while
the expansion is valid in the bulk phase of the front, it break down
at the tip of the front, where there are a very few particles per
lattice site. At least for a fluctuating ``pulled'' front such as the
stochastic Fisher equation, we already know that its properties are
tremendously sensitive to the fluctuation dynamics at the tip of the
front. For this reason alone, such Langevin type field theories has to
be applied with extreme caution, as we illustrate in the next two
paragraphs below.

For fluctuating ``pulled'' fronts, the reason why the results from the
chemists' side \cite{lemar1,lemar10,lemar2} contradict the $1/\ln^2N$
convergence of the front speed $v_N$ to $v^*$ from below is not
difficult to trace --- it lies in the fact that the form of the
stochastic terms that they consider amounts to an {\it additive
noise\/} to the deterministic equation. If we compare the resulting
evolution equation of the front with the microscopic
reaction-diffusion process X$\leftrightharpoons2$X for example, the
contrast between the physical implications of the stochastic
differential equation and what actually happens in the microscopic
process becomes really clear. In this microscopic process, lattice
sites on which there is no particle to start with cannot grow a
particle on its own, meaning that the first particle on any lattice
site has to diffuse from one of its nearest neighbours. When this idea
is applied to the lattice sites that lie ahead of the front, one
immediately recovers the idea of the i.f.o.l.s. (see footnote
\ref{ifols}). The lattice sites on the right of the i.f.o.l.s. cannot
generate particles on their own, and that  was the motivation of
Brunet and Derrida behind using a growth cutoff for $\phi<1/N$. The
growth cutoff then predicted that the leading edge of the front would
be of finite extent, and it also gave rise to the $1/\ln^2N$
convergence of the front speed $v_N$ to $v^*$ from below for
fluctuating ``pulled'' fronts.

An additive noise term totally changes this scenario. In that case,
{\it any place ahead of the actual front region\/} where $\phi$ value
is identically zero at a given time can generate a non-zero $\phi$
within an infinitesimal time interval simply due to
fluctuations. These fluctuations then have the capability of pulling
the rest of the front faster than even the leading edge of the
deterministic pulled fronts! It is therefore no wonder that these
models \cite{lemar1,lemar10,lemar2} do yield front speeds that are
{\it higher\/} than $v^*$.

\subsection{Fluctuating ``Pulled'' Fronts in Stochastic
Fisher-Kolmogorov-Petrovsky-Piscunov (sFKPP) Equation \label{sec4.1}}

From the discussions in the last few paragraphs above, it is clear
that as far as a stochastic differential (Fisher-type) equation is
concerned, additive noise to model fluctuating ``pulled'' fronts is
out of the question. The noise value has to be zero where the $\phi$
value is zero and this requirement then points the finger to
multiplicative noise.

While this argument is generally valid, question remains regarding
what form of multiplicative noise one should choose. Clearly, one
should certainly avoid a noise term that behaves
$\sim\tilde\varepsilon^{1/2}\phi$ for $\phi\rightarrow0$ --- for
example the case of fluctuating pulled fronts with multiplicative
noise as we encountered in Sec. \ref{sec3}. For such noise terms, no
matter howsoever small $\phi$ is, the noise amplitude is always an
order of $\tilde\varepsilon\ll1$ weaker than $\phi$, and one does not
expect such noise terms to yield a leading edge of a finite extent for
the front (although I have not seen any work proving or disproving it).

\begin{figure}[ht]
\begin{center}   
\includegraphics[width=0.55\linewidth,angle=-0.5]{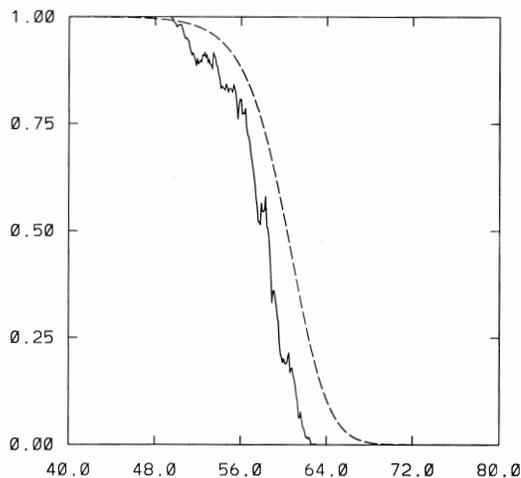}
\end{center}
\caption{A snapshot of the fluctuating ``pulled'' front with $D=1$ in
sFKPP equation (solid curve) demonstrating compact support property
\protect{\cite{doering21,doering22}}. The corresponding plot for the
deterministic pulled front in Fisher equation (\ref{e1}) [$n=2$] is
shown with dashed lines for comparison.}
\label{compactness}
\end{figure}
It turns out, however, that for the front in Fisher-type
(reaction-diffusion) equation, the most frequently used noise term
that does yield a leading edge of a finite extent in fact behaves
$\sim\tilde\varepsilon^{1/2}\sqrt{\phi}$ for $\phi\rightarrow0$ (of
this, the $\tilde\varepsilon\ll1$ and $\tilde\varepsilon\gg1$ limits
are respectively referred to as the {\it weak noise and the strong
noise limits\/}). The corresponding stochastic differential equation
is then known as the stochastic Fisher-Kolmogorov-Petrovsky-Piscunov,
or in short, the sFKPP equation
\cite{muller,levine,doering21,doering22}:
\begin{eqnarray}
\frac{\partial\phi}{\partial t}\,=\,D\frac{\partial^2\phi}{\partial
x^2}\,+\,\phi-\phi^2\,+\,\tilde\varepsilon^{1/2}\,\sqrt{\phi-\phi^2}\,\eta(x,t)\,.
\label{ee96}
\end{eqnarray}
Here $\eta(x,t)$ satisfies the properties that
$\langle\eta(x,t)\rangle_\eta=0$ and
$\langle\eta(x,t)\eta(x',t')\rangle_\eta=2\delta(x-x')\delta(t-t')$.
The stochastic term is interpreted in the It\^{o} sense. The factor of
$2$ in the noise correlation is really unnecessary, but it is
introduced to maintain consistency with the corresponding form
(\ref{e73}) in Sec. \ref{sec3}.

For the microscopic reaction-diffusion process X$\leftrightharpoons2$X
however, the sFKPP equation (\ref{ee96}) has been derived in terms of
a path integral formalism \cite{levine}, which avoids the subtleties
associated with very few particles per lattice site at the leading
edge of the front. In this derivation, one first rewrites the full
master equation in a path integral form, and then proceeds with the
idea that the fluctuations in the front around the mean front shape
are always small and retains the effects of the fluctuations only at
the leading order. The compact support property for sFKPP equation ---
not only that there exists an $x_r(t)$ for which $\phi[x>x_r(t),t]=0$,
but also there exists an $x_l(t)$ for which $\phi[x<x_l(t),t]=1$, and
that $[x_r(t)-x_l(t)]<\infty\,\,\forall t$ --- has been rigorously
derived too \cite{muller} (see Fig. \ref{compactness}). Although none
of these derivations will be discussed here any further, note that the
compact support property of the fronts in sFKPP equation is not a
surprise. Both around $\phi\sim\tilde\varepsilon$ and
$\phi\sim1-\tilde\varepsilon^{1/2}$, the strengths of the stochastic
term are respectively of the same order as $\phi$ and $1-\phi$, and
that helps to abruptly cut down the growth of $\phi$ ahead of and the
decay of $1-\phi$ behind the front region.

\begin{figure}[ht]
\begin{center}   
\includegraphics[width=0.6\linewidth]{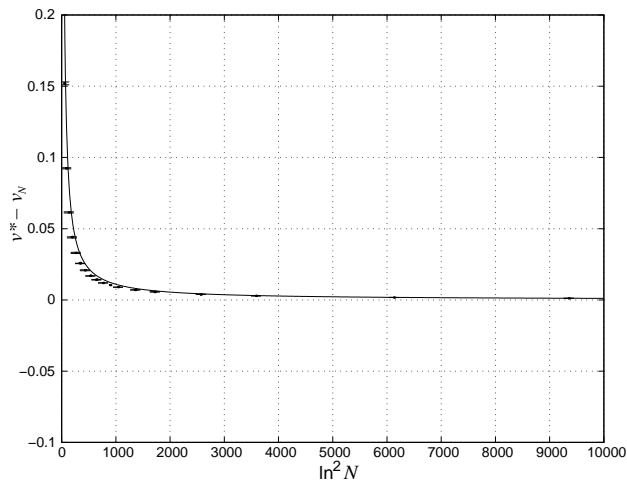}
\end{center}
\caption{Speed for the fluctuating ``pulled'' front in the sFKPP
equation on a discrete lattice with $D=1$ and $\tilde\varepsilon=1/2N$
\protect{\cite{levine}}. Circles: simulation data (with error bars),
solid line: the theoretical expression (\ref{e11}). The symbols have
been changed from the original to maintain notational consistency.}
\label{levine}
\end{figure}
In the weak noise limit, $\tilde\varepsilon$ is replaced by $1/2N$
(the factor of $2$ is put in to counter the $2$ appearing in the
expression of $\langle\eta(x,t)\eta(x',t')\rangle_\eta$) for
$N\rightarrow\infty$ to make contact with the reaction-diffusion
system X$\leftrightharpoons2$X. Even in this limit, due to  the
non-local nature of sFKPP equation, it is not possible to obtain a
theoretical expression for the speed of the corresponding fluctuating
``pulled'' front (although some simple local forms of it can be solved
exactly at all noise strengths \cite{levine}). Nevertheless, one can
put it in the computer \cite{levine}. For $N\rightarrow\infty$, the
speed of the fluctuating ``pulled'' front is then seen to confirm the
$1/\ln^2N$ convergence (\ref{e11}) of the front speed $v_N$ to $v^*$
from below (see Fig. \ref{levine}).\vfill

\subsubsection{Fronts and Duality between the Weak and the Strong
Noise Limits in the sFKPP Equation \label{4.1.1}}

The sFKPP equation, in fact, is more versatile than what appears from
above. What is required is that one puts it in appropriate perspective.

\begin{figure}[ht]
\begin{center}   
\includegraphics[width=0.5\linewidth]{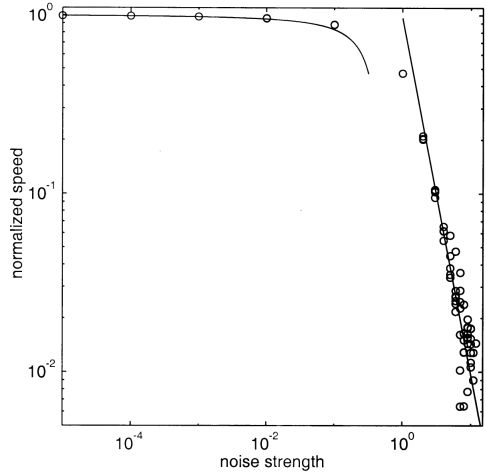}
\end{center}
\caption{Log-log plot of $v_N/v^*$ in the sFKPP equation on a discrete
lattice as a function of noise strength $\tilde\varepsilon^{1/2}$
\protect{\cite{doering21,doering22}}. Open circles: simulation data
for the sFKPP equation, solid lines: the theoretical expressions
(\ref{e11}) at weak noise $\tilde\varepsilon\ll1$ and (\ref{ee97}) at
strong noise $\tilde\varepsilon\gg1$.}
\label{muller1}
\end{figure}
Consider once again the reaction-diffusion process
X$\leftrightharpoons2$X. We had earlier seen, in Secs. \ref{sec2.3}
and \ref{sec2.6} that $N$, the number of particles per lattice site at
the stable phase of the front, dictates the strength of the
fluctuations. The smaller $N$ is, the (relatively) stronger the
fluctuations become (c.f. Sec. \ref{sec2.6}). On the other hand, since
$N$ is essentially the ratio between the rates of the forward and the
backward reactions in the X$\leftrightharpoons2$X process
\cite{breuer,deb11,doering21,doering22}, in Eq. (\ref{ee96}), the
noise strength $\tilde\varepsilon\sim1/N$ is decided by this ratio.

The formal connection between the noise strength and the ratio between
the forward vs. backward reactions in the reaction-diffusion process
X$\leftrightharpoons2$X has recently been rigorously proved through
the notion of duality \cite{doering21,doering22}. This connection then
enables one to predict the front speed at strong noise
$\tilde\varepsilon\gg1$ in the following way: the strong noise limit
reduces the X$\leftrightharpoons2$X system to the model of
Sec. \ref{sec2.6.5} in the $\varepsilon\rightarrow0$ and
$W\rightarrow\infty$ limit,\footnote{Although the
X$\leftrightharpoons2$X system formally allows more than one particles
per lattice site, effectively no lattice site ever gets occupied by
more than one particles when the backward reaction rate is much higher
than the forward reaction rate. The condition for the model of
Sec. \ref{sec2.6.5} that each lattice site can be occupied by at most
one particle at any given time is then automatically
satisfied.\label{restriction}} while keeping $D\sim{\cal O}(1)$. In
this limit $N=\varepsilon/(\varepsilon+W)\simeq\varepsilon/W$, and
then very easily, Eq. (\ref{e51}) predicts a front speed
\begin{eqnarray}
v_N\,=\,DN\,\sim\,\frac{D}{\tilde\varepsilon}\,.
\label{ee97}
\end{eqnarray}
The rescaled front speed $v_N/v^*=v_N/(2\sqrt{D\varepsilon})$ is
plotted as a function of the noise strength $\tilde\varepsilon^{1/2}$
in Fig. \ref{muller1} \cite{doering21,doering22}. The open circles
denote simulation results for the sFKPP equation and the solid lines
denote the corresponding theoretical results for weak
[Eq. (\ref{e11})] and strong [Eq. (\ref{ee97})] noise.

\subsection{The $D_f$ and $D_G$ Dilemma\label{sec4.2}}

From what we learnt in Sec. \ref{sec4.1}, it seems that the sFKPP
equation, interpreted in the It\^{o} sense, does provide a lot a of
input to the subject of fluctuating ``pulled'' fronts --- it correctly
predicts the front speed behaviour as a function of the noise strength
for the limiting cases of both the weak (very large values of $N$) and
strong noise ($N\ll1$).

There is, however, a missing piece of the puzzle in the success story
of the stochastic differential equations for internal fluctuations,
namely that the front diffusion coefficient in one spatial dimension
has been explored in neither for the strong and nor for the weak
noise. As the issue of front diffusion now becomes our focus, at the
very first step, it becomes a necessity that for general
reaction-diffusion systems, we reconcile $D_f$ of Sec. \ref{sec2.5}
and $D_G$ of Sec. \ref{sec3}, both introduced for the measure of front
diffusion.

\subsubsection{$D_f$ vs. $D_G$: What Exactly is the
Difference?\label{sec4.2.1}}

We start with the It\^{o} stochastic differential equation for
$\tilde\varepsilon\ll1$
\begin{eqnarray}
\frac{\partial\phi}{\partial t}\,=\,D\frac{\partial^2\phi}{\partial
x^2}\,+\,f(\phi)\,+\,\tilde\varepsilon^{1/2}\,R[\phi(x,t),x,t]
\label{e97}
\end{eqnarray}
[which is of the same form as Eq. (\ref{e76})]. Here
$R[\phi(x,t),x,t]=g[\phi(x,t)]\eta(x,t)$, and in the It\^{o}
interpretation,
\begin{eqnarray}
\langle
R[\phi(x,t),x,t]\rangle_\eta\,=\,g[\phi(x,t)]\,\langle\eta(x,t)\rangle_\eta\,=\,0\quad\mbox{and}\nonumber\\&&\hspace{-9cm}\langle
R[\phi(x,t),x,t]R[\phi(x',t'),x',t']\rangle_\eta=g[\phi(x,t)]g[\phi(x',t')]\langle\eta(x,t)\eta(x',t')\rangle_\eta\nonumber\\&&\hspace{-3.1cm}=2g^2[\phi(x,t)]\delta(x-x')\delta(t-t').
\label{e97aa}
\end{eqnarray}
In Eq. (\ref{e97}), $D_G$ is defined from $\dot{X}(t)$, the random
displacement of the Goldstone mode of the front [see
Eq. (\ref{e83})]. On the other hand, if one measures the position of
the front by its centre of mass location using the continuum version
of Eq. (\ref{srt}), then from Eq. (\ref{e81}) one obtains, for the
fluctuation in the speed of the centre of mass of the front,
\begin{eqnarray}
\dot{S}(t)\,=\,\dot{X}(t)\,+\,\int_{-\infty}^\infty
d\xi\,\delta\dot\phi(\xi,t)\,,
\label{e97a}
\end{eqnarray}
where {\it $\delta\dot\phi(t)$ is measured on the instantaneous
comoving frame of the Goldstone mode}.

Next, having simplified $\dot{S}(t)$ with the aid of Eqs. (\ref{e80})
and (\ref{e82}), and {\it with the  inherent assumption that
$\delta\phi(\xi,t)\rightarrow0$ for $\xi\rightarrow\pm\infty$} as
\begin{eqnarray}
\dot{S}(t)\,=\,\int_{-\infty}^\infty d\xi\,\frac{\delta
f(\phi)}{\delta\phi}\bigg|_{\phi^{(0)}}\,\delta\phi(\xi,t)\,\,+\,\tilde\varepsilon^{1/2}\int_{-\infty}^\infty
d\xi\,R[\phi(\xi,t),\xi,t]\,,
\label{e97b}
\end{eqnarray}
we immediately notice something interesting --- namely that through
Eq. (\ref{e97b}), we have essentially recovered
Eq. (\ref{diff1}). Remember that Eq. (\ref{diff1}) was obtained
through a heuristic argument for the clock model! For the uniformly
translating front solution $\phi^{(0)}$ of the deterministic
reaction-diffusion equation $\partial_t\phi=D\partial^2_x\phi+
f(\phi)$, the front speed is given by
$\displaystyle{\int_{-\infty}^\infty dx\,f[\phi^{(0)}]}$, and
naturally, when the distribution of $\phi(x,t)$ deviates from
$\phi^{(0)}$, at the leading order, the deviation in the front speed
is given by $\displaystyle{\int_{-\infty}^\infty\!\!dx\frac{\delta
f(\phi)}{\delta\phi}\bigg|_{\phi^{(0)}}}$ $\delta\phi(x,t)$ just like
we had seen in Eq. (\ref{clock2}). Secondly, in Eq. (\ref{e97b}), we
have not replaced $R[\phi(\xi,t),\xi,t]$ by
$R[\phi^{(0)}(\xi,t),\xi,t]$ as we did in Sec. \ref{sec3}. In fact,
the reason we had to do so in Sec. \ref{sec3} is to make the
stochastic term in Eq. (\ref{e75}) $\delta$-correlated both in space
and time in the Stratonovich interpretation. In the It\^{o}
interpretation (\ref{e97aa}) this replacement is no longer
necessary. The choice of not replacing $R[\phi(\xi,t),\xi,t]$ by
$R[\phi^{(0)}(\xi,t),\xi,t]$, however, implies that to obtain the
front diffusion coefficient, we would first have to average over the
noise realizations  (subscript $\eta$) and then average over an
ensemble of initial front realizations at time $t$ (subscript $t$),
just like in Eq. (\ref{kubo}):
\begin{eqnarray}
D_f=\frac{1}{2}\lim_{T\rightarrow\infty}\int_{0}^{T}dt'\,\langle\langle\dot{S}(t)\,\dot{S}(t+t')\rangle_\eta\rangle_t\,.
\label{kubo1}
\end{eqnarray}
Thirdly, with $\langle R[\phi(\xi,t),\xi,t]\rangle_\eta=0$, the
average of Eq. (\ref{e97b}) over all possible noise realizations at
time $t$ does not yield $\langle\dot{S}(t)\rangle_\eta=0$. Indeed,
this is the reason why we had a non-zero $\delta v_{\mbox{\scriptsize
r,mf}}(t)$ in Eq. (\ref{clock1}). Only when a further average over all
possible front realizations at time $t$ is taken, by virtue of
$\langle\delta\phi(\xi,t)\rangle_{t}=0$, one gets
$\langle\dot{S}(t)\rangle=0$. This is also the same as in
Eq. (\ref{diff1}).

Now to simplify Eq. (\ref{kubo1}). The r.h.s. of Eq. (\ref{e97b})
contains a sum of two terms, and the product of
$\dot{S}(t)\dot{S}(t+t')$ that appears in the Green-Kubo formula
(\ref{kubo1}) contains four terms. Of these, an average over all
possible noise realizations after time $t$ kills the
$\displaystyle{\int_{-\infty}^\infty d\xi\,\frac{\delta
f(\phi)}{\delta\phi}\bigg|_{\phi^{(0)}}\,\delta\phi(\xi,t)\int_{-\infty}^\infty
d\xi'\,R[\phi(\xi',t+t'),\xi',t+t']}$ but the rest three
survive. Among these three surviving terms, the easiest one to handle
is $\displaystyle{\int_{-\infty}^\infty d\xi
R[\phi(\xi,t),\xi,t]\int_{-\infty}^\infty
d\xi\,R[\phi(\xi,t+t'),\xi,\-t+t']}$. With Eq. (\ref{e97aa}), its
contribution to $D_f$ is seen to be
\begin{eqnarray}
D_f^{(1)}\,=\,\tilde\varepsilon\int_{-\infty}^\infty d\xi\,\langle
g^2[\phi(\xi,t)]\rangle_t\,.
\label{err}
\end{eqnarray}
To obtain the contribution $D_f^{(2)}$ of
$\displaystyle{\int_{-\infty}^\infty
\!\!\!d\xi\,R[\phi(\xi,t),\xi,t]\!\!\int_{-\infty}^\infty\!\!\!d\xi'\frac{\delta
f(\phi)}{\delta\phi}\bigg|_{\phi^{(0)}}\!\!\!\!\delta\phi(\xi'\!,t+t')}$,
we formally solve $\delta\phi(\xi,t+t')$ first by expanding it as in
Eq. (\ref{e85}), and then by using Eq. (\ref{e97c}) (without having
replaced $R[\phi(\xi,t''),\xi,t'']$ by
$R[\phi^{(0)}(\xi,t''),\xi,t'']$). Thereafter, Eq. (\ref{e97aa}) and
the integration over $t'$ yield
\begin{eqnarray}
D_f^{(2)}=\tilde\varepsilon\int_{-\infty}^{\infty}\!\!\!d\xi\!\!\int_{-\infty}^{\infty}\!\!\!d\xi'\sum_{m\neq0}\left[\tau_m\Psi_{m,L}(\xi)\Psi_{m,R}(\xi')\right]\langle
g^2[\phi(\xi,t)]\rangle_{t}\,\frac{\delta
f(\phi)}{\delta\phi}\bigg|_{\phi^{(0)}(\xi')}\!\!\!\!\!\!\!\!.
\label{e97d}
\end{eqnarray}
This is still not simple enough. To reduce Eq. (\ref{e97d}) further,
we return to Eqs. (\ref{e16}) and (\ref{e17}) and use the fact that
$\displaystyle{\int_{-\infty}^{\infty}\!\!\!d\xi'\Psi_{m,R}(\xi')\frac{\delta
f(\phi)}{\delta\phi}\bigg|_{\phi^{(0)}(\xi')}\!\!\!\!\!\!\!\!=\tau^{-1}_m\!\!\!\int_{-\infty}^{\infty}\!\!\!d\xi'\Psi_{m,R}(\xi')}$.
With the completeness relation (\ref{complete}), we then get a simple
expression:
\begin{eqnarray}
D_f^{(2)}\,=\,\tilde\varepsilon\int_{-\infty}^\infty d\xi\,\langle
g^2[\phi(\xi,t)]\rangle_t\,.
\label{df2}
\end{eqnarray}

Finally, proceeding exactly along the lines of Eqs. (\ref{e97d}) and
(\ref{df2}), $D^{(3)}_f$, the contribution of the
$\displaystyle{\int_{-\infty}^\infty d\xi\,\frac{\delta
f(\phi)}{\delta\phi}\bigg|_{\phi^{(0)}}\,\delta\phi(\xi,t)\!\!\!\int_{-\infty}^\infty
d\xi'\frac{\delta
f(\phi)}{\delta\phi}\bigg|_{\phi^{(0)}}\delta\phi(\xi',t+t')}$ term to
$D_f$ can also be simplified to
\begin{eqnarray}
D_f^{(3)}\,=\,\sum_{m,m'\neq0}\tau^{-1}_m\,\langle
c_m(t)\,c_{m'}(t)\rangle_t\int_{-\infty}^\infty
d\xi\int_{-\infty}^\infty d\xi'\,\Psi_{m,R}(\xi)\,\Psi_{m',R}(\xi')\,.
\label{e102}
\end{eqnarray}

To obtain $D_f=D^{(1)}_f+D^{(2)}_f+D^{(3)}_f$, one now has to
calculate the average of the quantities $\langle\rangle_t$ in
Eqs. (\ref{err}), (\ref{df2}) and (\ref{e102}), i.e., over an ensemble
of (initial) realizations at time $t$. When $R[\phi(\xi,t),\xi,t]$ is
replaced by $R[\phi^{(0)}(\xi,t),\xi,t]$ in the Stratonovich
interpretation of Sec. \ref{sec3}, this is an easy task: while
$\displaystyle{\int_{-\infty}^\infty d\xi\,g^2[\phi^{(0)}(\xi)]}$ is
very easy to evaluate directly, Eq. (\ref{e87}) effectively becomes a
Langevin equation (with additive noise) for a Brownian particle in a
viscous fluid. One can then simply use fluctuation-dissipation theorem
to evaluate $\langle c_m(t)c_{m'}(t)\rangle_t$. In the It\^{o}
interpretation, however, one does not require to replace
$R[\phi(\xi,t),\xi,t]$ by $R[\phi^{(0)}(\xi,t),\xi,t]$ anywhere
between Eqs. (\ref{e97a}) and (\ref{e102}), and then strictly
speaking, one cannot use the fluctuation-dissipation theorem for
multiplicative noise to evaluate $\langle c_m(t)c_{m'}(t)\rangle_t$
from Eq. (\ref{e87}) any longer --- is is a well-known problem in
Stochastic processes that there is no fluctuation-dissipation theorem
for  multiplicative noise (see, e.g., Ref. \cite{vankampen}).

Nevertheless, what is the problem if we simple-mindedly replace
$R[\phi(\xi,t),\xi,t]$ by $R[\phi^{(0)}(\xi,t),\xi,t]$? It certainly
makes calculations easier. As for the answer to this question, we note
that one really has to be very careful to do this. {\it The integrand
in Eq. (\ref{cm}) contains
$\Psi_{m,L}(\xi)=e^{v_\varepsilon\xi/(2D)}\psi_m(\xi)$, and due to the
presence of the exponential weight factor $e^{v_\varepsilon\xi/(2D)}$,
it is the fluctuations at the tip of the front that matter the
most}. This further implies that $c_m$'s are strong functions of time
in general --- after all, in case of internal fluctuations of fronts,
the fluctuations at the tip of the front are of the order of the front
value itself.\footnote{This actually depends on the functional form of
$g(\phi)$. For example, when $g(\phi)\sim\sqrt{\phi}$ for
$\phi\rightarrow0$, the fluctuations at the tip of the front are
certainly as strong as $\phi$.\label{fn70}} However, when the
nonlinearity $f(\phi)$  makes the front pushed, i.e., for fluctuating
pushed fronts, all fluctuation modes have a finite lifetime comparable
to the time scale set by $1/v_\varepsilon$. In that case, the
dependence of the precise value of $\delta\phi$ in one snapshot of a
front realization at any time $t$ at the tip of the front does not
matter so much on the precise noise realization that has been used to
update the front profile before $t$. This then justifies the
replacement of $R[\phi(\xi,t),\xi,t]$ by $R[\phi^{(0)}(\xi,t),\xi,t]$
[and further on the usage of fluctuation-dissipation theorem to obtain
$\langle c_m(t)c_{m'}(t)\rangle_t\sim{\cal
O}(\tilde\varepsilon)$]. For fluctuating ``pulled'' fronts however,
such as the sFKPP equation the fluctuation modes can have very long
decay times (c.f. Sec. \ref{sec2.4.2}). Then the the replacement of
$R[\phi(\xi,t),\xi,t]$ by $R[\phi^{(0)}(\xi,t),\xi,t]$ and the
subsequent usage of fluctuation-dissipation theorem to obtain $\langle
c_m(t)c_{m'}(t)\rangle_t$ is simply incorrect!

For fluctuating pushed fronts, when $R[\phi(\xi,t),\xi,t]$ is replaced
everywhere by $R[\phi^{(0)}(\xi,t),\xi,t]$, one can easily show that
$D^{(1)}_f=D^{(2)}_f=2D^{(3)}_f$ [after using fluctuation-dissipation
theorem in Eq. (\ref{e86}) and thereafter with the completeness
relation (\ref{complete})], yielding
$D_f=5\,\tilde\varepsilon\displaystyle{\int_{-\infty}^\infty
d\xi\,g^2[\phi^{(0)}(\xi)]}/2$. This expression is clearly different
from that of $D_G$ in Eq. (\ref{e84}).\footnote{For fluctuating pushed
fronts made of discrete particles on a lattice,
$\tilde\varepsilon\sim1/N$. This indicates that for these lattice
models, one does obtain a $D_f$ that scales asymptotically as
$1/N$. From Eq. (\ref{e84}), $D_G$ too is seen to scale as
$1/N$.\label{notepushed}} As for fluctuating ``pulled'' fronts, the
scalings of $D_f$ and $D_G$ will now be dealt with in
Sec. \ref{sec4.2.2}.

\subsubsection{$D_f$ and $D_G$ for the sFKPP Equation\label{sec4.2.2}}

From the point of view that the sFKPP equation belongs to the family
of Eq. (\ref{e97}), a question that naturally arises is whether it is
possible to apply our collected wisdom of Sec. \ref{sec4.2.1} to the
sFKPP equation.

A priori, one can expect such an attempt to fail, or at best expect it
to be full of subtleties, for by now, we have already learnt from
Sec. \ref{sec3.3} that in the limit of zero noise strength, whenever
Eq. (\ref{e97}) gives rise to a pulled front, $D_G$ [Eq. (\ref{e84})]
does not remain well defined any more. This simply has to do with the
fact that the integral in the denominator of Eq. (\ref{e84}) diverges
due to the semi-infinite leading edge of pulled fronts.

We can however ask ourselves the following question: from Brunet and
Derrida's analysis, we know that an effective description of
fluctuating ``pulled'' fronts in Fisher equation is in terms of a
growth cutoff below $\phi\simeq1/N$. So how about using Brunet and
Derrida's cutoff solution for $\phi^{(0)}$ in the whole formalism of
Sec. \ref{sec4.2.1}, assuming that the shape fluctuation of the front
in the sFKPP equation are indeed described by the shape fluctuation
modes $\{\Psi_m\}$?\footnote{Recall that in this case, $\{\Psi_m\}$
are localized within the front region and they vanish for
$\xi\rightarrow\pm\infty$ (c.f. Sec. \ref{sec2.4.2}). This is
precisely the compact support property of the front solution in the
sFKPP equation entails; and moreover, it is the same requirement that
underlies Eq. (\ref{e97a}). This means that all the results we
obtained in Sec. \ref{sec4.2.1} can be safely applied to this
case.\label{recall}} At least in that case, any complication
associated with the divergence of the denominator Eq. (\ref{e84}) is
safely avoided!

What we will show in this section is that such an simple-minded
approach works \cite{deb4}. It yields the asymptotic scalings
$D_G\sim1/\ln^6N$ and $D_f\sim1/\ln^3N$. Although at the time of
writing this review article, neither $D_G$ nor $D_f$ for the sFKPP
equation has been studied (analytically or numerically), we will later
argue that the asymptotic scaling $D_f\sim1/\ln^3N$ is a generic
property of fluctuating ``pulled'' fronts.

Between $D_G$ and $D_f$, the $1/\ln^6N$ asymptotic scaling for $D_G$
is the easiest to demonstrate. For simplicity, we set $D=1$. With
$\tilde\varepsilon=1/(2N)$ and  $g(\phi)=\sqrt{\phi-\phi^2}$ for the
sFKPP equation, from Eq. (\ref{e84}). we obtain
\begin{equation}
D_G\!=\!\!\left[\displaystyle{\int_{-\infty}^{\xi_c}\!\!\!d\xi\,e^{2v_N\xi}\!
\left(\frac{d\phi^{(0)}}{d\xi}\right)^{\!\!2}\!\frac{\langle\phi(\xi,t)-\phi^2(\xi,t)\rangle_t}{2N}}\right]\!\!\Bigg/\!\!\left[\int_{-\infty}^{\xi_c}\!
\!\!d\xi\,e^{v_N\xi}\!\left(\frac{d\phi^{(0)}}{d\xi}\right)^{\!\!2}\right]^2\!\!\!\!.\!
\label{e107}
\end{equation}
When one uses Eq. (\ref{e10}) [$\varepsilon=1/N$] in Eq. (\ref{e107}),
the exponential weight factor in the integrand of the denominator
drops out and the denominator itself asymptotically scales as
$1/\ln^6N$ at the leading order. Dealing with the integrand of the
numerator, however, is slightly more complicated --- there exists a
factor of $e^{v_N\xi}$ in the expanded expression of
$\displaystyle{e^{2v_N\xi}\,
\left(\frac{d\phi^{(0)}}{d\xi}\right)^2}$.\footnote{We are using, at
the leading order, $v_N\simeq v^*=2=2\lambda^*$.\label{fn73}} The
presence of this exponential weight factor in the integrand means that
mathematically, we need to consider
$\langle\phi(\xi,t)-\phi^2(\xi,t)\rangle_t$ pretty much only at the
tip of the front (i.e., around $\xi_0$, where the cutoff is
implemented; see Fig. \ref{simpschro}), and physically it means that
{\it the fluctuations at the tip of the front contribute the most to
$D_G$}. At this region, due to the fact $\phi^2(\xi,t)\ll\phi(\xi,t)$,
$\phi^2(\xi,t)$ can be dropped from Eq. (\ref{e107}). Thereafter, with
$\langle\phi(\xi,t)\rangle_t=\phi^{(0)}(\xi)$, the contribution of the
tip region to the numerator turns out to be of ${\cal O}(N)$ at the
leading order, which then cancels the $1/N$ prefactor.

The $1/\ln^6N$ asymptotic scaling of $D_G$ is finally obtained by
noticing that the region behind the tip cannot contribute enough to
cancel the $1/N$ prefactor.

The $1/\ln^3N$ asymptotic scaling of $D_f$, on the other hand, is a
much more involved process; but we can already considerably simplify
its expression with the results of Sec. \ref{sec4.2.1}. For example,
$D_f=D^{(1)}_f+D^{(2)}_f+D^{(3)}_f$ is a simplification. We also know
that {\it Eqs. (\ref{err}), (\ref{df2}) and (\ref{e102}) are exact
results for the It\^{o} differential equation (\ref{e97})}. Since
$D^{(1)}_f$ and $D^{(2)}_f$ for the sFKPP equation are easily seen to
scale $\sim1/N$ from Eqs. (\ref{err}) and (\ref{df2}), our focus
naturally concentrates on the asymptotic scaling of $D^{(3)}_f$.

Formally, it is not possible to simplify Eq. (\ref{e102}) any
further. We have also demonstrated in Sec. \ref{sec4.2.1} that one
cannot also simply replace $\phi(\xi,t)$ by $\phi^{(0)}(\xi)$
everywhere for fluctuating ``pulled'' fronts --- doing so yields the
wrong form $D^{(3)}_f$, namely that
$D^{(3)}_f=D^{(1)}_f\sim1/N$. Instead, to evaluate $D^{(3)}_f$, one
really has to compute the integrals and the sums in
Eq. (\ref{e102}). As it turns out, doing so is full of subtleties as
well.

The first difficulty we face is that we cannot calculate
$\displaystyle{\int_{-\infty}^{\infty}d\xi\,\Psi_{m,R}(\xi)}$
easily. While we do know the solutions of $\Psi_{m,R}(\xi)$ for the
low lying modes from Sec. \ref{sec2.4.2} --- {\it essentially all the
fluctuation modes are localized within the leading edge of the front}
--- 			   there is a danger in using them to evaluate
$\displaystyle{\int_{-\infty}^{\infty}d\xi\,\Psi_{m,R}(\xi)}$ right
away. This is easily demonstrated as follows: at the leading edge,
$\Psi_{m,R}(\xi)=A_m\sin[q_m(\xi-\xi_1)]e^{-\lambda^*\xi}$, where
$q_m=(m+1)\pi/\ln N$ and the nor\-m\-a\-l\-i\-z\-a\-t\-i\-o\-n
$A_m\simeq\sqrt{2}/\ln^{1/2}N$ is chosen in such a way that
$\displaystyle{\int_{-\infty}^{\infty}d\xi\,\Psi_{m,L}(\xi)\Psi_{m',R}(\xi)}=\delta_{m,m'}$.
When this expression of $\Psi_{m,R}(\xi)$ is used for not very high
values of $m$, one gets
$\displaystyle{\int_{-\infty}^{\infty}d\xi\,\Psi_{m,R}(\xi)}\sim(m+1)/\ln^{3/2}N$
at the leading order. The point to note here however, is that since
$\lambda^*=1$, {\it practically the only contribution to the integral
comes from the region around $\xi_1$, i.e., at the very left of the
leading edge within a distance of ${\cal O}(1/\lambda^*)={\cal
O}(1)$}, where the argument of the sin-function approaches zero [this
is what allowed us to replace $\sin[q_m(\xi-\xi_1)]$ by
$q_m(\xi-\xi_1)$]. Precisely at this region, the nonlinear term in
$f(\phi)$, too, becomes significant within a distance of ${\cal O}(1)$
(see Fig. \ref{simpschro}). Therefore, although in
Fig. \ref{simpschro}, $V(\xi)$ can be replaced by $V_0(\xi)$ to obtain
$\tau_m$, $\Psi_{m,R}(\xi)$, and the normalization constant $A_m$ with
a reasonable accuracy, $\Psi_{m,R}(\xi)$ cannot be used to evaluate
$\displaystyle{\int_{-\infty}^{\infty}d\xi\,\Psi_{m,R}(\xi)}$ so
simple-mindedly.

To circumvent this difficulty, we rewrite $D^{(3)}_f$ in a different
way:
\begin{eqnarray}
D^{(3)}_f\!=\!\frac{1}{2}\lim_{T\rightarrow\infty}\!\int_{t}^{t+T}\!\!dt'\!\!\int_{-\infty}^\infty
\!\!d\xi\!\!\int_{-\infty}^\infty\!\!
d\xi'\left\langle\!\!\left\langle\!\frac{\delta
f(\phi)}{\delta\phi}\bigg|_{\phi^{(0)}(\xi)}\!\!\!\!\!\!\!\delta\phi(\xi,t)\,\frac{\delta
f(\phi)}{\delta\phi}\bigg|_{\phi^{(0)}(\xi')}\!\!\!\!\!\!\!\delta\phi(\xi',t+t')\!\!\right\rangle_{\!\!\eta}\right\rangle_{\!\!t}\!\nonumber\\&&\hspace{-12.25cm}=\!\!\!\!\!\!\sum_{m,m'\neq0}\!\!\!\!\!\frac{\tau_m\langle
c_m(t)\,c_{m'}(t)\rangle_t}{2}\!\!\!\int_{-\infty}^\infty
\!\!\!d\xi\,\frac{\delta
f(\phi)}{\delta\phi}\bigg|_{\phi^{(0)}(\xi)}\!\!\!\!\!\!\!\!\!\!\!\!\Psi_{m,R}(\xi)\!\!\!\int_{-\infty}^\infty\!\!\!d\xi'\,\frac{\delta
f(\phi)}{\delta\phi}\bigg|_{\phi^{(0)}(\xi')}\!\!\!\!\!\!\!\!\!\!\!\!\!\Psi_{m',R}(\xi').
\label{e108}
\end{eqnarray}
In this form, $\displaystyle{\int_{-\infty}^\infty
\!d\xi\,\frac{\delta
f(\phi)}{\delta\phi}\bigg|_{\phi^{(0)}(\xi)}\!\!\!\!\!\!\Psi_{m,R}(\xi)}$
is the contribution of the $m$-th shape fluctuation mode  to $\dot{S}$
[Eq. (\ref{e97b})]. The rationale behind writing $D^{(3)}_f$ in this
form is to articulate the fact that for the uniformly translating
front solution $\phi^{(0)}$, the front speed is given by
$\displaystyle{\int_{-\infty}^\infty d\xi\,f[\phi^{(0)}(\xi)]}$, and
naturally, when the front shape $\phi(x,t)$ deviates from $\phi^{(0)}$
by an amount $\Psi_{m,R}(\xi)$, the contribution of the shape
fluctuation to the to the fluctuation in the front speed is given by
$\delta v_{\mbox{\scriptsize
mf},m}=\displaystyle{\int_{-\infty}^\infty \!d\xi\,\frac{\delta
f(\phi)}{\delta\phi}\bigg|_{\phi^{(0)}(\xi)}\!\!\!\!\!\!\Psi_{m,R}(\xi)}$
at the leading order [just like in Eq. (\ref{clock2}) for the clock
model; the notation $\delta v_{\mbox{\scriptsize mf},m}$, too, is
motivated by Eq. (\ref{clock2})].

From the paragraph above Eq. (\ref{e108}), we learnt that to evaluate
$\delta v_{\mbox{\scriptsize mf},m}$ exactly, we need the full
solution of Eq. (\ref{e17}) [this we do not have]. Nevertheless, what
we are really interested is how $\delta v_{\mbox{\scriptsize mf},m}$
scales with $\ln N$. To determine this, using
$A_m=\sqrt{2}/\ln^{1/2}N$, we first rewrite the front shape {\it
only\/} due to the shape fluctuation mode $\Psi_{m,R}$ at the
(linearized) leading edge as
\begin{eqnarray}
\phi(\xi,t)\,=\,\phi^{(0)}(\xi)\,+\,\Psi_{m,R}(\xi)\nonumber\\&&\hspace{-3.9cm}=\,\frac{\ln
N}{\pi}e^{-\lambda^*\xi}\bigg\{\sin[q_0\xi]+\sqrt{2}\pi\,\frac{\sin[q_m\xi]}{\ln^{3/2}N}\bigg\}\,.
\label{e109}
\end{eqnarray}
In this form, it becomes immediately clear that for any {\it pure
normalized\/} fluctuation mode, the fluctuations in the front field
are always an order $1/\ln^{3/2}N$ weaker than the front field
itself. Thereafter, as we further realize that for Brunet and
Derrida's cutoff solution,
$\displaystyle{\int_{-\infty}^{\infty}d\xi\,f[\phi^{(0)}(\xi)]}=v_N\simeq
v^*=2$ is actually of ${\cal O}(1)$, we conclude that the presence of
the pure normalized $m$-th shape fluctuation mode introduces a
fluctuation $\delta v_{\mbox{\scriptsize mf},m}$ in the speed of the
centre of mass of the front, such that
\begin{eqnarray}
\delta v_{\mbox{\scriptsize mf},m}\,=\,\frac{s_m}{\ln^{3/2}N}\,,
\label{e110}
\end{eqnarray}
at the leading order, and $s_m$ of ${\cal O}(1)$. Equations
(\ref{e108}-\ref{e110}) then imply
\begin{eqnarray}
D^{(3)}_f\,=\,\frac{1}{2}\sum_{m,m'\neq0}\tau_m\frac{s_ms_{m'}}{\ln^3N}\,\langle
c_m(t)\,c_{m'}(t)\rangle_t\,.
\label{e111}
\end{eqnarray}
Furthermore, since there are ${\cal O}(\ln N)$ number of bound states
within the potential well $V_0(\xi)$ of width $\ln N$ and depth ${\cal
O}(1)$ [see Fig. \ref{simpschro}], the sum over $m$ and $m'$ in
Eq. (\ref{e111}) runs from $1$ to $\ln N$.

The last tricky part now is how to obtain the dependence of $\langle
c_m(t)\,c_{m'}(t)\rangle_t$ on $\ln N$, and how to evaluate the sums
in Eq. (\ref{e111}). To this end, notice that for a given realization,
$c_m(t)$ is obtained from Eq. (\ref{cm}), and therein, the presence of
the factor $e^{\lambda^*\xi}$ in the $\Psi_{m,L}(\xi)$ implies that
$c_m(t)$ for a given realization is practically determined from the
fluctuation characteristics at the tip of the front. Thus, in
Eq. (\ref{cm}), we keep the integral over $\xi$ only over the leading
edge of the front and write
\begin{eqnarray}
c_m(t)\,=\,\frac{1}{\ln^{1/2}N}\int_{\xi_1}^{\xi_0}d\xi\,e^{\lambda^*\xi}\sin[q_m(\xi-\xi_1)]\,\delta\phi(\xi,t)\,.
\label{e112}
\end{eqnarray}
Although Eq. (\ref{e112}) yields $\langle c_m(t)\rangle_t=0$ as it
should, in the absence of any statistics of the shape fluctuations of
the front, one cannot obtain an expression of $\langle
c_m(t)\,c_{m'}(t)\rangle_t$ from it. Therefore, as far as any formal
derivation of the $1/\ln^3N$ asymptotic scaling of $D^{(3)}_f$ (or
$D_f$) is concerned, one simply cannot progress beyond
Eq. (\ref{e112}).

Nevertheless, we can still proceed with two approximations. The first
one stems from the fact that although it is clear from
Eq. (\ref{e112}) that $c_m(t)$ and $c_{m'}(t)$ ($m\neq m'$) are
correlated in general [after all, for a given realization, all the
$c_m(t)$'s are determined through the {\it same\/} $\delta\phi(t)$],
these fluctuation modes will have a finite correlation ``length'',
i.e., $\langle c_m(t)\,c_{m'}(t)\rangle_t$ will be negligibly small
[compared to both $\langle c^2_m(t)\rangle_t$ and $\langle
c^2_{m'}(t)\rangle_t$ values] when $|m-m'|$ exceeds a certain
threshold $a\ll\ln N$. Based on this observation, our approximation is
to  choose $a=0$ for the extreme (and unrealistic) case to simplify
the expression for $D^{(3)}_f$ to (see later for the discussion on
non-zero values of $a$)
\begin{eqnarray}
D^{(3)}_f\,=\,\frac{1}{2}\sum_{m\neq0}^{\ln
N}\tau_m\frac{s^2_m}{\ln^3N}\,\langle c^2_m(t)\rangle_t\,.
\label{e113}
\end{eqnarray}

Then the second approximation is that due to the presence of the
$e^{\lambda^*\xi}$ in the integrand of Eq. (\ref{e112}), only the
value of $\delta\phi$ within a distance $\sim1/\lambda^*=1$ of the tip
determines $c_m(t)$. This is argued in the following manner: typically
the magnitude of $\delta\phi(\xi,t)$ is of order
$\sqrt{\phi(\xi,t)/N}$; at the tip, $e^{\lambda^*\xi_0}\sim N$ cancels
$\delta\phi(\xi,t)\sim1/N$, but further behind, the $1/\sqrt{N}$
factor of $\sqrt{\phi(\xi,t)/N}$ can no longer be compensated by
$e^{\lambda^*\xi}$. We therefore use
\begin{eqnarray}
|c_m(t)|\,\sim\,\frac{1}{\ln^{1/2}N}\int_{\xi_0-1}^{\xi_0}d\xi\,|\sin[q_m(\xi-\xi_1)]|\sim\frac{q_m}{\ln^{1/2}N}\,=\,\frac{(m+1)\pi}{\ln^{3/2}N}\,.
\label{e114}
\end{eqnarray}
With $\tau_m=\ln^2N/[\pi^2\{(m+1)^2-1\}]$, the asymptotic scaling of
$D_f$ is finally obtained as
\begin{eqnarray}
D_f\simeq D^{(3)}_f\sim\frac{1}{2}\sum_{m\neq0}^{\ln
N}\frac{(m+1)^2\,s^2_m}{[(m+1)^2-1]\ln^4N}\sim\frac{1}{\ln^3N}\,.
\label{e115}
\end{eqnarray}
We end Sec. \ref{sec4.2.2} with four observations: (i) with more
reasonable assumption $a\neq0$, the expression for $D^{(3)}_f$
certainly gets more complicated. However, so long as $a\ll\ln N$,
which is what one expects in reality, the $1/\ln^3N$ asymptotic
scaling of $D^{(3)}_f$ (and hence of $D_f$ too) still continues to
hold. (ii) We have extensively used the construction for the left
eigenvector of the stability operator ${\cal L}_{v_N}$ for Fisher
equation (with a growth cutoff below $\phi=1/N$) all along
Sec. \ref{sec4.2.2}. For clock model
\cite{vanzon,vanzonarticles3,vanzon1,vanzonarticles1,vanzonarticles2},
or for the model that Brunet and Derrida considered \cite{bd3} to
demonstrate the $1/\ln^3N$ asymptotic scaling of $D_f$, construction
of the left eigenvector for the corresponding stability operator is
not easy. In view of that I do not know how to repeat the same
derivation for those two models with an equivalent amount of
rigour. Nevertheless, since all the arguments for the sFKPP equation
between Eqs. (\ref{e109}) and (\ref{e115}) are concentrated on the
leading edge or at the very tip of the front, they can be easily
repeated for these two models to derive the same $1/\ln^3N$ asymptotic
scaling of $D_f$. (iii) In the clock model, in terms of $\phi$, one
cannot create a localized fluctuation in the front shape --- any
fluctuation in the front shape is by definition non-local. The
``collisions'' between clocks are also non-local in nature [as
reflected in Eq. (\ref{clock1})]. From these complications, although a
priori it may seem that the clock model may have a different
asymptotic scaling of $D_f$, in reality, it does not happen. In fact,
Brunet and Derrida showed in a simplified version of their original
microscopic model (which closely resembles the clock model) that the
introduction of localized fluctuations in $\phi$ at the very tip of
the front is all that is needed for the $1/\ln^3N$ asymptotic scaling
of $D_f$. In that sense [also in view of point (ii)] the $1/\ln^3N$
asymptotic scaling of $D_f$ seems to be a generic property of
fluctuating ``pulled'' fronts, independent of the microscopic
model. (iv) Last but not the least, through the scaling of the
$c_m(t)$'s, the $1/\ln^3N$ asymptotic scaling of $D_f$ is seen to be
determined only from the tip of the front. This is in perfect
agreement with Brunet and Derrida's ``simplified model'' \cite{bd3}.

\subsection{Epilogue III\label{sec4.3}}

The existence of two different expressions for front diffusion, $D_f$
and $D_G$ seems confusing to say the least --- after all, any front
realization maintains its integrity at all times, and thus, at no
instant of time, the Goldstone mode and the centre of mass of any
given front realization can be infinitely far apart from each
other. This implies that whether measured through the Goldstone mode,
the center of mass, or the any place where the value of the front
field reaches any fixed value, say $\phi_0$, operationally (i.e., in a
computer simulation) one should always obtain the same front diffusion
coefficient!

In fact, the present situation can be compared to the a similar one
that occurs in gas mixtures (see for example, Chap. 11.2 of
Ref. \cite{degroot}) Therein, the expression (and the value) of the
the diffusion coefficient depends on its precise definition (although
operationally there is exactly one diffusion coefficient), but these
different expressions of the diffusion coefficient are (quite
non-trivially) related by means of the Onsager relations for the
diffusion coefficients. As for fronts too, it should not be surprising
that the precise values of $D_f$ and $D_G$ are not the same --- {\it
conceptually they are simply two entirely different
quantities}. Whether they could be related by any clever means or not
is left here for future investigation.

It is however not enough to stop at the above paragraph; one  has to
identify what physical quantities $D_f$ and $D_G$ respectively
correspond to. Although surprising it may seem in the first place, I
will argue in the next few paragraphs that {\it for fronts with
$\delta\phi\rightarrow0$ at $\xi\rightarrow\pm\infty$ and finite
$\tau_m$'s, $D_f$ is the long-time  front diffusion coefficient, while
$D_G$ is the short-time (or the instantaneous) diffusion coefficient
measured through the Goldstone mode of the front}.

The most important point to note is that from its very definition,
$D_G$ can be measured by tracking the spread in the position of the
Goldstone modes for an ensemble of front realizations over an
infinitesimally short interval of time; one does not at all need to
follow the front realizations over a long period of time. Through its
very definition, therefore, $D_G$ completely ignores the
time-correlations that exist in the speed fluctuations for individual
front realizations of the ensemble (precisely these correlations
exhibit themselves through the relaxation time scale $\tau_m$ of the
$m$-th mode of front shape fluctuation). These correlations do affect
the whole dynamics of the individual front realizations, and only
$D_f$ captures the full flavour of it (c.f. Sec. \ref{sec4.2}), albeit
one must pay a price of having to deal with more complicated
algebra. Thus, it should not come as a surprise that $D_f$ in general
yields an expression different from $D_G$. The best example that
supports this scenario --- rather an extreme case where $\tau_m$'s can
be arbitrarily large --- is the fluctuating ``pulled'' front
(Sec. \ref{sec4.2.2}). Therein, $D_G$ simply yield a wrong scaling for
the front diffusion as $N\rightarrow\infty$.

In view of existing simulation results on front diffusion, it is
worthwhile to revisit the examples for front diffusion in the
literature (reviewed in this article) with a critical eye to clarify
the above viewpoint.

\begin{itemize}
\item[]{1.} When $R(\phi,\xi,t)$ is replaced by $R(\phi^{(0)},\xi,t)$
in Sec. \ref{sec3.1.2}: with the $O(\tilde\varepsilon^{1/2})$
neglected in Eq. (\ref{e77}), in this case, there is no difference
between Eqs. (\ref{e75}) and (\ref{e97}). Then, with $D=1$, one obtains
\begin{eqnarray}
D_f\,=\,\frac{5\tilde\varepsilon}{2}\int_{-\infty}^{\infty}d\xi\,g^2[\phi^{(0)}]\,\neq
D_G\,=\,\tilde\varepsilon\,\,\frac{\displaystyle{\left[\,\int_{-\infty}^{\infty}d\xi\,\Phi_{G,L}^2(\xi)\,g^2[\phi^{(0)}]\right]}}{\displaystyle{\left[\,\int_{-\infty}^{\infty}d\xi\,e^{v_\varepsilon\xi}\,\Phi^2_{G,R}(\xi)\right]^2}}\,.
\label{add1}
\end{eqnarray}
The first point to note for this case is that just like $D_G$, $D_f$
also scales as $\tilde\varepsilon$. The second one is that for pushed
fronts, the numerical values of $D_f$ and $D_G$, in all likelihood,
are close by (i.e., one needs good data to distinguish between the
two). This is in fact easily seen from the following argument: pushed
fronts are strongly localized, and therefore, their Goldstone modes
$\Phi_{G,R}$ are localized in space, say around
$\xi=\xi_0$. Having used this information in Eq. (\ref{add1}), we see
that for $D_G$, the $e^{v_\varepsilon\xi}$ factors in the numerator
and the denominator cancel, and thus, effectively apart from some
numerical coefficients of $O(1)$, $D_G\simeq\tilde\varepsilon
g^2[\phi^{(0)}(\xi_0)]/\Phi^2_{G,R}(\xi_0)$, while
$D_f\simeq5\tilde\varepsilon g^2[\phi^{(0)}(\xi_0)]/2$. How their
numerical values compare depends on $\Phi^2_{G,R}(\xi_0)$.

To me therefore, the issue regarding the difference between $D_f$ and
$D_G$ for pushed fronts requires further numerical investigation. One
particular instance where the above argument breaks down is when the
actual front region is very large in size. This is precisely what
happens for fluctuating ``pulled'' (or weakly pushed fronts), i.e.,
when $\phi^{(0)}$ is the Brunet and Derrida's cutoff solution
(\ref{e11}). In that case, there is a clear difference between $D_f$
and $D_G$ through their respective scalings (also see 3 below).
\vspace{3mm}
\item[]{2.} Fluctuating pulled front with multiplicative noise in Sec.
\ref{sec3.3}: in this case, it has been clearly shown, numerically and
as well as via the Cole-Hopf transformation, that the front wandering
is subdiffusive. The first sign of the fact that there is a problem
associated with the separation of time scales in this particular
example (slow for front wandering and fast for front shape
fluctuations) is manifested through the breakdown of convergence of
the integrals in the definition of $D_G$. A priori, it seems strange
that $D_f$ [see Eq. (\ref{add1})] does not suffer from any similar
peculiarities.

The reason for the expression of $D_f$ not exhibiting peculiarities,
however, is not hard to trace. It is in fact hidden within the very
derivation of Eq. (\ref{add1}) --- recall that there have been
cancellations of factors of $\tau_m$ all along
Sec. \ref{sec4.2.1}. For a pulled front $\tau_m$'s are infinite, and
these cancellations are meaningless. The derivation of
Eq. (\ref{add1}), thus, simply breaks down for fluctuating pulled
fronts with multiplicative noise.

Nevertheless, since ${\cal W}(t)$ has been measured operationally via
the centre of mass in Ref. \cite{wim2}, it makes sense to  use $D_f$
to calculate ${\cal W}(t)$ as well. With the introduction of the upper
cutoff $\xi_c\sim\sqrt{4Dt}$, which makes the $\tau_m$'s finite, one
can very easily show [having $R(\phi,\xi,t)$ replaced by 
$R(\phi^{(0)},\xi,t)$] that
\begin{eqnarray}
D_f\,\sim\,\frac{4\tilde\varepsilon D^{5/2}}{\sqrt{t}}\neq
D_G\sim\frac{3\tilde\varepsilon}{v^2_\varepsilon\sqrt{\pi Dt}}
\label{add2}
\end{eqnarray}
for $t\gg1$. Interestingly enough, in this example, one gets away with
the usage of $D_G$ to obtain the theoretical curve (dashed line) for
${\cal W}(t)$ in Fig. \ref{wim2} --- both $D_f$ and $D_G$ scales as
$1/\sqrt{t}$ for $t\gg1$, and for Fig. \ref{wim2} that is all that
matters.
\vspace{3mm}
\item[]{3.} Fluctuating ``pulled'' fronts, such as the sFKPP equation:
although we have discussed this at length in Sec. \ref{sec4.2.2}, it
is still useful to articulately point out a few distinct
subtleties. First, if the noise term in Eq. (\ref{ee96}) is made
additive by replacing $\phi$ by $\phi^{(0)}$, where $\phi^{(0)}$ is
Brunet and Derrida's cutoff solution (\ref{e11}), $D_f$ scales as
$1/N$ [c.f. Eq. (\ref{add1})] for $N\rightarrow\infty$, as opposed to
$1/\ln^3N$. It is therefore important to take notice of this subtlety
for simulation purposes --- Brunet and Derrida properly take this
into account in their own simulations \cite{bd3}. For $D_G$ however,
it does not matter whether one uses $\phi$ or $\phi^{(0)}$ for the
noise term in Eq. (\ref{ee96}) [this is quite obvious from the
expressions of $D_G$ in Eqs. (\ref{e107}) and (\ref{add1})], one
always gets $D_G\sim1/\ln^6N$ for $N\rightarrow\infty$.

Secondly, since $D_G$ is the short-time (or the instantaneous)
diffusion coefficient of the front, if the diffusive spread
$\langle\Delta x^2_G(t)\rangle$ in the front positions measured
through the Goldstone modes are plotted as a function of time for an
ensemble of fluctuating ``pulled'' fronts, one would see that
$d\langle\Delta x^2_G(t)\rangle/dt=1/\ln^6N$  right from the very
start. On the other hand, when the front positions are tracked by the
positions of their centres of mass, one expects $d\langle\Delta
x^2_f(t)\rangle/dt$ to be smaller for small $t$, but it should
nevertheless converge to  $1/\ln^3N$ for $t\rightarrow\infty$. In
fact, one can further argue that $d\langle\Delta
x^2_f(t)\rangle/dt|_{t\rightarrow0}\sim1/\ln^4N$. In order to do so,
one simply has to observe that the $1/\ln^3N$ scaling of $D_f$ is
obtained from $\ln N$ number of modes, each of which contribute an
amount of $1/\ln^4N$ to $D_f$ [c.f. Eq. (\ref{e115})]. The shortest
and the longest surviving modes have lifetimes respectively of $O(1)$
and $O(\ln^2N)$. Thus, $d\langle\Delta
x^2_f(t)\rangle/dt|_{t\rightarrow0}$ captures only the contribution of
the shortest surviving modes and should behave as $\sim1/\ln^4N$,
while in order to observe the true $1/\ln^3N$ scaling of $D_f$, one
has to wait till $O(\ln^2N)$ time for the contributions of all the
$\ln N$ number of modes to $D_f$ to add up together.
\end{itemize}

I finally end Sec. \ref{sec4} (and this review article too) with the
comment that further simulations and research are necessary to confirm
the conjectures made in Sec. \ref{sec4.3}.

\section{Author's Note\label{sec5}}

This review article has been written from my own perspective, and it
is clearly reflected by the organization of the topics that I have
covered. After a brief introduction to the subject, I spent a longish
section on the discrete particle and lattice models of fluctuating
fronts. Needless to say, there is no unifying underlying structure for
the plethora of results in that section --- each problem or topic has
been dealt with a different approach in the existing literature. The
two sections \ref{sec3} and \ref{sec4} I spent on field-theory ---
therein, we find an organized theoretical backbone for studying the
effects of fluctuations on propagating fronts by means of stochastic
differential equations. Finally at the end, I tried to merge the
predictions of field-theories on internal fluctuations and the
corresponding results for the discrete particle and lattice models.

Unless existing derivations are too involved or too complex, I have
tried to outline the main line of procedure. The only exception has
been the issues of front diffusion in one spatial dimension, specially
for the fluctuating ``pulled'' fronts, where I have delved deep in the
dirty world of algebra. This was done with keeping in mind that front
diffusion in one spatial dimension is a relatively new topic. In any
case, I have also thrived to put in physical insights wherever
appropriate. At various places, open issues have been spelled out for
further investigations as well. As for myself, it would be gratifying
to see if the expert researchers, and as well as the beginners in this
field, both of whom are the target readers, find this review article
useful.

\section{Acknowledgement\label{sec6}}

A large part of my own work on this subject has been carried out in
close collaboration with Wim van Saarloos, whose insight has deeply
influenced me. I would also like to thank my one-time colleague (and
presently friend) Ramses van Zon for sharing his thoughts on many of
the topics discussed here. For many odds and ends of scientific
discussions, I am also grateful to (in alphabetical order) Henk van
Beijeren, Jaume Casademunt, Jean-Sebastien Caux, Charlie Doering,
Wouter Kager, Hans van Leeuwen,  Esteban Moro and Bernard Nienhuis. I
also wish to thank H.-P. Breuer, Charlie Doering, Wim van Saarloos,
Len Sander, Igor Sokolov and Ramses van Zon for giving me prompt
permission to use figures from their own papers. Thanks for all the
fish!

Financial support has been provided by the Dutch Research Organization
FOM (Fundamenteel Onderzoek der Materie).


\begin{thebibliography}{03}

%A
\bibitem{physics1} G. Ahlers and D. S. Cannell, {\it Vortex-front
propagation in rotating Couette-Taylor flow}, Phys. Rev. Lett. {\bf
50}, 1583 (1983).

\bibitem{bruno} B. And\`{o} and S. Graziani, {\it Stochastic Resonance
Theory and Applications}, (Kluwer Academic Publishers, Boston, 2000).

\bibitem{armero} J. Armero, J. Casademunt, L. Ram\'{\i}rez-Piscina,
and J.M. Sancho, {\it Ballistic and diffusive corrections to front
propagation in the presence of multiplicative noise}, Phys. Rev. E
{\bf 58}, 5494 (1998).

\bibitem{armeroprl} J. Armero, J. M. Sancho, J. Casademunt,
A. M. Lacasta, L. Ram\'{\i}rez-Piscina, F. Sagu\'es, {\it External
Fluctuations in Front Propagation}, Phys. Rev. Lett. {\bf 76}, 3045
(1996).

\bibitem{ba} D. ben-Avraham, {\it Fisher waves in the
diffusion-limited coalescence process A$+$A$\leftrightharpoons$A},
Physics Letters A {\bf 247}, 53 (1998).

\bibitem{ba15} D. ben-Avraham, {\it Inhomogeneous steady states of
diffusion-limited coalescence A$+$A$\leftrightarrow$A}, Physics
Letters A {\bf 249}, 415 (1998).

\bibitem{ba16} D. ben-Avraham, {\it Diffusion-limited coalescence,
A$+$A$\leftrightarrow$A, with a trap}, Phys. Rev. E {\bf 58}, 4351
(1998).

\bibitem{ba17} D. ben-Avraham, {\it Complete exact solution of
diffusion-limited coalescence A$+$A$\rightarrow$A},
Phys. Rev. Lett. {\bf 81}, 4756 (1998).

\bibitem{ba13} D. ben-Avraham, M. A. Burschka and C. R. Doering, {\it
Statics and dynamics of a diffusion-limited reaction: anomalous
kinetics, non-equilibrium self-ordering and a dynamic transition},
J. Stat. Phys. {\bf 60}, 695 (1990).

%********************************************************************

%B
\bibitem{barabasi} A.-L. Barabasi and H. E. Stanley, {\it Fractal
Concepts in Surface Growth\/}, (Cambridge Univ. Press, Cambridge,
1995).

\bibitem{becker1} A. Becker and L. Kramer, {\it Linear stability
analysis for bifurcations in spatially extended systems with
fluctuating control parameter}, Phys. Rev. Lett. {\bf 73}, 955 (1994).

\bibitem{bj} E. Ben-Jacob, H. R. Brand, G. Dee, L. Kramer,  and
J. S. Langer, {\it Pattern propagation in nonlinear dissipative
systems}, Physica D {\bf 14}, 348 (1985).

\bibitem{bramson} M. Bramson, {\it Convergence of solutions of the
Kolmogorov equation to travelling waves}, Mem. Am. Math. Soc. {\bf
44}, No. 285 (1983).

\bibitem{bramson1} M. Bramson, P. Calderoni, A. De Masi,
P. A. Ferrari, J. L. Lebowitz and R. H. Schonmann, {\it Microscopic
selection principle for a diffusion-reaction equation},
J. Stat. Phys. {\bf 45}, 905 (1986).

\bibitem{Levine1} E. Brener, H. Levine and Y. Tu, {\it Mean-field
theory for diffusion-limited aggregation in low dimensions},
Phys. Rev. Lett. {\bf 66}, 1978 (1991).

\bibitem{breuer} H. P. Breuer, W. Huber, and F. Petruccione, {\it
Fluctuation effects on wave propagation in a reaction-diffusion
process}, Physica D {\bf 73}, 259 (1994); Europhys. Lett. {\bf 30}, 69
(1995).

\bibitem{biology2} N. F. Britten, {\it Reaction-Diffusion Equations
and Their Applications to Biology\/} (Academic Press, New York).

\bibitem{parrondo2} C. van den Broeck, J. M. R. Parrondo and R. Toral
{\it Noise-Induced Nonequilibrium Phase Transition},
Phys. Rev. Lett. {\bf 73}, 3395 (1994).

\bibitem{bd1} E. Brunet and B. Derrida, {\it Shift in the velocity of
a front due to a cutoff}, Phys. Rev. E {\bf 56}, 2597 (1997).

\bibitem{bd2} E. Brunet and B. Derrida, {\it Microscopic models of
travelling wave equations}, Comp. Phys. Comm. {\bf 122}, 376 (1999).

\bibitem{bd3} E. Brunet and B. Derrida, {\it Effect of microscopic
noise on front propagation}, J. Stat. Phys. {\bf 103}, 269 (2001).

%********************************************************************

%C
\bibitem{oono3} L.-Y. Chen, N. Goldenfeld and Y. Oono, {\it
Renormalization group theory and variational calculations for
propagating fronts}, Phys. Rev. E {\bf 49}, 4502 (1994).

\bibitem{oono2} L.-Y. Chen, N. Goldenfeld, Y. Oono and G. C. Paquette,
{\it Selection, stability and renormalization}, Physica A {\bf 204},
111 (1994).

\bibitem{cook} J. Cook and B. Derrida, {\it Lyapounov exponents of
large, sparse random matrices and the problem ofdirected polymers with
random complex weights}, J. Stat. Phys. {\bf 61}, 961 (1990).

\bibitem{physics11} A. Couairon and J.-M. Chomaz, {\it Absolute and
convective instabilities, front velocities and global modes in
nonlinear systems}, Physica D {\bf 108}, 236 (1997).

\bibitem{ch} M. C. Cross and P. C. Hohenberg, {\it Pattern formation
outside of equilibrium}, Rev. Mod. Phys. {\bf 65}, 851 (1993).

%********************************************************************

%D
\bibitem{derrida1} B. Derrida, {\it Mean field theory of directed
polymers in a random medium}, Phys. Scr. {\bf T 38}, 6 (1991).

\bibitem{ds1} B. Derrida and H. Spohn, {\it Polymers on disordered
trees, spin glasses and travelling waves}, J. Stat. Phys. {\bf 51},
817 (1988).

\bibitem{dee} G. Dee and J. S. Langer, {\it Propagating pattern
selection}, Phys. Rev. Lett. {\bf 50}, 383 (1983).

\bibitem{physics10} S. J. Di Bartolo and A. T. Dorsey, {\it Velocity
selection for propagating fronts in superconductors},
Phys. Rev. Lett. {\bf 77}, 4442 (1996).

\bibitem{ba14} C. R. Doering, {\it Microscopic spatial correlations
induced by external noise in a reaction-diffusion system}, Physica A
{\bf 188}, 386 (1992).

\bibitem{ba11} C. R. Doering and D. ben-Avraham, {\it Interparticle
distribution functions and rate equations for diffusion-limited
reactions}, Phys. Rev. A {\bf 38}, 3035 (1988).

\bibitem{ba12} C. R. Doering and D. ben-Avraham, {\it
Diffusion-limited coalescence in the presence of particle input:exact
results in one dimensions}, Phys. Rev. Lett. {\bf 62}, 2563 (1989).

\bibitem{doering3} C. R. Doering, M. A. Burschka and W. Horsthemke,
{\it Fluctuations and correlations in a diffusion-reaction system:
exact hydrodynamics}, J. Stat. Phys. {\bf 65}, 953 (1991).

\bibitem{doering21} C. R. Doering, C. Mueller and P. Smerecka, {\it
Interacting particles, the stochastic
Fisher-Kolmogorov-Petrovsky-Piscunov equation, and duality}, in
Proceedings 3rd International Conference Unsolved Problems of Noise
(UPoN) and fluctuations in physics, biology and high technology
(2002); AIP Conference Proceedings {\bf 665}, 523 (2003).

\bibitem{doering22} C. R. Doering, C. Mueller and P. Smerecka, {\it
Interacting particles, the stochastic
Fisher-Kolmogorov-Petrovsky-Piscunov equation, and duality}, Physica A
{\bf 325}, 243 (2003).

%********************************************************************

%E
\bibitem{ebert1} U. Ebert and W. van Saarloos, {\it Breakdown of the
standard perturbation theory and moving boundary approximation},
Phys. Rep. {\bf 337}, 139 (2000).

\bibitem{ebert} U. Ebert and W. van Saarloos, {\it Front propagation
into unstable states: universal algebraic convergence towards
uniformly pulled fronts}, Physica D {\bf 146}, 1 (2000).

\bibitem{streamers} U. Ebert, W. van Saarloos and C. Caroli, {\it
Propagation and structure of planar streamer fronts}, Phys. Rev. E
{\bf 55}, 1530 (1997).

\bibitem{edwards1} S. Edwards and D. Wilkinson, {\it The surface
statistics of a granular aggregate}, Proc. R. Soc. London A {\bf 381},
17 (1982).

%********************************************************************

%F
\bibitem{biology1} P. C. Fife, {\it Mathematical Aspects of Reacting
and Diffusing Systems}, Lecture Notes in Biomathematics {\bf 28}
(Springer, Berlin, 1979).

\bibitem{physics4} J. Fineberg and V. Steinberg, {\it Vortex-front
propagation in Rayleigh-B\'enard convection}, Phys. Rev. Lett. {\bf
58}, 1332 (1987).

\bibitem{fisher} R. A. Fisher, {\it The wave of advance of
advantageous genes}, Ann. Eugenics {\bf 7}, 355 (1937).

%********************************************************************

%G
\bibitem{sancho1} J. Garc\'{\i}a-Ojalvo, A. Hern\'{a}ndez-Machado and
J. M. Sancho, {\it Effects of external noise on the Swift-Hohenberg
equation}, Phys. Rev. Lett. {\bf 71}, 1542 (1993).

\bibitem{gardiner} G. W. Gardiner, {\it Handbook of Stochastic
Methods\/} (Springer, Berlin, 1985).

\bibitem{degroot} S. R. de Groot and P. Mazur, {\it Non-equilibrium
Thermodynamics}, Dover Publications, New York (1984).
 
%********************************************************************

%H
\bibitem{halpin} T. J. Halpin-Healy and Y. C. Zhang, {\it Kinetic
roughening, stochastic growth, directed polymers and all that},
Phys. Rep. {\bf 254}, 215 (1995).

\bibitem{physics12} M. van Hecke, C. Storm and W. van Saarloos, {\it
Sources, sinks, and wavenumber selection in coupled CGL equations and
experimental implications for counter-propagating wave  equations},
Physica D {\bf 134}, 1 (1999).

\bibitem{chemistry21} D. Horv\'{a}th, V. Petrov, S. K. Scott and
K. Showalter, {\it Instabilities in propagating reaction-diffusion
fronts}, J. Chem. Phys. {\bf 98}, 6332 (1993).

\bibitem{chemistry22} D. Horv\'{a}th and A. T\'{o}th, {\it
Diffusion-driven front instabilities in the chlorite-tetrathionate
reaction}, J. Chem. Phys. {\bf 108}, 1447 (1998).

%********************************************************************

%I
%********************************************************************

%J
\bibitem{physics6} R. A. L. Jones, L. J. Norton, E. J. Kramer,
F. S. Bates and P. Wiltzius, {\it Surface-directed spinodal
decomposition}, Phys. Rev. Lett. {\bf 66}, 1326 (1991).

%********************************************************************

%K
\bibitem{vankampen} N. G. van Kampen, {\it Stochastic Processes in
Physics and Chemistry}, North-Holland (2003).

\bibitem{kpz} M. Kardar, G. Parisi and Y. C. Zhang, {\it Dynamic
scaling of growing interfaces}, Phys. Rev. Lett. {\bf 56}, 889 (1986).

\bibitem{kerstein1} A. R. Kerstein, {\it Computational study of
propagating fronts in a lattice-gas model}, J. Stat. Phys. {\bf 45},
921 (1986).

\bibitem{kerstein} A. R. Kerstein, {\it A two-particle representation
of front propagation in diffusion-reaction systems},
J. Stat. Phys. {\bf 53}, 703 (1988).

\bibitem{kns} D. A. Kessler, Z. Ner, and L. M. Sander, {\it Front
propagation: precursors, cutoffs, and structural stability},
Phys. Rev. E. {\bf 58}, 107 (1998).

\bibitem{kladko} K. Kladko, I. Mitkov and A. R. Bishop, {\it Universal
Scaling of Wave Propagation Failure in Arrays of Coupled Nonlinear
Cells}, Phys. Rev. Lett. {\bf 84}, 4505 (2000).

\bibitem{biology4} A. J. Koch and H. Meinhardt, {\it Biological
pattern formation: from basic mechanisms to complex structures},
Rev. Mod. Phys. {\bf 66}, 1481 (1994).

\bibitem{kpp} A. Kolmogorov, I. Petrovsky and N. Piscounov, {\it Study
of the diffusion equation with growth of the quantity of matter and
its application to a biology problem}, Bulletin de l'universit\'{e}
d'\'{e}tat \`{a} Moscou, Ser. int., Section A, {\bf 1}, 1 (1937).

\bibitem{krug3} J. Krug, {\it Origin of scale invariance in growth
processes}, Adv. Phys. {\bf 46}, 139 (1997).

\bibitem{krug2} J. Krug and H. Spohn, {\it Kinetic roughening of
growing surfaces} in ``Solids Far from Equilibrium'', C. Godr\`eche
Ed. (Cambridge Univ. Press, Cambridge 1992).

\bibitem{krug1} J. Krug and P. Meakin, {\it Columnar growth in oblique
incidence ballistic deposition: Faceting, noise reduction, and
mean-field theory}, Phys. Rev. A {\bf 43}, 900 (1991).

%********************************************************************

%L
\bibitem{physics2} J. S. Langer and H. M\"{u}ller-Krumbhaar, {\it Mode
selection in a dendritelike nonlinear system}, Phys. Rev. A {\bf 27},
499 (1983).

\bibitem{lee1} B. P. Lee, {\it Scaling of reaction zones in the
A$+$B$\rightarrow0$ diffusion-limited reaction}, Phys. Rev. E {\bf
50}, R3287 (1994).

\bibitem{lee2} B. P. Lee, {\it Renormalization group calculation for
the reaction $kA\rightarrow\emptyset$}, J. Phys. A {\bf 27}, 2633
(1994).

\bibitem{lemar32} A. Lemarchand, {\it Selection of an attractor in a
continuum of stable solutions: Descriptions of a wave front at
different scales}, J. Stat. Phys. {\bf 101}, 579 (2000).

\bibitem{lemar1} A. Lemarchand, A. Lesne, and M. Mareschal, {\it
Langevin approach to a chemical wave front: Selection of the
propagation velocity in the presence of internal noise}, Phys. Rev. E
{\bf 51}, 4457 (1995).

\bibitem{lemar10} M. A. Karzazi, A. Lemarchand and M. Mareschal, {\it
Fluctuation effects on chemical wave fronts}, Phys. Rev. E {\bf 54},
4888 (1996).
 
\bibitem{lemar11} A. Lemarchand and B. Nowakowski, {\it Perturbation
of a local equilibrium by a chemical wave front}, J. Chem. Phys. {\bf
109}, 7028 (1998).

\bibitem{lemar2} A. Lemarchand and B. Nowakowski, {\it Different
description levels of chemical wave front propagation speed
selection}, J. Chem. Phys. {\bf 111}, 6190 (1999).

%********************************************************************

%M
\bibitem{sokolov1} J. Mai, I. M. Sokolov and A. Blumen, {\it Front
propagation and local ordering in one-dimensional ireversible
autocatalytic reactions}, Phys. Rev. Lett. {\bf 77}, 4462 (1996).

\bibitem{sokolov3} J. Mai, I. M. Sokolov and A. Blumen, {\it
Discreteness effects on the front propagation in the
A$+$B$\rightarrow2$A reaction in $3$ dimensions}, Europhys. Lett. {\bf
44}, 7 (1998).

\bibitem{sokolov2} J. Mai, I. M. Sokolov and A. Blumen, {\it Front
propagation in one-dimensional autocatalytic reactions: the breakdown
of classical picture at small particle concentrations}, Phys. Rev. E
{\bf 62}, 141 (2000).

\bibitem{sokolov1a} J. Mai, I.M. sokolov, V. Kuzovkov and A. Blumen,
{\it Front form and velocity in a one-dimensional autocatalytic
A$+$B$\rightarrow2$A reaction}, Phys. Rev. E {\bf 56}, 4130 (1996).

\bibitem{chemistry3} A. Malevanets, R. Careta and R. Capral, {\it
Biscale chaos in propagating fronts}, Phys. Rev. E {\bf 52}, 4724
(1995).

\bibitem{manz} N. Manz, V. A. Davydov, V. S. Zykov and
S. C. M\"{u}ller, {\it Excitation fronts in a spatially modulated
light sensitive Belousov-Zhabotinsky system}, Phys. Rev. E {\bf 66},
036207 (2002).

\bibitem{marinari} E. Marinari, A. Pagnani and G. Parisi, {\it
Critical exponents of the KPZ equation via multi-surface coding
numerical simulations} J. Phys. A {\bf 33}, 8181 (2000).

\bibitem{meron} E. Meron, {\it Pattern formation in excitable media},
Phys. Rep. {\bf 218}, 1 (1992).

\bibitem{lutz1} A. S. Mikhailov, L. Schimansky-Geier and W. Ebeling,
{\it Stochastic motion of the propagating front in bistable media},
Phys. Lett. A {\bf 96}, 453 (1983).

\bibitem{moro} E. Moro, {\it Internal Fluctuations Effects on Fisher
Waves}, Phys. Rev. Lett., {\bf 87}, 238303 (2001).

\bibitem{moron} E. Moro, {\it Emergence of pulled fronts in fermionic
microscopic particle models}, Phys. Rev. E {\bf 68}, 025102 (2003).

\bibitem{biology3} J. A. Murray, {\it Mathematical Biology\/}
(Springer, Berlin, 1989).

\bibitem{muller} C. Mueller and R. B. Sowers, {\it Random travelling
waves for the KPP equation with noise}, J. Funct. Anal. {\bf 128}, 439
(1995).


%********************************************************************

%N
\bibitem{physics5} M. Niklas, M. L\"{u}cke and
H. M\"{u}ller-Krumbhaar, {\it Velocity of a propagating Taylor-vortex
front}, Phys. Rev. A {\bf 40}, 493 (1989).

\bibitem{novikov} E. A. Novikov, {\it Functionals and the random-force
method in turbulence theory}, JETP {\bf 20}, 1290 (1965).

%********************************************************************

%O
%********************************************************************

%P
\bibitem{deb12}  D. Panja, {\it Surprising aspects of fluctuating
pulled fronts}, in Proceedings 3rd International Conference Unsolved
Problems of Noise (UPoN) and fluctuations in physics, biology and high
technology 2002; AIP Conference Proceedings {\bf 661}, 539 (2003).

\bibitem{deb11} D. Panja and W. van Saarloos, {\it Fluctuating pulled
fronts: The origin and the effects of a finite particle cutoff},
Phys. Rev. E {\bf 66}, 036206 (2002).

\bibitem{deb2} D. Panja and W. van Saarloos, {\it Fronts with a growth
cutoff but speed higher than the linear spreading speed}, Phys. Rev. E
{\bf 66}, 015206(R) (2002).

\bibitem{deb3} D. Panja and W. van Saarloos, {\it The Weakly Pushed
Nature of ``Pulled'' Fronts with a Cutoff}, Phys. Rev. E {\bf 65},
057202 (2002).

\bibitem{deb5} D. Panja, G. Tripathy and W. van Saarloos, {\it Front
propagation and diffusion in the A$\leftrightarrow$A$+$A hard-core
reaction on a chain}, Phys. Rev. E 67, 046206 (2003).

\bibitem{deb4} D. Panja, {\it Asymptotic scaling of the diffusion
coefficient of fluctuating ``pulled'' fronts}, to appear in Rapid Comm. 
Phys. Rev. E; e-print arxiv cond-mat/0304371. 

\bibitem{oono1} G. C. Paquette, L.-Y. Chen, N. Goldenfeld and Y. Oono,
{\it Structural stability and renormalization group for propagating
fronts}, Phys. Rev. Lett. {\bf 72}, 76 (1994).

\bibitem{parrondo1} J. M. Parrondo, C. van den Broeck, J. Buceta and
F. J. de la Rubia, {\it Noise-induced spatial patterns}, Physica A
{\bf 224}, 153 (1996).

\bibitem{pasquale} F. de Pasquale, J. Gorecki and J. Popielawski, {\it
On the stochastic correlations in a randomly perturbed chemical
front}, J. Phys. A {\bf 25}, 433 (1992).

\bibitem{levine} L. Pechenik and  H. Levine, {\it Interfacial velocity
corrections due to multiplicative noise}, Phys. Rev. E {\bf 59}, 3893
(1999).

%********************************************************************

%Q
%********************************************************************

%R
\bibitem{doering1} J. Riordan, C. R. Doering, and D. ben-Avraham, {\it
Fluctuations and Stability of Fisher Waves}, Phys. Rev. Lett. {\bf
75}, 565 (1995).

\bibitem{jaume2} A. Rocco, J. Casademunt, U. Ebert, W. van Saarloos,
{\it Diffusion coefficient of propagating fronts with multiplicative
noise}, Phys. Rev. E {\bf 65}, 012102 (2002).

\bibitem{wim2} A. Rocco, U. Ebert and W. van Saarloos, {\it
Subdiffusive fluctuations of ``pulled'' fronts with multiplicative
noise}, Phys. Rev. E {\bf 62}, R13 (2000).

\bibitem{jaume3} A. Rocco, L. Ram\'{\i}rez-Piscina, J. Casademunt,
{\it Kinematic reduction of reaction-diffusion fronts with
multiplicative noise: derivation of stochastic sharp-interface
equations} Phys. Rev. E {\bf 65}, 056116 (2002).

%********************************************************************

%S
\bibitem{vs2} W. van Saarloos, {\it Front propagation into unstable
states. II. Linear versus nonlinear marginal stability and rate of
convergence}, Phys. Rev. A {\bf 39}, 6367 (1989).

\bibitem{vs1} W. van Saarloos, {\it Three basic issues concerning
interface dynamics in nonequilibrium pattern formation},
Phys. Rep. {\bf 301}, 9 (1998).

\bibitem{wimreview} W. van Saarloos, {\it Front propagation into
unstable states}, submitted to Physics Reports (2003).

\bibitem{physics9} W. van Saarloos, M. van Hecke and R. Holyst {\it
Front propagation into unstable and metastable states in smectic-$C^*$
liquid crystals:linear and nonlinear marginal stability analysis},
Phys. Rev. E {\bf 52}, 1773 (1995).

\bibitem{physics7} E. K. H. Salje, {\it On the kinetics of partially
conserved order parameters: a possible mechanism for pattern
formation}, J. Phys.: Condensed Matter {\bf 5}, 4775 (1992).

\bibitem{santos} M.-A. Santos and J. M. Sancho, {\it Noise-induced
fronts}, Phys. Rev. E {\bf 59}, 98 (1999).

\bibitem{santos1} M.-A. Santos and J. M. Sancho, {\it Front dynamics
in the presence of spatiotemporal structured noise}, Phys. Rev. E {\bf
64}, 016129 (2001).

\bibitem{lutz2} L. Schimansky-Geier, A. S. Mikhailov and W. Ebeling,
{\it Effect of fluctuation on plane front propagation in bistable
nonequilibrium systems}, Ann. der Physik (Leipzig) {\bf 40}, 277
(1983).

\bibitem{lutz3} L. Schimansky-Geier and Ch. Z\"{u}licke, {\it Kink
propagation induced by m\-u\-l\-t\-i\-plicative noise}, Z. Phys. {\bf
82}, 157 (1991).

\bibitem{schlogl1} F. Schl\"{o}gl and R. S. Berry, {\it Small
roughness fluctuations in the layer between two phases}, Phys. Rev. A
{\bf 21}, 2078 (1980).

\bibitem{schlogl2} F. Schl\"{o}gl, C. Escher and R. S. Berry {\it
Fluctuations in the interface between two phases}, Phys. Rev. A {\bf
27}, 2698 (1983).

\bibitem{jaume4} I. Sendi\~{n}a-Nadal, A. P. Mu\~{n}uzuri, D. Vives,
V. P\'{e}rez-Mu\~{n}uzuri, J, Casademunt, L. Ram\'{\i}rez-Piscina,
J. M. Sancho and F. Sagu\'{e}s, {\it Wave propagation in a medium with
disordered excitability}, Phys. Rev. Lett. {\bf 80}, 5437 (1998).

\bibitem{starobin} J. M. Starobin and C. F. Starmer, {\it
Boundary-layer analysis of waves propagating in an excitable medium:
Medium conditions for wave-front-obstacle separation}, Phys. Rev. E
{\bf 54}, 430 (1996).

%********************************************************************

%T
\bibitem{physics8} A. Torcini, P. Grassberger and A. Politi,
J. Phys. A, {\it Error propagation in extended chaotic systems}, {\bf
27}, 4533 (1995).

\bibitem{physics3} G. S. Triantafyllou, K. Kupfer and A. Bers, {\it
Absolute instabilities and self-sustained oscillations in the wake of
circular cylinders}, Phys. Rev. Lett. {\bf 59}, 1914 (1987).

\bibitem{jaume1} G. Tripathy, A. Rocco, J. Casademunt and W. van
Saarloos, {\it Universality Class of Fluctuating Pulled Fronts},
Phys. Rev. Lett. {\bf 86}, 5215 (2001).

\bibitem{goutam} G. Tripathy and W. van Saarloos, {\it Fluctuation and
Relaxation Properties of Pulled Fronts: A Scenario for Nonstandard
Kardar-Parisi-Zhang Scaling}, Phys. Rev. Lett. {\bf 85}, 3556 (2000).

\bibitem{tucross} Y. Tu and M. C. Cross, {\it Chaotic domain structure
in rotating convection}, Phys. Rev. Lett. {\bf 69}, 2515 (1992).

%********************************************************************

%U
%********************************************************************

%V
%********************************************************************

%W
\bibitem{ellak2} C. P. Warren, G. Mikus, E. Somfai and L. M. Sander,
{\it Fluctuation effects in an epidemic model}, Phys. Rev. E {\bf 63},
056103 (2001).

\bibitem{ellak1} C. P. Warren, E. Somfai and L. M. Sander, {\it
Velocity of front propagation in $1$-dimensional autocatalytic
reactions}, Braz. J. Phys. {\bf 30}, 157 (2000).

%********************************************************************

%X
%********************************************************************

%Y
%********************************************************************

%Z
\bibitem{chemistry1} Y. Zel'dovich, G. I. Braenblatt, V. B. Librovich
and G. M. Makhviladze, {\it The Mathematical Theory of Combustion and
Explosions\/} (Consultants Bureau, New York, 1985).

\bibitem{tabeling} G. Zocchi, P. Tabeling and M. Ben Amar, {\it
Saffman-Taylor plumes}, Phys. Rev. Lett. {\bf 69}, 601 (1992).

\bibitem{vanzonarticles3} R. van Zon, Ph.D. thesis, Univ. of Utrecht,
the Netherlands (2000).

\bibitem{vanzon1} R. van Zon and H. van Beijeren, {\it Front
propagation techniques to calculate the largest Lyapunov exponent of
dilute hard disk gases}, J. Stat. Phys. {\bf 109}, 641 (2002).

\bibitem{vanzon} R. van Zon, H. van Beijeren and Ch. Dellago, {\it
Largest Lyapunov Exponent for Many Particle Systems at Low Densities},
Phys. Rev. Lett. {\bf 80}, 2035 (1998).

\bibitem{vanzonarticles1} R. van Zon, H. van Beijeren and
J. R. Dorfman, {\it Kinetic theory of dynamical systems}, in
Proceedings NATO ASI on ``Dynamics: Models and Kinetic Methods for
Non-equilibrium Many-Body Systems'', NATO Science Series E, Vol. 317;
J. Karkheck Ed., 131 (Kluwer, 2000).

\bibitem{vanzonarticles2} R. van Zon, H. van Beijeren, and
J. R. Dorfman, {\it Kinetic theory estimates for the Kolmogorov-Sinai
entropy and the largest Lyapunov exponents for dilute, hard-ball gases
and for dilute, random Lorentz gases}, in ``Hard Ball Systems''
(editor D. Szasz), Mathematical Encyclopaedia of Mathematical
Sciences, Vol. 101 (Springer 2000).

\end{thebibliography}
\end{document}